\documentclass[aps,prb,twocolumn,nofootinbib,notitlepage,superscriptaddress,longbibliography,preprintnumbers]{revtex4-2}
\usepackage{times}
\usepackage{graphicx}
\usepackage[whole]{bxcjkjatype} % for Japanese

\usepackage[top=30truemm,bottom=30truemm,left=25truemm,right=25truemm]{geometry}

\usepackage[colorlinks=true,linktoc=page,citecolor=red,linkcolor=blue]{hyperref}	
\graphicspath{{./figs/}}  % pass for figs
\usepackage[usenames]{xcolor}
\usepackage{amsmath,amssymb,amsfonts}	
\usepackage{bm,braket,array}
\usepackage{multirow}
\usepackage[all]{xy}
\usepackage{amsthm}
\usepackage{amscd}
\usepackage{slashed} %for D slash
\usepackage[normalem]{ulem} % for edit
\theoremstyle{definition}

\theoremstyle{remark}

\def\im{{\rm Im\,}}
\def\pf{{\rm Pf\,}}

\def\tr{{\rm tr\,}}

\def\dim{{\rm dim\,}}
\def\p{\partial}

\def\wt{\widetilde}

\def\ker{{\rm Ker\,}}
\def\coker{{\rm Coker\,}}

\newcommand{\G}{\Gamma}

\newcommand{\Z}{\mathbb{Z}}

\newcommand{\bk}{{\bm{k}}}

\def\widebar{\accentset{{\cc@style\underline{\mskip10mu}}}} %widebar
\def\wideubar{\underaccent{{\cc@style\underline{\mskip10mu}}}} %wideubar

\begin{document}
\title{Atiyah-Hirzebruch Spectral Sequence in Band Topology: \\
General Formalism and Topological Invariants for 230 Space Groups }
\author{Ken Shiozaki}
\affiliation{Yukawa Institute for Theoretical Physics, Kyoto University, Kyoto 606-8502, Japan}
\author{Masatoshi Sato}
\affiliation{Yukawa Institute for Theoretical Physics, Kyoto University, Kyoto 606-8502, Japan}
\author{Kiyonori Gomi}
\affiliation{Department of Mathematics, Tokyo Institute of Technology, 
2-12-1 Ookayama, Meguro-ku, Tokyo 152-8551, Japan}
\date{\today}
\begin{abstract}
We study the Atiyah-Hirzebruch spectral sequence (AHSS) for equivariant $K$-theory in the context of band theory.
Various concepts in band theory, such as irreps at high-symmetry points, compatibility relations, topological gapless points and singularities, fit naturally into the AHSS.
As an application of the AHSS, we get the complete list of topological invariants for 230 space groups without time-reversal or particle-hole invariance. 
We find that many torsion topological invariants appear even for symmorphic space groups. 
\end{abstract}
\maketitle
%\tableofcontents
\parskip=\baselineskip

\section{Introduction}
\label{intro}

After the discovery of the quantum spin Hall effect by Kane and Mele,~\cite{KaneZ2} it has been realized that the rich topological nature inheres in band theory for crystalline materials.
The relations between topology and band theory go back to the celebrated TKNN formula,~\cite{TKNN} where they found that the band structure under a magnetic field shows a quantized Hall conductivity, and it was identified with the Chern number of the vector bundle.~\cite{Kohmoto} 
Kane and Mele pointed out that the time-reversal symmetry(TRS) plays an important role in the band topology; it gives the $\Z_2$-valued topological invariant in two space dimensions. 
For onsite symmetries such as TRS and particle-hole symmetry (PHS), the topological classification was summarized as the periodic table~\cite{RyuClass, Kitaev, RyuTenFold,PhysRevB.82.115120} for Altland-Zirnbauer (AZ) symmetry classes.~\cite{AZ} 
If the bulk has a nontrivial topological number, a gapless state localized at the boundary appears and is stable under perturbation. 
It is called bulk-boundary correspondence.~\cite{PhysRevLett.71.3697}
It was shown that crystalline symmetry also gives rise to new topological invariants of band structures and stabilizes gapless boundary states.~\cite{FuTCI, TaoMirror}
Such topological insulators and superconductors protected by crystalline symmetry are called topological crystalline insulators (TCIs) and superconductors (TCSCs). 
Shortly, the mathematical framework describing the topological classification of the band structure for arbitrary symmetries was formulated as the twisted equivariant $K$-theory by Freed and Moore.~\cite{FreedMoore, Thiang, Gomi17} 
The complete classification of TCIs and TCSCs for any magnetic space group symmetry has been called for. 
So far, much effort has been made around the world with various approaches such as topological invariants, Clifford algebra, and $K$-theory.~\cite{slager2013space,ChiuReflection, MorimotoClifford, Chern_Rotation, ArisSpin, SS14, LiuNonsymmorphic, ChenGlide, SSG15, SSG16, ArisHourGlass, SSG17,bradlyn2017topological,PhysRevB.98.024310,PhysRevB.99.075105,PhysRevResearch.3.013052,Ahn_2019,PhysRevB.105.094518}

The notion of compatibility relation in band theory~\cite{InuiGroup} has shed new light on the topological classification of band structures.~\cite{Kruthoff, Haruki230, TopoChem, Haruki1651}
The number of irreps (or called irreps) at a high-symmetry point is one of the topological invariants in the presence of space group symmetry. 
The compatibility relation measures how an irrep at a high-symmetry point is mapped to representations at a slightly off-symmetric line by the high-symmetric point. 
The set of solutions of the compatibility relation gives the combination of representations at high-symmetry points which can extend to the 1-dimensional subspace in the Brillouin zone (BZ) along the lines between high-symmetry points. 
As an application, in Refs.,~\cite{Haruki230,Haruki1651} they subtracted the atomic insulators by the set of solutions of compatibility relation to get the indicators for topological insulators (in the sense of band structures which have no description by a localized Wanner function) and Weyl semimetals, which is the comprehensive generalization of the Fu-Kane parity formula.~\cite{Fu-Kane} 

Toward the complete classification of band structures, in addition to irreps and the compatibility relation, we should take into account the following issues which are closely related to each other:  
\begin{itemize}
\item[(i)]
The compatibility relation does not ensure a uniform gap over the whole BZ. 
A higher-dimensional obstruction exists to extend a Bloch wave function on the 1-dimensional subspace to the whole BZ. 
For instance, the representation enforced Weyl semimetal~\cite{TurnerInversion} is nothing but the failure to glue the Bloch wave functions in a three-dimensional BZ. 
\item[(ii)]
An obstruction exists to glue a Bloch wave function defined over lines (planes) together on planes (volumes).
For instance, a Dirac point appearing in graphene with sublattice (chiral) symmetry is viewed as the obstruction to gluing a Bloch wave function over lines together on the plane enclosed by the lines. 
\item[(iii)]
Topological invariants of band structures are not limited to the number of irreps at high-symmetry points. 
There are various higher-dimensional topological invariants. 
For instance, the Chern number and the Kane-Mele $\Z_2$ invariant are examples of topological invariants defined over a two-dimensional subspace of BZ. 
\end{itemize}
The crucial point is that obstructions of type (i) and (ii) exist beyond the compatibility relation to glue Bloch wave functions. 

The purpose of this paper is to introduce the Atiyah-Hirzebruch spectral sequence (AHSS)~\cite{AHSS} as systematic machinery to deal with the above three issues (i), (ii), and (iii), as well as the compatibility relation. 
The AHSS is a mathematical tool calculating a generalized cohomology theory. 
We explain how the AHSS fits into band theory in detail.  

As an application, we report the complete classification of topological invariants for 230 space groups without the time-reversal or particle-hole invariance (i.e., A and AIII in AZ classes). 
We found that various torsion topological invariants (meaning cyclic Abelian groups like $\Z_2$) appear in the presence of space group symmetry even if they are symmorphic.
Picking a few symmetry classes, we show the explicit formulas of torsion invariants that have not been addressed in the literature.  

Throughout this paper, the classification of band structures means that in the sense of the $K$-theory, that is, every classification is an Abelian group, and it measures the classification between two different vector bundles stable under adding the same vector bundle. 

The organization of the paper is as follows. 
In Sec.~\ref{sec:AHSS_band}, before moving on to the mathematical detail of the AHSS, we give a brief overview of what the AHSS computes through the language of band theory.
The subsequent two sections are devoted to introducing the AHSS in detail for complex AZ classes (Sec.~\ref{sec:AHSS}) and general symmetry classes (Sec.~\ref{sec:pbar1}). 
Sec.~\ref{sec:e2} includes one of the main results of this paper, where we present the complete list of the topological invariants for 230 space groups in AZ symmetry classes A and AIII. 
We conclude in Sec.~\ref{sec:conc} with the outlook for future directions. 
Appendices are for the technical details to compute the AHSS. 

\section{Overview of the AHSS in band theory}
\label{sec:AHSS_band}
This section aims to illustrate the AHSS before moving on to the mathematical detail. 

\begin{figure}[!]
\centering
\includegraphics[width=\linewidth, trim=0cm 5cm 12cm 0cm]{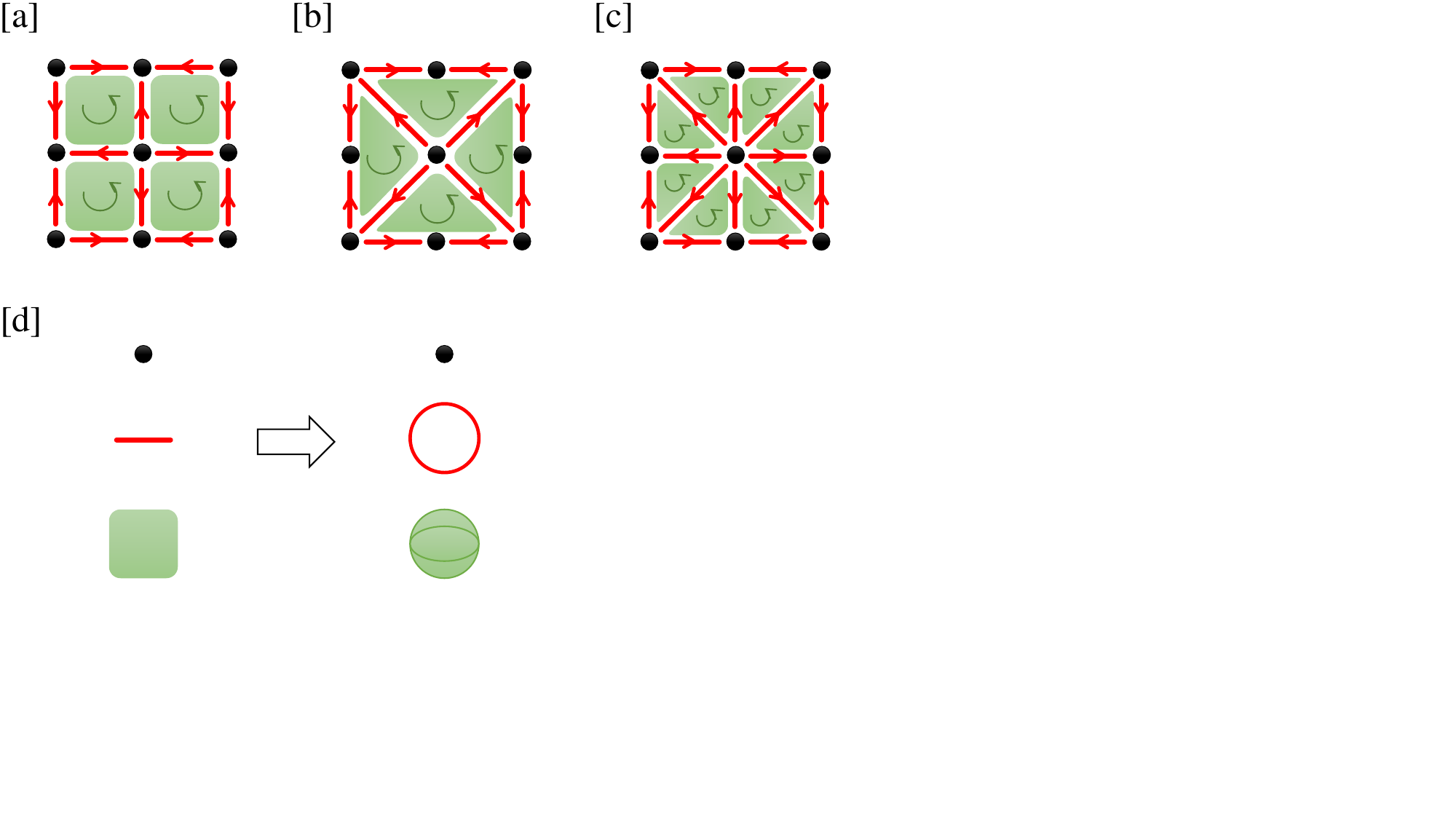}
\caption{[a,b,c] Cell decompositions of the BZ torus $T^2$ with $4$-fold rotation symmetry.
[d] The one-point compactification of the boundary of $p$-cell.}
\label{fig:f1}
\end{figure}

In the AHSS, we start with a cell decomposition of the BZ respecting symmetry. 
Also, we assign each cell an orientation symmetrically. 
For example, in two-dimensional systems with $4$-fold rotation symmetry, a decomposition of the BZ torus $T^2$ is given as Fig.~\ref{fig:f1} [a], which is composed of points, open lines, and open planes, and we call them $0$-cells, $1$-cells, and $2$-cells, respectively. 
We note that the cell decomposition is not unique: Fig.~\ref{fig:f1} [b] and [c] are other cell decompositions of the BZ torus $T^2$ with 4-fold rotation symmetry.

The next step is to assign each $p$-cell an Abelian group so that it possesses the information on band topology as a $p$-dimensional object. 
To do so, in the AHSS, we shrink the boundary of each p-cell to a point to get the $p$-dimensional sphere (or called $p$-sphere) as described in Fig.~\ref{fig:f1} [d].
The classification of band topology over a $p$-sphere for a given symmetry class is readily determined: 
It turns out that the classification is essentially computed by classifying irreps at a point inside the $p$-cell. 
The resulting data of Abelian groups assigned to the $p$-cells is called ``$E_1$-page'', denoted by $E_1=(E_1^{p,-n})$, where $p$ is the dimension of cells, and an integer $n$ indicates an AZ symmetry class which has the period $n = n+8$ ($n=n+2$) for real (complex) AZ classes. 

\begin{table}
\caption{$E_1$-page for complex AZ classes.}
\label{tab:e1_c}
$$
\begin{array}{cl|llll}
{\rm A} & n=0 & E_1^{0,0} & E_1^{1,0} & E_1^{2,0} & E_1^{3,0} \\
{\rm AIII} & n=1 & E_1^{0,-1} & E_1^{1,-1} & E_1^{2,-1} & E_1^{3,-1} \\
\hline 
& E_{1}^{p,-n} & p=0 & p=1 & p=2 & p=3 \\
\end{array}
$$
\end{table}

\begin{table}
\caption{$E_1$-page for real AZ classes.}
\label{tab:e1_r}
$$
\begin{array}{cl|llll}
{\rm AI} & n=0 & E_1^{0,0} & E_1^{1,0} & E_1^{2,0} & E_1^{3,0} \\
{\rm BDI} & n=1 & E_1^{0,-1} & E_1^{1,-1} & E_1^{2,-1} & E_1^{3,-1} \\
{\rm D} & n=2 & E_1^{0,-2} & E_1^{1,-2} & E_1^{2,-2} & E_1^{3,-2} \\
{\rm DIII} & n=3 & E_1^{0,-3} & E_1^{1,-3} & E_1^{2,-3} & E_1^{3,-3} \\
{\rm AII} & n=4 & E_1^{0,-4} & E_1^{1,-4} & E_1^{2,-4} & E_1^{3,-4} \\
{\rm CII} & n=5 & E_1^{0,-5} & E_1^{1,-5} & E_1^{2,-5} & E_1^{3,-5} \\
{\rm C} & n=6 & E_1^{0,-6} & E_1^{1,-6} & E_1^{2,-6} & E_1^{3,-6} \\
{\rm CI} & n=7 & E_1^{0,-7} & E_1^{1,-7} & E_1^{2,-7} & E_1^{3,-7} \\
\hline 
& E_{1}^{p,-n} & p=0 & p=1 & p=2 & p=3 \\
\end{array}
$$
\end{table}

It is useful to express the $E_1$-page as in Tables~\ref{tab:e1_c} and \ref{tab:e1_r}. 
$E_1^{p,-n}$ is the Abelian group of the classification of irreps over a point inside the $p$-cell with the AZ class $n$. 
Using an isomorphism of the $K$-theory, we find that $E_1^{p,-n}$ is also the Abelian group for the classification of topological insulators over the $p$-sphere for the AZ class $(n-p)$, where a representative Hamiltonian is described by the massive Dirac Hamiltonian $H= \sum_{\mu=1}^{p} k_{\mu} \gamma_{\mu} + (m-\epsilon k^2) \gamma_{p+1}$. 

Moreover, the $E_1$-page has a couple of other interpretations. 

The first one is topological gapless states. 
$E_1^{p,-n}$ is also the Abelian group for the classification of stable gapless points inside the $p$-cell for the AZ class $(n+1-p)$: 
Suppose that a topological gapless state in $p$-cell for the AZ class $(n+1-p)$ is described by the massless Dirac Hamiltonian $H'= \sum_{\mu=1}^{p} k_{\mu} \gamma_{\mu}'$. 
By adding a mass term, the Hamiltonian $H'$ can be viewed as the massive Dirac Hamiltonian $H = \sum_{\mu=1}^p k_{\mu} \gamma_{\mu} + (m-\epsilon k^2)\gamma_{p+1}$ that describes the topological insulator over the $p$-sphere with the shift of AZ class as $(n+1-p) \mapsto (n-p)$, i.e.\ the Abelian group $E_1^{p,-n}$. 

\begin{figure}[!]
	\begin{center}
	\includegraphics[width=\linewidth, trim=4cm 10cm 3cm 0cm]{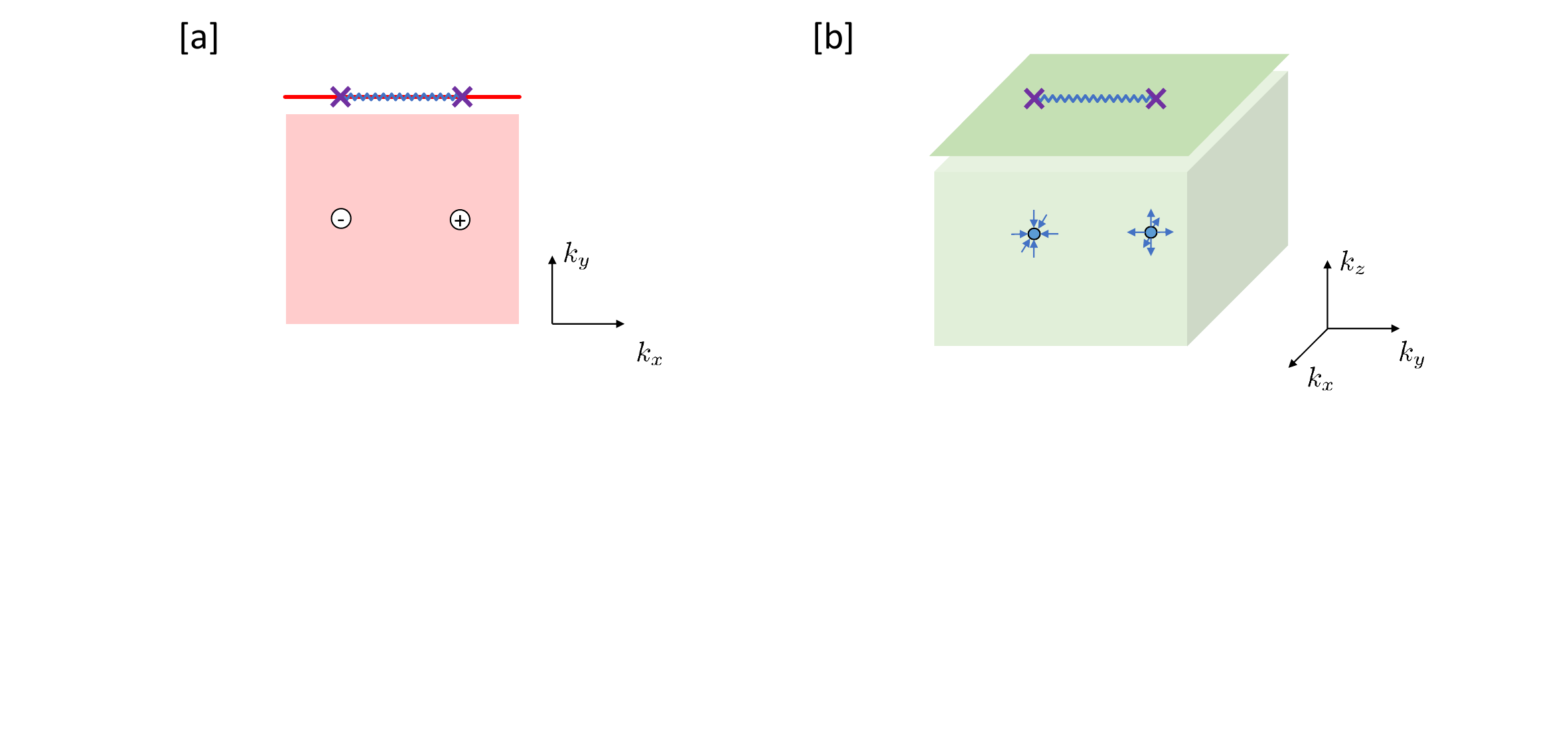}
	\end{center}
	\caption{Examples of topological singular points. 
	[a] The end point of the flat edge zero energy state with the chiral symmetry. 
	[b] The branching point of the Fermi arc in the surface state of the Weyl semimetal.}
	\label{fig:singular_pt_end_pt}
\end{figure}

$E_1^{p,-n}$ also represents the Abelian group for the classification of stable singular points inside $p$-cells for the AZ class $(n+2-p)$. 
Here, the singular point means a point in the BZ where the Hamiltonian is not single-valued. 
For example, the endpoint of the flat zero energy edge state for the zigzag edge boundary condition in the graphene with the chiral symmetry is an example of the singularity inside a 1-cell (Fig.~\ref{fig:singular_pt_end_pt} [a]). 
An example of the singular point in a 2-cell is the branching point of the Fermi arc appearing in the surface BZ of the Weyl semimetal (Fig.~\ref{fig:singular_pt_end_pt} [b]). 
In general, a singular point inside a $p$-cell appears as the end point of the massless Dirac line described by the Hamiltonian $H' = \sum_{\mu=1}^{p-1} k_{\mu} \gamma_{\mu}'$. 
Using this, we have the model Hamiltonian for the topological singular point 
\begin{align}
H''=\Im \ln \left[ k_{p} + i \sum_{\mu=1}^{p-1}k_{\mu} \gamma_{\mu}' \right], 
\label{eq:dirac_singular_pt}
\end{align}
where $\Im (z)$ is the imaginary part of $z$. 
On the $k_p$-axis, for $k_p>0$ the Hamiltonian $H''$ is recast as the massless Dirac Hamiltonian $H'= \sum_{\mu=1}^{p-1} k_{\mu} \gamma_{\mu}'$, whereas for $k_p<0$ the Hamiltonian $H''$ has a finite energy gap as $H'' \sim \pm \pi$. 
Therefore, the possibility of a topological singular point in a $p$-cell is equivalent to the existence of a $(p-1)$-dimensional topological gapless state. 
Using an isomorphism of the $K$-theory, we find that the latter is classified by the Abelian group $E_1^{p,-n}$. 
See Sec.~\ref{sec:complex_E_1-page} and Appendix~\ref{app:On the classification of singular points} for more details.

\begin{figure}[!]
	\begin{center}
	\includegraphics[width=\linewidth, trim=0cm 25cm 25cm 0cm]{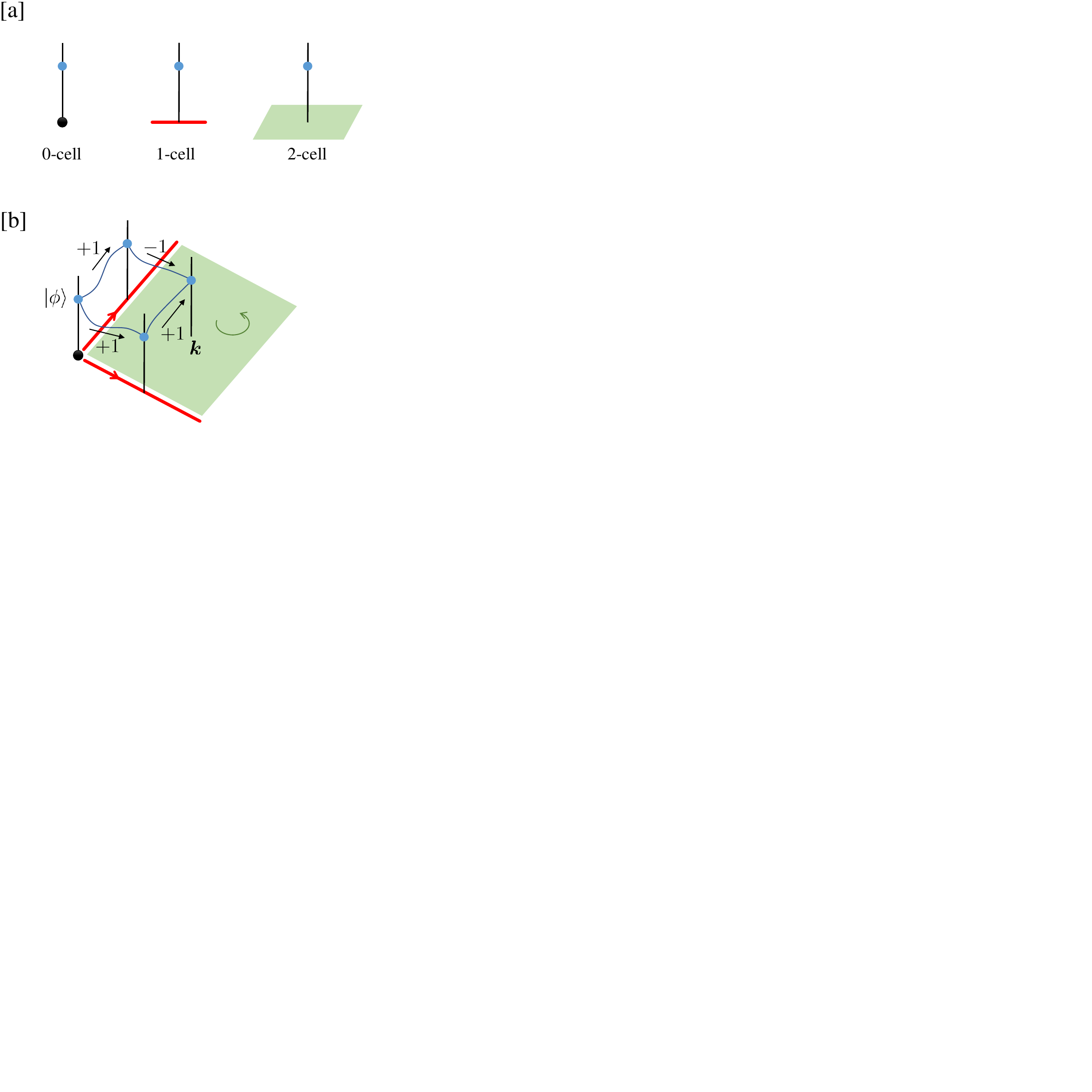}
	\end{center}
	\caption{
	[a] $E_1^{p,-n}$ as irreps at points $\bk$ inside $p$-cells for the AZ class $n$. 
	[b] Irrespective of adjacent intermediate 1-cells, a Bloch state $\ket{\phi}$ at a $0$-cell is connected to the same Bloch state at a point $\bk$ inside the $2$-cell. 
	}
	\label{fig:f2}
\end{figure}

In this way, there are four different interpretations of the $E_1$-page summarized as follows. 
\begin{itemize}
\item
\ [Irreps] 
$E_1^{p,-n}$ is the classification of irreps at points $\bk$ inside $p$-cells for the AZ class $n$. 
(See Fig.~\ref{fig:f2} [a]).
\item
\ [Topological insulators]
$E_1^{p,-n}$ is the classification of topological insulators over $p$-spheres for the AZ class $(n-p)$, where the $p$-sphere is defined by shrinking the boundary of the $p$-cell to one point. 
\item
\ [Topological gapless states]
$E_1^{p,-n}$ is the classification of topological gapless states inside $p$-cells for the AZ class $(n+1-p)$. 
\item
\ [Topological singular points]
$E_1^{p,-n}$ is the classification of topological singular points inside $p$-cells for the AZ class $(n+2-p)$. 
\end{itemize}
Table~\ref{tab:overview_e1} shows the correspondence between the latter three interpretations and the columns in the $E_1$-page in the view of a fixed AZ class $n$. 
In applying the AHSS to band theory, we should keep all the above four interpretations of the $E_1$-page in mind. 

\begin{table}
	\begin{center}
	\caption{The $E_1$-page for an AZ class $n$. In the table, 
	$E_1^{p,-(n+p)}, E_1^{p,-(n-1+p)}$, and $E_1^{p,-(n-2+p)}$ represent topological insulators, topological gapless states, and topological singular points, respectively, for the AZ class $n$. 
	}
	\label{tab:overview_e1}
	$$
\begin{array}{|c|c|c|c|c|}
\hline
\begin{minipage}{18mm}
\scalebox{0.5}{\includegraphics[width=35mm,clip]{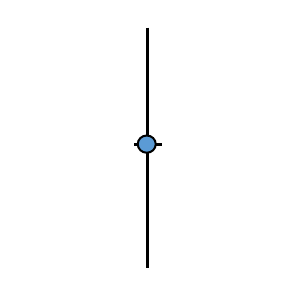}}
$E_1^{0,-(n-1)}$ 
\end{minipage}
&
\begin{minipage}{18mm}
\scalebox{0.5}{\includegraphics[width=35mm,clip]{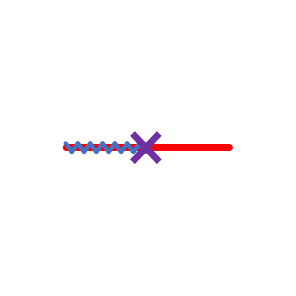}}
$E_1^{1,-(n-1)}$ 
\end{minipage}
& & 
\\
\hline
\begin{minipage}{18mm}
\scalebox{0.5}{\includegraphics[width=35mm,clip]{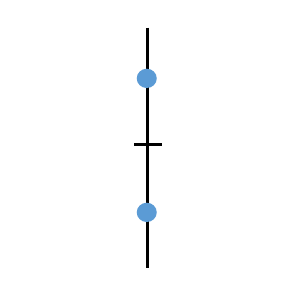}}
$E_1^{0,-n}$ 
\end{minipage}
& 
\begin{minipage}{18mm}
\scalebox{0.5}{\includegraphics[width=35mm,clip]{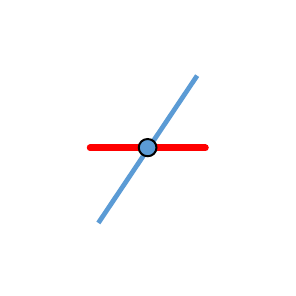}}
$E_1^{1,-n}$ 
\end{minipage}
&
\begin{minipage}{18mm}
\scalebox{0.5}{\includegraphics[width=35mm,clip]{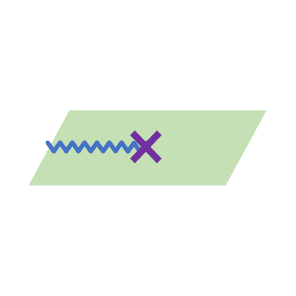}}
$E_1^{2,-n}$ 
\end{minipage}
& \\
\hline
& 
\begin{minipage}{18mm}
\scalebox{0.5}{\includegraphics[width=35mm,clip]{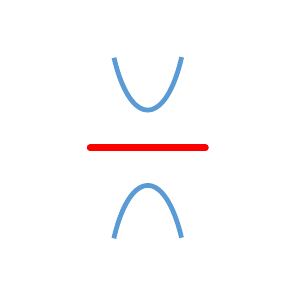}}
$E_1^{1,-(n+1)}$ 
\end{minipage}
& 
\begin{minipage}{18mm}
\scalebox{0.5}{\includegraphics[width=35mm,clip]{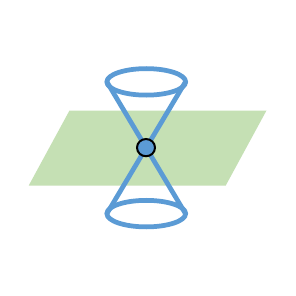}}
$E_1^{2,-(n+1)}$ 
\end{minipage}
&
\begin{minipage}{18mm}
\scalebox{0.5}{\includegraphics[width=35mm,clip]{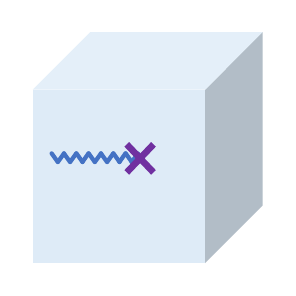}}
$E_1^{3,-(n+1)}$ 
\end{minipage}
\\
\hline
& & 
\begin{minipage}{18mm}
\scalebox{0.5}{\includegraphics[width=35mm,clip]{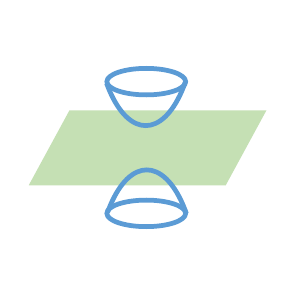}}
$E_1^{2,-(n+2)}$ 
\end{minipage}
& 
\begin{minipage}{18mm}
\scalebox{0.5}{\includegraphics[width=35mm,clip]{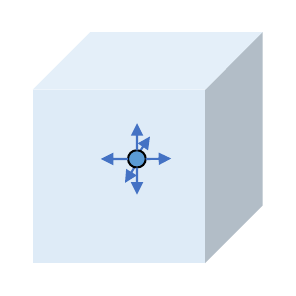}}
$E_1^{3,-(n+2)}$ 
\end{minipage}
\\
\hline
& & & 
\begin{minipage}{18mm}
\scalebox{0.5}{\includegraphics[width=35mm,clip]{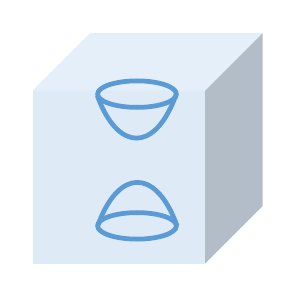}}
$E_1^{3,-(n+3)}$ 
\end{minipage}
\\
\hline
\mbox{0-cell}& 
\mbox{1-cell}&
\mbox{2-cell}&
\mbox{3-cell}\\
\hline
\end{array}
$$
	\end{center}
\end{table}

Generally, elements of groups $E_1^{p,-n}$ are dependent under the assumption of the continuity of the energy bands.
This constraint is known as compatibility relations,~\cite{bradley2010mathematical,Kruthoff,Haruki230} and is identified as the first differential 
\begin{align}
d_1^{p,-n}: E_1^{p,-n} \to E_1^{p+1,-n}
\end{align}
in the AHSS.
The first differential $d_1 = (d_1^{p,-n})$ can be viewed as the compatibility relation from $p$-cells to adjacent $(p+1)$-cells. 
The compatibility relation gives how an irrep $\rho^{p}_{\alpha}$ at a $p$-cell splits into representations $\rho^{p+1}_{\beta}$ at adjacent $(p+1)$-cells, 
\begin{align}
\rho^p_{\alpha} = \bigoplus_{\beta} n_{\alpha}^{\beta} \rho^{p+1}_{\beta}, 
\end{align}
which is characterized by the nonnegative integers $n_{\alpha}^{\beta}$. 
If the direction of the $p$-cell (dis)agrees with the adjacent $(p+1)$-cell, the nonnegative integer $n_{\alpha}^{\beta}$ contributes to the first differential $d_1^{p,-n}$ with the positive (negative) sign. 
We see that the first differential obeys $d_1 \circ d_1 =0$, i.e., \ taking the first differential twice is trivial. 
As shown in Fig.~\ref{fig:f2} [b], a Bloch state $\ket{\phi}$ at a $0$-cell is connected to, regardless of adjacent intermediate 1-cells, the same state at a point $\bk$ inside the $2$-cell, which is nothing but the relation $d_1 \circ d_1=0$. 
The relation $d_1 \circ d_1=0$ implies $\im (d_1^{p-1,-n}) \subset \ker (d_1^{p,-n})$ as an Abelian group.
One can take the cohomology of $d_1$ to get the $E_2$-page 
\begin{align}
E_2^{p,-n} := \ker (d_1^{p,-n})/\im (d_1^{p-1,-n}). 
\end{align}

\begin{figure}[!]
	\begin{center}
	\includegraphics[width=\linewidth, trim=0cm 14cm 22cm 0cm]{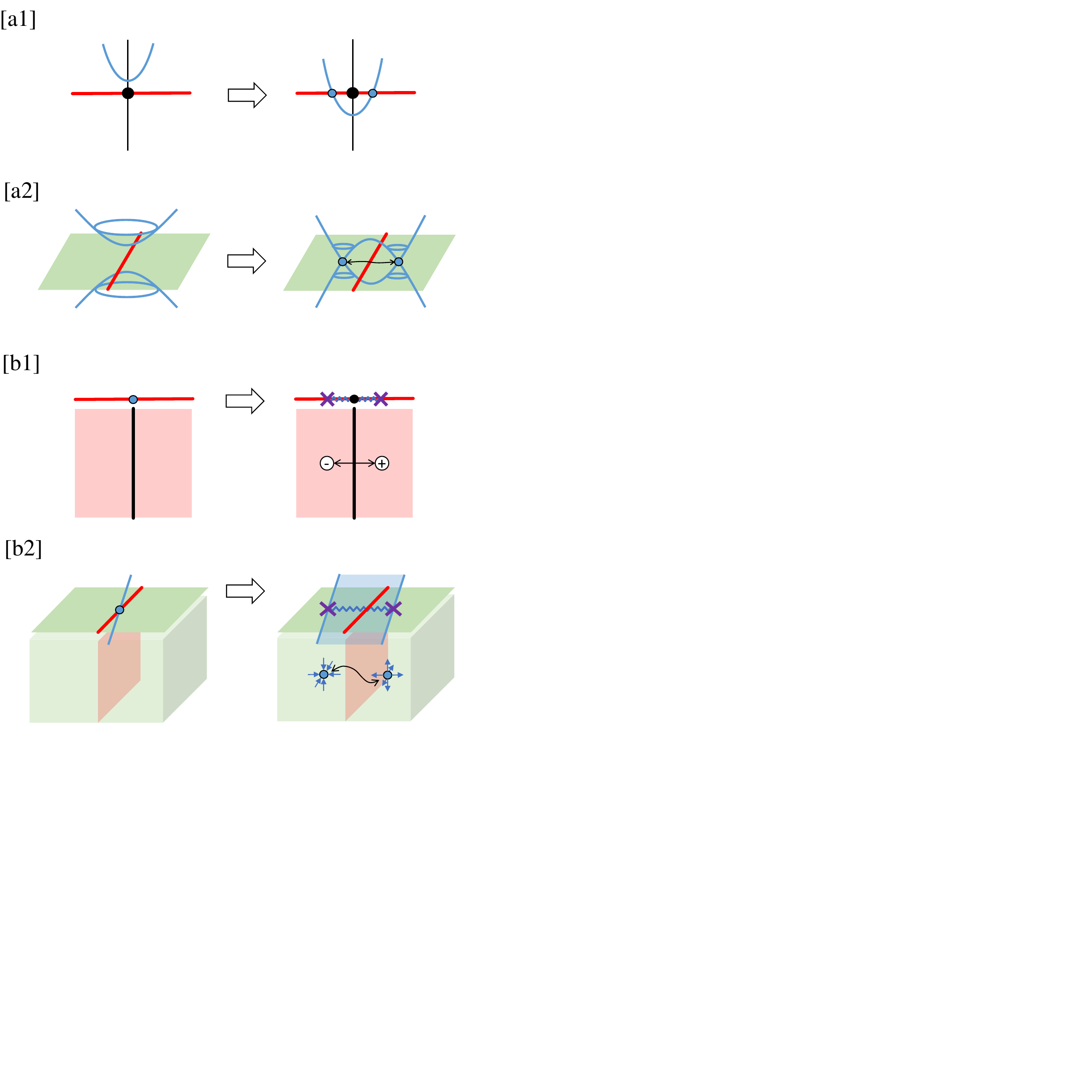}
	\end{center}
	\caption{
	[a1, a2] The first differential $d_1^{p-1,-(n-1+p)}$ can be viewed as changing the topological invariant over $(p-1)$-cells followed by creating gapless Dirac points to adjacent $p$-cells.
	[b1, b2] 
	The first differential $d_1^{p,-(n-1+p)}$ can be viewed as how gapless states inside $p$-cells are continuously extended with endpoints of topological singularities in adjacent $(p+1)$-cells. 
	The corresponding bulk semimetal phases with a pair of gapless points created from the  $p$-cell are also shown. 
	}
	\label{fig:f3}
\end{figure}

Interpreting the $E_2$-page as topological gapless states clarifies the meaning of the $E_2$-page. 
The $E_1$-page $E_1^{p,-(n-1+p)}$ is the candidate for an anomalous gapless state for the AZ class $n$ in the sense that it can not be realized as a lattice system where the number of bands is finite. 
Not every element in the $E_1$-page corresponds to a genuine anomalous gapless state because of the following two reasons. 
The first reason is that a topological gapless state inside a $p$-cell may be trivialized by the creation/annihilation of Dirac points from adjacent $(p-1)$-cells, which corresponds to the image of the first differential $d_1$: 
\begin{align}
\im \left[ d_1^{p-1,-(n-1+p)}: E_1^{p-1,-(n+p-1)} \to E_1^{p,-(n-1+p)} \right].
\end{align}
Recall that $E_1^{p-1,-(n+p-1)}$ gives the classification of topological insulators over $(p-1)$-spheres for the AZ class $n$. 
We see that $d_1^{p-1,-(n-1+p)}$ describes changing the topological invariant over $(p-1)$-cells followed by creating gapless Dirac points to adjacent $p$-cells, as shown in Fig.~\ref{fig:f3} [a1] and [a2].
The second reason is that a gapless state in the $E_1$-page may be singular in adjacent $(p+1)$-cells. 
The topological gapless states in $p$-cells that can extend the adjacent $(p+1)$-cells without a singularity are represented by the kernel 
\begin{align}
\ker \left[ d_1^{p,-(n-1+p)}: E_1^{p,-(n-1+p)} \to E_1^{p+1,-(n-2+p+1)} \right].
\end{align}
Recall that $E_1^{p+1,-(n-2+p+1)}$ gives the classification of topological singular points inside $(p+1)$-cells for the AZ class $n$. 
We see that $d_1^{p,-(n-1+p)}$ describes how gapless states inside $p$-cells are continuously extended with endpoints of topological singularities in adjacent $(p+1)$-cells, as shown in Fig.~\ref{fig:f3} [b1] and [b2].
The kernel of $d_1^{p,-(n-1+p)}$ implies that the topological singular points created by the first differential $d_1^{p,-(n-1+p)}$ cancel out, i.e.,  the absence of a singularity. 
In sum, we have that 
\begin{itemize}
\item
The $E_2$-page $E_2^{p,-(n-1+p)}$ is the classification of topological gapless states inside $p$-cells for the AZ class $n$ which can not be trivialized by creation of topological Dirac points from adjacent $(p-1)$-cells and can extend to adjacent $(p+1)$-cells without a singularity. 
\end{itemize}

\begin{figure}[!]
	\begin{center}
	\includegraphics[width=\linewidth, trim=0cm 21cm 16cm 0cm]{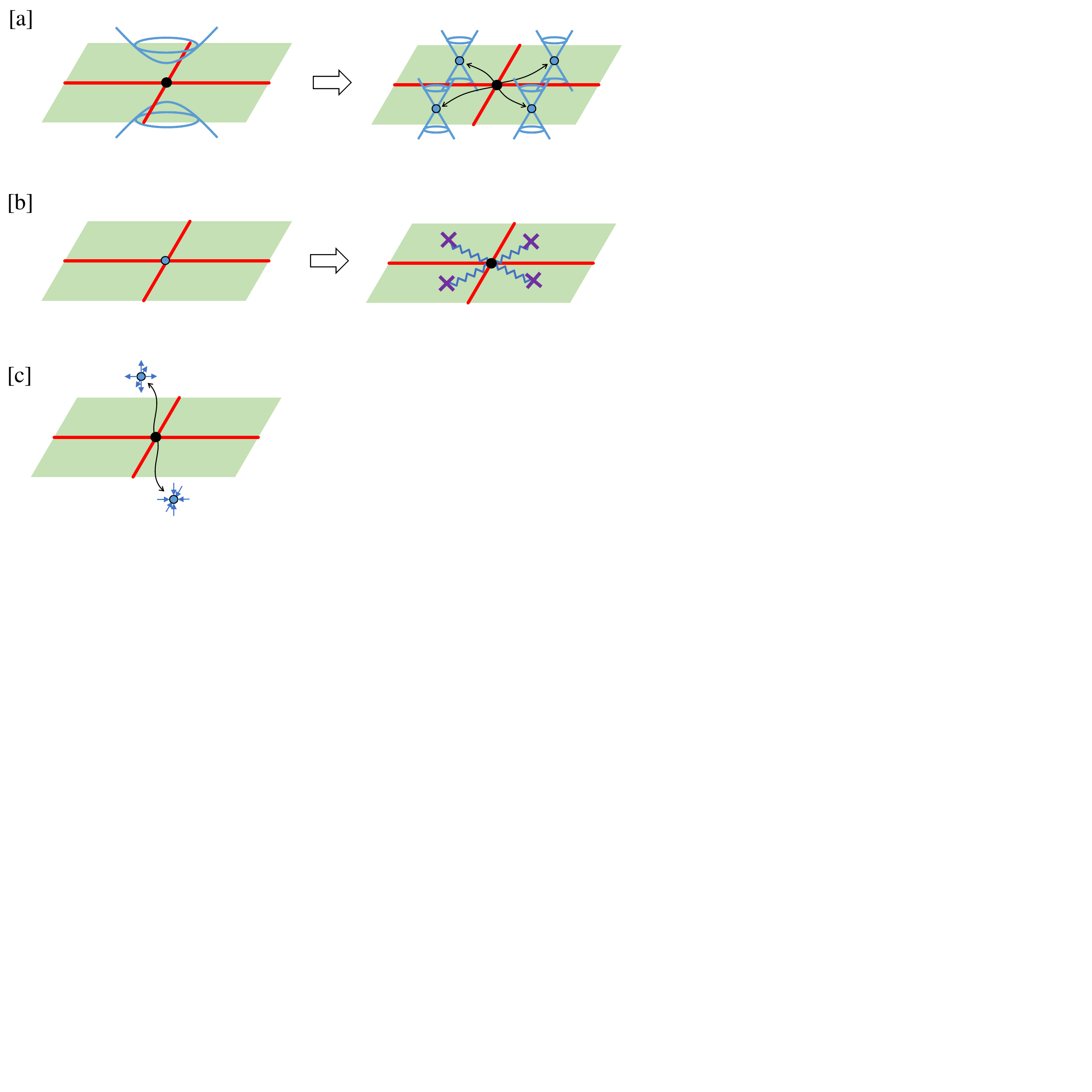}
	\end{center}
	\caption{
	[a]
	The second differential $d_2^{p-2,-(n+p-2)}$ represents a band inversion at a $(p-2)$-cell followed by creating gapless Dirac points in the $p$-cells nearby the $(p-2)$-cell. 
	[b]
	The second differential $d_2^{p,-(n-1+p)}$ represents how topological gapless states in $p$-cells extend into adjacent $(p+2)$-cells in the form of lines with singular points as  endpoints. 
	[c]
	The third differential $d_3^{p-3,-(n+p-3)}$ represents a band inversion at a $(p-3)$-cell followed by creating gapless Dirac points in the $p$-cells nearby the $(p-3)$-cell. 
	}
	\label{fig:f4}
\end{figure}

Here is not the end of the story. 
The topological gapless states described by the $E_2$-page may be further trivialized. 
The band inversion at a $(p-2)$-cell may create gapless Dirac points in the $p$-cells nearby the $(p-2)$-cell, as shown in Fig.~\ref{fig:f4} [a], and this defines the second differential 
\begin{align}
d_2^{p-2,-(n+p-2)}: E_2^{p-2,-(n+p-2)} \to E_2^{p,-(n-1+p)}.
\end{align}
The created Dirac points in $p$-cells can trivialize the topological gapless states left in the $E_2$-page $E_2^{p,-(n-1+p)}$, i.e.\ the trivialization by the image $\im (d_2^{p-2,-(n+p-2)})$. 
The second differential also represents how topological gapless states in $p$-cells create topological singular points in adjacent $(p+2)$-cells with branch cut lines, as shown in Fig.~\ref{fig:f4} [b], and this is represented by the second differential 
\begin{align}
d_2^{p,-(n-1+p)}: E_2^{p,-(n-1+p)} \to E_2^{p+2,-(n+p)}. 
\end{align}
In the same way as the first differential, taking the kernel of $d_2^{p,-(n-1+p)}$ leads to the subspace of topological gapless states in $E_2^{p,-(n-1+p)}$ so that the created singular points inside $(p+2)$-cells cancel out. 
We see that the second differential also obeys $d_2 \circ d_2=0$. 
The $E_3$-page is defined as the cohomology of the second differential 
\begin{align}
E_3^{p,-n}:= \ker (d_2^{p,-n})/\im (d_2^{p-2,-(n-1)}).
\end{align}
It is clear that the $E_3$-page has the following meaning as topological gapless states
\begin{itemize}
\item
The $E_3$-page $E_3^{p,-(n-1+p)}$ is the classification of topological gapless states inside $p$-cells for the AZ class $n$ which can not be trivialized by creation of Dirac points from adjacent $(p-1)$- and $(p-2)$-cells and can extend to adjacent $(p+1)$- and $(p+2)$-cells without a singularity. 
\end{itemize}

In the same way, the topological gapless states inside $p$-cells contained in the $E_3$-page may be further trivialized by the band inversion at a $(p-3)$-cell followed by the creation of Dirac points to adjacent $p$-cells, which defines the third differential 
\begin{align}
d_3^{p-3,-(n+p-3)}: E_3^{p-3,-(n+p-3)} \to E_3^{p,-(n-1+p)}. 
\end{align}
For example, the third differential $d_3^{0,-n}$ represents the band inversion and creating the Dirac points inside 3-cells, as shown in Fig.~\ref{fig:f4} [c].
Similarly, the third differential $d_3^{p,-(n-1+p)}: E_3^{p,-(n-1+p)} \to E_3^{p+3,-(n+1+p)}$ gives the creation of topological singular points inside $(p+3)$-cells from the gapless point in $p$-cells. 
We find that the third differential obeys that $d_3 \circ d_3=0$. 
The $E_4$-page is defined as the cohomology of $d_3$ 
\begin{align}
E_4^{p,-n}:= \ker (d_3^{p,-n}) / \im (d_3^{p-3,-(n+2)}).
\end{align}
In three space dimensions, there is no further trivialization and the compatibility relation for the absence of a singular point. 
We get the limiting page $E_{\infty} = E_4$. 
(In two space dimensions, the $E_3$-page becomes the limit $E_{\infty}=E_3$.)

The limiting page $E_{\infty}^{p,-(n-1+p)}$ represents the topological gapless states in $p$-cells for the AZ class $n$ which have no singularity and can not be trivialized by the creation of Dirac points from any adjacent low-dimensional cells. 
Therefore, elements of the $E_{\infty}$-page $E_{\infty}^{p,-(n-1+p)}$ are genuine anomalous gapless phases, i.e., gapless states which can not be realized as standalone lattice systems. 
Since the classification of anomalous gapless phases is equivalent to the classification of bulk gapped phases over the same BZ with a shift of the AZ class (this is the bulk-boundary correspondence), we find that the $E_{\infty}$-page $E_{\infty}^{p,-(n+p)}$ represents bulk gapped phases for the AZ class $n$. 
To be precise, the $E_{\infty}$-page approximates the classification of the gapped (and gapless) phases with the set of exact sequences. 
See eq.(\ref{eq:e_infty_appro}) below. 

\section{The AHSS in the absence of antiunitary symmetry}
\label{sec:AHSS}

Spectral sequences are mathematical tools to calculate (co)homology groups.
(For an introductory exposition, see Ref.\cite{BottTu}.) 
In particular, the Atiyah-Hirzebruch spectral sequence (AHSS) calculates a generalized cohomology theory, of which the $K$ theory version was first introduced by Atiyah and Hirzebruch.~\cite{AHSS} 
In the context of physics, the AHSS has been applied to string theories.~\cite{Maldacena} 
The AHSS in this section is defined by applying the general recipe~\cite{Tamaki} to the twisted equivariant $K$ theory. 
In the following, we explain the resulting AHSS along with the setup of our interest.

\subsection{Formulation}
\label{sec:ahss_com_for}

\begin{figure}[h]
	\begin{center}
	\includegraphics[width=\linewidth, trim=0cm 5cm 3cm 0cm]{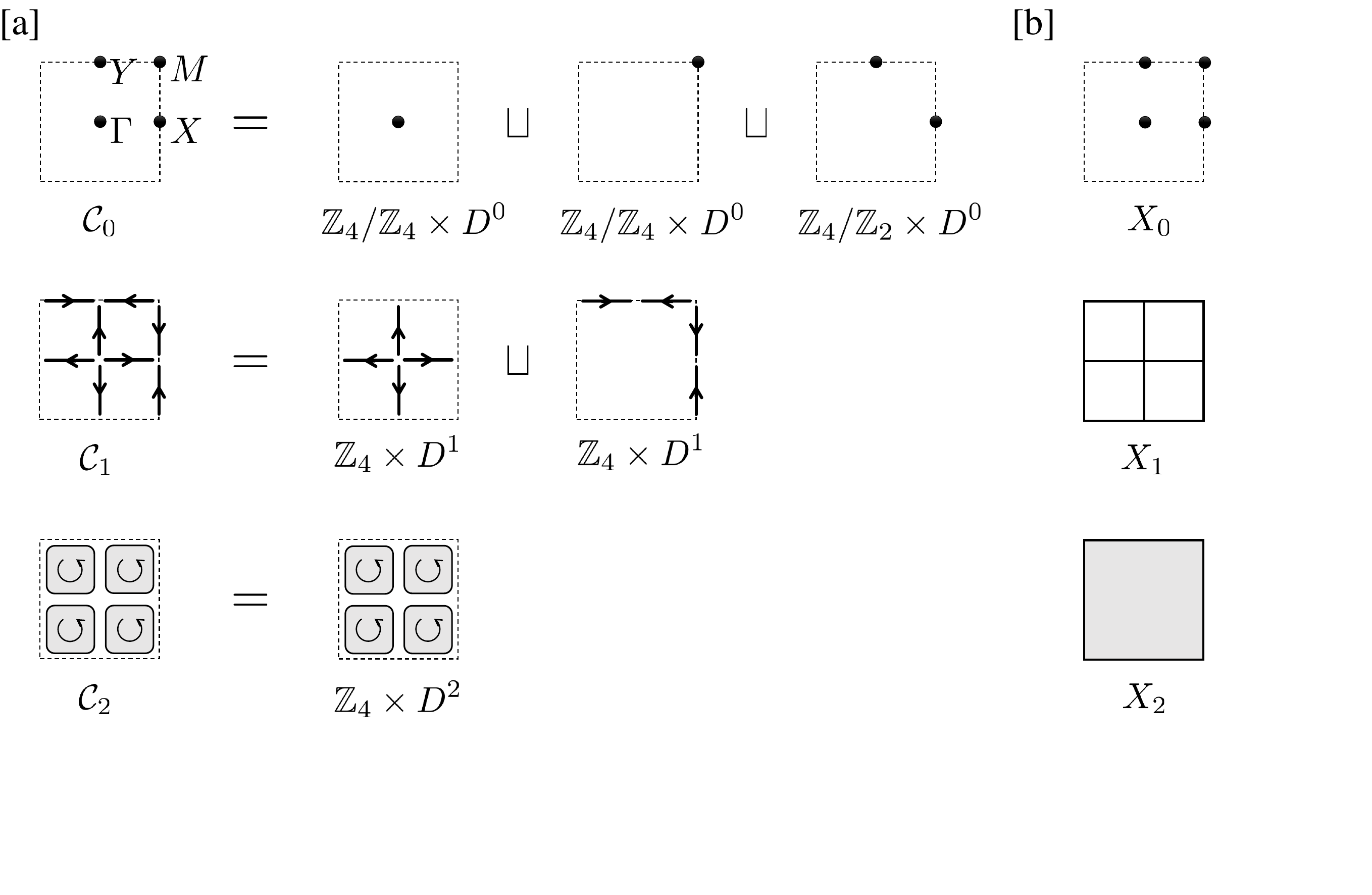}
	\end{center}
	\caption{
	An example of $\Z_4$-symmetric filtration of $T^2$ with $4$-fold rotation symmetry. 
	[a] $0$-, $1$- and $2$-cells. Arrows represent directions of $p$-cells which are $\Z_4$-symmetrically assigned. 
	The r.h.s.\ shows the orbits $(\Z_4/G_{\bk_j}) \times D^p (p=0,1,2)$.
	[b] $0$-, $1$- and $2$-skeletons. The 2-skeleton $X_2$ is the $2$-torus $X_2 = T^2$ itself. 
	}
	\label{fig:cell}
\end{figure}

Let $T^3$ be a three-dimensional BZ torus and $G$ a point group. 
The group $G$ acts on $T^3$ associatively, i.e., it holds that $g (h \bk) =(g h) \bk$ for $\bk \in T^3$ and $g,h \in G$.
The first step is to take a series of subspaces of $T^3$,  called $G$-symmetric filtration of $T^3$,
\begin{align}
X_0 \subset X_1 \subset X_2 \subset X_3= T^3, 
\end{align}
where $X_p$ is a $p$-dimensional subspace closed under the $G$-symmetry. 
We here take a particular filtration of $T^3$ associated with a $G$-CW decomposition,~\cite{MayBook} in which the subspace $X_p$, called the $p$-skeleton, is given in the following manner. 

\subsubsection{Cell decomposition \label{sec:cell_deco}}

We first divide $T^3$ by cells with dimensions lower or equal to 3,
i.e., points (0-cells), open line segments (1-cells), open polygons (2-cells), and open polyhedrons (3-cells):  
Each $p$-cell, which is isomorphic to a $p$-dimensional open disk $D^p$, is assigned an orientation, and its boundary, $\partial D^p=\overline{D^p}-D^p$, consists of $(p-1)$-cells.
The whole set of the oriented $p$-cells, which we denote ${\cal C}_p$, should form a set of $G$-symmetric cells (called $G$-equivariant cells or $G$-cells), where each $G$-cell consists of the $p$-cells that are obtained by applying $G$ on a $p$-cell.
In other words, a $G$-cell is an orbit of a $p$-cell $D^p$ under the $G$-action, $(G/G_{D^p})\times D^p$, where $G_{D^p}$ is the little group of the $p$-cell $D^p$ (that is $G_{D^p}:=\{g\in G|g{\bm k}={\bm k}, \mbox{for $^\forall$${\bm k}\in D^p$} \}$). 
We also require that all the $p$-cells in the orbit should be different if $(G/G_{D^p_j})\neq 1$, and the orientations of the $p$-cells are consistent with the $G$-action.
The former requirement implies that ${\cal C}_p$ contains all $p$-dimensional high-symmetry regions (points for $p=0$, lines for $p=1$ and planes for $p=2$) under $G$. 
The number of $p$-cells contained in the orbit $(G/G_{D^p}) \times D^p$ is $|G/G_{D^p}|$. 
The orbit $(G/G_{D^p}) \times D^p$ is homotopic to the set of points $(G/G_{\bk}) \times \{ \bk \}$ ($\bk\in D^p$), which is known as the ``star'' in the literature,~\cite{InuiGroup} since the $p$-cell $D^p$ can shrink to a point ${\bm k}\in D^p$ smoothly.
Keeping in mind the requirements for the orbit in the above, we write ${\cal C}_p$ as the direct sum of orbits of $p$-cells 
\begin{align}
{\cal C}_p = \coprod_{j \in I^p_{\rm orb}} (G/G_{D_j^p}) \times D_j^p, 
\end{align}
with $I^p_{\rm orb}$ a label set of orbits in ${\cal C}_p$, $D_j^p$ a representative $p$-cell of the $j$-th orbit, and $G_{D^p_j}$ the little group of the $p$-cell $D_j^{p}$.

The 0-skeleton $X_0$ is given by the set of 0-cells ${\cal C}_0$.
Then, for $p > 0$, the $p$-skeleton $X_p$ is defined inductively by gluing each orbit $(G/G_{\bk_j})\times D_j^p$ in ${\cal C}_p$ to the $(p-1)$-skeleton $X_{p-1}$: Using the obvious map $(G/G_{\bk_j}) \times \partial D^p \to X_{p-1}$, we have 
\begin{align}
X_p = X_{p-1} \cup \coprod_{j \in
I^p_{\rm orb}} (G/G_{D^p_j}) \times D_j^p.
\label{eq:XpXp-1}
\end{align}
If the resultant $X_3$ satisfies $X_3=T^3$, we have a $G$-symmetric filtration, and if not, we add (or remove) a proper orbit to (from) ${\cal C}_p$ and repeat the same procedure until $X_3$ coincides with $T^3$.
In any case, we can obtain a $G$-symmetric filtration. 
For illustration, we provide an example of a $\Z_4$-symmetric filtration of a 2-torus $T^2$ with $4$-fold rotation symmetry in Fig.~\ref{fig:cell}.
(A filtration of $T^2$ is defined similarly.)

\subsubsection{$E_1$-page}
Now consider a space group with the point group $G$ and the factor system $\tau$.
(For the definition of the factor system, see Sec.~\ref{sec:e1_wigner}).
Using the $G$-symmetric filtration above,  we introduce the AHSS.
The AHSS consists of  a collection of two sequences, i.e. pages $E_r$ and differentials $d_r$ ($r=1,2,\dots$).
According to the general recipe, the first page (called $E_1$-page)  is given by the twisted equivariant $K$-group $K_G^{\tau-n}$,~\cite{FreedMoore, SSG17}  
 \begin{align}
E_1^{p,-n} 
&= K^{\tau-(n-p)}_G(X_p, X_{p-1}) \nonumber \\
&\cong 
\left\{\begin{array}{ll}
\prod_{j \in I^p_{\rm orb}} K^{\tau|_{D^p_j}+0}_{G_{D^p_j}}( D^p_j ) & (n \in {\rm even}), \\
0 &  (n \in {\rm odd}).  \\
\end{array}\right.
\label{eq:e1}
\end{align}
where $X_{-1}=\emptyset$ and $\tau|_{D^p_j}$ is the factor system at $D^p_j$.~\footnote{
For derivation of Eq.(\ref{eq:e1}), see also Sec.\ref{sec:ahss_real}.}
Here only the parity of the degree $n \in \Z$ matters because of the Bott periodicity $K^{\tau-n}_G\cong K^{\tau-n+2}_G$.
In the AZ classification scheme, the even degree and the odd one are referred to as class A and class AIII, respectively. 
In the context of band theory, an element of the $K$-group $K^{\tau|_{D^p_j}-0}_{G_{D_j^p}}({D^p_j})$ corresponds to  a set of numbers 
\begin{align}
(n_{\rho_1(D^p_j)}, n_{\rho_2(D_j^p)}, n_{\rho_3(D_j^p)}, \dots )
\end{align}
in which $n_{\rho_{a}(D_j^p)}$ counts (occupied) states in the irrep $\rho_{\alpha}(D_j^p)$ of $G_{D_j^p}$ on the $p$-cell $D_j^p$. 
(More precisely, $n_{\rho_\alpha(D_j^p)}$ denotes the difference between the number of occupied states and the number of empty ones in $\rho_\alpha(D_j^p)$.)
It should be noted here that any Bloch state on the $p$-cell $D_j^p$ is a representation of $G_{D^p}$ since the little group is good symmetry on $D_j^p$.
The $K$-group is an Abelian group, where the addition (the subtraction) is defined obviously as an increase (decrease) of states in the corresponding representations. 
Therefore, $E_1^{p,0}$ is also an Abelian group.
It holds that $E_1^{p,-1} = 0$ because the chiral symmetry in class AIII enforces that occupied and empty states are the same in number, so  $n_{\Gamma_a}=0$.

\subsubsection{First differential $d_1$}

The first differential $d_1$ in the AHSS is given as a series of homomorphisms among $E_1$ pages,
 \begin{align}
E_1^{0,-n} \xrightarrow{d_1^{0,-n}} E_1^{1,-n} \xrightarrow{d_1^{1,-n}} E_1^{2,-n} \xrightarrow{d_1^{2,-n}} E_1^{3,-n},
\end{align}
where the differential satisfies $d_1 \circ d_1 = 0$.
In the present case, only $d_1^{p,0}$ is nontrivial since $E_1^{p,-1}=0$.
From the general recipe~\cite{Tamaki}, the first differential $d_1^{p,-n}$ is defined by the composition 
\begin{align}
    d_1^{p,-n}: &K^{\tau-(n-p)}_G(X_p,X_{p-1})
    \xrightarrow{i^*} 
    K^{\tau-(n-p)}_G(X_p) \nonumber\\
    &\xrightarrow{d}
    K^{\tau-(n-p-1)}_G(X_{p+1},X_{p}), 
\end{align}
where $i^*$ is induced map of the inclusion $i:X_p \to (X_p,X_{p-1})$, and $d$ is the coboundary map. 
We present here the physical meaning of $d_1$ in terms of band theory.
As we explained above, an element of $E^{p,0}_1$ specifies a particular set of irreps in each $p$-cell in $T^3$.
Furthermore, $d_1$ should be defined locally like an ordinary differential operator.
These properties suggest that $d^{p,0}_1$ is a map from representations in a $p$-cell to those in an adjacent $(p+1)$-cell.
In other words, $d^{p,0}_1$ gives a relation between the representation of a state at a $p$-cell and those of the same state at an adjacent $(p+1)$-cell:
In general, the representation $\rho_\alpha(D^p)$ at a $p$-cell $D^p$ splits into a set of representations $\rho_\beta(D^{p+1}) $ at an adjacent $(p+1)$-cell $D^{p+1}$, 
\begin{align}
\rho_{\alpha}(D^p) = \bigoplus_{\beta \in {\rm irreps}} n_{\alpha}^{\beta} \,\rho_{\beta}(D^{p+1}),  
\label{eq:compatibility}
\end{align}
since the little group $G_{D^{p+1}}$ is a subgroup of $G_{D^p}$ when the $(p+1)$-cell is adjacent to the $p$-cell.
Here $\beta$ runs over the irreps at the $(p+1)$-cell, and $n_{\alpha}^{\beta}$ is determined by the characters $\chi_{\alpha}(g)$ ($\chi_{\beta}(g)$) of $g$ in the representation $\rho_\alpha(D^{p})$ ($\rho_\beta(D^{p+1})$), 
\begin{align}
n_{\alpha}^{\beta} = \frac{1}{|G_{D^{p+1}}|} \sum_{g \in G_{D^{p+1}}} \chi_{\beta}(g)^* \chi_{\alpha}(g).
\end{align}
This relation (\ref{eq:compatibility}), which is known as compatibility relation in band theory, defines a required map from $E_1^{p,0}$ to $E_1^{p+1,0}$.  
For the map to be a differential, it also needs to satisfy $d_1\circ d_1=0$, but this can be met by generalizing the compatibility relation slightly:
In the original compatibility relation, the coefficients $\{n_{\alpha}^\beta\}$ in Eq.(\ref{eq:compatibility}) are non-negative integers, but we assign the sign to the coefficients according to orientations of the $p$- and $(p+1)$-cells.
If the orientation of the $p$-cell is the same as that of the boundary of the adjacent $(p+1)$-cell, we retain the non-negative integers, but if not, we assign the minus sign to them. 
It can be shown that this simple modification leads to $d_1\circ d_1=0$.
In this sense, the first differential $d_1$ compactly encodes the compatibility relations for representations in cells with the additional information on orientations. 

\subsubsection{Higher differentials}

The second and higher pages $E_r$ ($r=2,3,\dots)$ are introduced as follows.
First, the $E_2$-page is an Abelian group given as the cohomology of $d_1$,
\begin{align}
E_2^{p,-n} := \ker(d_1^{p,-n})/\im(d_1^{p-1,-n}),
\end{align}
which is well-defined since $\im(d_1^{p-1,-n}) \subset \ker(d_1^{p,-n})$ due to $d_1\circ d_1=0$.~\footnote{This is interpreted as a twisted version of the Bredon equivariant cohomology.~\cite{Bre}}
For the $E_2$-page, the second differential $d_2$ is defined as a homomorphism from $E_2^{p,-n}$ to $E_2^{p+2, -(n+1)}$, i.e. $E_2^{p,-n} \xrightarrow{d_2^{p,-n}} E_2^{p+2,-(n+1)}$, but it holds that $d_2=0$ since $d_2$ changes the parity of the degree $n$ and we have $E_2^{p,-n}\subset E_1^{p,-n}=0$ for an odd $n$.
Next, the  $E_3$-page is defined as
\begin{align}
E_3^{p,-n} := \ker(d_2^{p,-n})/\im(d_2^{p-2,-(n-1)}),
\end{align}
which reduces to $E_2^{p,-n}$ since $d_2=0$, and the third differential $d_3$ is a homomorphism from $E_3^{p,-n}$ to $E_3^{p+3, -(n+2)}$, $E_3^{p,-n} \xrightarrow{d_3^{p,-n}} E_3^{p+3.-(n+2)}$.
Below, we show possible nontrivial parts of the $E_3$-page for $T^3$,
\begin{align}
\begin{array}{c|ccccc}
n=0 & E_3^{0,0} & E_3^{1,0} & E_3^{2,0} & E_3^{3,0} & 0 \\ 
n=1 & 0&0&0&0& 0\\ 
n=2 & E_3^{0,-2} & E_3^{1,-2} & E_3^{2,-2} & E_3^{3,-2} & 0 \\ 
\hline
E_3^{p,-n} & p=0 & p=1 & p=2 & p=3 & p=4\\
\end{array}, 
\end{align}
which implies that the only possibly nontrivial third differential is $d_3^{0,0}: E_3^{0,0} \to E_3^{3,-2}$.
Then,  the $E_4$ page is defined by 
\begin{align}
E_4^{p,-n}:= \ker(d_3^{p,-n})/\im(d_3^{p-3,-(n-2)}). 
\end{align}
The fourth differential $d_4$ is given as the homomorphism, $E_4^{p,-n}\xrightarrow{d_4^{p,-n}}E_4^{p+4,-(n+3)}$, but from the dimensional reason, $d_4$ is trivial for $T^3$.  
In a similar manner, for $r\ge 5$, the $E_r$-page is defined by
\begin{align}
E_r^{p,-n}:= \ker(d_{r-1}^{p,-n})/\im(d_{r-1}^{p-r+1,-(n-r+2)}), 
\end{align}
and the $r$-th differential $d_r$ is given as the homomorphism, $E_r^{p,-n}\xrightarrow{d_r^{p,-n}}E_r^{p+r,-(n+r-1)}$, but $d_r$ is also trivial for the same dimensional reason, which means that 
\begin{align}
E_4=E_5=E_6=\dots,
\end{align}
hence the $E_4$ page gives the limit $E_{\infty} = E_4$. 
All the higher pages are also Abelian groups.
As discussed in detail below, the higher pages and higher differentials also have their physical meanings, like the $E_1$-page and the first differential $d_1$.

\subsubsection{Limiting page $E_\infty$}

In mathematics, the limiting page $E_{\infty}$ approximates the $K$-group $K^{\tau-q}_G(T^3)$:
For spatial dimensions lower than or equal to $3$, the following short exact sequences hold true 
\begin{align}
0 \to E_{\infty}^{2,0}\to & K^{\tau-0}_G(T^3)\to E_{\infty}^{0,0}\to 0, \label{eq:exseq_k0}\\
0 \to E_{\infty}^{3,0} \to &K^{\tau-1}_G(T^3) \to E_{\infty}^{1,0}\to 0. \label{eq:exseq_k1}
\end{align}
In the present case, $E_{\infty}^{0,0}$ is found to be a free Abelian group, and thus Eq.(\ref{eq:exseq_k0}) splits.~\footnote{As an $R(G)$-module, the exact sequence (\ref{eq:exseq_k0}) does not split in general.~\cite{SSG17} 
The nontriviality of the extension of Eq. (\ref{eq:exseq_k0}) as an $R(G)$-module implies the data of reps.\ of $E_{\infty}^{0,0}$ is constrained by the $2d$ topological invariants $E_{\infty}^{2,0}$.} 
Therefore, we obtain
\begin{align}
K^{\tau-0}_G(T^3) \cong E_{\infty}^{2,0} \oplus E_{\infty}^{0,0}.
\end{align}
On the other hand, $E_{\infty}^{1,0}$ contains a torsion in general, so the extension of Eq.(\ref{eq:exseq_k1}) is not unique. 
The nontrivial extension of Eq. (\ref{eq:exseq_k1}) implies that the torsion part of $E_{\infty}^{1,0}$ is determined by $E_{\infty}^{3,0}$.

It should be noted that the $G$-symmetric filtration is not unique. 
A different choice of $G$-filtration leads to a different $E_1$-page, the first differential $d_1^{p,-n}$, and $\coker (d_1^{p,-n})$. 
On the other hand, as we will see soon later, we take the cohomology of the first differential $d_1$ to get the $E_2$-page, which means that choices of $G$-filtration do not matter to the $E_2$-page, provided that $G$-filtration is constructed in the manner in Sec.~\ref{sec:cell_deco}. 

In the following,  we sketch the physical implications of $E_r$ and $d_r$ in order. 

\subsection{Physical interpretation of $E_r$ and $d_r$}

This section explains the physical meaning of $E_r$-pages and the differentials $d_r$ in band theory.
Each issue is numbered as (i), (ii), and so on. 

\subsubsection{$E_1$-page}
\label{sec:complex_E_1-page}

As we mentioned above, $E_1^{p,0}$ represents the space of representations of band electrons at $p$-cells.
However, there is another interpretation, which follows from the following relation in the $K$-theory,
\begin{align}
E_1^{p,-n}&=K^{\tau-(n-p)}_{G}(X_p, X_{p-1})
\nonumber\\
&\cong\prod_{j \in I^p_{\rm orb}} K^{\tau|_{D^p_j}-(n-p)}_{G_{D_j^p}}(D_j^p, \partial D_j^p)
\nonumber\\
&\cong\prod_{j \in I^p_{\rm orb}} \wt K^{\tau|_{D^p_j}-(n-p)}_{G_{D_j^p}}(S_j^p),
\label{eq:EKKK}
\end{align}
where $S^p_j$ is a $p$-dimensional sphere (or called $p$-sphere) obtained from the $p$-cell $D_j^p$ by identifying its boundary $\partial D_j^p$ to a point, and $\wt K^{\tau|_{D^p_j}-(n-p)}_{G_{D_j^p}}(S_j^p)$ is the reduced $K$-theory. 
The $K$-group $\wt K_{G|_{D_j^p}}^{\tau|_{D_j^p}-(n-p)}(S_j^p)$ in Eq.(\ref{eq:EKKK}) specifies a class $(n-p)$ topological insulator on the $p$-sphere $S_j^p$ with additional point group symmetry $G_{D_j^p}$.
Since $G_{D_j^p}$ is a little group on $S_j^p$,  the topological insulator splits into irreps of $G_{D_j^p}$, each of which also belongs to class $(n-p)$.
$E_1^{p,-n}$ represents space of such $p$-dimensional topological insulators.  

Furthermore, the bulk-boundary correspondence leads to an interpretation of $E_1^{p,-n}$ as space of gapless states.
We can regard $\wt K_{G|_{D_j^p}}^{\tau|_{D_j^p}-(n-p)}(S_j^p)$ as the $K$-group for  gapless states on the boundary $S_j^p$ of  class $(n+1-p)$ topological insulators on $S_j^p\times S^1$, where the gapless states are representations of $G_j^p$.
This correspondence enables us to interpret $E_1^{p,-n}$ as space of $p$-dimensional gapless states in class $(n+1-p)$.  

\begin{figure*}[t]
	\begin{center}
	\includegraphics[width=\linewidth, trim=0cm 9cm 0cm 0cm]{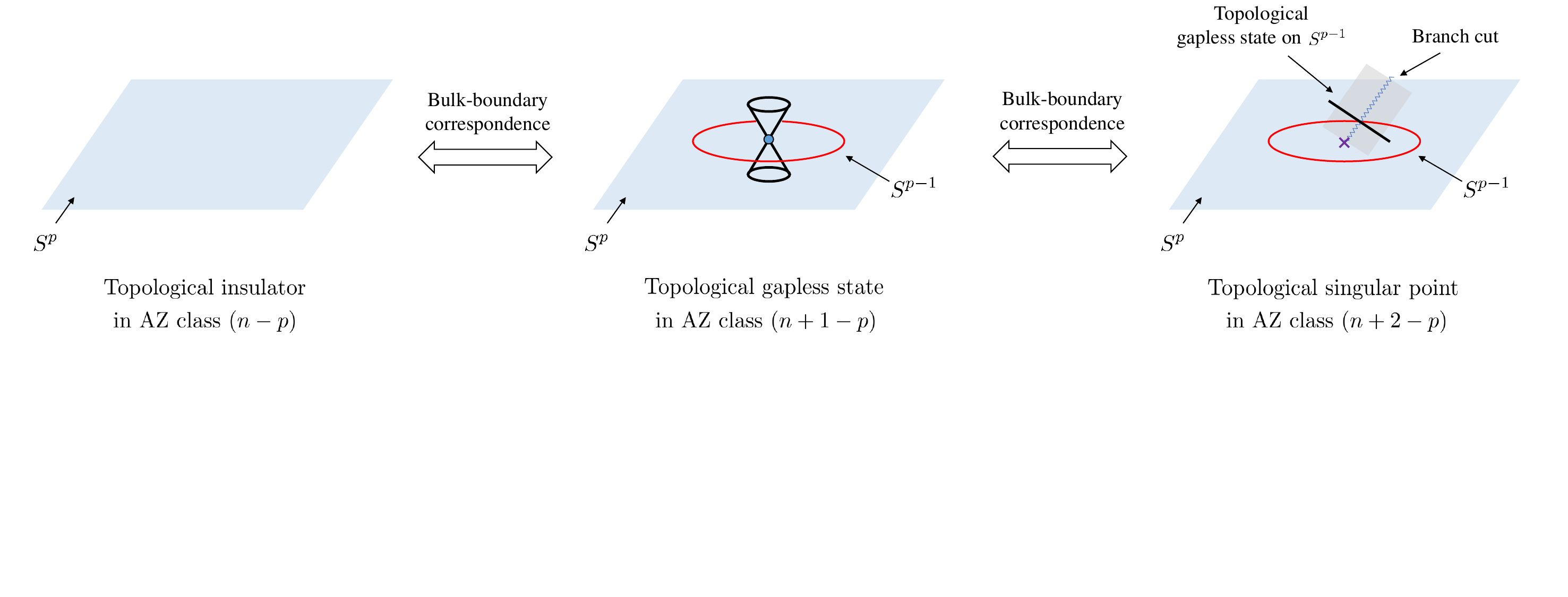}
	\end{center}
	\caption{
	The relationship among topological insulators, topological gapless states, and topological singular points over the $p$-sphere ($p\geq 1$). 
	These different interpretations of $E_1^{p,0}$ are related via the bulk-boundary correspondence. 
	}
	\label{fig:bulk_to_singular}
\end{figure*}

Finally, $E_1^{p,-n}$ also can be interpreted as space of singular points.
As mentioned above, $\wt K_{G|_{D_j^p}}^{\tau|_{D_j^p}-(n-p)}(S_j^p)$ in $E_1^{p,-n}$ describes topological gapless states in class $(n+1-p)$. 
For $p \geq 1$, an explicit topological invariant characterizing the gapless states is given by the isomorphism 
\begin{align}
\wt K_{G|_{D_j^p}}^{\tau|_{D_j^p}-(n-p)}(S_j^{p}) \cong \wt K_{G|_{D_j^p}}^{\tau|_{D_j^p}-(n+1-p)}(S^{p-1}), 
\label{eq:bbc2}
\end{align}
where $S^{p-1}$ is a $(p-1)$-sphere surrounding the gapless points. 
Since the system is gapful on $S_j^{p-1}$, the topological invariant of the gapless states is calculated as the topological invariant of topological insulators in the right-hand side of Eq.(\ref{eq:bbc2}). 
Applying the bulk-boundary correspondence again to $S^{p-1}$, the right-hand side of Eq.(\ref{eq:bbc2}) also represents class $(n+2-p)$ topological gapless states over the $(p-1)$-sphere $S^{p-1}$. 
The existence of a topological gapless state on $S^{p-1}$ implies that $S^{p-1}$ can not shrink to a point without a singularity. 
Therefore, we conclude that there must be a topological singular point in the original $p$-cell $D_j^p$,  which forms a branch cut with the gapless point on $S^{p-1}$. 
In Fig.~\ref{fig:bulk_to_singular}, we summarize these different interpretations of $E_1^{p,-n}$ and illustrate how they are related to each other by the bulk-boundary correspondence. 

Below, we describe the details of $E_1^{p,0}$ for each $p$.

\begin{figure}[!]
	\begin{center}
\includegraphics[width=\linewidth, trim=0cm 4cm 2cm 0cm]{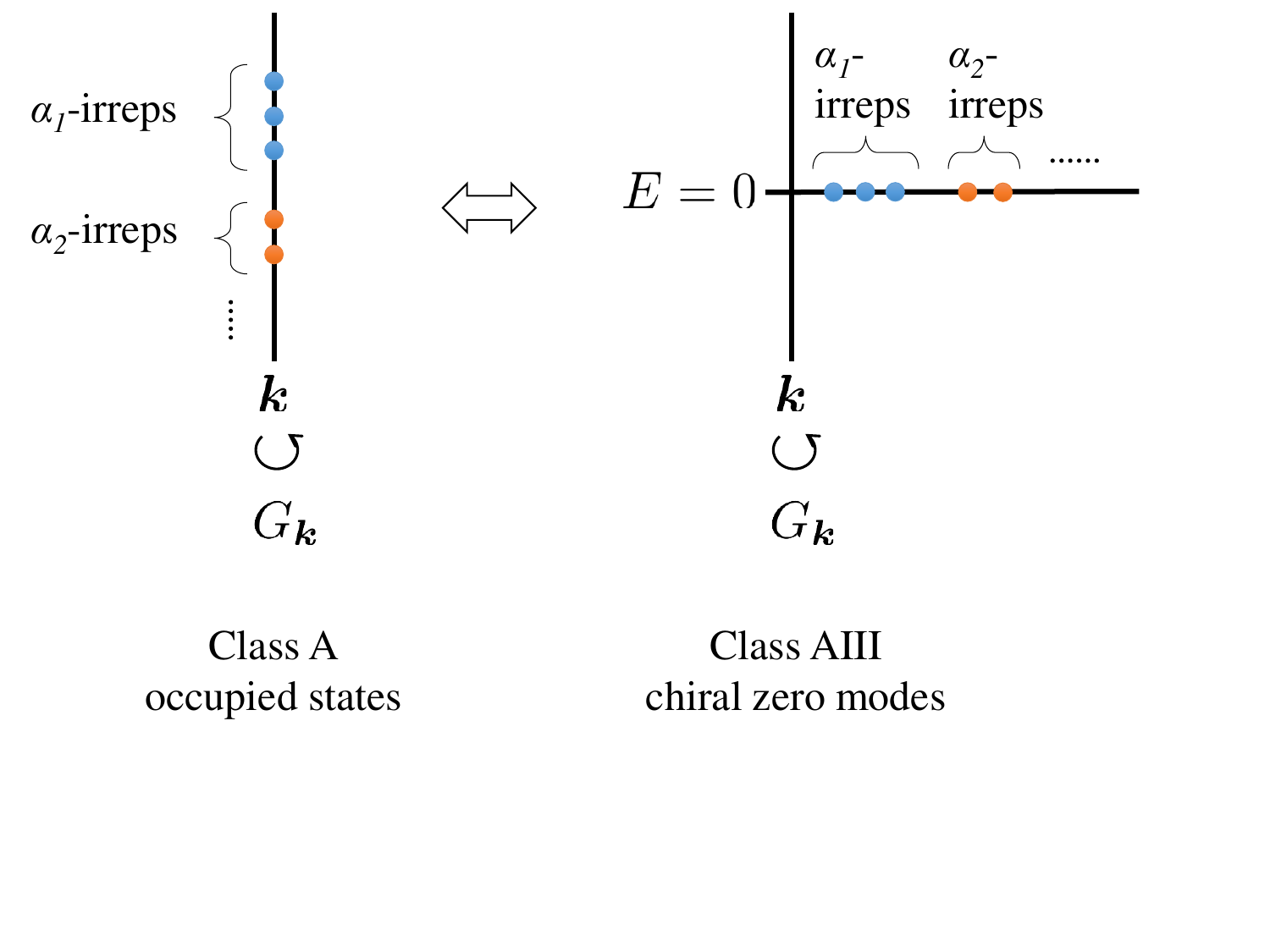}
	\end{center}
	\caption{Two interpretations of $E_1^{0,0}$.
	}
	\label{fig:E_1^0}
\end{figure}

\noindent
(i)---
$E_1^{0,0}$ gives a space of irreps in class A topological insulators on $0$-cells.
At the same time, it gives a space of irreps of class AIII zero mode on $0$-cells.
In the latter interpretation, an element $(n_{\rho_1(D^0_j)}, n_{\rho_2(D_j^0)}, \dots ) \in K^{\tau|_{D^0_j}-0}_{G_{D_j^0}}({D^0_j})$ represents the chirality of zero modes for each representation. 
More precisely, $n_{\rho_\alpha(D^0_j)}$ indicates the difference between the number of zero modes with positive chirality and that of negative one for the irrep  $\rho_\alpha(D^0_j)$. 
See Fig.~\ref{fig:E_1^0}

\begin{figure*}[!]
	\begin{center}
\includegraphics[width=\linewidth, trim=0.5cm 8cm 0.5cm 0cm]{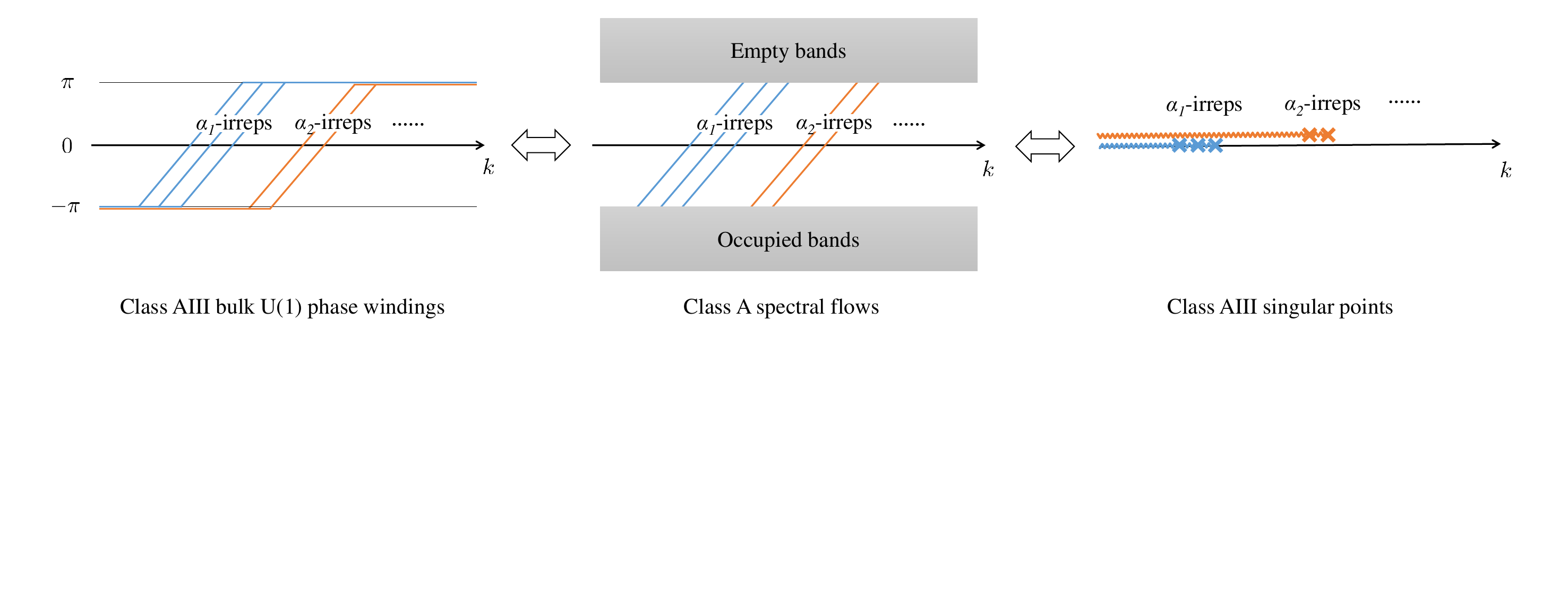}
	\end{center}
	\caption{Three interpretations of $E_1^{1,0}$.
	}
	\label{fig:E_1^1}
\end{figure*}

\noindent
(ii)---
As illustrated in Fig.\ref{fig:E_1^1}, $E_1^{1,0}$ has three interpretations: a) 1-dim class AIII topological insulators, b) 1-dim class A gapless states, and c) class AIII singular points on 1-cells.  
Correspondingly, an element $(n_{\rho_1(D^1_j)}, n_{\rho_2(D_j^1)}, \dots ) \in K^{\tau|_{D^1_j}-0}_{G_{D_j^1}}({D^1_j})$ indicates a set of a) 1-dim winding numbers for class AIII topological insulators,  b) spectral flows for 1-dim class A gapless states, and c) the numbers of  brunch cuts for class AIII singular points, all of which split into irreps $\{\rho_\alpha(D_j^1)\}_{\alpha=1,2,\dots}$ of $G_{D_j^1}$.

As explained above, these interpretations come from isomorphism in the $K$-theory, but we can also reproduce the same interpretations in terms of Hamiltonians.
Let us start with 1-dim class A gapless states (i.e., spectral flows) in  the irrep $\alpha$,
\begin{align}
\mbox{b)} \quad H_{\rm TGS}(k)=k {\bf 1}_{\alpha},
\label{eq:1dA}
\end{align}
where ${\bf 1}_{\alpha}$ is the identity matrix in the space of the irrep $\alpha$,  ${\bf 1}_{\alpha} = \sum_{i} \ket{\alpha,i} \bra{\alpha,i}$ with $\ket{\alpha,i}$ a basis of $\alpha$. 
Then, doubling the degrees of freedom and adding a mass term with UV cut-off into Eq.(\ref{eq:1dA}), we get a class AIII Hamiltonian 
\begin{align}
\mbox{a)}\quad 
H_{\rm TI}(k)=(k \sigma_x + (m-\epsilon k^2) \sigma_y) \otimes {\bf 1}_{\alpha},
\end{align}
with the chiral operator $\Gamma=\sigma_z$ .
We also have 1-dim class AIII topological singular points as
\begin{align}
\mbox{c)}\quad 
H_{\rm TSP}(k)=
\left\{\begin{array}{ll}
0 \times {\bf 1}_{\alpha}   & (\mbox{for $k<0$}), \\
\emptyset & (\mbox{for $k>0$}), \\
\end{array}\right.
\quad \Gamma={\bm 1}_\alpha,
\label{eq:def_1d_AIII_TSP}
\end{align}
where $\emptyset$ means the absence of states, so the $k=0$ point behaves as a singularity. 

\begin{figure*}[!]
	\begin{center}
\includegraphics[width=\linewidth, trim=0cm 10cm 0cm 0cm]{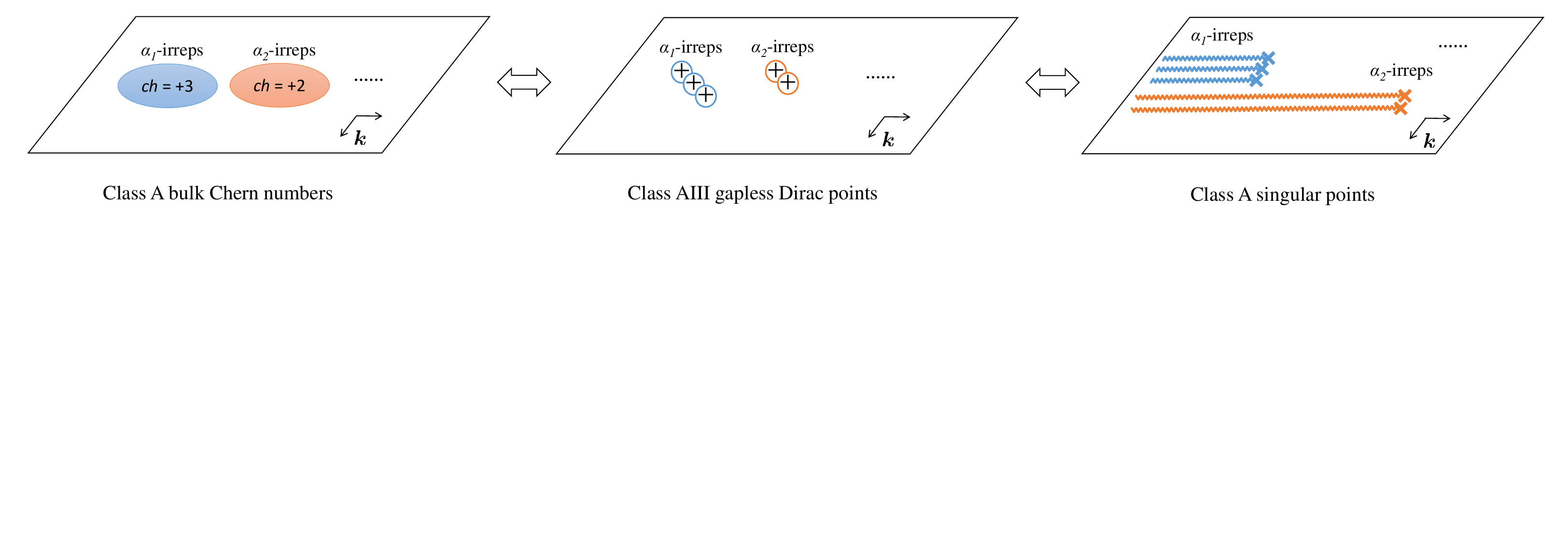}
	\end{center}
	\caption{Three interpretations of $E_1^{2,0}$.
	}
	\label{fig:E_1^2}
\end{figure*}

\noindent
(iii)---
Fig.~\ref{fig:E_1^2} illustrates the three interpretations for $E_1^{2,0}$:
The left is class A bulk Chern insulators on a 2-cell, the center is 2-dim class AIII topological gapless Dirac points, and the right is class A singular points on a 2-cell.
In the left interpretation, an element $(n_{\rho_1(D^2_j)}, _{\rho_2(D_j^2)}, \dots ) \in K^{\tau|_{D^2_j}-0}_{G_{D_j^2}}({D^2_j})$ represents the Chern numbers for irreps of $G_{D_j^2}$, then in the center, the same element specifies the topological numbers of class AIII Dirac points in the irreps. 
In the latter class AIII case, the topological number of a Dirac point is given by the 1-dim winding number on a circle enclosing the Dirac point: On the diagonal basis of the chiral operator,  the class AIII Hamiltonian in each irrep has an off-diagonal form
$H(\bk) = \begin{pmatrix}
0 & q(\bk) \\
q(\bk)^{\dag} & 0 \\
\end{pmatrix}$, where $q(\bk)$ has a vortex in the presence of a Dirac point.
The topological number of the Dirac point is nothing but the $U(1)$ phase winding of the vortex. 
Moreover, by applying the bulk-boundary correspondence to the circle surrounding the class AIII Dirac point, we get a class A gapless edge mode on the circle.  
Extending the gapless mode consistently in the entire region of a 2-cell, we have a singular point with a branch cut, as illustrated in the right of Fig.~\ref{fig:E_1^2}. 
Here the number of the branches in the $\alpha$-th irrep corresponds to $n_{\rho_\alpha(D_j^2)}$ in the above.

Again, these relations can be understood in terms of Hamiltonians. 
Gapless Dirac points with an irrep $\alpha$ in the center of Fig.\ref{fig:E_1^2} are described by 
\begin{align}
&H_{\rm TGS}(k_1,k_2)
= 
(k_1 \sigma_x + k_2 \sigma_y) \otimes {\bf 1}_{\alpha}, \label{eq:H_TGS(k_1,k_2)} \\
&\Gamma = \sigma_z. 
\end{align} 
Then, adding a mass term with UV cut-off to Eq.(\ref{eq:H_TGS(k_1,k_2)}), we have a class A Chern insulator in the left,
\begin{align}
H_{\rm TI}(k_1,k_2)
= 
(k_1 \sigma_x + k_2 \sigma_y + (m-\epsilon k^2) \sigma_z) \otimes {\bf 1}_{\alpha}.
\end{align}
Finally, from the the off-diagonal part $q(\bk)=k_1-ik_2$ of Eq.(\ref{eq:H_TGS(k_1,k_2)}), the Hamiltonian for the right of Fig.\ref{fig:E_1^2} is obtained as the imaginary part of logarithm of $q(k_1,k_2)$, $\Im \ln \left[ - q(k_1,k_2) \right]$, 
\begin{align}
H_{\rm TSP}(k_1,k_2)
= 
\Im \ln (-k_1+i k_2) \otimes {\bf 1}_{\alpha}.
\label{eq:Imk_1k_2}
\end{align}

\begin{figure*}[!]
	\begin{center}
\includegraphics[width=\linewidth, trim=0cm 10cm 0cm 0cm]{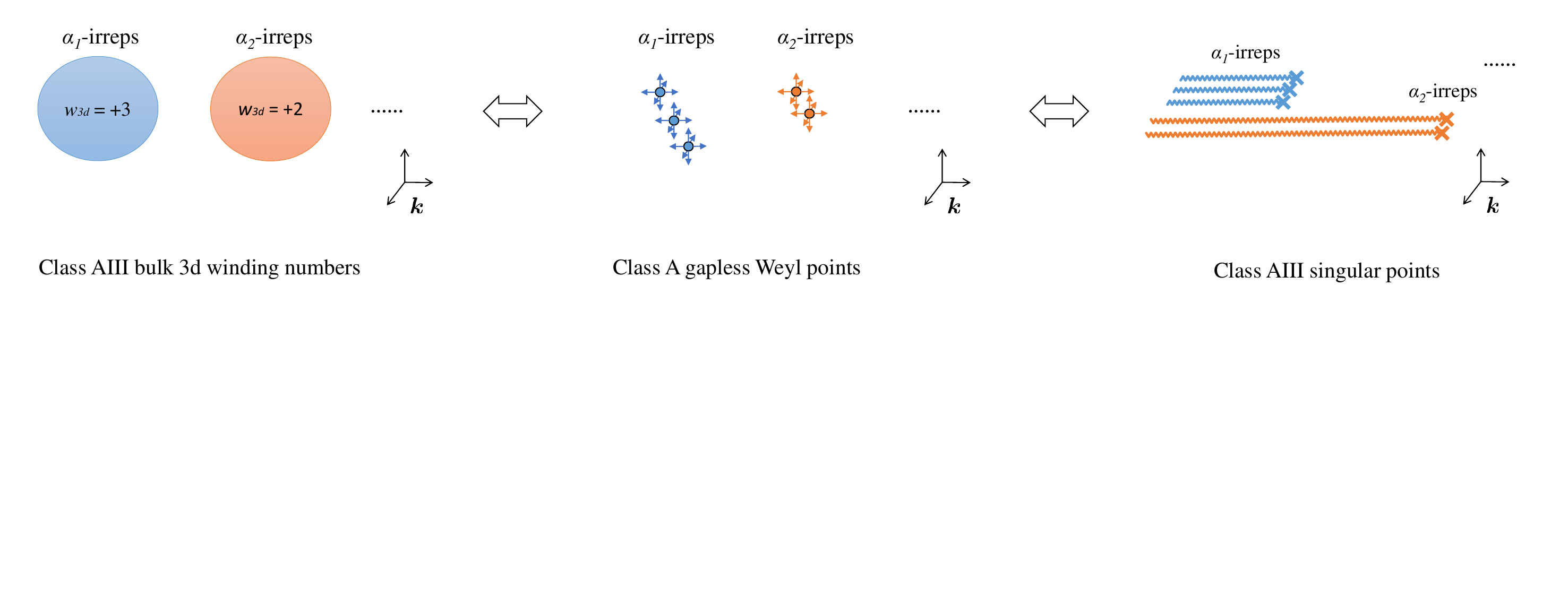}
	\end{center}
	\caption{Three interpretations of $E_1^{3,0}$.
	}
	\label{fig:E_1^3}
\end{figure*}

\noindent
(iv)---
The interpretations of $E_1^{3,0}$ are summarized in Fig.\ref{fig:E_1^3}.
An element $(n_{\rho_1(D^3_j)}, n_{\rho_2(D_j^3)}, \dots ) \in K^{\tau|_{D^3_j}-0}_{G_{D_j^3}}({D^3_j})$ represents a set of class AIII 3-dim winding numbers (left), class A Weyl charges (center), and class AIII 3-dim singular points (right) for irreps $\rho_\alpha(D^3_j)$.
In the Hamiltonian description, a class A Weyl point of an irrep $\alpha$ in the center is given by 
\begin{align}
H_{\rm TGS}(k_1,k_2,k_3)
= 
(k_1 \sigma_x + k_2 \sigma_y + k_3 \sigma_z) \otimes {\bf 1}_{\alpha}. 
\end{align} 
By doubling the degrees of freedom and adding a mass term, this Hamiltonian gives a class AIII Hamiltonian with a non-zero $3d$ winding number on the left of Fig.\ref{fig:E_1^3},
\begin{align}
&H_{\rm TI}(k_1,k_2,k_3)
= \big[ (k_1 \sigma_x + k_2 \sigma_y + k_3 \sigma_z) \otimes \sigma_x
\nonumber \\
&\qquad \qquad+ (m-\epsilon k^2) {\bf 1} \otimes \sigma_y \big] \otimes {\bf 1}_{\alpha}, \nonumber \\
&\Gamma = {\bf 1} \otimes \sigma_z.
\end{align}
Similarly, the class A topological singular point is described as 
\begin{align} 
&H_{\rm TSP}(k_1,k_2,k_3) = 
\Im \ln \left[ -k_1 + i (k_2 \sigma_x + k_3 \sigma_y) \right] \otimes {\bf 1}_{\alpha}, \nonumber \\
&\Gamma = \sigma_z,
\label{eq:H_TSP(k_1,k_2,k_3)}
\end{align}
where the branch cut extends from $\bk = \bm{0}$ to the negative region of the $k_1$-axis. 
One can see that around the branch cut the Hamiltonian (\ref{eq:H_TSP(k_1,k_2,k_3)}) is recast into that of a class AIII Dirac point, $H_{\rm TSP}(k_1<0,k_2,k_3) \sim k_2 \sigma_x + k_3 \sigma_y$, while around the positive region of the $k_1$-axis the Hamiltonian has a finite energy gap of $2 \pi$.

\subsubsection{First differential $d_1$}

\begin{figure}[!]
	\begin{center}
\includegraphics[width=\linewidth, trim=0cm 3cm 0cm 0cm]{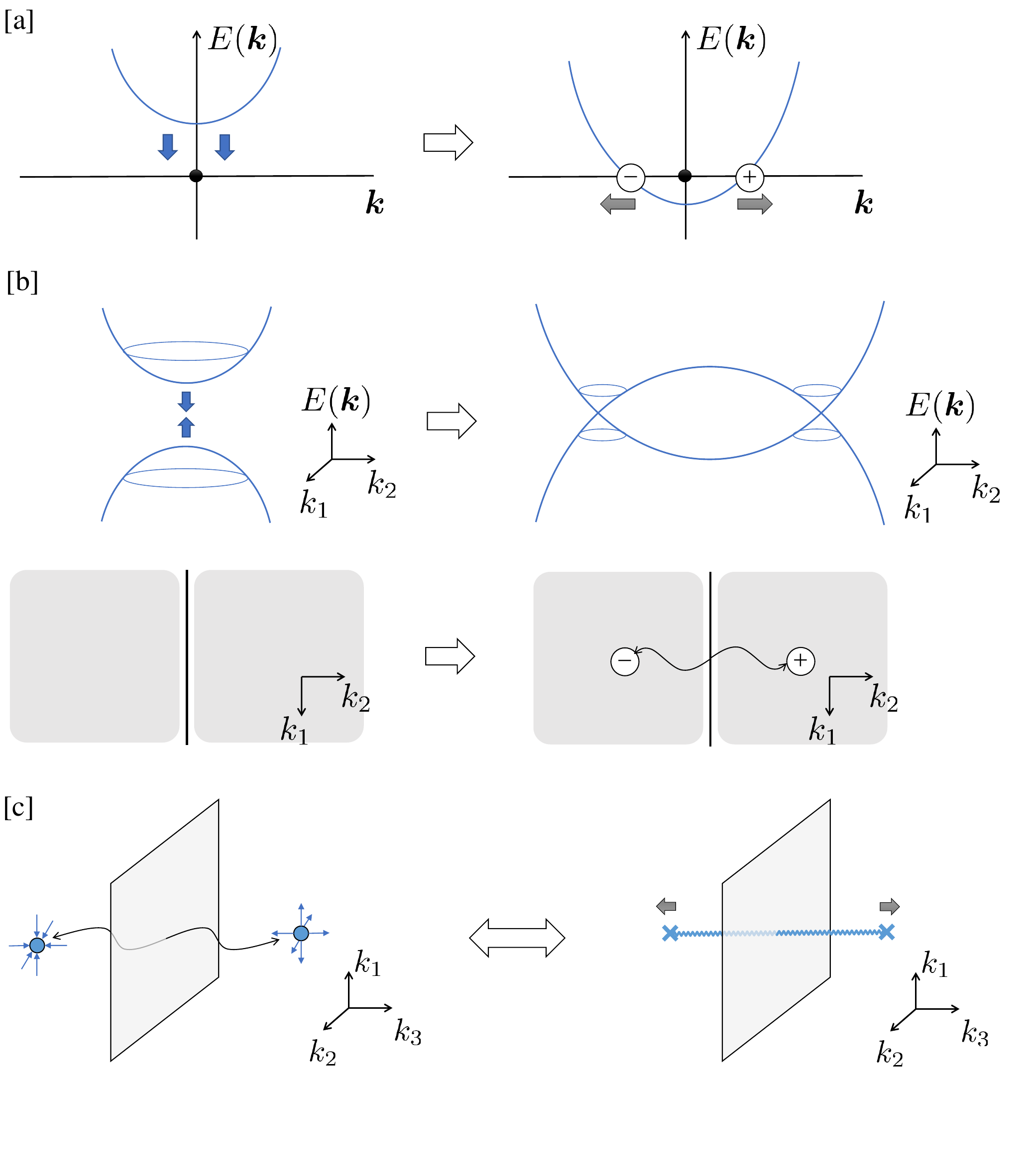}
	\end{center}
	\caption{
	[a]
	$d_1^{0,0}$ represents a class A topological phase transition at 0-cell creating a class A gapless state in adjacent 1-cells. 
	[b]
	$d_1^{1,0}$ represents the creation of class AIII Dirac points on 2-cells by a class AIII topological transition at a 1-cell. 
	[c]
	$d_1^{2,0}$ represents the creation of class A Weyl points in 3-cells by a class A topological phase transition on a 2-cell.
	}
	\label{fig:d1}
\end{figure}

The different interpretations of $E_1^{p,-n}$ above lead to different interpretations of  the first differential $d_1$:
As illustrated below, the first differential $d_1^{p,-n}$ relates class $(n-p)$ topological insulators on $p$-cells to class $(n-p)$ gapless states on their adjacent $(p+1)$-cells.
Moreover, for $p\geq 1$, the first differential $d_1^{p,-n}$ also relates class  $(n+1-p)$ topological gapless states on $p$-cells to class $(n+1-p)$ singular points on their adjacent $(p+1)$-cells.

\noindent
(v)---
$d_1^{0,0}: E_1^{0,0} \to E_1^{1,0}$ represents a class A topological phase transition at 0-cell creating a class A gapless state in adjacent 1-cells. 
This process can be modeled by the Hamiltonian 
\begin{align}
{\rm Class\ A}:\qquad  
H_{\rm A}(\bk) = (k^2-\mu) {\bf 1}_{\alpha},
\label{eq:ham_0to1}
\end{align}
with the change of the sign of $\mu$.
Indeed, by changing the sign of $\mu$ from negative to positive, an occupied state of the $\alpha$-th irrep is added to the 0-cell at $k=0$, and at the same time, gapless states appear on the adjacent 1-cells along $k>0$ and $k<0$. 
See Fig.~\ref{fig:d1} [a].

Like Eq.(\ref{eq:def_1d_AIII_TSP}), we can also obtain a class AIII Hamiltonian from the Hamiltonian (\ref{eq:ham_0to1}), 
\begin{align}
&{\rm Class\ AIII}:\nonumber \\
&H_{\rm AIII}(k)=
\left\{\begin{array}{ll}
0 \times {\bf 1}_{\alpha} & (k^2<\mu), \\
\emptyset & (k^2>\mu), \\
\end{array}\right.,
\quad \Gamma={\bm 1}_\alpha
\end{align}
which leads to an alternative interpretation of $d_1^{0,0}$:
When $\mu$ changes the sign from positive to negative, the above Hamiltonian hosts a class AIII zero mode at $k=0$, which is accompanied by singular points at $k=\pm \sqrt{\mu}$ with a branch cut between them.  
See the left figure in Fig.~\ref{fig:f1} [b1].
This means that $d_1^{0,0}$ also represents the creation of class AIII singularities with a branch cut on 1-cells by the creation of a class AIII zero mode on a 0-cell.

\noindent
(vi)---
$d_1^{1,0}: E_1^{1,0} \to E_1^{2,0}$ represents creation of class AIII Dirac points on 2-cells by class AIII topological transition at a 1-cell:
The Hamiltonian describing $d_1^{1,0}$ is 
\begin{align}
&{\rm Class\ AIII}:\nonumber \\
&H_{\rm AIII}(k_{\parallel},k_{\perp}) = \left[ (k_2^2-\mu) \sigma_x + k_1 \sigma_y \right] \otimes {\bf 1}_{\alpha}, \\
&\Gamma = \sigma_z, \nonumber
\label{eq:ham_1to2}
\end{align}
where $k_1$ ($k_2$) is the wave vector parallel (perpendicular) to the 1-cell.
When $\mu$ changes the sign from negative to positive, the class AIII winding number of the 1-cell on the $k_1$-axis jumps by 1, and there appear Dirac points at $(k_1, k_2)=(0,\pm \sqrt{\mu})$ in 2-cells, as illustrated Fig.~\ref{fig:d1} [b].

The first differential $d_1^{1,0}$ is also interpreted as creating class A singular points in 2-cells by creating a class A gapless point in 1-cell. 
In a manner similar to Eq.(\ref{eq:Imk_1k_2}),  the corresponding class A model Hamiltonian is derived from Eq. (\ref{eq:ham_1to2}) as
\begin{align}
{\rm Class\ A}:\quad
H_{\rm A}(k_1,k_2) = \Im \ln \left[ (k_2^2-\mu) + i k_1 \right] \otimes {\bf 1}_{\alpha}.
\end{align} 
See the right figure in Fig.~\ref{fig:f3} [b1].

\noindent
(vii)---
$d_1^{2,0}: E_1^{2,0} \to E_1^{3,0}$ represents creation of class A Weyl points in 3-cells by class A topological phase transition on a 2-cell.
The Hamiltonian describes this process is 
\begin{align}
&{\rm Class\ A}:\nonumber \\
&H_{\rm A}(\bk ) = \left[ (k_{3}^2-\mu) \sigma_x + k_{1} \sigma_y + k_{2} \sigma_z \right] \otimes {\bf 1}_{\alpha}, 
\label{eq:ham_2to3}
\end{align}
where $(k_1, k_2)$  ($k_3$) are parallel (is perpendicular) to the 2-cell. 
When $\mu$ passes zero, the Chern number on the 2-cell jumps by 1, and a pair of Weyl points is created in 3-cells.
See Fig.~\ref{fig:d1} [c]. 

Also, $d_1^{2,0}$ is interpreted as pair creation of class AIII singular points from 2-cells. 
See the right figure in the above.
The model Hamiltonian is 
\begin{align}
&{\rm Class\ AIII}:\nonumber \\
&H_{\rm AIII}(\bk) 
= 
\Im \ln \left[ (k_3^2-\mu) + i (k_{1} \sigma_x + k_{2} \sigma_y) \right] \otimes {\bf 1}_{\alpha}, \\ &\Gamma=\sigma_z.\nonumber
\end{align}

From the above interpretations of $d_1$, the image of $d_1^{p,-n}$ (denoted by $\im (d_1^{p,-n})$) gives a set of class $(n-p)$ gapless states (class $(n+1-p)$ singularities) on $(p+1)$-cells that are created by class $(n-p)$ topological phase transitions of topological insulators (gapless states) at adjacent $p$-cells. 
Therefore,  the complement $\coker(d_1^{p,-n})=E_1^{p+1,-n}/\im (d_1^{p,-n})$ has the following physical meanings.

\begin{figure}[!]
	\begin{center}
\includegraphics[width=\linewidth, trim=0cm 3cm 0cm 0cm]{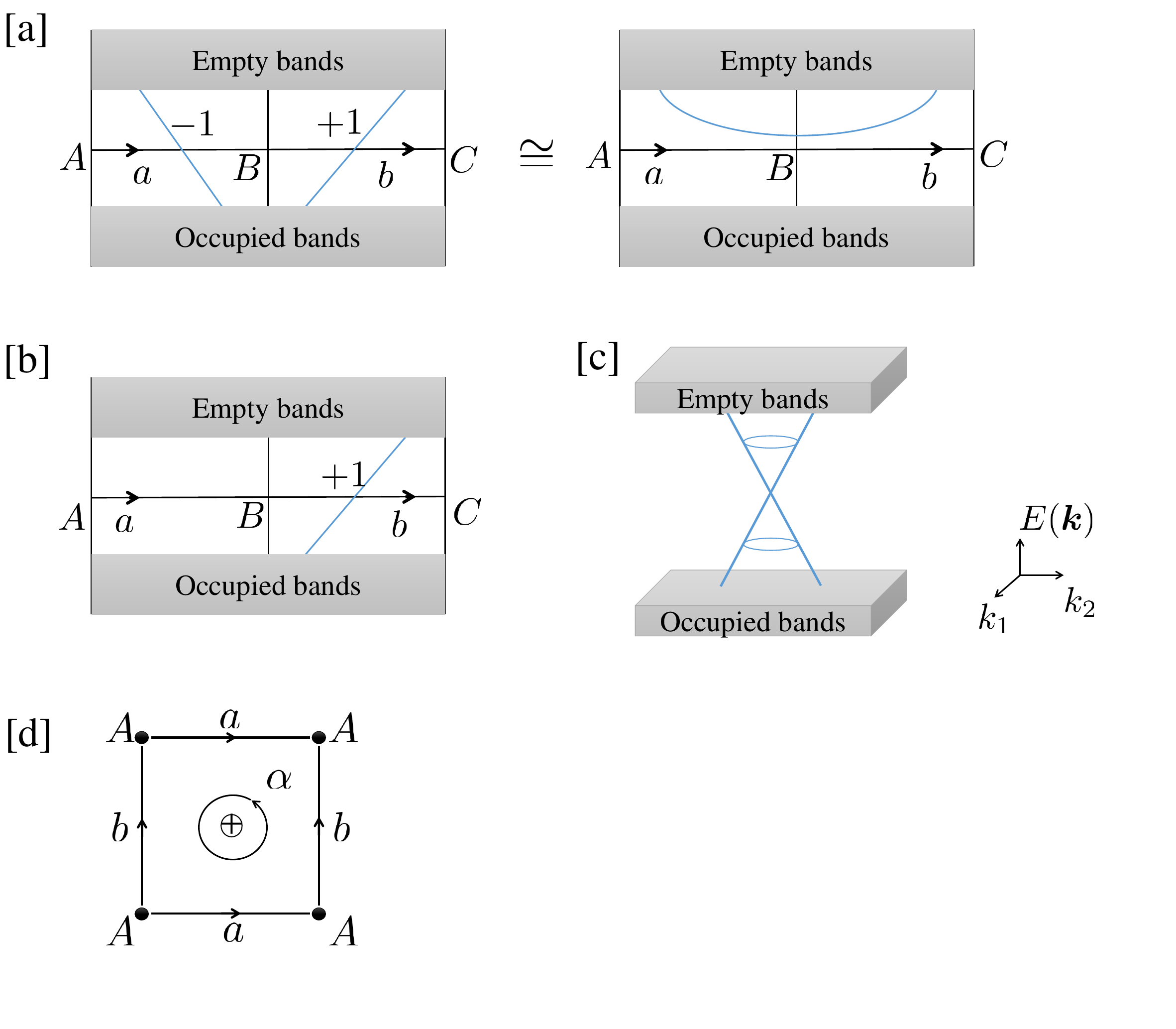}
	\end{center}
	\caption{
	[a]
	A pair annihilation of spectral flows. 
    [b] A genuine spectral flow that is not removed by a 0-cell. 
	[c]
	A single class AIII Dirac cone in $2d$, which is realized as a boundary state of the $3d$ class AIII topological insulator.
	[d]
	2-dimensional BZ torus $T^2$ with a single AIII Dirac point.  
	}
	\label{fig:coker_d1}
\end{figure}

\noindent
(viii)---
$\coker (d_1^{0,0}) = E_1^{1,0}/\im (d_1^{0,0})$ represents class A gapless states on 1-cells that can not be pair-annihilated at 0-cells. 
For example, consider class A gapless states on the 1-cells $a$ and $b$ as shown in the left of Fig.~\ref{fig:coker_d1} [a]. 
When they have opposite charges (i.e., spectral flows), the pair $(-1,+1)$ of the spectral flows can be trivialized. 
In this case, the pair $(-1,1)$ is nothing but the image of $d_1^{0,0}: 1 \mapsto (-1,1)$ from the 0-cell $B$ to 1-cells $a$ and $b$, resulting in no spectral flows as shown in the right of Fig.~\ref{fig:coker_d1} [a].
On the other hand, the spectral flow with the charge $(0,1)$ can not be trivialized. 
See Fig.~\ref{fig:coker_d1} [b].

Since no class A stable zero mode is possible at 0-cells, $\coker (d_1^{0,0})$ fully characterizes class A gapless modes on the whole 1-skeleton $X_1$. 
Thus, we have the relation, $\coker (d_1^{0,0}) = K^{\tau|_{X_1}-1}_G(X_1)$. 
The bulk-boundary correspondence also implies that $\coker(d_1^{0,0})$ gives the topological classification of class AIII gapped Hamiltonians over the 1-skeleton $X_1$. 

\noindent
(ix)---
$\coker (d_1^{1,0}) = E_1^{2,0}/\im (d_1^{1,0})$ represents class AIII Dirac points (or class A singularities) inside 2-cells which can not be pair-annihilated at 1-cells. 
A typical example is a single class AIII Dirac cone in $2d$ BZ, which is realized as a boundary state of the $3d$ class AIII topological insulator as illustrated in Fig.~\ref{fig:coker_d1} [c].
$\coker (d_1^{1,0})$ is the origin of $2d$ bulk class A topological invariants (such as the Chern number). 
To see this, as an example, let us consider the 2-torus $T^2$ with the cell decomposition composed by a 0-cell $\{ A \}$, 1-cells $\{ a,b\}$, and a 2-cell $\{\alpha\}$ as shown in Fig.~\ref{fig:coker_d1} [d].
We find that $\coker (d_1^{1,0}) = \Z$, and this is generated by a $U(1)$ phase winding of the transition function between patches of $T^2$, namely, the Chern number. 
Moreover, since the 0-dimensional topological invariants have no torsion in class A, $\coker (d_1^{1,0})$ coincides with the Abelian group structure of the $2d$ class A topological invariants defined over the 2-skeleton $X_2$.~\footnote{This is because the short exact sequence (\ref{eq:exseq_k0}) splits.}
We should note that, in general, the explicit definition of $2d$ class A topological invariants needs a correction from 0- and 1-cells in addition to the Berry curvature in 2-cells to make the topological invariant well-defined. 
The glide $\Z_2$ invariant~\cite{ChenGlide, SSG15}, the $\Z_2$ 1st Chern class on the real projective plane~\cite{Freed86,SSG17}, and the $\Z_2$ invariant appearing in the space group $F222$ introduced in Sec.~\ref{sec:f222} are such examples. 

\noindent
(x)---
$\coker(d_1^{3,0}) = E_1^{3,0}/\im (d_1^{2,0})$ represents class A Weyl points (class AIII singularities) inside 3-cells which can not be pair-annihilated at adjacent 2-cells. 
As discussed later,  the remaining Weyl points or singularities in $E_2^{3,0}$ may be pair-annihilated at 0-cells.  
See the issue (xv). 

\subsubsection{$E_2$-page}

Next, we discuss the meanings of the $E_2$-page in terms of band theory.
Since $E_2^{p,-n}$ is obtained from $\ker(d_1^{p,-n})$ by removing the trivial part of $\im(d_1^{p-1, -n})$, it specifies class $(n-p)$ topological insulators on $p$-cells that are consistently extended to nearby $(p+1)$-cells without gapless states.
At the same time, it also gives space of class $(n+1-p)$ gapless states on $p$-cells that are compatible with $(p+1)$-cells without class $(n+1-p)$ singularities.
In the latter interpretation, the gapless states on $p$-cells should not be annihilated at  $(p-1)$-cells because $E_2^{p,-n}\subset \coker (d_1^{p-1,-n})$.  

\noindent
(xi)---
$E_2^{0,0} = \ker(d_1^{0,0})$ represents a set of irreps at high-symmetry points which can be glued together on the 1-skeleton $X_1$ with keeping a gap.
Therefore,  $E_2^{0,0}$ is the class A $K$-group $K^{\tau|_{X_1}-0}_G(X_1)$ over the 1-skeleton $X_1$.
We note that $E_2^{0,0}$ for the 230 space groups reproduces $\Z^{d_{\rm BS}}$ in Ref.~\cite{Haruki230}.

\begin{figure*}[!]
	\begin{center}
\includegraphics[width=\linewidth, trim=0cm 0cm 0cm 0cm]{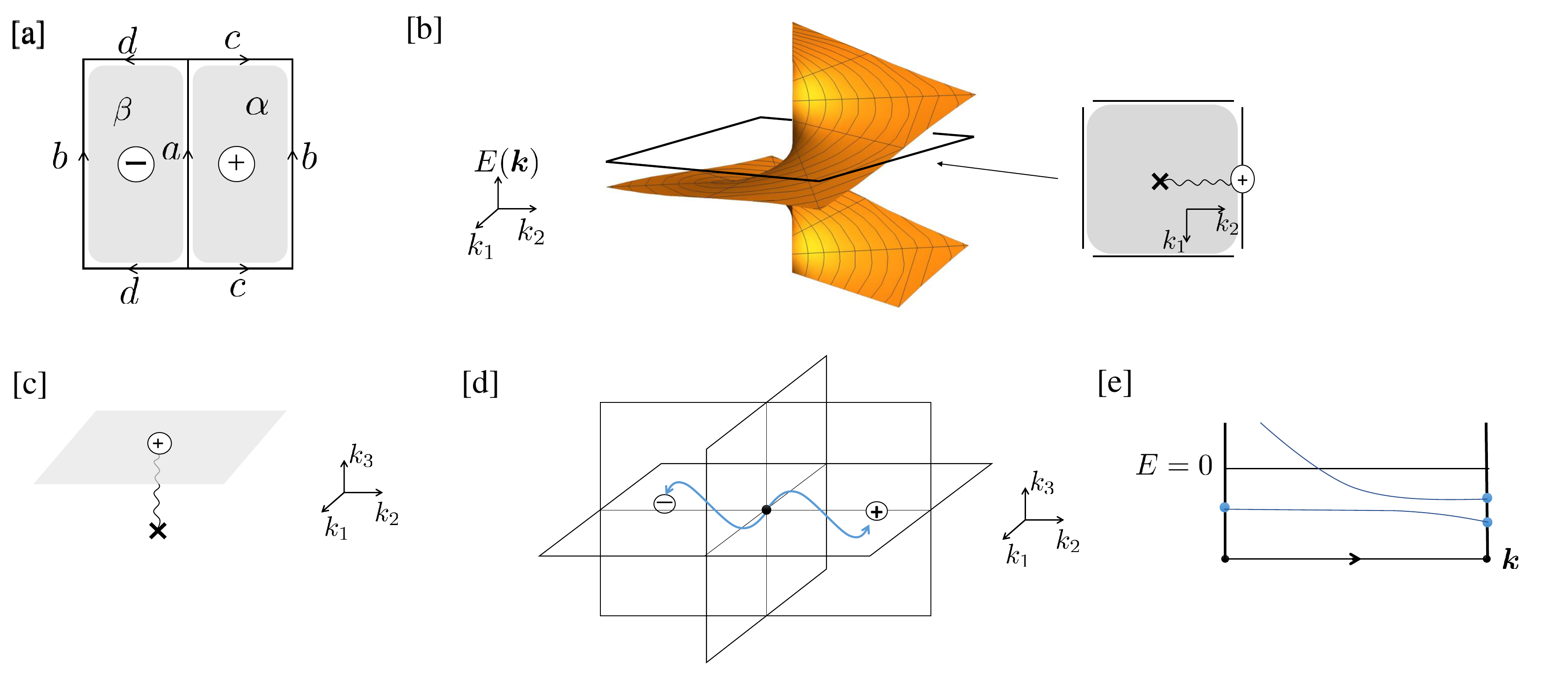}
	\end{center}
	\caption{
	[a]
	2-dimensional BZ $T^2$ with the  1-skeleton composed of 1-cells $a,b,c$, and $d$. 
	$\alpha$ and $\beta$ are 2-cells. 
	The circles with signs represent the class AIII Dirac points. 
	[b]
	A class A singular point inside a 2-cell. 
	The wavy line in the right figure represents the Fermi line ending at the singularity. 
	[c]
	A class AIII singular point inside a 3-cell. 
	The wavy line represents the Dirac nodal line ending at the singularity. 
	[d]
	Band inversion and creating a pair of Weyl points. 
	[e]
	The mismatch between the number of occupied bands results in a Fermi point. 
	}
	\label{fig:er}
\end{figure*}

\noindent
(xii)---
$E_2^{1,0}$ represents class AIII gapped Hamiltonians on 1-cells that can be consistently extended to 2-cells with keeping a gap.
Because there is no two-dimensional topological invariant in class AIII, such an extension is unique. 
Therefore, we also obtain $E_2^{1,0} = K^{\tau|_{X_2}-1}_G(X_2)$. 

As illustration, consider the 1-skeleton $X_1$ of $T^2(=X_2)$ composed by 1-cells $\{a,b,c,d\}$ (See Fig.~\ref{fig:er}) [a].
Here no space groups are assumed.
In this case, $\coker (d_1^{0,0})$ and $E_2^{1,0}$ are found to be $\Z^3$ and $\Z^2$, respectively. 
As explained in (viii), $\coker(d_1^{0,0})$ gives class AIII topological insulators on $X_1$.  
However, such a class AIII topological insulator allows a gapless point with a non-zero winding number on the closed loop $b-c-a+c$.  
On the other hand, such a gapless point is not allowed for $E_2^{1,0}$.
This difference gives the difference between $\coker (d_1^{0,0})$ and $E_2^{1,0}$.
   
$E_2^{1,0}$ also represents class A gapless states on 1-cells that are compatible with the presence of 2-cells. 
The compatibility with 2-cells, which comes from $\ker (d_1^{1,0})$, forbids a class A branch cut like the Fig.~\ref{fig:er} [b].

\noindent
(xiii)---
$E_2^{2,0}$ can be viewed as space of class A topological insulators on 2-cells which can extend to the whole 3-dimensional BZ without gapless states. 
$E_2^{2,0}$ also represents class AIII gapless Dirac points inside 2-cells without singularities in 3-cells. 
The compatibility with 3-cells comes from $\ker (d_1^{2,0})$ in the definition of $E_2^{2.0}$. In the latter interpretation, the compatibility forbids a class AIII gapless state terminated by a monopole singularity inside 3-cells. 
See Fig.~\ref{fig:er} [c].

\noindent
(xiv)---
$E_2^{3,0}$ coincides with $\coker(d_1^{3,0}) = E_1^{3,0}/\im (d_1^{2,0})$. See (x) for interpretation.

\subsubsection{Second differential $d_2$ and $E_3$-page}

The second differential $d_2^{p,-n}$ maps an element of $E_2^{p,-n}$ to that of $E_2^{p+2,-n-1}$. 
As explained above,  we can regard $E_2^{p,-n}$ as class $(n-p)$ insulators on $p$-cells that do not have gapless states on $(p+1)$-cells, and $E_2^{p+2,-n-1}$ as class $(n-p)$ gapless states on $(p+2)$-cells that do not host singularities on $(p+3)$-cells.
$d_2^{p,-n}$ relates such $E_2^{p,-n}$ insulators on $p$-cells to $E_2^{p+2, -n-1}$ gapless states on $(p+2)$-cells. 
In a manner similar to $d_1$, $\ker(d_2^{p,-n})$ gives $E_2^{p,-n}$ insulators on $p$-cells that can be extended to $(p+2)$-cells without gap closing.
In this sense, $d_2$ provides a ``2-dimensional compatibility relation'', which measures obstructions to extending the Bloch wave function to two-higher dimensional regions continuously. 
For $T^3$, we have $d_2^{p,-n}=0$ for $p\ge 2$.

From the meaning of $\ker(d_2^{p,-n})$, $E_3^{p,-n}$ represents the $E_2^{p,-n}$ insulators on $p$-cells  that can be extended to $(p+2)$-cells without gap closing. 
For $p\ge 1$, $E_3^{p,-n}$ also has another interpretation as gapless states as in the case of previous pages:
$E_3^{p,-n}$ represents the $E_2^{p,-n}$ gapless states on $p$-cells that can be extended continuously to $(p+2)$-cells without singularities.

For complex AZ classes, it holds that $d_2^{p,-n}=0$.  
The absence of obstruction by $d_2$ is understood as the absence of stable gapless lines (Weyl points) in class A (class AIII) systems. 
Since there is no obstruction by $d_2$,  the $E_3$-page reduces to the $E_2$-page in this case. 

\subsubsection{Third differential $d_3$ and $E_4$-page}

In a manner similar to the above, $d_3$ measures obstructions to extend the Bloch wave function to three-higher dimensional regions, and $E_4^{p,-n}$ represents the $E_3^{p,-n}$ insulators on $p$-cells that can be extended continuously to $(p+3)$-cells without gap closing. 
For $T^3$, from the dimensional reason,  $d_3^{p,-n}=0$ for $p\ge 1$.  In addition, we have $E_i^{p,1}=0$ in the present case, so only $d_3^{0,0}$ can be non-trivial.

\noindent
(xv)---
An element of $E_3^{0,0}$ is a set of irreps at 0-cells that can be continuously extended to the 2-skeleton $X_2$ without gap closing, and $d_3^{0,0}$ maps it to an element of $E_3^{3,0}$ which describes Weyl points in 3-cells.  
Therefore, a nontrivial third differential $d_3^{0,0}$ is identified with representation enforced Weyl semimetals discussed in Ref.\cite{TurnerInversion}.
Furthermore, by changing an element of $E_3^{0,0}$ by band inversion at $0$-cells, one can change the number of class A Weyl points in 3-cells. 
Therefore, $d_3^{0,0}$ also can be interpreted as the band inversion at 0-cells followed by pair-creation (or pair-annihilation) of class A Weyl points in 3-cells, which is illustrated in Fig.~\ref{fig:er} [d].
See Sec.~\ref{sec:pbar1} for the explicit Hamiltonian describing this process in the presence of inversion symmetry. 

We have the $E_4$ page by $E_4^{p,-n}:= \ker(d_3^{p,-n})/\im(d_3^{p-3,-(n-2)})$. 
The triviality of $d_3^{p,-n}$ for $p\ge 1$ implies that  $E_4^{1,-n}=E_3^{1,-n}$ and $E_4^{2,-n}=E_3^{2,-n}$.

\noindent
(xvi)---
$E_4^{0,0}$ is the space of class A representations at 0-cells which can be extended to the whole $3$-dimensional BZ without any gapless point. 

\noindent
(xvii)---
$E_4^{3,0}$ provides a subset of possible class AIII topological insulators on  $T^3$: 
As discussed above, an element of $E_1^{3,0}$ gives a set of class AIII topological insulators on $3$-cells, but those given by images of differentials become topologically trivial in the whole BZ. 
From the definition of $E_4^{3,0}$,
$E_4^{3,0} 
=\{[E_1^{3,0}/\im (d_1^{2,0})]/\im (d_2^{1,1})\}/\im (d_3^{0,2})$, 
such topologically trivial combinations of 3-cells are completely removed in $E_4^{3,0}$.
Therefore, $E_4^{3,0}$ gives class AIII topological insulators on $T^3$.
Note that $E_4^{3,0}$ does not fully characterize class AIII topological insulators on $T^3$ in general because it only contains topological information captured by 3-cells.
In the interpretation of gapless states, $E_4^{3,0}$ also represents class A Weyl points inside 3-cells that can not be trivialized.

The following coset spaces also have definite physical meanings. 

\noindent
(xviii)---
$E_1^{0,0}/E_2^{0,0} = K^{\tau|_{X_0}-0}_G(X_0)/K^{\tau|_{X_1}-0}_G(X_1)$ represents class A bulk gapless phases (i.e.\ metals) enforced by representations at 0-cells. 
This is because $E_1^{0,0}/E_2^{0,0}$ expresses the failure to glue irreps in $E_1^{0,0}$ along the whole 1-skeleton $X_1$. 
A combination of irreps at high-symmetry points belonging to $E_1^{0,0}/E_2^{0,0}$ implies the existence of a Fermi surface inside a 1-cell. 
See Fig.~\ref{fig:er} [e].

\noindent
(xix)---
$\coker (d_1^{0,0})/E_2^{1,0} = K^{\tau|_{X_1}-1}_G(X_1)/K^{\tau|_{X_2}-1}_G(X_2)$ represents class AIII bulk gapless phases enforced by topological invariants on the 1-skeleton $X_1$. 
Typically, such gapless phases have nodal lines in the 3-dimensional BZ. 
This is the class AIII analog of the representation enforced metal in class A defined in (xviii). 

\noindent
(xx)---
$\coker(d_1^{1,0}) / E_2^{2,0}$ represents Weyl semimetals enforced by $2$-dimensional class A topological invariants. 
A typical example is Weyl semimetals enforced by a mismatch of weak Chern numbers. 

\noindent
(xxi)---
$E_2^{0,0}/E_4^{0,0}$ represents class A Weyl semimetals enforced by representations at $0$-cells.

\subsection{Case studies}
\label{sec:exs}
In this section, we illustrate the computation of the AHSS for complex AZ classes. 
The complete list of the $E_{\infty}$-pages for all the 230 space groups is in Sec.~\ref{sec:e2}.  
We pick examples of torsion topological invariants which have been overlooked in the literature.

\subsubsection{Spinless $P222$}
\label{sec:16}
\begin{figure}[!]
	\begin{center}
	\includegraphics[width=0.7\linewidth, trim=5cm 14cm 5cm 0cm]{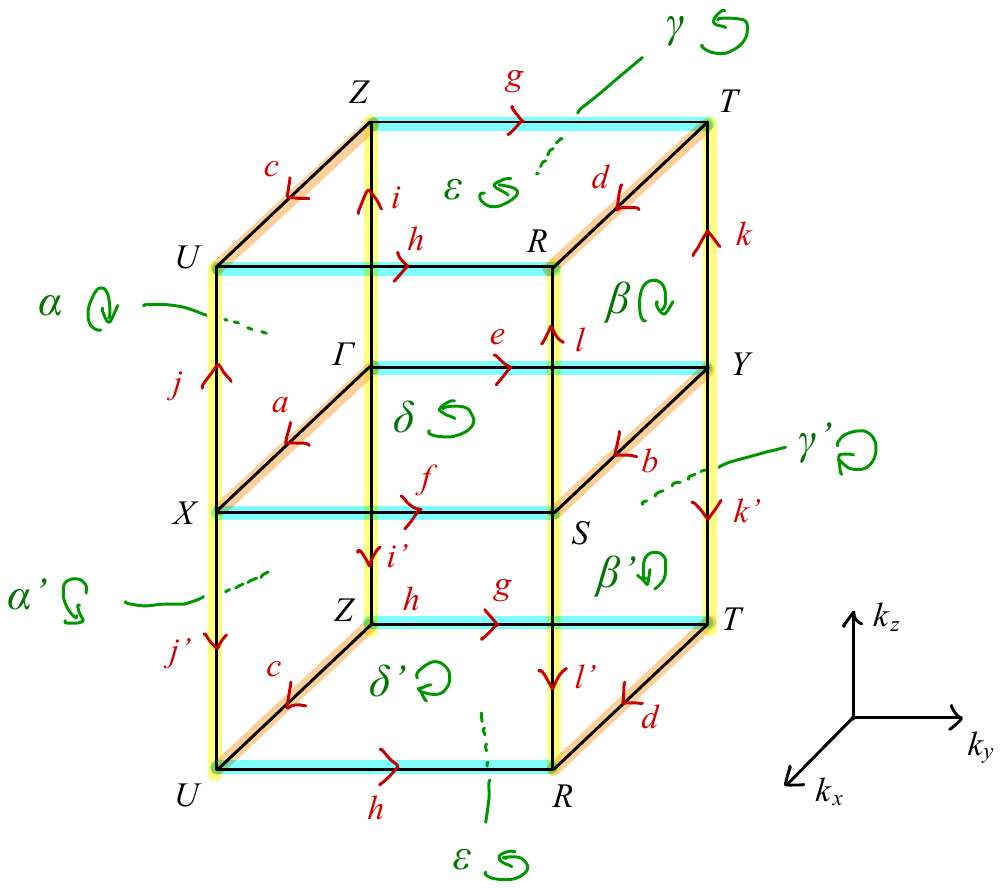}
	\end{center}
	\caption{A $D_2$-equivariant cell decomposition.}
	\label{fig:16}
\end{figure}
The first example is the space group P222 (No.$16$), which is symmorphic. 
The Bravais lattice is primitive, and the point group is $D_2 (\cong \Z_2 \times \Z_2)$, which is generated by two-fold rotations along $x$ and $y$ axes. 
We here consider spin integer electrons; namely, 2-fold rotations commute with each other. 
A $D_2$-equivariant cell decomposition is shown in Fig.~\ref{fig:16}. 
It is sufficient to draw an independent region in the BZ, a quarter of the whole BZ 3-torus. 
The $p$-cells $(p=0,1,2,3)$ are composed as 
\begin{align*}
&\mbox{0-cells} = \{\Gamma,X,Y,S,Z,U,T,R\}, \\
&\mbox{1-cells} = \{a,b,c,d,e,f,g,h,i,g,k,\ell\}, \\
&\mbox{2-cells} = \{\alpha,\beta,\gamma,\delta,\varepsilon\}, \\
&\mbox{3-cells} = \{ vol\ (\mbox{$\frac{1}{4}$BZ shown in Fig.~\ref{fig:16}} ) \}. 
\end{align*}
The little groups are $D_2$ itself on the $0$-cells, a $\Z_2$ subgroup on $1$-cells, and the trivial group on 2 and 3 cells. 
The $E_1$-pages, which are defined to be the space of irreps, are given as 
\begin{widetext}
\begin{align}
&E_1^{0,0}
=K_{D_2}^0(\mbox{0-cells})
=
\underbrace{\Z^4}_{\Gamma}
\oplus 
\underbrace{\Z^4}_{X}
\oplus 
\underbrace{\Z^4}_{Y}
\oplus 
\underbrace{\Z^4}_{S}
\oplus 
\underbrace{\Z^4}_{Z}
\oplus 
\underbrace{\Z^4}_{U}
\oplus 
\underbrace{\Z^4}_{T}
\oplus 
\underbrace{\Z^4}_{R}, \\
&E_1^{1,0}
=K_{D_2}^0(\mbox{1-cells})
=
\underbrace{\Z^2}_{a}
\oplus 
\underbrace{\Z^2}_{b}
\oplus 
\underbrace{\Z^2}_{c}
\oplus 
\underbrace{\Z^2}_{d}
\oplus 
\underbrace{\Z^2}_{e}
\oplus 
\underbrace{\Z^2}_{f}
\oplus 
\underbrace{\Z^2}_{g}
\oplus 
\underbrace{\Z^2}_{h}
\oplus 
\underbrace{\Z^2}_{i}
\oplus 
\underbrace{\Z^2}_{j}
\oplus 
\underbrace{\Z^2}_{k}
\oplus 
\underbrace{\Z^2}_{\ell}, \\
&E_1^{2,0}
=K_{D_2}^0(\mbox{2-cells})
=
\underbrace{\Z}_{\alpha}
\oplus 
\underbrace{\Z}_{\beta}
\oplus 
\underbrace{\Z}_{\gamma}
\oplus 
\underbrace{\Z}_{\delta}
\oplus 
\underbrace{\Z}_{\varepsilon}, \\
&E_1^{3,0}
=K_{D_2}^0(\mbox{3-cells})
=
\underbrace{\Z}_{vol}.
\end{align}
The first differential $d_1^{p,0}: E_1^{p,0}\to E_1^{p+1,0}$ is defined to be the compatibility relation. 
The matrices of the first differentials, 
\begin{align}
\Z^{32} \xrightarrow{d_1^{0,0}}
\Z^{24} \xrightarrow{d_1^{1,0}}
\Z^{5} \xrightarrow{d_1^{2,0}}
\Z, 
\end{align}
are given as follows. 
\begin{align}
d_1^{0,0}
= 
{\scriptsize
\begin{array}{|c@{}c@{}c@{}c@{}|c@{}c@{}c@{}c@{}|c@{}c@{}c@{}c@{}|c@{}c@{}c@{}c@{}|c@{}c@{}c@{}c@{}|c@{}c@{}c@{}c@{}|c@{}c@{}c@{}c@{}|c@{}c@{}c@{}c@{}|cc@{}}
\Gamma&&&&
X&&&&
Y&&&&
S&&&&
Z&&&&
U&&&&
T&&&&
R&&&\\
A&B_1&B_2&B_3&
A&B_1&B_2&B_3&
A&B_1&B_2&B_3&
A&B_1&B_2&B_3&
A&B_1&B_2&B_3&
A&B_1&B_2&B_3&
A&B_1&B_2&B_3&
A&B_1&B_2&B_3\\
\hline
1&0&0&1&-1&0&0&-1&&&&&&&&&&&&&&&&&&&&&&&&&A&a\\
0&1&1&0&0&-1&-1&0&&&&&&&&&&&&&&&&&&&&&&&&&B&\\
\hline
&&&&&&&&1&0&0&1&-1&0&0&-1&&&&&&&&&&&&&&&&&A&b\\
&&&&&&&&0&1&1&0&0&-1&-1&0&&&&&&&&&&&&&&&&&B&\\
\hline
&&&&&&&&&&&&&&&&1&0&0&1&-1&0&0&-1&&&&&&&&&A&c\\
&&&&&&&&&&&&&&&&0&1&1&0&0&-1&-1&0&&&&&&&&&B&\\
\hline
&&&&&&&&&&&&&&&&&&&&&&&&1&0&0&1&-1&0&0&-1&A&d\\
&&&&&&&&&&&&&&&&&&&&&&&&0&1&1&0&0&-1&-1&0&B&\\
\hline
1&0&1&0&&&&&-1&0&-1&0&&&&&&&&&&&&&&&&&&&&&A&e\\
0&1&0&1&&&&&0&-1&0&-1&&&&&&&&&&&&&&&&&&&&&B&\\
\hline
&&&&1&0&1&0&&&&&-1&0&-1&0&&&&&&&&&&&&&&&&&A&f\\
&&&&0&1&0&1&&&&&0&-1&0&-1&&&&&&&&&&&&&&&&&B&\\
\hline
&&&&&&&&&&&&&&&&1&0&1&0&&&&&-1&0&-1&0&&&&&A&g\\
&&&&&&&&&&&&&&&&0&1&0&1&&&&&0&-1&0&-1&&&&&B&\\
\hline
&&&&&&&&&&&&&&&&&&&&1&0&1&0&&&&&-1&0&-1&0&A&h\\
&&&&&&&&&&&&&&&&&&&&0&1&0&1&&&&&0&-1&0&-1&B&\\
\hline
1&1&0&0&&&&&&&&&&&&&-1&-1&0&0&&&&&&&&&&&&&A&i\\
0&0&1&1&&&&&&&&&&&&&0&0&-1&-1&&&&&&&&&&&&&B&\\
\hline
&&&&1&1&0&0&&&&&&&&&&&&&-1&-1&0&0&&&&&&&&&A&j\\
&&&&0&0&1&1&&&&&&&&&&&&&0&0&-1&-1&&&&&&&&&B&\\
\hline
&&&&&&&&1&1&0&0&&&&&&&&&&&&&-1&-1&0&0&&&&&A&k\\
&&&&&&&&0&0&1&1&&&&&&&&&&&&&0&0&-1&-1&&&&&B&\\
\hline
&&&&&&&&&&&&1&1&0&0&&&&&&&&&&&&&-1&-1&0&0&A&\ell\\
&&&&&&&&&&&&0&0&1&1&&&&&&&&&&&&&0&0&-1&-1&B&\\
\hline
\end{array}
}
\end{align}
\begin{align}
d_1^{1,0}
= 
\begin{array}{|c@{}c@{}|c@{}c@{}|c@{}c@{}|c@{}c@{}|c@{}c@{}|c@{}c@{}|c@{}c@{}|c@{}c@{}|c@{}c@{}|c@{}c@{}|c@{}c@{}|c@{}c@{}|c@{}c@{}}
a&&b&&c&&d&&e&&f&&g&&h&&i&&j&&k&&\ell&&\\
A&B&A&B&A&B&A&B&A&B&A&B&A&B&A&B&A&B&A&B&A&B&A&B&&\\
\hline
1&1&&&-1&-1&&&&&&&&&&&-1&-1&1&1&&&&&\alpha \\
\hline
&&1&1&&&-1&-1&&&&&&&&&&&&&-1&-1&1&1&\beta\\
\hline
&&&&&&&&1&1&&&-1&-1&&&-1&-1&&&1&1&&&\gamma\\
\hline
&&&&&&&&&&1&1&&&-1&-1&&&-1&-1&&&1&1&\delta\\
\hline
&&&&1&1&-1&-1&&&&&-1&-1&1&1&&&&&&&&&\varepsilon\\
\hline
\end{array}
\end{align}
\begin{align}
d_1^{2,0}
= 
\begin{array}{|c|c|c|c|c|c}
\alpha&\beta&\gamma&\delta&\varepsilon&\\
\hline
0&0&0&0&0&vol\\
\hline
\end{array}
\end{align}
\end{widetext}
Here, $\{A,B_1,B_2,B_3\}$ and $\{A,B\}$ represent irreps of $D_2$ and $\Z_2$ groups, respectively. 
It is straightforward to see that $d_1^{1,0}d_1^{0,0}=d_1^{2,0}d_1^{1,0}=0$. 
The $E_2$-page is found as 
\begin{align}
&E_2^{0,0}=\ker(d_1^{0,0})=\Z^{13}, \\
&E_2^{1,0}=\ker(d_1^{1,0})/\im(d_1^{0,0})=\Z_2, \\
&E_2^{2,0}=\ker(d_1^{2,0})/\im(d_1^{1,0})=0, \\
&E_2^{3,0}=\coker(d_1^{2,0})=\Z. 
\end{align}

Since $E_3^{3,0}=E_2^{3,0}=\Z$ is nonzero, the third differential $d_3^{0,2}:E_3^{0,2}\to E_3^{3,0}$ can be nontrivial. 
One can find that $d_3^{0,2}$ is trivial: 
We note that $E_2^{3,0} = E_1^{3,0}$ arises from the 3-cell $vol$ which has no symmetry left, which implies $E_2^{3,0} = \Z$ is the trivial irrep $A$ under $D_2$. 
On the other hand, as pointed out in Sec.~\ref{sec:ahss_com_for}, the third differential $d_3^{0,0}$ should accompany a band inversion between a pair of different irreps, which means the inverse image $(d_3^{0,0})^{-1}$ should be a nontrivial irrep under $D_2$, i.e.\ $B_1$, $B_2$ or $B_3$. 
However, we do not have a homomorphism from irreps $B_1$, $B_2$, and $B_3$ to the trivial irrep $A$. 
Therefore, $d_3^{0,0}$ is the zero map, and $E_2$-page is the limit $E_{\infty} = E_2$.

It should be noticed that $E_2^{1,0} = \Z_2$ means the appearance of a $\Z_2$ invariant $(-1)^{\nu} \in \{\pm 1\}$ defined on the 1-skeleton $X_1$ in class AIII, of which the construction is described in Sec.~\ref{sec:16z2}. 
We find that the $\Z_2$ invariant $(-1)^{\nu}$ is the parity of the half of the 3d winding number: 
The $K$-group of degree -1 fits into the exact sequence
\begin{align}
&0 \to \underbrace{\Z}_{E_2^{3,0}, w_{3d}=4}\to K^{-1}_{D_2}(T^3) \to \underbrace{\Z_2}_{E_2^{1,0},\nu\equiv 1} \to 0. 
\label{eq:exseq16aiii}
\end{align}
Here, $E_2^{3,0}=\Z$ is generated by a Hamiltonian with the 3d winding number $w_{3d}$ of $4$ since the fundamental region of BZ is $\frac{1}{4}BZ$ and each quarter contributes by 1 to the 3d winding number. 
On the one hand, the $K$-group $K^{-1}_{D_2}(T^3)$ is computed by the Clifford algebra~\cite{MorimotoClifford, SS14}, and we find that $K^{-1}_{D_2}(T^3) = \Z$, and $K^{-1}_{D_2}(T^3)$ is generated by a Hamiltonian with the 3d winding number $w_{3d} = 2$.~\footnote{
Let $H = k_x \gamma_x + k_y \gamma_y + k_z \gamma_z + M$ be the Dirac Hamiltonian, $\Gamma$ a chiral operator, and $U_x$ and $U_y$ the two-fold rotation operators along $x$ and $y$-axes, respectively. 
The classification of symmetry-respecting mass $M$ is equivalent to the classification of the extension of the complex Clifford algebra $\{\gamma_x,\gamma_y,\gamma_z,\Gamma\} \otimes Cl_2 \to \{\gamma_x,\gamma_y,\gamma_z,\Gamma,M\} \otimes Cl_2$, where $Cl_2$ is the complex Clifford algebra generated by $\{U_x\gamma_y\gamma_z, U_y\gamma_x\gamma_z\}$. 
The classification is recast as that for $3d$ class AIII without symmetry, i.e.\ $\Z$. 
Also, due to the commuting algebra $Cl_2$, the $3d$ winding number $w_{3d}$ should be an even integer. 
}
This implies that the extension (\ref{eq:exseq16aiii}) is nontrivial. 
Otherwise, the $K$ group $K^{-1}_{D_2}(T^3)$ becomes $\Z \oplus \Z_2$. 
Therefore, there is the relation 
\begin{align}
    (-1)^{\nu} = (-1)^{w_{3d}/2}.
    \label{eq:rel_nu_w3d}
\end{align}
The $\Z_2$ invariant $\nu$ is an analog of the symmetry indicator: 
If a band structure in class AIII is fully gapped and the 3d winding number is zero, then $\nu \equiv 0$. 
Therefore, if $\nu \equiv 1$ then the system is deformable to a gapped band structure with a finite 3d winding number $w_{3d} \equiv 2 \ ({\rm mod\ }4)$ without closing a gap on the 1-skeleton. 

\paragraph{Construction of $\Z_2$ invariant}
\label{sec:16z2}
Let $q(\bk)$ be the off-diagonal part of the Hamiltonian $H(\bk) = \begin{pmatrix}
0 & q(\bk) \\
q(\bk)^{\dag} & 0 \\
\end{pmatrix}$ in the basis so that the chiral operator is $\Gamma = \begin{pmatrix}
1 & 0 \\
0 & -1 
\end{pmatrix}$.
WLOG, $q(\bk)$ is assumed to be a unitary matrix. 
The $D_2$ symmetry matrices in spinless systems are written as 
\begin{align}
\left\{\begin{array}{ll}
U_x(\bk)q(\bk)=q(C_{2x}\bk)U_x(\bk), \\ U_x(C_{2x}\bk)U_x(\bk)=1, \\
U_y(\bk)q(\bk)=q(C_{2y}\bk)U_y(\bk), \\ U_y(C_{2y}\bk)U_y(\bk)=1, \\
U_x(C_{2y}\bk)U_y(\bk)=U_y(C_{2x}\bk)U_x(\bk).
\end{array}\right.
\end{align}
On the symmetric lines (1-cells), the matrix $q(\bk)$ becomes a block-diagonal form as 
\begin{align}
q(\bk)=\begin{pmatrix}
q_A(\bk) &  \\
& q_B(\bk)\\
\end{pmatrix}, \qquad  (\bk \in \mbox{1-cells}), 
\end{align}
according to the symmetry of the little group $G_{\bk}$. 
In the same way, at the high-symmetry points (0-cells), the matrix $q(\bk)$ is decomposed as 
\begin{align}
&q(\bk) = \begin{pmatrix}
q_A(\bk) &&&\\
&q_{B_1}(\bk)&&\\
&&q_{B_2}(\bk)&\\
&&&q_{B_3}(\bk)\\
\end{pmatrix}, \nonumber \\
&(\bk \in \mbox{0-cells}).
\end{align}
Let us focus on the determinant of the matrix $q(\bk)$ within the subsectors. 
We denote $e^{i \theta_{A/B}(\bk)} = \det q_{A/B}(\bk)$ for $\bk \in $1-cells and $e^{i \phi_{A/B_i}(\bk)} = \det q_{A/B_i}(\bk)$ $(i=1,2,3)$ for $\bk \in$ 0-cells. 
The origin of the $\Z_2$ invariant is the fact that the $U(1)$ phases $e^{i \theta_{A/B} (\bk)}$ on 1-cells do not fully determine the $U(1)$ phases $e^{i \phi_{A/B_i}(\bk)}$ at 0-cells: 
For instance, around the $\Gamma$ point, the compatibility relation reads 
\begin{align}
\left\{\begin{array}{ll}
e^{i \theta_B(\bk \in a)}|_{\bk \to \Gamma} = e^{i (\phi_{B_1}(\Gamma) + \phi_{B_2}(\Gamma))}, \\
e^{i \theta_B(\bk \in e)}|_{\bk \to \Gamma} = e^{i (\phi_{B_1}(\Gamma) + \phi_{B_3}(\Gamma))}, \\
e^{i \theta_B(\bk \in i)}|_{\bk \to \Gamma} = e^{i (\phi_{B_2}(\Gamma) + \phi_{B_3}(\Gamma))}, 
\end{array}\right.
\end{align}
from which we have 
\begin{align}
\left\{\begin{array}{ll}
e^{2 i \phi_{B_1}(\Gamma)} = e^{i \{ \theta_B(\bk \in a) +\theta_B(\bk \in e) - \theta_B(\bk \in i)\}}|_{\bk \to \Gamma}, \\
e^{2 i \phi_{B_2}(\Gamma)} = e^{i \{ \theta_B(\bk \in a) -\theta_B(\bk \in e) + \theta_B(\bk \in i)\}}|_{\bk \to \Gamma}, \\
e^{2 i \phi_{B_3}(\Gamma)} = e^{i \{ -\theta_B(\bk \in a) +\theta_B(\bk \in e) + \theta_B(\bk \in i)\}}|_{\bk \to \Gamma}, 
\end{array}\right.
\label{eq:16u1rel}
\end{align}
where $a,e,i$ indicate $1$-cells shown in Fig.~\ref{fig:16}. 
Thus, $U(1)$ phases $\phi_{B_i}(\Gamma)$ are fixed by the $U(1)$ phases on 1-cells up to a $\pi$-phase.
Similar relations hold true for other $0$-cells. 
From these $\Z_2$ ambiguities, one can define a $\Z_2$ invariant 
\begin{align}
&(-1)^{\nu} \nonumber\\
&:= \exp\Big[\frac{i}{2}\int_{a-b-c+d-e+f+g-h-i+j+k-\ell} d \theta_B(\bk) \nonumber\\
&+i \sum_{\bk \in \Gamma,S,U,T} \phi_{B_3}(\bk)-i\sum_{\bk \in X,Y,Z,R} \phi_{B_3}(\bk)
\Big]. 
\end{align}
The constraint (\ref{eq:16u1rel}) (and same ones for other 0-cells) leads to the $\Z_2$ quantization $\{ (-1)^{\nu} \}^2 = 1$. 
We have checked that a model Hamiltonian with 3d winding number $w_{3d}=2$ gives $(-1)^{\nu}=-1$.~\footnote{
From the Clifford algebra, a model Hamiltonian is given by 
$H=\sin k_x \sigma_x \tau_x \mu_0+ \sin k_y \sigma_y \tau_x \mu_0+ \sin k_z \sigma_z \tau_x \mu_0+ (m+\cos k_x+\cos k_y+\cos k_z) \sigma_0 \tau_y \mu_0$, 
$\Gamma = \sigma_0 \tau_z \mu_0$, $U_x=\sigma_x \tau_0 \mu_x, U_y = \sigma_y \tau_0 \mu_y$, where $\sigma_{i}, \tau_i, \mu_i (i=0,x,y,z)$ are Pauli matrices. 
}

\subsubsection{$F222$}
\label{sec:f222}
\begin{figure}[!]
	\begin{center}
	\includegraphics[width=0.6\linewidth, trim=5cm 15cm 4cm 0cm]{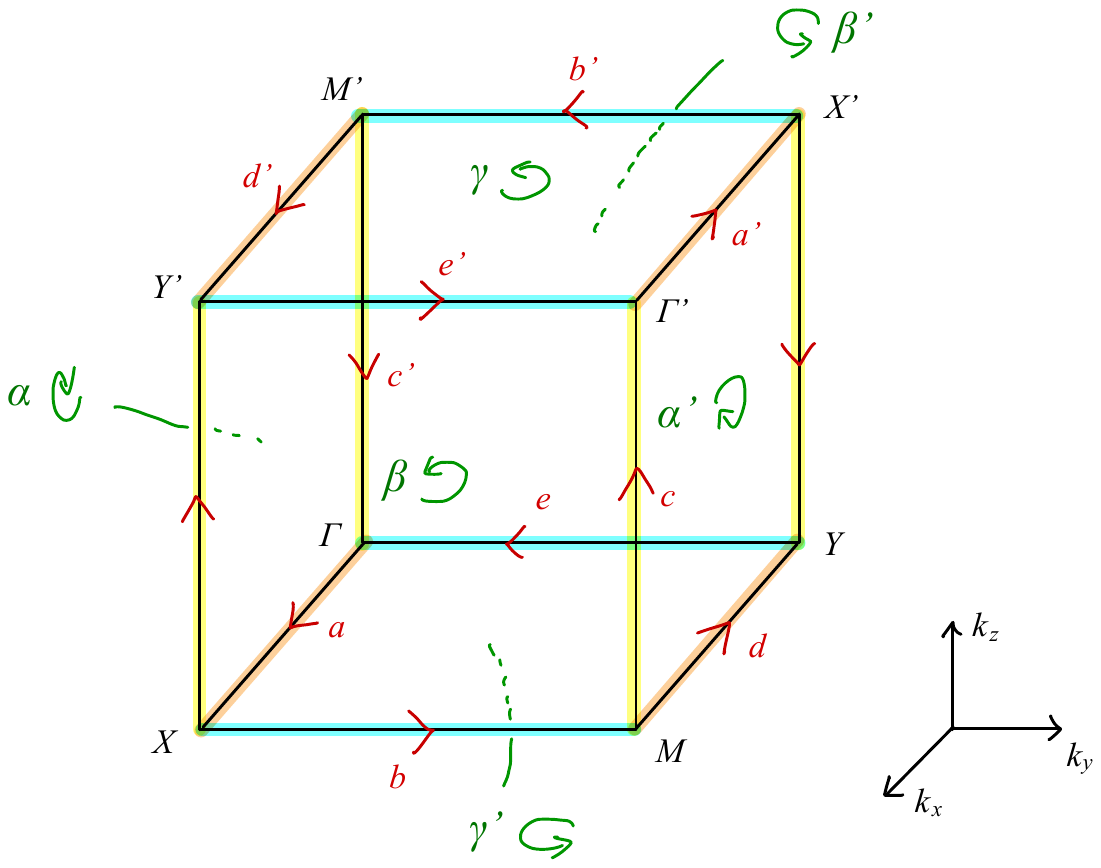}
	\end{center}
	\caption{A $D_2$-equivariant cell decomposition of BZ for the space group $F222$. The figure shows the a quarter of BZ. 
	}
	\label{fig:22}
\end{figure}
The second example is the space group $F222$ (No.\ $22$, symmorphic).
The Bravais lattice is face-centered and the reciprocal lattice vectors (in the unit of $2 \pi$) are $\bm{b}_1=(-1,1,1), \bm{b}_2=(1,-1,1)$ and $\bm{b}_3=(1,1,-1)$. 
The point group is $D_2$ again and is generated by $C_{2x}$ and $C_{2y}$. 
A $D_2$-equivariant cell decomposition is shown in Fig.~\ref{fig:22}. 
The list of $p$-cells ($p=0,1,2,3$) is as follows. 
\begin{align*}
&\mbox{0-cells} = \{\Gamma = (0,0,0), X=(1,0,0), \\
&\qquad Y=(0,1,0), M=(1,1,0)\}, \\
&\mbox{1-cells} = \{a,b,c,d,e,f\}, \\
&\mbox{2-cells} = \{\alpha,\beta,\gamma\}, \\
&\mbox{3-cells} = \{ vol\ (\mbox{$\frac{1}{4}$BZ shown in Fig.~\ref{fig:22}} ) \}. 
\end{align*}
The 2-cells $\alpha', \beta'$ and $\gamma'$ are equivalent to $\alpha$, $\beta$ and $\gamma$, respectively: 
$\alpha' = C_{2y} \alpha + \bm{b}_1+\bm{b}_2+\bm{b}_3$, 
$\beta' = C_{2x} \beta + \bm{b}_1$ and 
$\gamma' = C_{2z}\gamma+\bm{b}_3$. 
The similar equivalence relations hold true for 1- and 0-cells. 
An interesting feature is found in $E_2^{2,0}$, the $2d$ class A topological invariant. 
For both factor systems of spinful and spinless electrons, the first differentials $d_1^{1,0}$ and $d_1^{2,0}$ are given by 
\begin{align}
&d_1^{1,0}
= 
\begin{array}{|c@{}c@{}|c@{}c@{}|c@{}c@{}|c@{}c@{}|c@{}c@{}|c@{}c@{}|c@{}}
a&&b&&c&&d&&e&&f&\\
A&B&A&B&A&B&A&B&A&B&A&B&\\
\hline
1&1&&&1&1&-1&-1&&&1&1&\alpha\\
\hline
&&1&1&1&1&&&-1&-1&-1&-1&\beta\\
\hline
1&1&1&1&&&1&1&1&1&&&\gamma\\
\hline
\end{array}, \\
&d_1^{2,0}
= 
\begin{array}{|c|c|c|c}
\alpha&\beta&\gamma&\\
\hline
0&0&0&vol\\
\hline
\end{array}, 
\end{align}
where $A$ ($B$) represents the trivial (nontrivial) irrep of $\Z_2$. 
It is found that $E_2^{2,0}=\ker(d_1^{2,0})/\im(d_1^{1,0}) = \Z_2$, which means that the existence of a $\Z_2$-valued class A topological invariant defined on the 2-skeleton $X_2$. 

The appearance of the $\Z_2$ invariant is understood from that the 2-dimensional boundary $\p(\frac{1}{4} BZ)$ of the quarter of BZ has the same structure as the real projective plane $RP^2$ owing to the identification of the 2-cells. 
Similar to the formula for the torsion part of the first Chern class $c_1 \in H^2(RP^2,\Z)=\Z_2$~\cite{Freed86, SSG17}, the $\Z_2$ invariant for $F222$ is defined by
\begin{align}
(-1)^{\nu}
:= 
\exp \left[ \int_{a+b+c} \tr {\cal A} - \frac{1}{2} \int_{\alpha+\beta+\gamma} \tr {\cal F} \right], 
\end{align}
with ${\cal A}$ and ${\cal F}$ the Berry connection and curvature, respectively. 
Notice that the path $a+b+c$ is closed since $\Gamma'$ is equivalent to $\Gamma$. 
The $\Z_2$-quantization follows from the Stokes' theorem applied to the boundary $\p (\alpha+\beta+\gamma) = a+b+c+a'+b'+c'$.
We have checked that an atomic insulator generates a $K$-group with $(-1)^{\nu}=-1$.~\footnote{See \cite{KS_homology} for the detail. K.S.\ thanks Judith H\"oller for pointing out this fact.}
This example implies that topological invariants defined on a two-dimensional surface in the momentum space do not necessarily yield a nontrivial topological insulator with a surface state.

\subsubsection{$P3_1$ and torsion invariant}
\label{sec:screw}
This section gives an example of the AHSS for a nonsymmorphic space group and shows that the $n$-fold screw axis leads to a $\Z_n$ torsion invariant in class AIII.

\begin{figure}[!]
	\begin{center}
	\includegraphics[width=0.7\linewidth, trim=3cm 18cm 4cm 0cm]{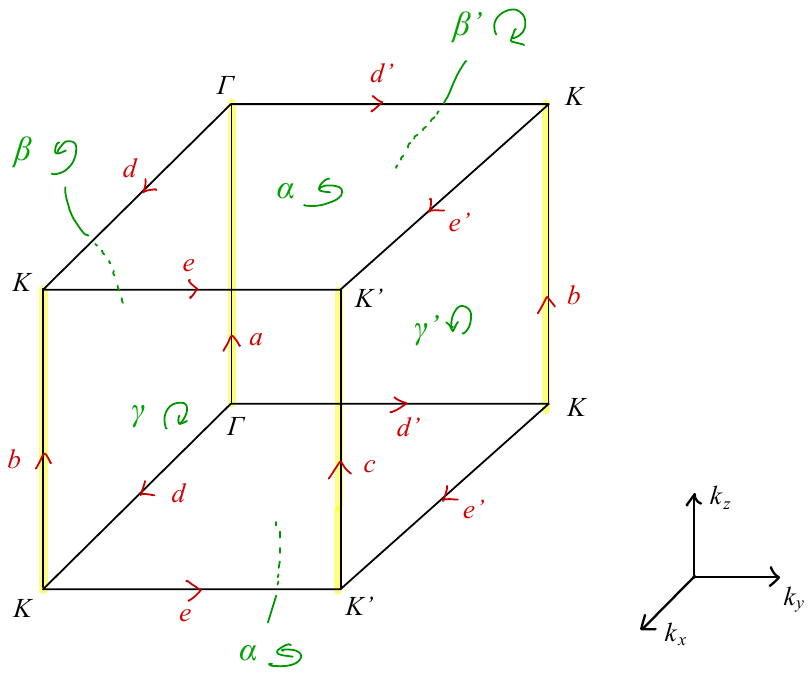}
	\end{center}
	\caption{A $C_3$-equivariant cell decomposition of BZ. 
	The figure shows one third of the BZ. 
	}
	\label{fig:144}
\end{figure}
Let us consider the space group $P3_1$ (No.144) in spinless systems. 
The Bravais lattice is primitive and the point group is $C_3$, and the 3-fold rotation $C_{3z}$ is accompanied by the nonprimitive lattice translation $\frac{1}{3} \hat z$.
The $C_3$ group acts on the $k$-space with the twist $U(C_{3z}^2\bk)U(C_{3z}\bk)U(\bk)=e^{-i k_z}$. 
At the screw axes, the three irreps are labeled by eigenvalues $\lambda \in \{ e^{-i k_z/3}, \omega e^{-i k_z/3}, \omega^2 e^{-i k_z/3}\}$ ($\omega = e^{-2 \pi i/3}$) which are cyclically permuted by the shift $k_z \mapsto k_z + 2 \pi$. 
\begin{figure}[!]
	\begin{center}
	\includegraphics[width=\linewidth, clip, trim=2cm 20cm 2cm 0cm]{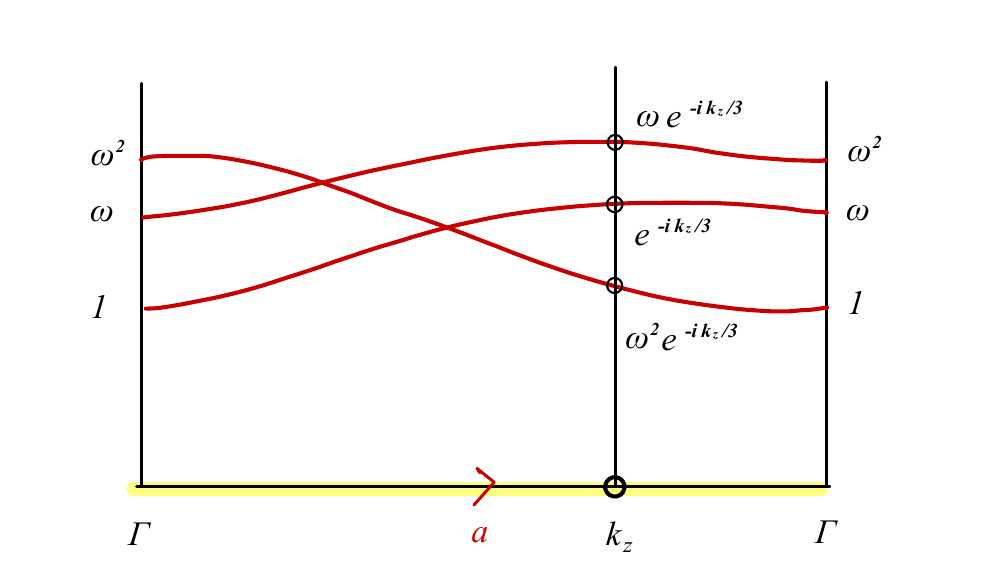}
	\end{center}
	\caption{The structure of Bloch states on a $3$-fold screw axis.}
	\label{fig:screw}
\end{figure}
A $C_3$-equivariant cell decomposition is given as in Fig.~\ref{fig:144}. 
The list of $p$-cells ($p=0,1,2,3$) is as follows. 
\begin{align*}
&\mbox{0-cells} = \{\Gamma, K, K'\}, \\
&\mbox{1-cells} = \{a,b,c,d,e\}, \\
&\mbox{2-cells} = \{\alpha,\beta,\gamma\}, \\
&\mbox{3-cells} = \{ vol\ (\mbox{$\frac{1}{3}$BZ shown in Fig.~\ref{fig:144}} ) \}. 
\end{align*}
We should take care of the compatibility relation on the screw axes.
See Fig.~\ref{fig:screw}. 
An irrep on the 1-cell $a$ is connected to the different irreps at $k_z=0$ and $k_z=2\pi$. 
The block matrix from $\Gamma$ to $a$ of the differential $d_1^{0,0}$ is given by
\begin{align}
d_1^{0,0}|_{\Gamma \to a} = 
\begin{array}{|ccc|cc}
\Gamma &&&&\\
1&\omega&\omega^2&&\\
\hline 
1&-1&0&e^{-i k_z/3}&a \\
0&1&-1&\omega e^{-i k_z/3}&\\
-1&0&1&\omega^2 e^{-i k_z/3}&\\
\hline 
\end{array}
\end{align}
Taking all contributions into account, we obtain 
\begin{align}
&E_2^{0,0}=\Z, \quad 
E_2^{1,0}=\Z \oplus \Z_3^2, \nonumber\\
&E_2^{2,0}=\Z, \quad 
E_2^{3,0}=\Z.
\end{align}
The torsion in $E_2^{1,0}$ means that there are two $\Z_3$ invariants defined on the 1-skeleton $X_1$. 
On the one hand, the class AIII $K$-group is found to be 
$K^{\tau+1}_{\Z_3}(T^3) = \Z^2 \oplus \Z_3$ from the Mayer-Vietoris sequence,~\footnote{
Proof. 
Applying the Mayer-Vietoris sequence to the $k_z$-direction, we have 
\begin{align*}
0 
\to 
K^{\tau+0}_{\Z_3}(T^3) 
\to
K\oplus K
\xrightarrow{\Delta} 
K\oplus K
\to 
K^{\tau+1}_{\Z_3}(T^3) 
\to 
0, 
\end{align*}
where $K=K^0_{\Z_3}(T^2) \cong R(\Z_3) \oplus R(\Z_3) \oplus (1-t)$ is the class A $K$-group with the $C_3$ rotation symmetry.~\cite{SSG17}
Here, $R(\Z_3)=\Z[t]/(1-t^3)$ is the representation ring of $\Z_3$ and $(1-t)$ is an $R(\Z_3)$-ideal. 
The homomorphism $\Delta$ is given by $(x,y) \mapsto (x-y,x-ty)$. 
Then, we have $K^{\tau+0}_{\Z_3}(T^3) \cong \ker \Delta \cong \Z^2$, 
$K^{\tau+1}_{\Z_3}(T^3) \cong \coker \Delta \cong \Z^2 \oplus \Z_3$.
}
which implies that one of the two $\Z_3$ invariants is determined by the 3d winding number $w_{3d}$. 

\paragraph{The construction of the screw $\Z_n$ invariant}
In general, a pair of $n$-fold screw axes gives rise to a $\Z_n$ invariant in class AIII, which is the generalization of the $\Z_2$ invariant in $2d$ class AIII with the glide symmetry.~\cite{SSG17}
Let $q(k_z)$ be the off-diagonal part of the Hamiltonian $H(k_z) = \begin{pmatrix}
0 & q(k_z) \\
q(k_z)^{\dag} & 0 \\
\end{pmatrix}$ on a $n$-fold screw axis. 
The matrix $q(k_z)$ splits into subsectors as 
\begin{align}
q(k_z) = q_0(k_z) \oplus q_1(k_z) \oplus \cdots \oplus q_{n-1}(k_z) 
\end{align}
with respect to the eigenvalues $\omega^j e^{- i k_z/n} (j=0,1,\dots,n-1)$ with $\omega = e^{-2 \pi i/n}$. 
Because of the twist, $q_j(k_z)$ are cyclically permuted as $q_j(k_z+2 \pi) = q_{j+1}(k_z)$. 
We introduce the following $U(1)$-valued quantity \begin{widetext}
\begin{align}
e^{i \phi(k_z)}:= \det q_0(k_z) \cdot \exp \left[ \sum_{j=0}^{n-2} \frac{n-j-1}{n} \int_{k_z}^{k_z+2 \pi} d k_z' \p_{k_z'} \log \det q_j(k_z') \right] 
\end{align}
so that its $n$th power becomes the total determinant $e^{i n \phi(k_z)} = \det q(k_z)$. 
Suppose that there are two $n$-fold screw axes at $(k_x,k_y) = X$ and $Y$. 
The $\Z_n$ invariant $e^{2 \pi i \nu/n}$ is defined to be 
\begin{align}
e^{2 \pi i \nu/n}
:= \exp\left[ i \phi(X,k_z) - i \phi(Y,k_z) - \frac{1}{n} \int_{X \to Y} d \bk \cdot \bm{\nabla} \log \det q(\bk,k_z) \right].
\end{align}
\end{widetext}
It is easy to show that $\{ e^{2 \pi i \nu/n} \}^n = 1$, i.e.\ $e^{2 \pi i \nu/n}$ takes values in $\Z_n$ quantized $U(1)$ phases. 

For the space group $P3_1$, a nontrivial model showing $\nu=1$ is given by putting a single SSH chain along the $x$-direction and extending to the $3d$ lattice by group elements of the space group $P3_1$.

\subsubsection{$P\bar 1$}
\label{sec:pbar1}
The final example is the space group $P \bar 1$ (No.2). 
We illustrate how the representation enforced Weyl semimetal~\cite{TurnerInversion} appears in the AHSS. 

\begin{figure}[!]
	\begin{center}
	\includegraphics[width=0.05\linewidth, trim=10cm 18cm 10cm 0cm]{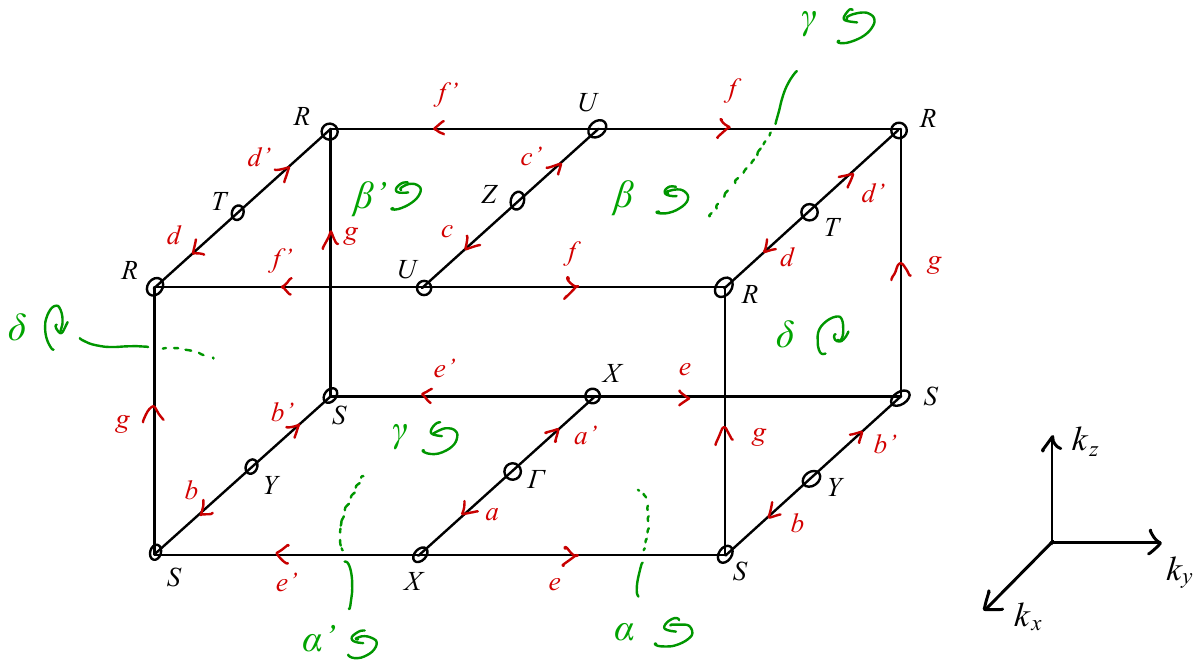}
	\end{center}
	\caption{A $C_i$-equivariant cell decomposition of the BZ. The figure shows a half of the BZ. 
	}
	\label{fig:2}
\end{figure}
The Bravais lattice is primitive and the point group is $C_i (\cong \Z_2)$ generated by the inversion $I$. 
Fig.~\ref{fig:2} shows a $C_i$-equivariant cell decomposition, where the $p$-cells are given as 
\begin{align*}
&\mbox{0-cells} = \{\Gamma, X, Y, S, Z, U, T, R\}, \\
&\mbox{1-cells} = \{a,b,c,d,e,f,g\}, \\
&\mbox{2-cells} = \{\alpha,\beta,\gamma,\delta\}, \\
&\mbox{3-cells} = \{ vol\ (\mbox{$\frac{1}{2}$BZ shown in Fig.~\ref{fig:2}} ) \}. 
\end{align*}
An important point is that the orientation of the $2$-cell $\alpha (\beta)$ is the same as the equivalent 2-cell $\alpha' (\beta')$ thereof since the inversion $I$ acts on the $k_z=0 (\pi)$ plane as the $2$-fold rotation $C_{2z}$. 
Then, the first differential $d_1^{2,0}: \Z^4 \to \Z$ is 
\begin{align}
d_1^{2,0}
= 
\begin{array}{|c|c|c|c|c}
\alpha&\beta&\gamma&\delta&\\
\hline
2&-2&0&0&vol\\
\hline
\end{array}
\end{align}
Apparently, we have $E_2^{3,0} = E_1^{3,0}/\im(d_1^{2,0}) =\Z/2\Z=\Z_2$. 
This means that odd numbers of Weyl points inside $\frac{1}{2}$BZ region can not be trivialized from the pair creation of Weyl points from 2-cells. 

From a straightforward calculation, we find that $E_2^{0,0} = \Z^9$. 
The third differential $d_3^{0,2}: \Z^9 \to \Z_2$ can be nontrivial, and one can show this is the case.
The explicit form of $d_3^{0,2}$ is given by~\cite{TurnerInversion}
\begin{align}
&d_3^{0,2}:\Z^9 \to \Z_2, \nonumber\\
&(n, \{ n_-^{\bk}\}_{\bk \in {\rm 0-cells}}) \mapsto 
(-1)^{\nu}:= (-1)^{\sum_{\bk \in {\rm 0-cells}} n^{\bk}_-}, 
\label{eq:2weyl}
\end{align}
where $n$ is the filling number and $n^{\bk}_-$ is the number of irreps with $I=-1$.~\footnote{
A quick derivation of (\ref{eq:2weyl}) is as follows. 
On the $k_z = 0$ and $k_z=\pi$ planes, the parity of the Chern number is constrained as $(-1)^{ch|_{k_z=0}} = (-1)^{\sum_{\bk \in \Gamma,X,Y,S} n^{\bk}_-}$ and $(-1)^{ch|_{k_z=\pi}} = (-1)^{\sum_{\bk \in Z,U,R,T} n^{\bk}_-}$. 
If the band structure is fully gapped, the Chern number should be uniform $ch|_{k_z=0} = ch|_{k_z=\pi}$, which implies $(-1)^{\nu}=1$. 
}

Alternatively, as we seen in Sec.~\ref{sec:AHSS}, the third differential represents the band inversion resulting in the creation of Weyl points from $0$-cells. 
Around the $\Gamma$ point, a model Hamiltonian is given by 
\begin{align}
H(k_x,k_y,k_z) = (k^2-\mu) \sigma_z + k_x \sigma_x + k_y \sigma_y, \nonumber\\
I = \sigma_z.
\end{align}
It is clear that when $\mu$ passes zero, the band inversion between $I=1$ and $I=-1$ occurs, and a pair of Weyl points is pumped on the $k_z$ axis.

\section{The AHSS with antiunitary symmetry}
\label{The AHSS with antiunitary symmetry}
In this section, we formulate the AHSS for band theory in the presence of TRS and/or PHS. 
First, we give the mathematical detail of the AHSS in Sec.~\ref{sec:ahss_real}.
In Sec.~\ref{sec:2daz}, as a warm-up, we calculate the AHSS for 2-dimensional systems without space group symmetry and reproduce the classification table for 2-dimensions.~\cite{RyuClass}
In Sec.~\ref{sec:e1_wigner}, we describe how to determine the $E_1$-page in general. 
Although one can readily determine the first differential $d_1$ using the compatibility relation, it is not straightforward to give the higher-differentials $d_2$ and $d_3$. 
We sketch the Hamiltonian formalism for $d_2$ and $d_3$ in Sec.~\ref{sec:d2d3}. 
We present some examples of the AHSS in Sec.~\ref{sec:casestudy_au}. 
Some technical details relevant to this section are in Appendices \ref{app:wigner} and \ref{sec:factor}. 

\subsection{Formulation of the AHSS for general symmetry classes}
\label{sec:ahss_real}
In this section, we formulate the AHSS for general symmetry classes including antiunitary symmetry. 
Since, the relationship among the sequences of $E_r$-pages and $r$-th differentials $d_r$ is almost the same as Sec.~\ref{sec:ahss_com_for}, we briefly sketch the mathematical formulation. 

Let $G$ be the symmetry group and $X_0\subset X_1 \subset X_2 \subset X_3 = T^3$ be a $G$-filtration of the BZ torus associated to a cell decomposition. 
Let $(\phi,c,\tau)$ be the data of a symmetry class, where $\phi,c: G \to \Z_2$ indicate whether $g \in G$ is unitary or antiunitary and symmetry or antisymmetry,~\footnote{$g \in G$ is said antisymmetry if $g$ anticommutes with the Hamiltonian like the particle-hole symmetry and the chiral symmetry.} respectively, and $\tau = \tau_{g,h}(\bk) (g,h \in G, \bk \in T^3)$ is the factor system for a given magnetic space group.
$\tau_{g,h}(\bk)$ also depends on representations of the superconducting gap function of the point group. 
Let ${}^{\phi}K^{(\tau,c)-n}_G(T^3)$ be the twisted equivariant $K$-group of the BZ torus $T^3$,~\cite{FreedMoore, SSG17, Gomi17} where $n \in \Z$ is the integer grading defined by adding chiral symmetries (see Sec.\ IV in Ref.\cite{SSG17}). 
The Bott periodicity ${}^{\phi}K^{(\tau,c)-n}_G(T^3) \cong {}^{\phi}K^{(\tau,c)-n+8}_G(T^3)$ holds true. 
In the absence of antiunitary symmetry (meaning the case where $\phi: G \to \Z_2$ is trivial), the period of the Bott periodicity reduces to two, $K^{(\tau,c)-n}_G(T^3) \cong K^{(\tau,c)-n+2}_G(T^3)$. 

The $E_1$-page of the AHSS is defined to be the $K$-group of the pair $(X_p,X_{p-1})$, 
\begin{align}
E_1^{p,-n} := {}^{\phi}K^{(\tau,c)-(n-p)}_G(X_p,X_{p-1}). 
\end{align} 
Here, the $K$-group over the pair $(X_p,X_{p-1})$ means the classification of gaped Hamiltonians over the $p$-skeleton $X_p$ which are constant on the $(p-1)$-skeleton $X_{p-1}$. 
From the definition of $p$-skeletons, this $K$-group is recast as the direct sum of $K$-groups of orbits of $p$-cells, 
\begin{widetext}
\begin{align}
E_1^{p,-n} 
&\cong {}^{\phi}K^{(\tau,c)-(n-p)}_G(\coprod_{j \in I^p_{\rm orb}} G/G_{D^p_j} \times D^p_j, \coprod_{j \in I^p_{\rm orb}} G/G_{D^p_j} \times \p D^p_j)  \nonumber \\
&\cong \prod_{j \in I^p_{\rm orb}} {}^{\phi|_{D^p_j}}K^{(\tau,c)|_{D^p_j}-(n-p)}_{G_{D^p_j}}(D^p_j,\p D^p_j). 
\end{align}
\end{widetext}
Here, $D^p_j$ is a representative $p$-cell for the orbit $j$, and $(\phi,c,\tau)|_{D^p_j} = (\phi|_{D^p_j},c|_{D^p_j}.\tau|_{D^p_j})$ means the data $(\phi,c,\tau)$ of the symmetry class restricted to a $p$-cell $D^p_j$, i.e.\ the data $(\phi,c,\tau)$ for the little group $G_{D^p_j}$. 
We find that the following four interpretations of the group $E_1^{p,-n}$. 

\noindent
(I)---
$E_1^{p,-n}$ is the direct sum of the $K$-groups over orbits of $p$-cells for topological insulators $H_{\rm TI}(\bk)$ with the symmetry class of the integer grading $(n-p)$. 
On each $p$-cell $D^p_j$, a gapped Hamiltonian obeys the boundary condition so that it is constant on the boundary $\p D^p_j$, which implies that $H_{\rm TI}(\bk)$ should be a massive Dirac Hamiltonian $H_{\rm TI}(\bk) \sim \sum_{\mu=1}^p k_{\mu} \gamma_{\mu} + (m-\epsilon k^2) \gamma_{p+1}$.  

By the use of the bulk-boundary correspondence,~\cite{SSG17} $E_1^{p,-n}$ is also identified with the group classifying topological gapless states with a shift of integer grading: 

\noindent
(II)---
$E_1^{p,-n}$ is the direct sum of the $K$-groups over orbits of $p$-cells for topological gapless states $H_{\rm TGS}(\bk)$ with the symmetry class of integer grading $(n-p+1)$. 
On each $p$-cell $D^p_j$, the spectrum is gapped on the boundary, which implies that $H_{\rm TGS}(\bk)$ should be a massless Dirac Hamiltonian $H_{\rm TGS}(\bk) \sim \sum_{\mu=1}^p k_{\mu} \gamma_{\mu}$. 

In the same way as Sec.~\ref{sec:complex_E_1-page}, for $p \geq 1$, applying the bulk-boundary correspondence to the $(p-1)$-dimensional sphere $S^{p-1}$ surrounding the Dirac point of the topological gapless state in the $p$-cell, we find that the $E_1$-page is viewed as the group classifying topological singular points in $p$-cells: 

\noindent
(III)---
$E_1^{p,-n}$ is the direct sum of the $K$-groups over orbits of $p$-cells for topological singular points $H_{\rm TSP}(\bk)$ with the symmetry class of integer grading $(n-p+2)$. 

The $E_1$-page is further deformed as follows. 
By shrinking the boundary on each $p$-cell, the pair $(D^p_j,\p D^p_j)$ is considered as the $p$-dimensional sphere $D^p_j/\p D^p_j \cong S^p_j$. 
Using the Thom isomorphism, we have  
\begin{align}
E_1^{p,-n} 
&\cong \prod_{j \in I^p_{\rm orb}} {}^{\phi|_{D^p_j}} \wt K^{(\tau,c)|_{D^p_j}-(n-p)}_{G_{D^p_j}}(D^p_j/\p D^p_j) \nonumber \\
&\cong \prod_{j \in I^p_{\rm orb}} {}^{\phi|_{D^p_j}} K^{(\tau,c)|_{D^p_j}-n}_{G_{D^p_j}}(D^p_j), \label{eq:e1_real_az}
\end{align} 
where ${}^{\phi|_{D^p_j}} \wt K^{(\tau,c)|_{D^p_j}-(n-p)}_{G_{D^p_j}}(D^p_j/\p D^p_j)$ meant the reduced $K$-theory. 
${}^{\phi|_{D^p_j}} K^{(\tau,c)|_{D^p_j}-n}_{G_{D^p_j}}(D^p_j)$ is the $K$-group over the $p$-cell $D^p_j$. 
Therefore, we arrived at the following formula to give the $E_1$-page: 

\noindent
(IV)---
$E_1^{p,-n}$ is the direct sum of the $K$-groups for the representations on $p$-cells. 
Each $K$-group ${}^{\phi|_{D^p_j}} K^{(\tau,c)|_{D^p_j}-n}_{G_{D^p_j}}(D^p_j)$ represents the space of representations on the $p$-cell $D^p_j$ with the symmetry class of the integer grading $n$. 

This enables us to compute the $E_1$-page quickly by looking at what the symmetry class realized at the $p$-cell $D^p_j$ is (See Sec.~\ref{sec:e1_wigner}).

\begin{figure}[!]
	\begin{center}
	\includegraphics[width=\linewidth, trim=0cm 6cm 0cm 0cm]{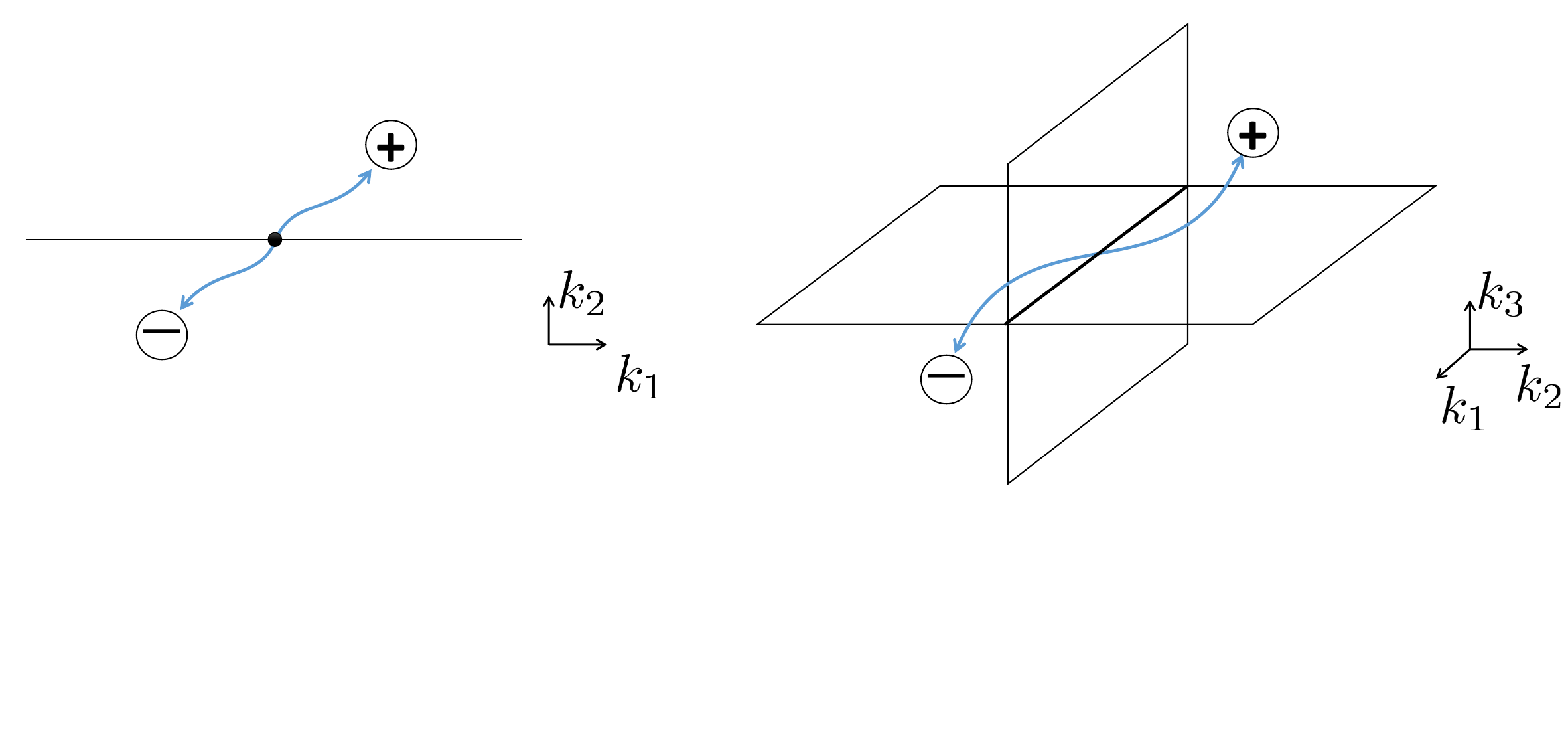}
	\end{center}
	\caption{
	The second differential $d_2^{p,-n}$ can be viewed as the creation of gapless points from $p$-cells to adjacent $(p+2)$-cells. 
	}
	\label{fig:d2}
\end{figure}

We define the first differential 
\begin{align}
d_1^{p,-n}: E_1^{p,-n} \to E_1^{p+1,-n}
\end{align}
in the same way as in Sec.~\ref{sec:ahss_com_for}.
The homomorphism $d_1^{p,-n}$ describes how irreps at $p$-cells are mapped to representations at adjacent $(p+1)$-cells. 
From the meaning (II) of the $E_1$-page, $d_1^{p,-n}$ can be also viewed as the creation of stable gapless points in $(p+1)$-cells from $p$-cells. 
It holds that $d_1\circ d_1=0$. 
The $E_2$-page is defined by the cohomology of $d_1$, 
\begin{align}
E_2^{p,-n} := \ker(d_1^{p,-n})/\im(d_1^{p-1,-n}).
\end{align}
We have the second differential in the $E_2$-page 
\begin{align}
d_2^{p,-n}: E_2^{p,-n} \to E_2^{p+2,-(n+1)}. 
\end{align}
This expresses the creations of topological gapless points from a $p$-cell to adjacent $(p+2)$-cells, as shown in Fig.~\ref{fig:d2}. 

Similarly, $d_2 \circ d_2 = 0$ holds true and the $E_3$-page is defined to be \begin{align}
E_3^{p,-n}:= \ker (d_2^{p,-n})/\im (d_2^{p-2,-(n-1)}), 
\end{align}
and we define the third differential 
\begin{align}
d_3^{p,-n}: E_3^{p,-n} \to E_3^{p+3,-(n+2)}
\end{align}
which represents the creation of topological gapless points from 0-cells to adjacent $3$-cells. 
We also have the $E_4$-page 
\begin{align}
E_4^{p,-n}:= \ker (d_3^{p,-n})/\im (d_3^{p-3,-(n-2)}). 
\end{align}
From the dimensional reason, the fourth differential $d_4^{p,-n}: E_4^{p,-n} \to E_4^{p+4,-(n+3)}$ is trivial for 3-space dimensions. 
This means that the $E_4$ page is the limit $E_{\infty} = E_4$. 

The limiting page $E_{\infty} = E_{\infty}^{p,-n}$ approximates the $K$-group ${}^{\phi} K^{(\tau,c)-n}_G(T^3)$. 
The topological invariants characterizing the $K$-group ${}^{\phi} K^{(\tau,c)-n}_G(T^3)$ are given by ``sticking'' the local contributions $E_{\infty}^{p,-(n+p)}$ arising from $p$-cells together appropriately. This is an extension problem in the algebraic point of view. 
The precise relationship among the $K$-group and the local contributions is described as 
\begin{equation}\begin{split}
&E_\infty^{0, -n} 
\cong
{}^{\phi} K^{(\tau,c)-n}_G(T^3)/F^1K^{-n}, \label{eq:e_infty_appro} \\
&E_\infty^{1, -(n+1)} 
\cong
F^{1}K^{-n}/F^{2}K^{-n}, \\
&E_\infty^{2, -(n+2)} 
\cong
F^{2}K^{-n}/E_\infty^{3, -n},
\end{split}\end{equation}
or equivalently, the short exact sequences 
\begin{widetext}
\begin{align}
\begin{CD}
0 @>>> F^{1}K^{-n} @>>> {}^{\phi} K_G^{(\tau,c)-n}(T^3) @>>> E_\infty^{0, -n} @>>> 0, \\
0 @>>> F^{2}K^{-n} @>>> F^{1}K^{-n} @>>> E_\infty^{1, -(n+1)} @>>> 0, \\
0 @>>> E_\infty^{3, -(n+3)} @>>> F^{2}K^{-n} @>>> E_\infty^{2, -(n+2)} @>>> 0, \\
\end{CD}\label{eq:e_inf}
\end{align}
in terms of the intermediate subgroups of ${}^{\phi} K^{(\tau,c)-n}_G(T^3)$
\begin{align}
{}^{\phi} K_G^{(\tau,c)-n}(T^3) = 
F^{0}K^{-n} \supset 
F^{1}K^{-n} \supset 
F^{2}K^{-n} \supset 
F^{3}K^{-n} = E_\infty^{3, -(n+3)}
\label{eq:ahss_f_inclusion}
\end{align}
given by
\begin{align}
F^{p}K^{-n}
&= \mathrm{Ker}\left[ \mathrm{res} : {}^{\phi} K^{(\tau,c)-n}_G(T^3) \to {}^{\phi} K^{(\tau,c)|_{X_{p-1}}-n}_G(X_{p-1})\right].
\label{eq:ahss_f_def}
\end{align}
Here ${\rm res}$ is the restriction homomorphism induced by inclusion
$X_{p-1}\subset T^3$.
When the hierarchical extension problem (\ref{eq:e_inf}) has a unique solution, the $K$-group ${}^{\phi} K_G^{(\tau,c)-n}(T^3)$ is fixed as an Abelian group. 
However, the extension problem (\ref{eq:e_inf}) has multiple solutions in general. 
In such cases, one can not evaluate the $K$-group only by the data $E_{\infty}^{n,-(n+p)}$ (, although the rank of ${}^{\phi} K^{(\tau,c)-n}_G(T^3)$ is the sum of those of $E_\infty^{p, -(n+p)}$).
A brute force approach to determine the $K$-group in such cases is finding an explicit formula of topological invariants compatible with the exact sequences (\ref{eq:e_inf}) and collecting the ``compatibility relation'' among the topological invariants ((\ref{eq:rel_nu_w3d}) is an example of such relations). 
Other exact sequences such as the Mayer-Vietoris and Gysin sequences in the $K$-theory can help us to determine the $K$-group.~\cite{SSG16,SSG17}

Similarly, in two space dimensions, the $K$-group ${}^{\phi} K_G^{(\tau,c)-n}(T^2)$ fits into the short exact sequences 
\begin{align}
\begin{CD}
0 @>>> F^{1}K^{-n} @>>> {}^{\phi} K_G^{(\tau,c)-n}(T^2) @>>> E_\infty^{0, -n} @>>> 0, \\
0 @>>> E_{\infty}^{2, -(n+2)} @>>> F^{1}K^{-n} @>>> E_\infty^{1, -(n+1)} @>>> 0. \\
\end{CD}\label{eq:e_inf_2d}
\end{align}
Also, in one space dimension, the $K$-group ${}^{\phi} K_G^{(\tau,c)-n}(S^1)$ obeys the short exact sequence 
\begin{align}
\begin{CD}
0 @>>> E_{\infty}^{1, -(n+1)} @>>> {}^{\phi} K_G^{(\tau,c)-n}(S^1) @>>> E_\infty^{0, -n} @>>> 0. 
\end{CD}\label{eq:e_inf_1d}
\end{align}
\end{widetext}

\subsubsection{On the exact sequences (\ref{eq:e_inf})}
Here, we sketch why the exact sequences (\ref{eq:e_inf}) hold true. 
We employ the interpretation of the $K$-group ${}^{\phi} K^{(\tau,c)-n}_G(T^3)$ as topological gapless states over $T^3$ for symmetry class $(n+1)$. 
(The sketch with the interpretation of the $K$-group ${}^{\phi} K^{(\tau,c)-n}_G(T^3)$ as topological insulators for symmetry class $n$ is parallel.)

Recall that $E_{\infty}^{p,-(n+p)}$ has the following meaning: 
\begin{itemize}
\item
$E_{\infty}^{p,-(n+p)}$ is the space of topological gapless states in $p$-cells for the symmetry class $(n+1)$ which can extend to all the adjacent higher-dimensional cells without a singularity, and can not be trivialized by the creation of topological gapless points from any adjacent low-dimensional cells.
\end{itemize}
The definition (\ref{eq:ahss_f_def}) of $F^{p,-n}$ implies that 
\begin{itemize}
\item
$F^{p}K^{-n}$ is the space of topological gapless states over $T^3$ for the symmetry class $(n+1)$ which have a finite energy gap on the $(p-1)$-skeleton $X_{p-1}$. 
\end{itemize}

Obviously, we have the injection $F^{p+1}K^{-n} \hookrightarrow F^{p}K^{-n}$, since the existence of a finite energy gap over $X_p$ implies that there is also a finite energy gap over $X_{p-1}$, which leads to the sequence of inclusions (\ref{eq:ahss_f_inclusion}). 

The homomorphism $F^pK^{-n} \to E_\infty^{p,-(n+p)}$ is defined by restricting gapless states into the  $p$-cells. 
Therefore, the group $\ker[F^{p}K^{-n} \to E^{p,-(n+p)}_{\infty}]$ represents gapless states with a finite energy gap on $X_{p-1}$ that also have a finite energy gap over the $p$-cells. 
On the other hand, the group $\im [F^{p+1}K^{-n} \to F^{p}K^{-n}]$ represents gapless states with a finite gap on $X_{p-1}$ that also have a finite energy gap on $X_p$. 
Since $X_p \backslash X_{p-1}$ is the set of $p$-cells, two groups $\ker[F^{p}K^{-n} \to E^{p,-(n+p)}_{\infty}]$ and $\im [F^{p+1}K^{-n} \to F^{p}K^{-n}]$ are equivalent. 

The finial step is to show that the homomorphism $F^{p}K^{-n} \to E^{p,-(n+p)}_{\infty}$ is surjectve. 
This follows from the definition of $E_{\infty}^{p,-(n+p)}$: 
It is possible to extend a topological gapless state in an orbit $G/G_{D^p_j} \times D^p_j$ of $p$-cells to the neighborhood of the orbit $G/G_{D^p_j} \times D^p_j$ in the BZ $T^3$ without affecting the Hamiltonian outside the neighborhood. 
This means that there must be a representative topological gapless state in $F^{p}K^{-n}$ for a topological gapless state of $E^{p,-(n+p)}$.

\subsection{Warm-up: real AZ class in two space dimensions}
\label{sec:2daz}

\begin{figure*}[!]
	\begin{center}
	\includegraphics[width=\linewidth, trim=0cm 0cm 0cm 0cm]{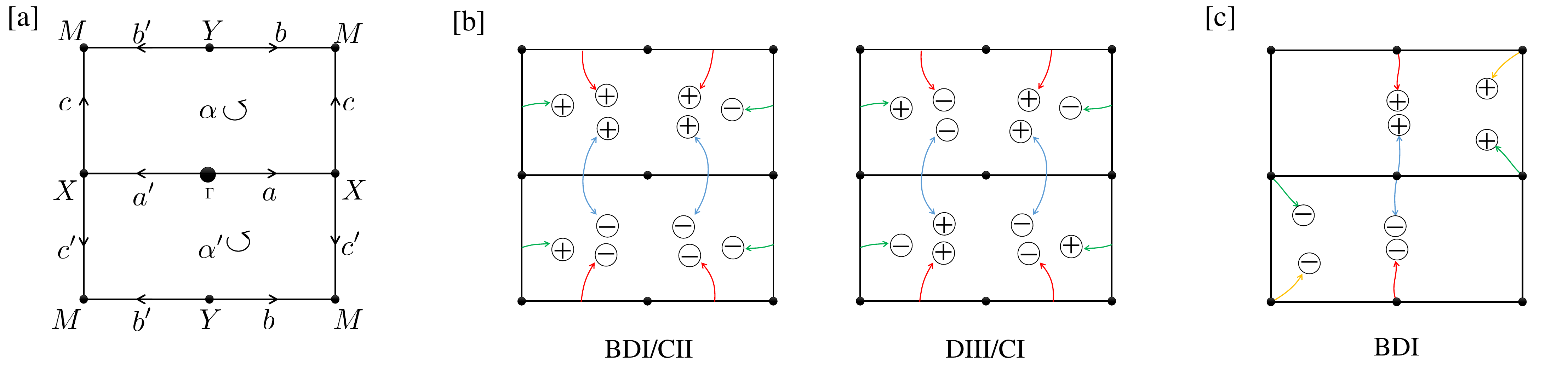}
    \end{center}
	\caption{
	[a]
	A cell decomposition of the 2d BZ 
	with TRS $T: (k_x,k_y) \mapsto (-k_x,-k_y)$. 
	Here, $a' = T(a), b'=T(b), \alpha'=T(\alpha)$ represent equivalent $p$-cells. 
	The first differential $d_1^{1,{\rm even}}$. 
	The signs $\pm$ in the circles represent the charges of the gapless Dirac point with chiral symmertry. 
	[c]
	The second differential $d_2^{0,-1}$. 
	The band inversion at 0-cells creates a pair of Dirac points. 
	}
	\label{fig:2d}
\end{figure*}

Let us compute the AHSS for $2d$ systems with real AZ symmetry classes. 
As the symmetry class for $n=0$, we consider the $2d$ spinless systems with TRS. 
I.e., class AI in the AZ classes. 
We denote the TRS operator by $T$. 
The symmetry group is $G = \Z_2 = \{1,T\}$ which acts on the $2d$ BZ torus by $T: (k_x,k_y) \mapsto (-k_x,-k_y)$. 
(Here, we denoted the group element of TRS by the same symbol $T$.)
The factor system is trivial $T^2 = 1$. 
We use the cell-decomposition of the BZ torus shown in Fig.~\ref{fig:2d} [a]. 

From the formula (\ref{eq:e1_real_az}), the $E_1$-page is the collection of the $K$-groups over $p$-cells with symmetry class shifted by $n \in \Z$. 
We have the following $E_1$-page. 
\begin{align}
\begin{array}{cc|cccc}
{\rm AI} & n=0 & \Z^4 & \Z^3 & \Z \\
{\rm BDI} & n=1 & \Z_2^4 & 0 & 0 \\
{\rm D} & n=2 & \Z_2^4 & \Z^3 & \Z \\
{\rm DIII} & n=3 & 0 & 0 & 0 \\
{\rm AII} & n=4 & \Z^4 & \Z^3 & \Z \\
{\rm CII} & n=5 & 0 & 0 & 0 \\
{\rm C} & n=6 & 0 & \Z^3 & \Z \\
{\rm CI} & n=7 & 0 & 0 & 0 \\
\hline 
& E_1^{p,-n} & p=0 & p=1 & p=2 
\end{array}.
\end{align}
The first differential $d_1^{p,-n}: E_1^{p,-n} \to E_1^{p+1,-n}$ is computed by the compatibility relation incorporating PHS as in 
\begin{align}
&d_1^{0,0} = \begin{array}{cccc|c}
\Gamma & X & Y & M & \\
\hline
1&-1&0&0&a\\
0&0&1&-1&b\\
0&1&0&-1&c\\
\end{array}, \\
&d_1^{0,-4} = \begin{array}{cccc|c}
\Gamma & X & Y & M & \\
\hline
2&-2&0&0&a\\
0&0&2&-2&b\\
0&2&0&-2&c\\
\end{array}, \\
&d_1^{1,-2} = d_1^{1,-6} = \begin{array}{ccc|c}
a&b&c& \\
\hline
2&-2&0&\alpha\\
\end{array}, 
\end{align}
and $d_1^{p,-n}=0$ for other $(p,-n)$s. 
We should be careful about the PHS. 
For $n=2$ and $6$, the 1-cell $a$ ($b$) changes to $a'$ ($b'$) with the particle-hole transformation $C$. 
Then, an occupied state $\ket{\phi}$ at $a$ is sent to an empty state $C \ket{\phi}$ at $a'$, which results in the nontrivial first differentials $d_1^{1,-2}$ and $d_1^{1,-6}$. 

\begin{figure*}[!]
	\begin{center}
	\includegraphics[width=\linewidth, trim=0cm 0cm 0cm 0cm]{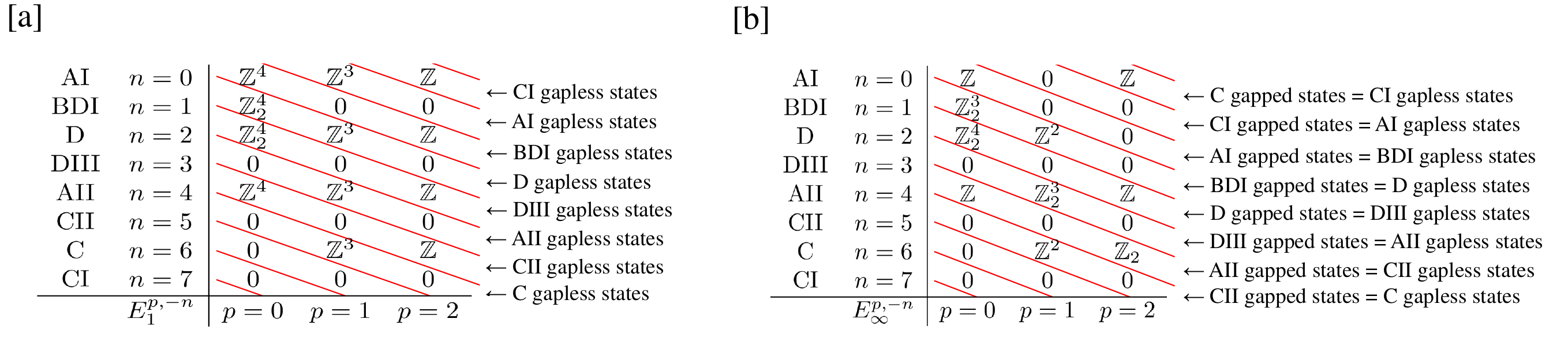}
    \end{center}
	\caption{
	[a]$E_1$-page as topological gapless states. 
	[b] Relationship among the groups $E_\infty^{p,-n}$ and AZ classes for gapped and gapless states. 
	}
	\label{fig:2d_e1}
\end{figure*}

Before moving on to the second differential $d_2$, it is worth understanding the first differential in terms of gapless Dirac points.
According to the meaning (II) of the $E_1$-page in Sec.~\ref{sec:ahss_real}, $E_1^{p,-n}$ represents topological gapless points inside $p$-cells with the symmetry class $(n-p+1)$ as illustrated in Fig.~\ref{fig:2d_e1} [a]. 
$E_1^{2,{\rm even}} = \Z$ comes from that the chiral symmetry stabilizes a Dirac point in a 2-cell. 
The first differentials $d_1^{1,{\rm even}}: E_1^{1,{\rm even}} \to E_1^{2,{\rm even}}$ means how gapless Dirac points in the 2-cell are absorbed by the pair creation of the Dirac points from 1-cells. 
For class BDI/CII, a time-reversal symmetric pair of Dirac points has the opposite charge because of the algebra $T \Gamma = \Gamma T$ between the TRS $T$ and the chiral symmetry $\Gamma$, whereas for class DIII/CI, the charge is the same because of the algebra $T \Gamma =-\Gamma T$.
See Fig.~\ref{fig:2d} [b]. 
This accounts for $d_1^{1,(-2)/(-6)}=(2,-2,0)$ and $d_1^{1,0/(-4)} = (0,0,0)$.

Taking the cohomology group of $d_1$, we have the $E_2$-page 
\begin{align}
\begin{array}{cc|cccc}
{\rm AI} & n=0 & \Z & 0 & \Z \\
{\rm BDI} & n=1 & \Z_2^4 & 0 & 0 \\
{\rm D} & n=2 & \Z_2^4 & \Z^2 & \Z_2 \\
{\rm DIII} & n=3 & 0 & 0 & 0 \\
{\rm AII} & n=4 & \Z & \Z_2^3 & \Z \\
{\rm CII} & n=5 & 0 & 0 & 0 \\
{\rm C} & n=6 & 0 & \Z^2 & \Z_2 \\
{\rm CI} & n=7 & 0 & 0 & 0 \\
\hline 
& E_2^{p,-n} & p=0 & p=1 & p=2 
\end{array}.
\end{align}
In the $E_2$-page, the only second differential $d_2^{0,-1}: E_2^{0,-1} \to E_2^{2,-2}$ can be nontrivial, and we find that $d_2^{0,-1}$ is surjective: 
We label 0-, 1-, and 2-cells as in Fig.~\ref{fig:2d} [a]. 
Notice that the even/odd parity of the class BDI $1d$ winding number $w_{1d}[-a'+a]$ along the loop $-a'+a$ is the product of the $\Z_2$ invariants at $\Gamma$ and $X$ as in  $(-1)^{w_{1d}[-a'+a]} = (-1)^{\nu(\Gamma)} (-1)^{\nu(X)}$.
In the same way, it holds that $(-1)^{w_{1d}[-b'+b]} = (-1)^{\nu(Y)} (-1)^{\nu(M)}$ for the $1d$ winding number along the loop $-b'+b$. 
Therefore, the product $(-1)^{\nu}:= \prod_{\bk \in \{\Gamma,X,Y,Z\}} (-1)^{\nu(\bk)}$ is the $\Z_2$ indicator to detect an odd number of class BDI Dirac points inside the 2-cell $\alpha$ and $(-1)^{\nu}$ is nothing but the second differential $d_2^{0,-1}$. 

Alternatively, one can evaluate the second differential $d_2^{0,-1}$ by the pair creation of the Dirac points from 0-cells. 
The class BDI symmetry permits creating a pair of Dirac points from 0-cells that removes the $\Z_2$ remainder of $E_2^{2,-2}=\Z_2$, the odd charges of Dirac points in the 2-cell. 
(See Fig.~\ref{fig:2d} [c]).
Explicitly, such a process can be modeled as 
\begin{align}
&H(\bk) = (|\bk-\bk_0|^2-\mu) \tau_z + (\bk-\bk_0) \cdot \bm{n} \tau_y, \nonumber\\
&T=K, \qquad C=\tau_x, 
\end{align}
around a 0-cell $\bk_0$. 
It is clear that when $\mu$ passes zero, the band inversion occurs with the change of the $\Z_2$ invariant $(-1)^{\nu(\bk_0)}$ at the $0$-cell $\bk_0$, resulting in a pair creation of of Dirac points direction perpendicular to $\bm{n}$. 
This can contrast well with the case of class CII, where the pair creation of Dirac points from 0-cells should be doubly degenerate due to the TRS with Kramers degeneracy.  
As a result, there are no class CII topological invariants at 0-cells ($E_1^{0,-5}=0$), which means that 0-cells can not be a new source of Dirac points, i.e., Dirac points arising from a 0-cell are recast as ones from 1-cells with continuous deformation. 

Taking the cohomology of $d_2$, we arrive at the limiting page $E_{\infty} = E_3$, 
\begin{align}
\begin{array}{cc|cccc}
{\rm AI} & n=0 & \Z & 0 & \Z \\
{\rm BDI} & n=1 & \Z_2^3 & 0 & 0 \\
{\rm D} & n=2 & \Z_2^4 & \Z^2 & 0 \\
{\rm DIII} & n=3 & 0 & 0 & 0 \\
{\rm AII} & n=4 & \Z & \Z_2^3 & \Z \\
{\rm CII} & n=5 & 0 & 0 & 0 \\
{\rm C} & n=6 & 0 & \Z^2 & \Z_2 \\
{\rm CI} & n=7 & 0 & 0 & 0 \\
\hline 
& E_{\infty}^{p,-n} & p=0 & p=1 & p=2 
\end{array}
\end{align}
The data $\{ E_{\infty}^{0,-n}, E_{\infty}^{1,-(n+1)}, E_{\infty}^{2,-(n+2)} \}$ approximate the $K$-group ${}^{\phi} K_{\Z_2}^{-n}(T^2)$ in the sense of the exact sequences (\ref{eq:e_inf_2d}). 
The relationship among columns of the $E_{\infty}$-page, AZ symmetry classes of $2d$ bulk insulators and $2d$ gapless states should be kept in mind, which is shown in Fig.~\ref{fig:2d_e1} [b]. 
The dimension $p$ of $E_{\infty}^{p,-n}$ indicates the skeleton $X_p$ on which the topological invariant is defined.~\footnote{
The band structures representing the group $E_\infty^{p,-n}$ are insufficient to define the topological invariant, since $E_\infty^{p,-n}$ represent only band structures inside $p$-cells that are trivial over the boundary of the $p$-cells.}
The exact sequences (\ref{eq:e_inf_2d}) are recast as 
\begin{align}
&{\rm AI}:&&{}^{\phi} K_{\Z_2}^{0}(T^2) = \Z, \\
&{\rm BDI}:&&0\to \Z^2 \to {}^{\phi} K_{\Z_2}^{-1}(T^2) \to \Z_2^3 \to 0, \label{eq:2d_bdi}\\
&{\rm D}:&&0\to \Z \to {}^{\phi} K_{\Z_2}^{-2}(T^2) \to \Z_2^4 \to 0, \label{eq:2d_d}\\
&{\rm DIII}:&&{}^{\phi} K_{\Z_2}^{-3}(T^2) = \Z_2^3, \\
&{\rm AII}:&&{}^{\phi} K_{\Z_2}^{-4}(T^2) = \Z+\Z_2, \\
&{\rm CII}:&&{}^{\phi} K_{\Z_2}^{-5}(T^2) = \Z^2, \\
&{\rm C}:&&{}^{\phi} K_{\Z_2}^{-6}(T^2) = \Z, \\
&{\rm CI}:&&{}^{\phi} K_{\Z_2}^{-7}(T^2) = 0.
\end{align}
These agree with the literature.~\cite{RyuClass}
Especially, $E_{\infty}^{2,-6} = \Z_2$ corresponds to the Kane-Mele $\Z_2$ topological invariant.~\cite{KaneZ2}
The short exact sequences (\ref{eq:2d_bdi}) and (\ref{eq:2d_d}) show nontrivial group extensions.~\footnote{
Class BDI: 
The $1d$ winding numbers $w_{1d}^{x}, w_{1d}^{y}$ along the $k_x$ and $k_y$ directions, respectively, give the constraints on the $\Z_2$ invariants defined at 0-cells as $(-1)^{w_{1d}^x} = (-1)^{\nu(\Gamma)} (-1)^{\nu(X)} = (-1)^{\nu(Y)} (-1)^{\nu(M)}$ and $(-1)^{w_{1d}^y} = (-1)^{\nu(\Gamma)} (-1)^{\nu(Y)} = (-1)^{\nu(X)} (-1)^{\nu(M)}$. 
Class D: 
The parity of the Chern number $C$ is related to the $\Z_2$ invariants (Pfaffian invariants) at 0-cells as $(-1)^C = \prod_{\bk \in \Gamma,X,Y,M} (-1)^{\nu(\bk)}$.
}
We find that ${}^{\phi}K^{-1}_{\Z_2}(T^2) = \Z^2+\Z_2^2$ and ${}^{\phi}K^{-2}_{\Z_2}(T^2) = \Z+\Z_2^3$.

\subsection{$E_1$-page for general symmetry}
\label{sec:e1_wigner}
In this section we describe how to compute the $E_1$-page for general symmetry classes realized in lattice systems. 
Let $G$ be the symmetry group (magnetic point group) and $\phi, c: G \to \Z_2 = \{\pm 1\}$ be the indicators for unitary/antiunitary and symmetry/antisymmetry, respectively. 
We denote the symmetry operator for $g \in G$ by $U_g(\bk)$. 
For the Hamiltonian $H(\bk)$ over BZ, the constraint relation is 
\begin{align}
&U_g(\bk)H(\bk)U_g(\bk)^{-1}=c(g)H(g\bk),\\
&U_g(\bk)i=\phi(g)iU_g(\bk),
\end{align}
with $i$ the imaginary unit.
The (magnetic) space group is specified by the point group action $p_g$ on the real space and the possibly nonprimitive lattice translation $\{g|\bm{a}_g\}: \bm{x} \mapsto p_g \bm{x}+\bm{a}_g$ associated with group elements $g \in G$.
Due to antiunitarity, $g \in G$ with $\phi(g)=-1$ acts on the momentum space by $\bk \mapsto -p_g \bk$. 
To make the notation simple, we denote the group action on the momentum space by $\bk \mapsto g \bk$, i.e.\ $g = \phi(g) p_g$ on $\bk$.
The nonsymmorphic part of the factor system is described by a 2-cocycle $\bm{\nu} \in Z^2(G,BL)$, where $BL \cong \Z^3$ is the Bravais lattice translation group, which is given as 
\begin{align}
&\{g|\bm{a}_g\} \{h|\bm{a}_h\}
= \{e|\bm{\nu}_{g,h}\}\{gh|\bm{a}_{gh}\}, \\
&\bm{\nu}_{g,h} = p_g \bm{a}_h + \bm{a}_g - \bm{a}_{gh} \in BL.
\end{align}
Using  $\bm{\nu}$, one can write down the factor system explicitly as 
\begin{align}
z_{g,h} e^{-i g h \bm{k} \cdot \bm{\nu}_{g,h}} U_{gh}(\bk)
= 
U_g(h \bk) U_h(\bk)
\end{align}
where $z_{g,h}$ represents the factor system of fundamental degrees of freedom under the point group.
For instance, $z_{g,h}\equiv 1$ for spinless electrons, and $z_{g,h} \in \{\pm 1\}$ for spinful electrons. 
With the twisting $e^{i \tau_{g,h}(\bk)} = z_{g,h} e^{-i \bm{k} \cdot \bm{\nu}_{g,h}}$, the $K$-group ${}^{\phi}K^{(\tau,c)-0}_G(T^3)$ is defined.
For finite integer gradings $n>0$, the $K$-group ${}^{\phi}K^{(\tau,c)-n}_G(T^3)$ is defined by adding chiral symmetries $\Gamma_i$, $\{ \Gamma_i, H(\bk)\}=0 $, with the following algebra:~\cite{SSG17}
\begin{align}
&\{\Gamma_i, \Gamma_j\} = 2 \delta_{ij}, \nonumber\\
&
\Gamma_i U_g(\bk) = c(g) U_g(\bk) \Gamma_i, \qquad (g \in G).
\label{eq:shift_n}
\end{align}
One can show the Bott periodicity ${}^{\phi}K^{(\tau,c)-(n+8)}_G(T^3) \cong {}^{\phi}K^{(\tau,c)-n}_G(T^3)$.
In Appendix~\ref{sec:factor}, we summarize the list of factor systems for $n>0$ much relevant to the condensed matter physics. 

Let us move on to the $E_1$-page. 
Let $X_0 \subset X_1 \subset X_2 \subset X_3 = T^3$ be $G$-filtration associated to a cell decomposition described in Sec.~\ref{sec:cell_deco}.
From (\ref{eq:e1_real_az}), the $E_1$-page is given by the space of representations incorporating TRS and PHS at $p$-cells 
\begin{align}
E_1^{p,-n} = \prod_{j \in I^p_{\rm orb}} {}^{\phi|_{D^p_j}} K^{(\tau,c)|_{D^p_j}-n}_{G_{D^p_j}}(D^p_j).
\end{align}
The problem is recast to finding irreps at a point in the $p$-cell $D^p_j$, which can be systematically solved by using the Wigner criteria~\cite{InuiGroup} and the generalization thereof in the presence of PHS (see Appendix \ref{app:wigner}). 
The little group $G_{\bk}$ at $\bk$ is the subgroup of $G$ so that $g \in G_{\bk}$ fixes the point $\bk$, i.e.\ $G_{\bk} = \{g \in G | g \bk = \bk\}$.
The little group $G_{\bk}$ splits into the disjoint union of left cosets as 
\begin{align}
G_{\bk} 
&= \underbrace{G_{\bk}^0}_{\rm unitary\ symmetries} \nonumber\\
&\sqcup \underbrace{a G_{\bk}^0}_{\rm magnetic\ symmetries} \nonumber\\
&\sqcup \underbrace{b G_{\bk}^0}_{\rm particle-hole\ symmetries} \nonumber\\
&\sqcup \underbrace{ab G_{\bk}^0}_{\rm magnetic\ particle-hole\ symmetries}, 
\label{eq:G_deco}
\end{align}
where $G^0_{\bk} = \{g \in G_{\bk} | \phi(g)=c(g)=1\}$ is the subgroup of unitary symmetries, 
$a \in G$ $(\phi(a)=-c(a)=-1)$ is a magnetic symmetry group element, 
$b \in G$ $(\phi(b)=c(b)=-1)$ is a particle-hole symmetry group element, 
and $ab \in G$ $(\phi(ab)=-c(ab)=1)$ is a magnetic particle-hole symmetry group element.
In the Wigner criteria, we first determine irreps of the subgroup $G_{\bk}^0$. 
The factor system of the little group $G_{\bk}^0$ is given by 
\begin{align}
z_{g,h}^{\bk} = z_{g,h} e^{-i \bm{k} \cdot \bm{\nu}_{g,h}},
\quad g,h \in G_{\bk}^0. 
\end{align}
Let $\alpha,\beta,\dots,$ be irreps of $G_{\bk}^0$ with the factor system $z_{g,h}^{\bk}$. 
We introduce the following integer-valued quantities on each irrep
\begin{align}
&W^T_{\alpha} := \frac{1}{|G_{\bk}^0|} \sum_{g \in G_{\bk}^0} z^{\bk}_{ag,ag} \chi_{\alpha}((ag)^2) \in \{\pm1, 0\}, \label{eq:wt_trs}\\
&W^C_{\alpha} := \frac{1}{|G_{\bk}^0|} \sum_{g \in G_{\bk}^0} z^{\bk}_{bg,bg} \chi_{\alpha}((bg)^2) \in \{\pm 1,0\}, \label{eq:wt_phs}\\
&W^{\Gamma}_{\alpha}:=\frac{1}{|G_0|} \sum_{g \in G_0} \nonumber\\
&\left[\frac{z^{\bk}_{g,ab}}{z^{\bk}_{ab,(ab)^{-1}gab}} \chi_{\alpha}((ab)^{-1}gab)\right]^* \chi_{\alpha}(g)  \in \{1,0\} \label{eq:wt_cs}, 
\end{align}
which we call the Wigner indices. 
Here, $\chi_{\alpha} (g \in G_{\bk}^0)$ is the character of the irrep $\alpha$. 
See Appendix~\ref{app:wigner} for the detail.
The datum $(W^T_{\alpha}, W^C_{\alpha}, W^{\Gamma}_{\alpha})$ gives the emergent AZ class realized on the irrep $\alpha$, which determines the $E_1^{p,0}$ terms.
Table~\ref{tab:wigner_az} summarizes the relationship among Wigner indices $(W^T_{\alpha}, W^C_{\alpha}, W^{\Gamma}_{\alpha})$, emergent AZ classes, and corresponding band structures. 

\begin{table*}
	\begin{center}
	\caption{The relationship among Wigner indices $(W^T_{\alpha}, W^C_{\alpha}, W^{\Gamma}_{\alpha})$, emergent AZ classes, and band structures.}
	\label{tab:wigner_az}
\begin{align*}
\begin{array}{cccccc}
W^T_{\alpha} & {\rm AZ} &{\rm Classification} & {\rm Band\ str.}\\
\hline \hline
1 & {\rm AI} & \Z & 
\begin{minipage}{20mm}
        \scalebox{0.5}{\includegraphics[width=40mm,clip]{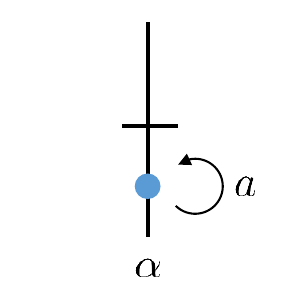}}
    \end{minipage}\\
\hline
-1 & {\rm AII} & \Z & 
\begin{minipage}{20mm}
        \scalebox{0.5}{\includegraphics[width=40mm,clip]{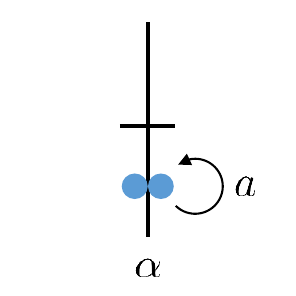}}
    \end{minipage}\\
    \hline
0 & {\rm A} & \Z & 
\begin{minipage}{20mm}
        \scalebox{0.5}{\includegraphics[width=40mm,clip]{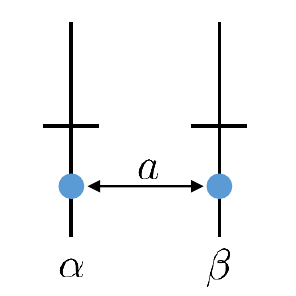}}
    \end{minipage}\\
\hline \hline
\medskip \\
W^C_{\alpha} & {\rm AZ} &{\rm Classification} & {\rm Band\ str.}\\
\hline \hline
1 & {\rm D} & \Z_2 & 
\begin{minipage}{20mm}
        \scalebox{0.5}{\includegraphics[width=40mm,clip]{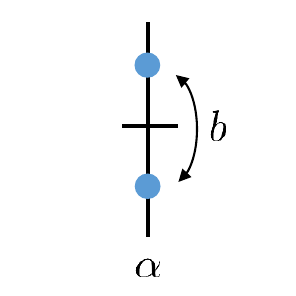}}
    \end{minipage}\\
\hline
-1 & {\rm C} & 0 & 
\begin{minipage}{20mm}
        \scalebox{0.5}{\includegraphics[width=40mm,clip]{d}}
    \end{minipage}\\
    \hline
0 & {\rm A} & \Z & 
\begin{minipage}{20mm}
        \scalebox{0.5}{\includegraphics[width=40mm,clip]{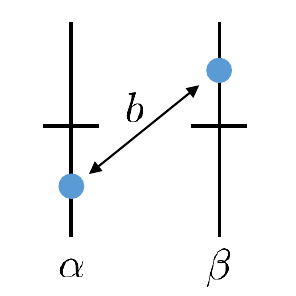}}
    \end{minipage}\\
\hline \hline
\medskip \\
W^{\Gamma}_{\alpha} & {\rm AZ} &{\rm Classification} & {\rm Band\ str.}\\
\hline \hline
1 & {\rm AIII} & 0 & 
\begin{minipage}{20mm}
        \scalebox{0.5}{\includegraphics[width=40mm,clip]{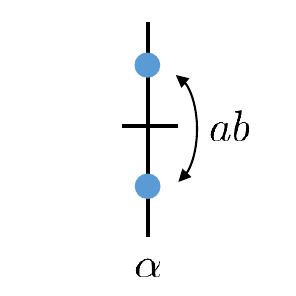}}
    \end{minipage}\\
\hline
0 & {\rm A} & \Z & 
\begin{minipage}{20mm}
        \scalebox{0.5}{\includegraphics[width=40mm,clip]{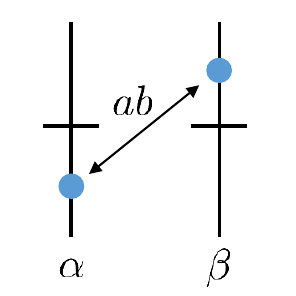}}
    \end{minipage}\\
\hline \hline
\medskip \\
\medskip \\
\ \\
\end{array} && && 
\begin{array}{ccccccccc}
W^T_{\alpha} & W^C_{\alpha} & W^{\Gamma}_{\alpha} & {\rm AZ} &{\rm Classification} & {\rm Band\ str.}\\
\hline \hline
0&0&0&{\rm A}& \Z & 
\begin{minipage}{25mm}
        \scalebox{0.5}{\includegraphics[width=55mm,clip]{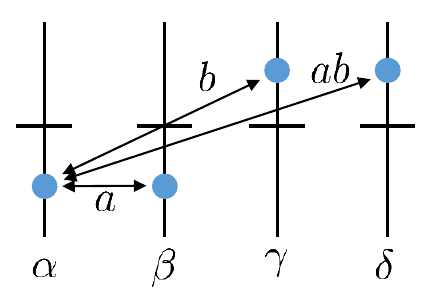}}
    \end{minipage}\\ 
    \hline
0&0&1&{\rm AIII}& 0 & 
\begin{minipage}{20mm}
        \scalebox{0.5}{\includegraphics[width=40mm,clip]{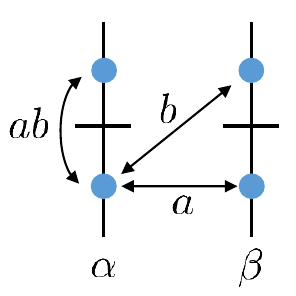}}
    \end{minipage}\\
    \hline
1&0&0&{\rm AI}& \Z & 
\begin{minipage}{20mm}
        \scalebox{0.5}{\includegraphics[width=40mm,clip]{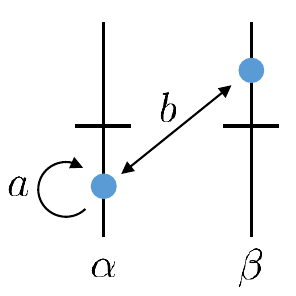}}
    \end{minipage}\\
    \hline
1&1&1&{\rm BDI}& \Z_2 & 
\begin{minipage}{20mm}
        \scalebox{0.5}{\includegraphics[width=40mm,clip]{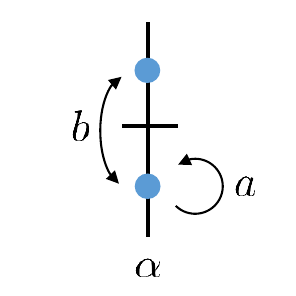}}
    \end{minipage}\\
    \hline
0&1&0&{\rm D}& \Z_2 & 
\begin{minipage}{20mm}
        \scalebox{0.5}{\includegraphics[width=40mm,clip]{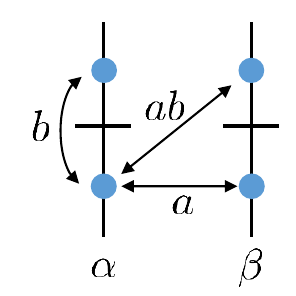}}
    \end{minipage}\\
    \hline
-1&1&1&{\rm DIII}& 0 & 
\begin{minipage}{20mm}
        \scalebox{0.5}{\includegraphics[width=40mm,clip]{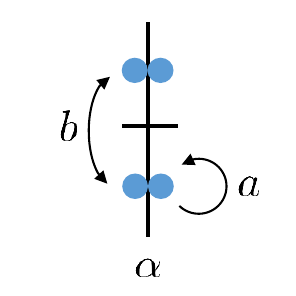}}
    \end{minipage}\\
    \hline
-1&0&0&{\rm AII}& \Z & 
\begin{minipage}{20mm}
        \scalebox{0.5}{\includegraphics[width=40mm,clip]{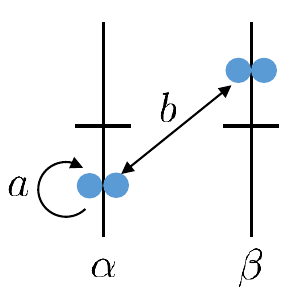}}
    \end{minipage}\\
    \hline
-1&-1&1&{\rm CII}& 0 &
\begin{minipage}{20mm}
        \scalebox{0.5}{\includegraphics[width=40mm,clip]{d3_both}}
    \end{minipage}\\
    \hline
0&-1&0&{\rm C}& 0 & 
\begin{minipage}{20mm}
        \scalebox{0.5}{\includegraphics[width=40mm,clip]{d_both}}
    \end{minipage}\\
    \hline
1&-1&1&{\rm CI}& 0 & 
\begin{minipage}{20mm}
        \scalebox{0.5}{\includegraphics[width=40mm,clip]{bd1_both}}
    \end{minipage}\\
\hline \hline
\end{array}
\end{align*}
	\end{center}
\end{table*}

Once we have determined the emergent AZ classes for the grading $n=0$, from the definition (\ref{eq:shift_n}), the emergent AZ class and the band structure for other grading $n>0$ follow the shift of symmetry classes as in 
\begin{align}
\to {\rm A} \to {\rm AIII} \to {\rm A} \to, 
\end{align}
for emergent complex AZ classes, and 
\begin{align}
&\to {\rm AI} \to {\rm BDI} \to {\rm D} \to {\rm DIII} \nonumber\\
&\qquad \to {\rm AII} \to {\rm CII} \to {\rm C} \to {\rm CI} \to {\rm AI} \to,
\end{align}
for emergent real AZ classes.

\subsection{Construction of higher differentials $d_2$ and $d_3$}
\label{sec:d2d3}
As seen in Sec.~\ref{sec:AHSS}, the construction of the matrix $d_1^{p,-n}: E_1^{p,-n} \to E_1^{p+1,-n}$ is straightforward. 
$d_1^{p,-n}$ is determined by the compatibility relation (incorporating PHS in addition to TRS) among irreps. 
On the other hand, there are no simple formulas for higher-differentials $d_r (r\geq 2)$. 
However, as a case-by-case problem, they can be constructed through a model Hamiltonian. 

As a preliminary step, consider the Hamiltonian description of the first differential $d_1^{0,-n}:E_1^{0,-n} \to E_1^{1,-n}$ as in 
\begin{align}
H(k) = 
\left\{\begin{array}{ll}
(k^2-\mu)  & ({\rm without\ PHS}), \\
(k^2-\mu)\tau_z & ({\rm with\ PHS}, \, C=\tau_x K). \\
\end{array}\right.
\label{eq:fermi_surface_d1}
\end{align}
Here, $k$ is the distance from the 0-cell we are considering. 
The degeneracy of the irrep, if any, is implicit. 
This Hamiltonian describes how gapless points appear on adjacent 1-cells when the irrep at the 0-cell passes the zero energy.
By identifying the irreps at the 0-cell and the gapless points on the 1-cells with elements of $E_1^{0,-n}$ and $E_1^{1,-n}$, respectively, and taking into account the signs related to the orientations of these cells at the same time, the model Hamiltonian (\ref{eq:fermi_surface_d1}) provides matrix elements of $d_1^{0,-n}$.

In a similar manner, we may construct the second differential $d_2^{0,-n}: E_2^{0,-n} \to E_2^{2,-(n+1)}$ by using a model Hamiltonian.
Since $E_2^{0,-n} = \ker (d_1^{0,-n})$, $E_2^{0,-n}$ represents sets of irreps at 0-cells that are smoothly extended to the 1-skeleton $X_1$ without gapless states on it. 
Some elements of $E_2^{0,-n}$ can give rise to a gapless Dirac point in 2-cells, which is detected by $d_2^{0,-n}$.     
To describe this process, we generalize Eq. (\ref{eq:fermi_surface_d1}) to 2-dimensions. 
Below, we construct model Hamiltonians for two examples. 

\medskip 
\begin{figure}[!]
	\begin{center}
	\includegraphics[width=\linewidth, trim=5cm 8cm 5cm 0cm]{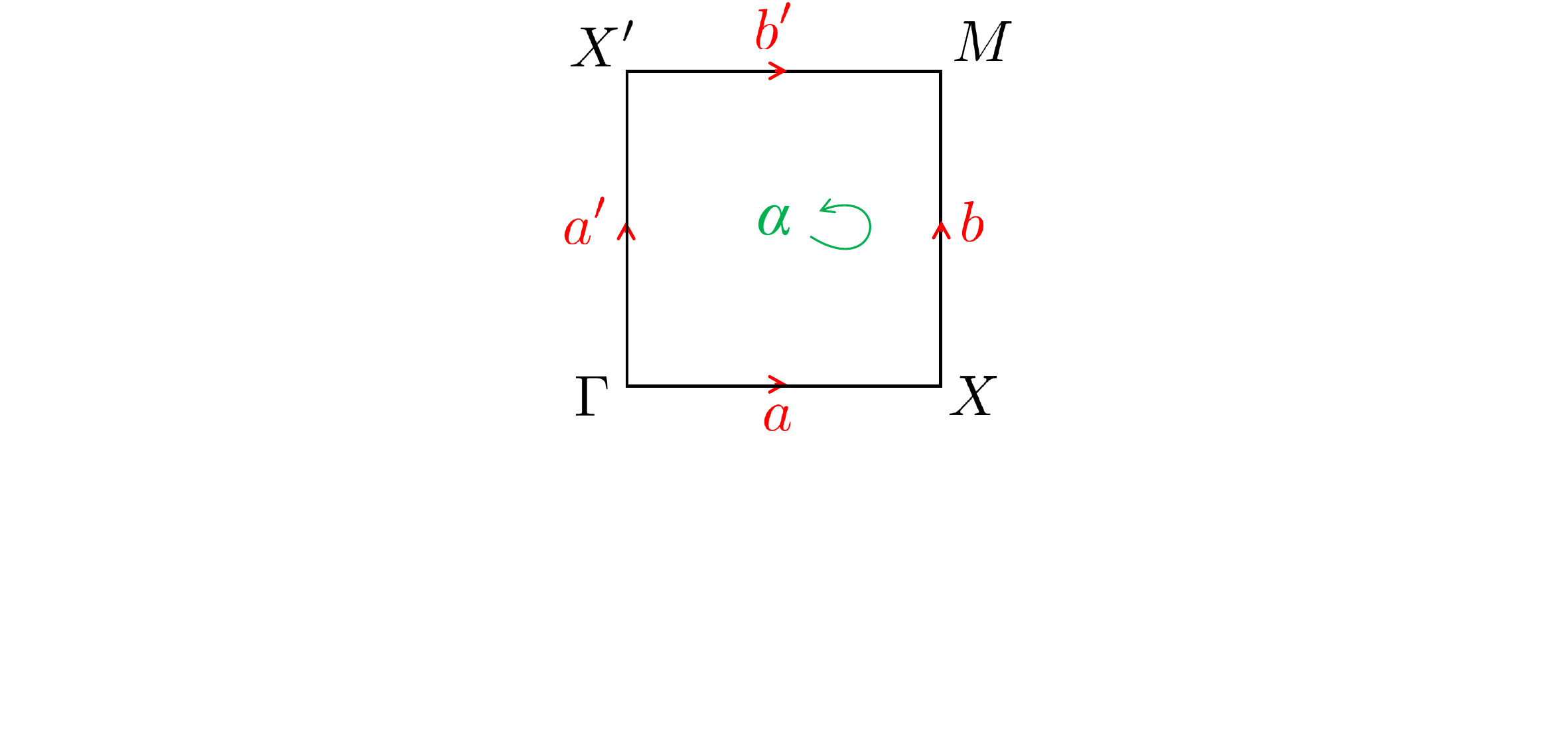}
	\end{center}
	\caption{A $C_4$ symmetric cell decomposition of 2-torus. 
	The figure shows one fourth of the BZ.}
	\label{fig:c4}
\end{figure}

\noindent 
{\it $2d$ class AI with $C_4$ rotation symmetry (spinless electrons with the wallpaper group $p41'$)---}
The TRS $T$ and 4-fold rotation $C_4$ acts on the 2d BZ as $T: (k_x,k_y) \mapsto (-k_x,-k_y)$ and $C_4: (k_x,k_y)\mapsto (-k_y,k_x)$, respectively. 
The factor system is $T^2=1, C_4^4=1$ and $T C_4 = C_4 T$. 
We employ a $C_4$ symmetric cell decomposition as in Fig. \ref{fig:c4}.
A gapless Dirac point in the 2-cell is stabilized by a $\pi$-Berry phase introduced by the symmetry operator $TC_2$ with $(TC_2)^2 =1$. 
First, consider the creation of Dirac points from the $C_4$-invariant 0-cell, say the $\Gamma$ point (or the $M$ point). 
To describe a band inversion, we consider a 2d generalization of  Eq.(\ref{eq:fermi_surface_d1}),  $H=(\bk^2-\mu)^2$.
We also need another term in order to remove gapless point on adjacent 1-cells. 
For this purpose, we add the $d_{x^2-y^2}$-density wave $\propto (k_x^2-k_y^2)$. 
Since $(k_x^2-k_y^2)$ gives a $(-1)$ sign under $C_4$, to make the Hamiltonian $C_4$-symmetric, an orbital degrees of freedom compensating the $(-1)$ sign is necessary.
Finally, the Hamiltonian for $d_2^{0,0}$ around the $\G/M$ point is given by
\begin{align}
&H_{\Gamma/M}(k_x,k_y)
= 
(\bk^2-\mu)\sigma_z + (k_x^2-k_y^2) \sigma_x, \nonumber\\ 
&T=K, \quad C_4= \sigma_z,
\end{align}
where $\sigma_i$ is the Pauli matrix in the orbital space.
This Hamiltonian describes the process that when $\mu$ changes the sign the band inversion between the $C_4=1$ and $c_4=-1$ orbits occurs at $\G/M$, and a quartet of Dirac points appear in the 2-cells. 
Similarly, the creation of Dirac points from the $C_2$-invariant 0-cell (the $X$ point) is described by 
\begin{align}
&H_{X}(k_x,k_y)
= 
(\bk^2-\mu)\sigma_z + (k_x+k_y) \sigma_x, \nonumber\\
&T= K, \quad C_2=\sigma_z.
\end{align}
These Hamiltonians specify the second differentials $d_2^{0,0}$ from $\Gamma$, $M$ and $X$ to the 2-cell $\alpha$. 

\medskip 

\noindent
{\it $2d$ class AII with $C_4$ rotation symmetry (spinful electrons with the wallpaper group $p41'$)---}
It is instructive to compare the previous example with the case of class AII systems with $C_4$ symmetry.
For spinful electrons, the factor system is given by $T^2=-1, C_4^4=-1$ and $T C_4 = C_4 T$. 
A gapless Dirac point in the 2-cell again has the quantized $\pi$-Berry phase protected by the symmetry $TC_2$ with $(TC_2)^2 = 1$. 
However, in contrast to the class AI case, irreps at $\Gamma$, $M$ are 2-dimensional, and thus a band inversion can create only an even number of Dirac points, which are not protected by the $\pi$-Berry phase.
Also, at the $X$ point, no band inversion may occur because the irrep is unique. 
Thus, we can conclude that $d_2^{0,0}$ is trivial.

\medskip

The third differential $d_3^{0,-n}: E_3^{0,-n} \to E_3^{3,-(n+2)}$ may be constructed similarly. 
$d_3$ describes how Weyl points are created in 3-cells by band inversion at 0-cells. 
Here we give an example. 

\medskip

\noindent
{\it $3d$ class D (3d TRS-breaking superconductors)---}
The PHS $C$ acts on 3d BZ as $C:(k_x,k_y,k_z)\mapsto (-k_x,-k_y,-k_z)$. 
The factor system is $C^2=1$.
We set $C=\tau_x K$, where $\tau_{\mu\in \{x,y,z\}}$ is the Pauli matrices in the Nambu space. 
We consider Fig.\ref{fig:2} as a particle-hole symmetric cell decomposition.
Each high symmetry point $\bk_0$ of PHS is a fixed point of $C$, and the $\Z_2$-valued Pfaffian invariant $\pf[\tau_xH(\bk_0)]$ is defined. 
A band inversion at a high symmetric point may change the $\Z_2$ invariant, and creates two Weyl points in the opposite direction at the same time.
This process is described by the Hamiltonian 
\begin{align}
&H(k_x,k_y,k_z)
=
(\bk^2-\mu) \tau_z + k_x \tau_x + k_y \tau_y,\nonumber\\
&C = \tau_x K,
\end{align}
where the origin of $\bk$ denotes a high symmetry point. 
We conclude that the corresponding $d_3^{0,0}$ is nontrivial.

\subsection{Indicators for gapless phases}
\label{sec:ind}
From the definition of the differentials $d_1^{0,-n}$, $d_2^{0,-n}$ and $d_3^{0,-n}$, it is clear that these kernels serve as the indicators detecting bulk gapless phases characterized by topological invariants at 0-cells. 
Recall that $E_1^{0,-n}$ is the data of topological invariants at 0-cells. 
In $E_1^{0,-n}$, elements that satisfy the compatibility relation, namely $E_2^{0,-n}=\ker (d_1^{0,-n}) \subset E_1^{0,-n}$, can glue together in the 1-skeleton $X_1$. 
The Abelian group $E_1^{0,-n}/E_2^{0,-n}$ describes bulk gapless phases with a gapess point inside a 1-cell. 
In the same way, $E_2^{0,-n}/E_3^{0,-n}$ and $E_3^{0,-n}/E_4^{0,-n}$ describe bulk gapless phases thereof, which is summarized as follows:  
\begin{itemize}
\item 
$E_1^{0,-n}/E_2^{0,-n}$ is the indicator for the existence of a topological gapless point inside a 1-cell under the assumption that the Hamiltonian is gapped at all 0-cells. 
\item 
$E_2^{0,-n}/E_3^{0,-n}$ is the indicator for the existence of a topological gapless point inside a 2-cell under the assumption that the Hamiltonian is gapped on the 1-skeleton $X_1$. 
\item 
$E_3^{0,-n}/E_4^{0,-n}$ is the indicator for the existence of a topological gapless point inside a 3-cell under the assumption that the Hamiltonian is gapped on the 2-skeleton $X_2$. 
\end{itemize}

\subsection{Case studies}
\label{sec:casestudy_au}
We give four examples of the AHSS with TRS and/or PHS. 

\subsubsection{$2d$ time-reversal symmetric spinless systems with two-fold rotation symmetry}
\begin{table*}
	\begin{center}
	\caption{
	The factor systems and the homomorphism $c:G\to \Z_2$ for integer grading $n\geq 0$ for spinless systems with TRS and $C_2$ symmetry. 
	}
	\label{tab:fac_2d_p21'}
$$
\renewcommand{\arraystretch}{1.5}
\begin{array}{l|l|l|c|l|ll}
{\rm AZ} & n & z_{T,T}, z_{C,C} & z_{T,g}/z_{g,T}, z_{C,g}/z_{g,C} & z_{g,h} & c(g) \\
\hline
{\rm AI} & n=0
&T^2=1&&&\\
\cline{1-3}
{\rm BDI} & n=1
&T^2=1, C^2=1&&&\\
\cline{1-3}
{\rm D} & n=2
&C^2=1&T C_2(\bk) = C_2(-\bk) T&&\\
\cline{1-3}
{\rm DIII} & n=3 
&T^2=-1, C^2=1&{\rm and/or}&C_2(-\bk) C_2(\bk)=1 & C_2(\bk) H(\bk) = H(-\bk) C_2(\bk)\\
\cline{1-3}
{\rm AII} & n=4
&T^2=-1&C C_2(\bk) = C_2(-\bk) C&&\\
\cline{1-3}
{\rm CII} & n=5 
&T^2=-1,C^2=-1&  & &\\
\cline{1-3}
{\rm C} & n=6
&C^2=-1&& &\\
\cline{1-3}
{\rm CI} & n=7
&T^2=1,C^2=1&& &\\
\hline
\end{array}
\renewcommand{\arraystretch}{1}
$$
\end{center}
\end{table*}
Let us consider, as the initial grading $n=0$, 2-dimensional spinless insulators with TRS $T$ and the two-fold rotation symmetry $C_2$, where both the generators act on the 2d BZ torus as an inversion $T, C_2:(k_x,k_y) \mapsto (-k_x,-k_y)$. 
The symmetry group is $G=\Z_2 \times \Z_2^T$. 
A $G$ symmetric cell decomposition is given by the same one in Fig.~\ref{fig:2d} [a]. 
The factor systems for $n=0$ as well as $n>0$ are summarizes as in Table.~\ref{tab:fac_2d_p21'}. 
At a 0-cell $\bk_0 \in \{\Gamma, X, Y, M\}$, the unitary subgroup introduced in (\ref{eq:G_deco}) is $G_{\bk_0}^0 = \Z_2 = \{e,C_2\}$ and the Wigner criterion for each irrep $C_2=\pm 1$ is given by $W_{C_2=\pm 1}^{T} =1$, namely, the emergent AZ class is AI. 
On 1- and 2-cells, the little group is $G_{\bk} = \{e,TC_2\}$ and the emergent AZ class is AI because $(TC_2(\bk))^2=1$.
On the basis of the emergent AZ classes realized in $p$-cells, we get the $E_1$-page 
\begin{align}
\begin{array}{cc|cccc}
{\rm AI} & n=0 & \Z^2+\Z^2+\Z^2+\Z^2 & \Z+\Z+\Z & \Z \\
{\rm BDI} & n=1 & \Z_2^2+\Z_2^2+\Z_2^2+\Z_2^2 & \Z_2+\Z_2+\Z_2 & \Z_2 \\
{\rm D} & n=2 & \Z_2^2+\Z_2^2+\Z_2^2+\Z_2^2& \Z_2+\Z_2+\Z_2 & \Z_2 \\
{\rm DIII} & n=3 & 0 & 0 & 0 \\
{\rm AII} & n=4 & \Z^2+\Z^2+\Z^2+\Z^2 & \Z+\Z+\Z & \Z \\
{\rm CII} & n=5 & 0 & 0 & 0 \\
{\rm C} & n=6 & 0 & 0 & 0 \\
{\rm CI} & n=7 & 0 & 0 & 0 \\
\hline 
& E_1^{p,-n} & \{\Gamma,X,Y,M\} & \{ a,b,c\} & \{ \alpha\} \\ 
& & p=0 & p=1 & p=2 
\end{array}
\end{align}
The first differential $d_1^{p,-n}$ is straightforwardly given by the compatibility relation, i.e.\ how irreps at $p$-cells split into representations at adjacent $(p+1)$-cells. 
We get the followings.
\begin{align}
&d_1^{0,0} = d_1^{0,-4}\nonumber\\
&=\begin{array}{|cc|cc|cc|cc|c}
\Gamma && X && Y && M && \\
1 & -1 & 1 & -1 & 1 & -1 & 1 & -1 & \\
\hline
1&1&-1&-1&&&&&a\\
&&&&1&1&-1&-1&b\\
&&1&1&&&-1&-1&c\\
\hline
\end{array}, 
\end{align}
$d_1^{0,-2} = d_1^{0,-1} = d_1^{0,0}$ (mod 2), and other first differentials are zeros. 
We have the $E_2$-page $E_2^{p,-n} = \ker (d_1^{p,-n}) / \im (d_1^{p-1,-n})$ 
\begin{align}
=
\begin{array}{cc|cccc}
{\rm AI} & n=0 & \Z^5 & 0 & \Z \\
{\rm BDI} & n=1 & \Z_2^5 & 0 & \Z_2 \\
{\rm D} & n=2 & \Z_2^5 & 0 & \Z_2 \\
{\rm DIII} & n=3 & 0 & 0 & 0 \\
{\rm AII} & n=4 & \Z^5 & 0 & \Z \\
{\rm CII} & n=5 & 0 & 0 & 0 \\
{\rm C} & n=6 & 0 & 0 & 0 \\
{\rm CI} & n=7 & 0 & 0 & 0 \\
\hline 
& E_2^{p,-n} & p=0 & p=1 & p=2 \\
\end{array}.
\end{align}

We find that the second differential $d_2^{0,0}$ is nontrivial: 
$E_1^{2,-1} = \Z_2$ indicates the class AI point node in the 2-cell $\alpha$ described by the Hamiltonian 
\begin{align}
H(k_x,k_y) = k_x \sigma_z +k_y \sigma_x, \qquad 
TC_2(\bk)=K, 
\label{eq:2d_ai_point_node}
\end{align}
where $K$ is the complex conjugate. 
The point node (\ref{eq:2d_ai_point_node}) is stabilized by the $\pi$-Berry phase, and can be removed by the band inversion at $0$-cells described by the Hamiltonian 
\begin{align}
&d_2^{0,0}|_{\Gamma,X,Y,M \to \alpha}:\nonumber\\
&\left\{\begin{array}{ll}
H(k_x,k_y)=(k^2-\mu)\sigma_z+\bk \cdot \bm{n} \sigma_x, \\
T=\sigma_z K, \\
C_2=\sigma_z, 
\end{array}\right.
\end{align}
where $\bm{n}$ is a unit vector. 
Similarly, the second differential $d_2^{0,-1}$ is nontrivial: 
The class BDI point node of $E_1^{2,-2} = \Z_2$ in the 2-cell $\alpha$ is described as 
\begin{align}
&H(k_x,k_y) = k_x \tau_z +k_y \sigma_y \tau_y, \nonumber\\
&TC_2(\bk)=K, \qquad 
\Gamma=\tau_x. 
\label{eq:2d+c2_bdi_gapless}
\end{align}
This point node is removed by a band inversion at 0-cells followed by creating a point node to the 2-cell $\alpha$ described by the Hamiltonian 
\begin{align}
&d_2^{0,-1}|_{\Gamma,X,Y,M \to \alpha}:\nonumber\\
&\left\{\begin{array}{ll}
H(k_x,k_y) = (k^2-\mu) \tau_z + \bk \cdot \bm{n} \sigma_y \tau_y, \\
TC_2(\bk)=K, \\
\Gamma=\tau_x, \\
C_2=\sigma_z, 
\end{array}\right.
\end{align}
with $\bm{n}$ a unit vector. 
We arrive at the $E_3 (=E_{\infty})$-page 
\begin{align}
\begin{array}{cc|cccc}
{\rm AI} & n=0 & \Z^5 & 0 & \Z \\
{\rm BDI} & n=1 & \Z_2^4 & 0 & 0 \\
{\rm D} & n=2 & \Z_2^5 & 0 & 0 \\
{\rm DIII} & n=3 & 0 & 0 & 0 \\
{\rm AII} & n=4 & \Z^5 & 0 & \Z \\
{\rm CII} & n=5 & 0 & 0 & 0 \\
{\rm C} & n=6 & 0 & 0 & 0 \\
{\rm CI} & n=7 & 0 & 0 & 0 \\
\hline 
& E_3^{p,-n} & p=0 & p=1 & p=2 \\
\end{array}, 
\end{align}
and the $K$-groups read as 
\begin{equation}\begin{split}
&{}^{\phi}K^{-0}_{\Z_2 \times \Z_2^T}(T^2) = \Z^5, \\
&{}^{\phi}K^{-1}_{\Z_2 \times \Z_2^T}(T^2) = \Z_2^5, \\
&0 \to \Z \to {}^{\phi}K^{-2}_{\Z_2 \times \Z_2^T}(T^2) \to \Z_2^5 \to 0, \\
&{}^{\phi}K^{-3}_{\Z_2 \times \Z_2^T}(T^2) = 0, \\
&{}^{\phi}K^{-4}_{\Z_2 \times \Z_2^T}(T^2) = \Z^5, \\
&{}^{\phi}K^{-5}_{\Z_2 \times \Z_2^T}(T^2) = 0, \\
&{}^{\phi}K^{-6}_{\Z_2 \times \Z_2^T}(T^2) = \Z, \\
&{}^{\phi}K^{-7}_{\Z_2 \times \Z_2^T}(T^2) = 0. 
\end{split}\end{equation}
These are consistent with the $K$-groups computed in Ref.~\cite{SS14}, where ${}^{\phi}K^{-2}_{\Z_2 \times \Z_2^T}(T^2) = \Z + \Z_2^4$.

We, here, illustrate the different interpretations (I)-(IV) of the $E_1$-page in Sec.~\ref{sec:ahss_real} for the group $E_1^{2,-2} = \Z_2$.
Let us denote the subgroup $\{e,T C_2\} \subset \Z_2 \times \Z_2^T$ by $\Z_2^{TC_2}$. 
In the followings, $R_s$ represents the classifying space of the emergent AZ class $s$.

\noindent
(IV)---
$E_1^{2,-2} \cong {}^{\phi} K^{-2}_{\Z_2^{TC_2}}(D^2_{\alpha}) \cong \pi_0(R_2) \cong \Z_2$ is the space of representations (incorporating PHS) of the $2$-cell $\alpha$ with the symmetry class $n=2$ (class D). 

\noindent
(I)---
$E_1^{2,-2} \cong {}^{\phi} K^{-0}_{\Z_2^{TC_2}}(D^2_{\alpha},\p D^2_{\alpha})$ is the classification of topological insulators on the 2-cell $\alpha$ for symmetry class $n=0$ (class AI) with the  condition that the Hamiltonian is constant on the boundary $\p D^2_{\alpha}$. 
Using the isomorphism ${}^{\phi} K^{-0}_{\Z_2^{TC_2}}(D^2_{\alpha},\p D^2_{\alpha}) \cong {}^{\phi} \wt K^{-0}_{\Z_2^{TC_2}}(S^2_{\alpha}) \cong \pi_2(R_0) = \Z_2$, we confirm that $E_2^{2,-2}=\Z_2$.
A model Hamiltonian is given by
\begin{align}
&H(k_x,k_y) = k_x \sigma_x \tau_x + k_y \sigma_z \tau_x + (m-\epsilon k^2)\tau_z,\nonumber\\
&TC_2(\bk) = K.	
\end{align}

\noindent
(II)---
Using the isomorphism ${}^{\phi} \wt K^{-0}_{\Z_2^{TC_2}}(S^2_{\alpha}) \cong {}^{\phi} \wt K^{-1}_{\Z_2^{TC_2}}(S^1_{\alpha}) \cong \pi_1 (R_1)= \Z_2$, we can interpret $E_2^{2,-2}$ as the $\Z_2$ topological invariant defined on a circle $S^1_{\alpha}$ enclosing the topological gapless point for symmetry class $n=1$ (class BDI), which we already showed in (\ref{eq:2d+c2_bdi_gapless}). 

\begin{figure*}[!]
	\begin{center}
	\includegraphics[width=0.8\linewidth, trim=0cm 1cm 0cm 0cm]{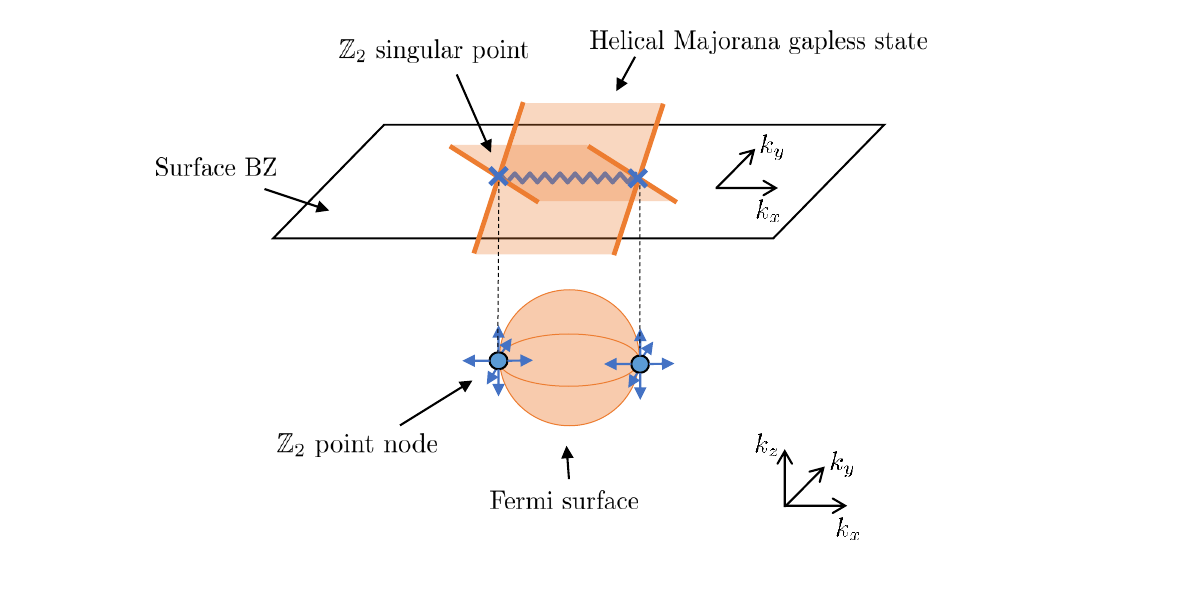}
	\end{center}
	\caption{Interpretation of $E_1^{2,-2} = \Z_2$ as topological singular points in class D superconductors with $C_2$ rotation symmetry. 
	The figure shows the corresponding topological point nodes in the 3-dimensional superconductor (\ref{eq:2d+c2_b_singular_bulk}). }
	\label{fig:2daz_ai_c2}
\end{figure*}
\noindent
(III)---
Applying the bulk-boundary correspondence to the circle $S^1_{\alpha}$ enclosing the topological gapless point in the interpretation (II), we have a gapless helical Majorana mode on $S^1_{\alpha}$ with the symmetry class $n=2$ (class D).
The existence of the topological gapless states on $S^1_{\alpha}$ implies that there must be a singular point inside the circle $S^1_{\alpha}$. 
Such a singular point appears on the surface of 3-dimensional gapless class D superconductors with $C_2^z$ rotation symmetry. 
A model Bogoliubov de Gennes (BdG) Hamiltonian is given by 
\begin{align}
&H(k_x,k_y,k_z) = \left( \frac{k^2}{2 m} - \mu \right) \tau_z + k_y \sigma_z \tau_x + k_z \tau_y\nonumber\\
&C=\tau_x K, \qquad C_2^z=\sigma_z
\label{eq:2d+c2_b_singular_bulk}
\end{align}
around the $\Gamma$ point, where the gap function is $\Delta(\bk) = (k_y \sigma_x + k_z \sigma_y) i \sigma_y$. 
The combined symmetry $C C_2: (k_x,k_y,k_z) \mapsto (k_x,k_y,-k_z)$ defines the $\Z_2$ invariant $(-1)^{\nu}$ on a plane $\Sigma = S^1_{xy} \times S^1_{z}$ with $S^1_{xy}$ a circle on the $k_xk_y$-plane and $S^1_{z}$ the circle along the $k_z$-direction.~\cite{SS14}
We see that the gapless points at $\bk = (\pm \sqrt{2 m \mu},0,0)$ have nontrivial $\Z_2$ charge with respect to $(-1)^{\nu}$, i.e., when a plane $\Sigma$ passes the gapless point the $\Z_2$ invariant $(-1)^{\nu}$ is flipped. 
Between the two gapless points, $-\sqrt{2 m \mu} < k_x < \sqrt{2 m \mu}$, there appears the $\Z_2$ gapless helical Majorana state on the surface BZ. 
(Here, the surface was defined to be perpendicular to the $z$-axis in order to preserve the $C_2^z$ rotation symmetry.)
The projection of the $\Z_2$ topological gapless points in bulk to the surface BZ becomes the singular points, each of which is described by $E_1^{2,-2}=\Z_2$. 
Fig.~\ref{fig:2daz_ai_c2} illustrates the relation between the surface singular points and bulk gapless points of the Hamiltonian (\ref{eq:2d+c2_b_singular_bulk}).

\subsubsection{$2d$ time-reversal symmetric superconductors with the half lattice translation symmetry}
Let us consider $2d$ spinful time-reversal symmetric superconductors with the half lattice translation symmetry: $L_{\hat x/2}: (x,y) \mapsto (x+1/2,y)$.  
We assume the gap function is odd under the half lattice translation,
$L_{\hat x/2}(\bk) \Delta(\bk)L_{\hat x/2}^T(-\bk)=-\Delta(\bk)$, 
while the normal Hamiltonian ${\cal E}(\bk)$ is invariant, $L_{\hat x/2}(\bk){\cal E}(\bk) L^{-1}_{\hat x/2}(\bk)={\cal E}(\bk)$.
For the Nambu space of the BdG Hamiltonian
\begin{align}
H(\bk)=\left(
\begin{array}{cc}
{\cal E}(\bk) &\Delta(\bk)\\
\Delta^{\dagger}(\bk) & -{\cal E}^t(-\bk)
\end{array}
\right),
\end{align}
the half lattice translation is given by
\begin{align}
\tilde{L}_{\hat x/2}(\bk)=\left(
\begin{array}{cc}
L_{\hat x/2}(\bk) & 0 \\
0 &-L_{\hat x/2}^*(-\bk)
\end{array}
\right),
\label{eq:htodd}
\end{align}
which yields the following symmetry for the BdG Hamiltonian,
\begin{align}
\tilde{L}_{\hat x/2}(\bk) H(\bk) \tilde{L}^{-1}_{\hat x/2}(\bk)=H(\bk).
\end{align}
Here $\tilde{L}_{\hat x/2}(\bk)$ satisfies  $C\tilde{L}_{\hat x/2}(\bk)=-L_{\hat x/2}(-\bk)C$, because of the minus sign in Eq.(\ref{eq:htodd}), which reflects the odd representation of the gap function under $L_{\hat x/2}$.
In general, the representation of the gap function under the symmetry group is encoded in the twist between PHS and unbroken symmetries in the Nambu space.
Once the twist is identified, only symmetry operators in the Nambu space are needed.  
Therefore, we only use $\tilde{L}_{\hat x/2}$  below, and we omit the symbol of ``tilde" in the notation for simplicity.
It has been known that there is a $\Z_4$ invariant for this symmetry class.~\cite{SSG16}

The symmetry group is $\Z_2 \times \Z_2^{T}$, which is generated by the half lattice translation ($\Z_2$) and TRS ($\Z_2^T$). 
The factor system reads~\footnote{
The half translation symmetry in the $xy$-plane can be naturally realized as the glide reflection $G_z: (x,y,z) \mapsto (x+1/2,y,-z)$ with $z=0$ (glide plane).
Since the glide operator satisfies $G_z^2=-e^{-ik_x}$, we impose $L^2_{\hat x/2}(\bk)=-e^{-i k_x}$ on 
the factor systems in Eq. (\ref{eq:2d_TRSC_half_lattice_tr_fact_sys}).
}
\begin{align}
\left\{\begin{array}{ll}
T^2=-1, \qquad C^2=1, \\
L^2_{\hat x/2}(\bk)=-e^{-ik_x}, \\
TL_{\hat x/2}(\bk)=L_{\hat x/2}(-\bk)T, \\ CL_{\hat x/2}(\bk)=-L_{\hat x/2}(-\bk)C, 
\end{array}\right. 
\label{eq:2d_TRSC_half_lattice_tr_fact_sys}
\end{align}
for the grading $n=3$ (DIII). 
According to Table~\ref{tab:shift_sym} in Appendix~\ref{sec:factor}, the factor systems for other AZ classes are shown in  Table~\ref{tab:fac_2d_ht}.

\begin{table*}
	\begin{center}
	\caption{
	The factor systems and the homomorphism $c:G\to \Z_2$ for integer grading$n\geq 0$ for time-reversal symmetric superconductors with the half lattice translation symmetry.
	Here, we assume the square of the half-lattice translation is the fermion parity, as for the glide plane of spinful electrons. 
	}
	\label{tab:fac_2d_ht}
$$
{\small
\renewcommand{\arraystretch}{1.5}
\begin{array}{l|l|l|l|l|l}
{\rm AZ} & n & z_{T,T}, z_{C,C} & z_{T,g}/z_{g,T}, z_{C,g}/z_{g,C} & z_{g,h} & c(g) \\
\hline
{\rm AI} & n=0
&T^2=1&TL_{\hat x/2}(\bk)=L_{\hat x/2}(-\bk)T& &L_{\hat x/2}(\bk)H(\bk)=-H(\bk)L_{\hat x/2}(\bk)\\
\cline{1-4}\cline{6-6} 
{\rm BDI} & n=1
&T^2=1&TL_{\hat x/2}(\bk)=-L_{\hat x/2}(-\bk)T& &L_{\hat x/2}(\bk)H(\bk)=H(\bk)L_{\hat x/2}(\bk)\\
&&C^2=1&CL_{\hat x/2}(\bk)=L_{\hat x/2}(-\bk)C&&\\
\cline{1-4}\cline{6-6} 
{\rm D} & n=2
&C^2=1&CL_{\hat x/2}(\bk)=L_{\hat x/2}(-\bk)C& &L_{\hat x/2}(\bk)H(\bk)=-H(\bk)L_{\hat x/2}(\bk)\\
\cline{1-4}\cline{6-6} 
{\rm DIII} & n=3 
&T^2=-1&TL_{\hat x/2}(\bk)=L_{\hat x/2}(-\bk)T& &L_{\hat x/2}(\bk)H(\bk)=H(\bk)L_{\hat x/2}(\bk)\\
&&C^2=1&CL_{\hat x/2}(\bk)=-L_{\hat x/2}(-\bk)C&L^2_{\hat x/2}(\bk)=-e^{-ik_x} &\\
\cline{1-4}\cline{6-6} 
{\rm AII} & n=4
&T^2=-1&TL_{\hat x/2}(\bk)=L_{\hat x/2}(-\bk)T& &L_{\hat x/2}(\bk)H(\bk)=-H(\bk)L_{\hat x/2}(\bk)\\
\cline{1-4}\cline{6-6} 
{\rm CII} & n=5 
&T^2=-1&TL_{\hat x/2}(\bk)=-L_{\hat x/2}(-\bk)T& &L_{\hat x/2}(\bk)H(\bk)=H(\bk)L_{\hat x/2}(\bk)\\
&&C^2=-1&CL_{\hat x/2}(\bk)=L_{\hat x/2}(-\bk)C&&\\
\cline{1-4}\cline{6-6} 
{\rm C} & n=6
&C^2=-1&CL_{\hat x/2}(\bk)=L_{\hat x/2}(-\bk)C& &L_{\hat x/2}(\bk)H(\bk)=-H(\bk)L_{\hat x/2}(\bk)\\
\cline{1-4}\cline{6-6} 
{\rm CI} & n=7
&T^2=1&TL_{\hat x/2}(\bk)=L_{\hat x/2}(-\bk)T& &L_{\hat x/2}(\bk)H(\bk)=H(\bk)L_{\hat x/2}(\bk)\\
&&C^2=-1&CL_{\hat x/2}(\bk)=-L_{\hat x/2}(-\bk)C&&\\
\hline
\end{array}
\renewcommand{\arraystretch}{1}
}
$$
\end{center}
\end{table*}

As a $(\Z_2 \times \Z_2^T)$ symmetric cell decomposition of the 2-torus, we use Fig.~(\ref{fig:2d}) [a] again. 
The emergent AZ class for each $p$-cell at $n=0$ is readily obtained. 
The $E_1$-page is given by 
\begin{align}
\begin{array}{cc|cccc}
{\rm AI} & n=0 & 0+\Z_2+0+\Z_2 & 0 & 0 \\
{\rm BDI} & n=1 & \Z+\Z_2+\Z+\Z_2 & \Z+\Z+\Z & \Z \\
{\rm D} & n=2 & \Z_2+0+\Z_2+0 & 0 & 0 \\
{\rm DIII} & n=3 & \Z_2+\Z+\Z_2+\Z & \Z+\Z+\Z & \Z \\
{\rm AII} & n=4 & 0 & 0 & 0 \\
{\rm CII} & n=5 & \Z+0+\Z+0 & \Z+\Z+\Z & \Z \\
{\rm C} & n=6 & 0 & 0 & 0 \\
{\rm CI} & n=7 & 0+\Z+0+\Z & \Z+\Z+\Z & \Z \\
\hline 
& E_1^{p,-n} & \{\Gamma,X,Y,M\} & \{ a,b,c\} & \{ \alpha\} \\ 
& & p=0 & p=1 & p=2 
\end{array}.
\end{align}
From the compatibility relation, the first differentials are given as 
\begin{align}
&d_1^{0,-1}=\begin{array}{|cccc|c}
\Gamma&X&Y&M&\\
\hline
1&0&0&0&a\\
0&0&1&0&b\\
0&0&0&0&c\\
\hline
\end{array}, \\
&d_1^{0,-3}=\begin{array}{|cccc|c}
\Gamma&X&Y&M&\\
\hline
0&-2&0&0&a\\
0&0&0&-2&b\\
0&2&0&-2&c\\
\hline
\end{array}, \\
&d_1^{0,-5}=\begin{array}{|cc|c}
\Gamma&Y&\\
\hline
2&0&a\\
0&2&b\\
0&0&c\\
\hline
\end{array}, \\
&d_1^{0,-7}=\begin{array}{|cc|c}
X&M&\\
\hline
-1&0&a\\
0&-1&b\\
1&-1&c\\
\hline
\end{array}, \\
&d_1^{1,{\rm -1/-5}}=\begin{array}{|ccc|c}
a&b&c\\
\hline
0&0&2&\alpha\\
\hline
\end{array}, \\
&d_1^{1,{\rm -3/-7}}=\begin{array}{|ccc|c}
a&b&c\\
\hline
2&-2&2&\alpha\\
\hline
\end{array}, 
\end{align}
and others are trivial. 
The $E_2$-page reads as 
\begin{align}
\begin{array}{cc|cccc}
{\rm AI} & n=0 & \Z_2^2 & 0 & 0 \\
{\rm BDI} & n=1 & \Z_2^2 & 0 & \Z_2 \\
{\rm D} & n=2 & \Z_2^2 & 0 & 0 \\
{\rm DIII} & n=3 & \Z_2^2& \Z_2^2& \Z_2 \\
{\rm AII} & n=4 & 0 & 0 & 0 \\
{\rm CII} & n=5 & 0 & \Z_2^2 & \Z_2 \\
{\rm C} & n=6 & 0 & 0 & 0 \\
{\rm CI} & n=7 & 0 & 0 & \Z_2 \\
\hline 
& E_2^{p,-n} & p=0 & p=1 & p=2 
\end{array}.
\end{align}
We find that the only $d_2^{0,-0}$ and $d_2^{0,-2}$ are nontrivial in the second differentials: 
As seen in Sec.~\ref{sec:2daz}, since the emergent AZ class at $X$ and $M$ ($\Gamma$ and $Y$) for $n=0$ ($n=2$) is BDI, Dirac points can be created at these points, by which we can pair-annihilate Dirac points in the adjacent 2-cells. 
As a result, we obtain the $E_3$(=$E_\infty$)-page 
\begin{align}
\begin{array}{cc|cccc}
{\rm AI} & n=0 & \Z_2 & 0 & 0 \\
{\rm BDI} & n=1 & \Z_2^2 & 0 & 0 \\
{\rm D} & n=2 & \Z_2 & 0 & 0 \\
{\rm DIII} & n=3 & \Z_2^2& \Z_2^2& 0 \\
{\rm AII} & n=4 & 0 & 0 & 0 \\
{\rm CII} & n=5 & 0 & \Z_2^2 & \Z_2 \\
{\rm C} & n=6 & 0 & 0 & 0 \\
{\rm CI} & n=7 & 0 & 0 & \Z_2 \\
\hline 
& E_3^{p,-n} & p=0 & p=1 & p=2 
\end{array}.
\end{align}
The datum $\{E_3^{0,-n}, E_3^{1,-(n+1)}, E_3^{-2,-(n+2)}\}$ approximates the $K$-group ${}^{\phi}K_{\Z_2 \times \Z_2^T}^{(\tau,c)-n}(T^2)$ according to the exact sequences (\ref{eq:e_inf_2d}). 
Let us focus on the class DIII $K$-group, which fits into the exact sequence 
\begin{align}
0\to \underbrace{\Z_2}_{E_3^{2,-5}} \to {}^{\phi}K_{\Z_2 \times \Z_2^T}^{(\tau,c)-3}(T^2) \to \underbrace{\Z_2^2}_{E_3^{0,-3}} \to 0. 
\end{align}
We find that ${}^{\phi}K_{\Z_2 \times \Z_2^T}^{(\tau,c)-3}(T^2) = \Z_4+\Z_2$ from the explicit construction of the $\Z_4$ invariant.~\cite{SSG16}

\subsubsection{$3d$ spinful insulators with TRS and 4-fold screw rotation symmetry ($P4_11'$)}
\begin{table*}
	\begin{center}
	\caption{
	The factor systems and the homomorphism $c:G\to \Z_2$ for integer gradings $n\geq 0$ for insulators in spinful electrons with TRS and 4-fold screw rotatioan symmetry.
	}
	\label{tab:fac_3d_s4}
$$
{\small
\renewcommand{\arraystretch}{1.5}
\begin{array}{l|l|l|l|l|l|l}
{\rm AZ} & n & z_{T,T} & z_{C,C} & z_{T,g}/z_{g,T} & z_{C,g}/z_{g,C} & z_{g,h} \\
\hline 
{\rm AI} & n=0 & T^2=1 & & TS(\bk) = S(-\bk)T & & \\
\cline{1-6}
{\rm BDI} & n=1 & T^2=1 & C^2=1& TS(\bk) = S(-\bk)T & CS(\bk) = S(-\bk)C & \\
\cline{1-6}
{\rm D} & n=2 & & C^2=1 & & CS(\bk) = S(-\bk)C & \\
\cline{1-6}
{\rm DIII} & n=3 & T^2=-1 & C^2=1& TS(\bk) = S(-\bk)T & CS(\bk) = S(-\bk)C & S(c_4^3\bk)S(c_4^2\bk)S(c_4\bk)S(\bk)=-e^{-i k_z} \\
\cline{1-6}
{\rm AII} & n=4 & T^2=-1 & & TS(\bk) = S(-\bk)T & & \\
\cline{1-6}
{\rm CII} & n=5 & T^2=-1 & C^2=-1& TS(\bk) = S(-\bk)T & CS(\bk) = S(-\bk)C &\\
\cline{1-6}
{\rm C} & n=6 & & C^2=-1 & & CS(\bk) = S(-\bk)C &  \\
\cline{1-6}
{\rm CI} & n=7 & T^2=1 & C^2=-1& TS(\bk) = S(-\bk)T & CS(\bk) = S(-\bk)C &  \\
\hline
\end{array}
\renewcommand{\arraystretch}{1}
}
$$
\end{center}
\end{table*}

\begin{figure}[!]
	\begin{center}
	\includegraphics[width=\linewidth, trim=5cm 2cm 5cm 0cm]{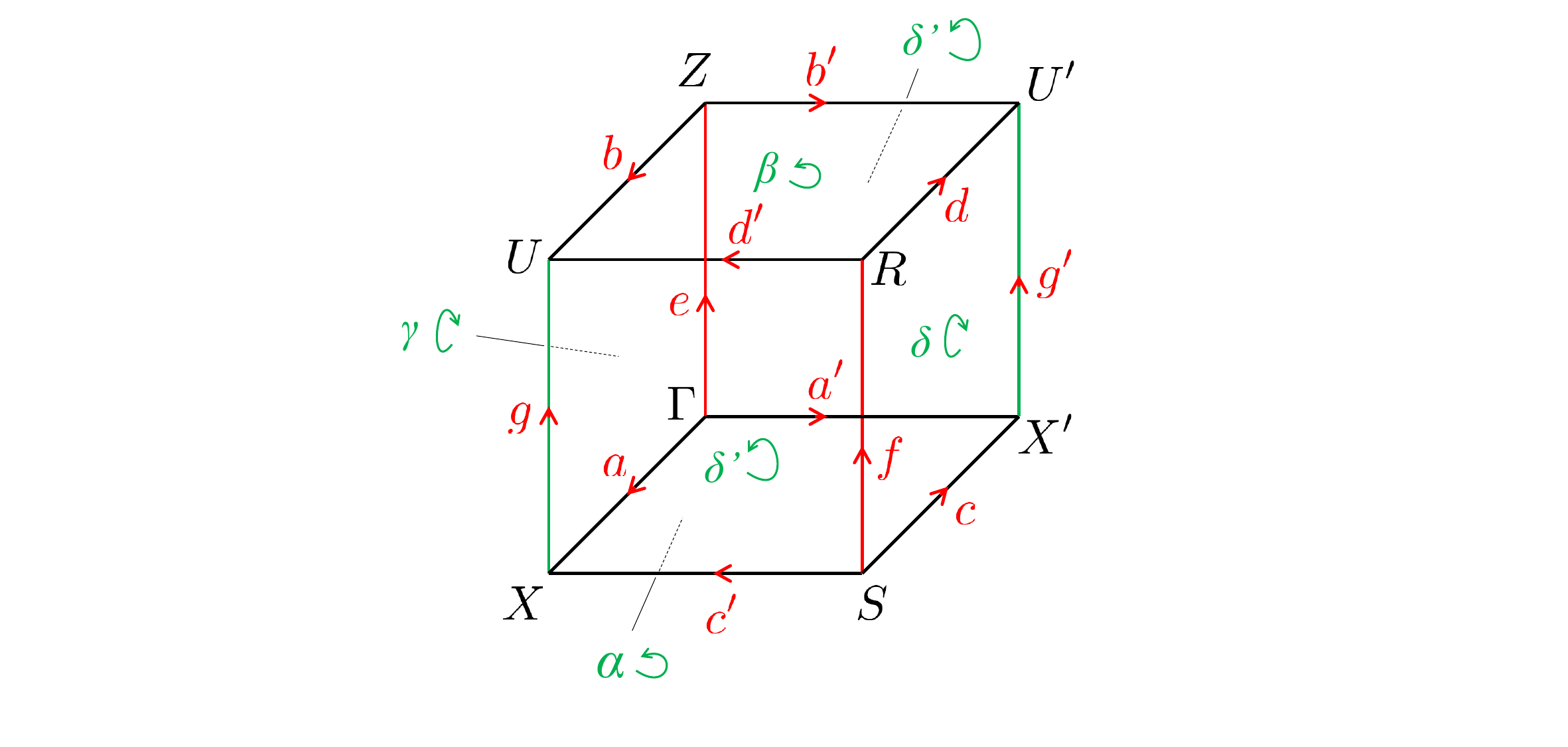}
	\end{center}
	\caption{A cell decomposition for the magnetic space group $P4_11'$.
	The figure shows one-eighth of BZ.}
	\label{fig:p4_11'}
\end{figure}

Let us consider 3d spinful systems with TRS $T$ and $4$-fold screw rotation symmetry $S: (x,y,z) \mapsto (-y,x,z+1/4)$.
The symmetry group is $\Z_4 \times \Z_2^T$ which is generated by $4$-fold screw ($\Z_4$) and time-reversal ($\Z_2^T$). 
The factor system for class AII ($n=4$) is 
\begin{align}
&T^2=-1, \nonumber\\
&S(c_4^3\bk)S(c_4^2\bk)S(c_4\bk)S(\bk)=-e^{-i k_z}, \nonumber\\
&TS(\bk)=S(-\bk)T, 
\end{align}
where $c_4\bk= (-k_y,k_x,k_z)$. 
According to Table~\ref{tab:shift_sym} in Appendix~\ref{sec:factor}, the factor systems for general $n$ are given in Table~\ref{tab:fac_3d_s4}.
Especially, the factor system for class DIII ($n=3$) is that for time-reversal symmetric superconductors with the trivial representation $S(\bk)\Delta(\bk)S(-\bk)^T=\Delta(c_4\bk)$ of the gap function under the 4-fold screw rotation. 
We use a $(\Z_4 \times \Z_2^{T})$-symmetric cell decomposition as in Fig.~\ref{fig:p4_11'}. 

It is straightforward to get the $E_1$-page 
\begin{widetext}
\begin{align}
\begin{array}{cc|ccccc}
{\rm AI} & n=0 & \Z^2+\Z+\Z^2+\Z^3+\Z^2+\Z^3& \Z+\Z+\Z+\Z+\Z^4+\Z^4+\Z^2&\Z+\Z+\Z+\Z&\Z\\
{\rm BDI} & n=1 & 0+0+0+\Z_2^2+\Z_2^2+\Z_2^2& 0+\Z_2+0+\Z_2+0+0+0&0+\Z_2+0+0&0\\
{\rm D} & n=2 & \Z^2+\Z+\Z^2+(\Z+\Z_2^2)+\Z_2^2+(\Z+\Z_2^2)& 0+\Z_2+0+\Z_2+\Z^4+\Z^4+\Z^2&0+\Z_2+\Z+\Z&\Z\\
{\rm DIII} & n=3 & 0& 0&0&0\\
{\rm AII} & n=4 & \Z^2+\Z+\Z^2+\Z^3+\Z^2+\Z^3 & \Z+\Z+\Z+\Z+\Z^4+\Z^4+\Z^2 & \Z+\Z+\Z+\Z & \Z \\
{\rm CII} & n=5 & 0 & \Z_2+0+\Z_2+0+0+0+0& \Z_2+0+0+0&0\\
{\rm C} & n=6 & \Z^2+\Z+\Z^2+\Z+0+\Z & \Z_2+0+\Z_2+0+\Z^4+\Z^4+\Z^2&\Z_2+0+\Z+\Z&\Z\\
{\rm CI} & n=7 & 0&0&0&0\\
\hline 
& E_1^{p,-n} & \{\Gamma,X,S,Z,U,R\} & \{a,b,c,d,e,f,g\} & \{\alpha,\beta,\gamma,\delta\} & \{vol\} \\ 
& & p=0 & p=1 & p=2 & p=3
\end{array}.
\end{align}
The first differential $d_1^{p,-n}: E_1^{p,-n} \to E_1^{p+1,-n}$ is also straightforwardly given. 
For example, $d_1^{0,-4}$ is 
\begin{align}
d_1^{0,-4}
= 
{\scriptsize
\begin{array}{|cc|c|cc|ccc|cc|ccc|cc}
\Gamma&&X&S&&Z&&&U&&R&&&\\
(\lambda,\lambda^*)&(i\lambda,-i\lambda^*)&
(i,-i)&
(\lambda,\lambda^*)&(i\lambda,-i\lambda^*)&
1&(i,-i)&-1& 
1&-1&
1&(i,-i)&-1\\
\hline
2&2&-2&&&&&&&&&&&
a\\
\hline
&&&&&1&1&1&-1&-1&&&&
b\\
\hline
&&-2&2&2&&&&&&&&&
c\\
\hline
&&&&&&&&-1&-1&1&1&1&
d\\
\hline
1&0&&&&-2&0&0&&&&&&
\lambda e^{-\frac{ik_z}{4}}&e\\
0&1&&&&0&-1&0&&&&&&
i \lambda e^{-\frac{ik_z}{4}}&\\
0&1&&&&0&0&-2&&&&&&
-\lambda e^{-\frac{ik_z}{4}}&\\
1&0&&&&0&-1&0&&&&&&
-i \lambda e^{-\frac{ik_z}{4}}&\\
\hline
&&&1&0&&&&&&-2&0&0&
\lambda e^{-\frac{ik_z}{4}}&f\\
&&&0&1&&&&&&0&-1&0&
i \lambda e^{-\frac{ik_z}{4}}&\\
&&&0&1&&&&&&0&0&-2&
-\lambda e^{-\frac{ik_z}{4}}&\\
&&&1&0&&&&&&0&-1&0&
-i\lambda e^{-\frac{ik_z}{4}}&\\
\hline
&&1&&&&&&-2&0&&&&
ie^{-\frac{ik_z}{2}}&g\\
&&1&&&&&&0&-2&&&&
-ie^{-\frac{ik_z}{2}}&\\
\hline
\end{array}
}
\end{align}
with $\lambda=e^{\pi i/4}$. 
We get the $E_2$-page 
\begin{align}
\begin{array}{cc|ccccc}
{\rm AI} & n=0 & \Z& \Z_2&\Z&0\\
{\rm BDI} & n=1 & \Z_2^4& 0&\Z_2&0\\
{\rm D} & n=2 & \Z_2^4& \Z&\Z_2&\Z\\
{\rm DIII} & n=3 & 0& 0&0&0\\
{\rm AII} & n=4 & \Z & \Z_8+\Z_4+\Z_2& \Z & 0 \\
{\rm CII} & n=5 & 0 & \Z_2&0&0\\
{\rm C} & n=6 & 0 & \Z+\Z_2^2&\Z_2&\Z\\
{\rm CI} & n=7 & 0&0&0&0\\
\hline 
& E_2^{p,-n} & p=0 & p=1 & p=2 & p=3
\end{array}
\end{align}

In class AII ($n=4$), the $E_2$-page is already the limiting page. 
The exact sequences (\ref{eq:e_inf}) imply that 
\begin{align}
{}^{\phi}K_{\Z_4 \times \Z_2^T}^{\tau-4}(T^3) = \underbrace{\Z}_{\rm Filling\ number}+F^{1,-5}, 
\end{align}
where the group $F^{1,-5}$ filts into the exact sequence 
\begin{align}
    0\to\underbrace{\Z_2}_{E_{\infty}^{2,-6}}\to F^{1,-5}\to\underbrace{\Z_2}_{E_{\infty}^{1,-5}}\to 0.
\end{align}
The AHSS alone can not determine $F^{1,-5}$. 
Nevertheless, we find that either $\Z_4$ or $\Z_2+\Z_2$ topological invariant should be defined on the $2$-skeleton. 

Interestingly, there appears a $\Z_8$ topological invariant in class DIII ($n=3$). 
Since $E_2^{0,-4}=\Z$ represents the filling number, $d_3^{0,-4}:E_3^{0,-4}\to E_3^{3,-6}$ is trivial. 
Then, the $K$-group fits into the short exact sequence 
\begin{align}
0\to \underbrace{\Z}_{3d {\rm\ winding\ number}} \to {}^{\phi}K_{\Z_4 \times \Z_2^T}^{\tau-3}(T^3) \to \underbrace{\Z_8+\Z_4+\Z_2}_{1d{\rm\ topological\ invariants}} \to 0.
\end{align}
The construction of the $\Z_8$ invariant is similar to the screw $\Z_n$ invariant discussed in Sec.~\ref{sec:screw}.
Let $q(\bk)$ be the off-diagonal part of the flattened Hamiltonian ${\rm sgn} [H(\bk)] = \begin{pmatrix}
0 & q(\bk) \\
q(\bk)^{\dag} & 0 \\
\end{pmatrix}$ in the basis so that the chiral operator is $\Gamma = \begin{pmatrix}
1 & 0 \\
0 & -1 \\
\end{pmatrix}$. 
The symmetry of class DIII is written by $\sigma_y q(\bk)^* = q(-\bk)^{\dag} \sigma_y$, $S(\bk) q(\bk) = q(c_4 \bk) S(\bk)$, $\sigma_y S(\bk)^* = S(-\bk) \sigma_y$. 
The matrix $q(k_z)$ on the line $(k_x,k_y) = \Gamma$ and $M$ splits into the screw eigensectors as $q(k_z) = q_0(k_z)\oplus q_1(k_z) \oplus q_2(k_z) \oplus q_3(k_z), k_z \in [-\pi,\pi],$ with the eigenvalues $e^{i m \pi /4} e^{-i k_z/4}$, $m \in \{0,1,2,3\}$. 
Moreover, at the high-symmetry points $(k_x,k_y,k_z) = (\Gamma,-\pi)$ and $(M,-\pi)$ on the zone boundary, 
the TRS is closed inside the eigensector with $m=0$ (and $m=2$), then, the Pfaffian ${\rm Pf} [\sigma_y q_0(-\pi)]$ is well-defined. 
The $\Z_8$ invariant $e^{\pi i \nu/4}, \nu \in \{0,1,\dots, 7\}$ is defined as 
\begin{equation}\begin{split}
&e^{i \pi \nu/4}\\
&:= 
\frac{
{\rm Pf}[\sigma_y q_0(\Gamma,-\pi)] \times 
\exp \left[ 
\frac{3}{8} \int_{-\pi}^{\pi} d \log \det q_0(\Gamma,k_z) 
+\frac{2}{8} \int_{-\pi}^{\pi} d \log \det q_1(\Gamma,k_z) 
+\frac{1}{8} \int_{-\pi}^{\pi} d \log \det q_2(\Gamma,k_z) \right]
}{
{\rm Pf}[\sigma_y q_0(M,-\pi)] \times 
\exp \left[ 
\frac{3}{8} \int_{-\pi}^{\pi} d \log \det q_0(M,k_z) 
+\frac{2}{8} \int_{-\pi}^{\pi} d \log \det q_1(M,k_z) 
+\frac{1}{8} \int_{-\pi}^{\pi} d \log \det q_2(M,k_z) 
\right]
} \\
&
\qquad \times 
\exp \left[ \frac{1}{8} \int_{\Gamma \to M} d \bk \cdot \bm{\nabla} \log \det q(\bk,-\pi)
\right]. 
\end{split}\end{equation}
Using ${\rm Pf}(A)^2 = \det (A)$, one can show $(e^{i \pi \nu/4})^8 = 1$. 
\end{widetext}

\subsubsection{$2d$ superconductors in spinful systems with the space group $P4/m$ and with the $B_g$ representation}

\begin{table*}
	\begin{center}
	\caption{
	The factor systems and the homomorphism $c:G\to \Z_2$ for integer gradings $n\geq 0$ for superconductors in spinful systems with $P4/m$ symmetry and with the  $B_g$ representation for the gap function.  
	We have omitted the symbol of tilde in symmetry operators $\tilde I(\bk)$ and $\tilde C_4(\bk)$ in the Nambu space.
	}
	\label{tab:fac_exam4}
$$
{\scriptsize
\renewcommand{\arraystretch}{1.5}
\begin{array}{l|l|l|l|l|l|l}
{\rm AZ} & n & z_{T,T}, z_{C,C} & z_{T,c_4}/z_{c_4,T}, z_{c_4,g}/z_{g,c_4} & z_{T,I}/z_{I,T}, z_{I,g}/z_{g,I} & z_{g,h} & c(g) \\
\cline{1-7}
{\rm AI} & n=0
&T^2=1&TC_4(\bk)=C_4(-\bk)T&TI(\bk)=I(-\bk)T&&C_4(\bk)H(\bk)=-H(\bk)C_4(\bk)\\
&&&&&&I(\bk)H(\bk)=H(-\bk)I(\bk)\\
\cline{1-5}\cline{7-7}
{\rm BDI} & n=1 
&T^2=1&TC_4(\bk)=-C_4(-\bk)T&TI(\bk)=I(-\bk)T&&C_4(\bk)H(\bk)=H(\bk)C_4(\bk)\\
&&C^2=1&CC_4(\bk)=C_4(-\bk)C&CI(\bk)=I(-\bk)C&&I(\bk)H(\bk)=H(-\bk)I(\bk)\\
\cline{1-5}\cline{7-7}
{\rm D} & n=2 
&C^2=1&CC_4(\bk)=C_4(-\bk)C&CI(\bk)=I(-\bk)C&&C_4(\bk)H(\bk)=-H(\bk)C_4(\bk)\\
&&&&&&I(\bk)H(\bk)=H(-\bk)I(\bk)\\
\cline{1-5}\cline{7-7}
{\rm DIII} & n=3 
&T^2=-1&TC_4(\bk)=C_4(-\bk)T&TI(\bk)=I(-\bk)T&C_4(c_4^3\bk)C_4(c_4^2\bk)C_4(c_4\bk)C_4(\bk)=-1&C_4(\bk)H(\bk)=H(\bk)C_4(\bk)\\
&&C^2=1&CC_4(\bk)=-C_4(-\bk)C&CI(\bk)=I(-\bk)C&I(-\bk)I(\bk)=1&I(\bk)H(\bk)=H(-\bk)I(\bk)\\
\cline{1-5}\cline{7-7}
{\rm AII} & n=4 
&T^2=-1&TC_4(\bk)=C_4(-\bk)T&TI(\bk)=I(-\bk)T&I(c_4\bk)C_4(\bk)=C_4(-\bk)I(\bk)&C_4(\bk)H(\bk)=-H(\bk)C_4(\bk)\\
&&&&&&I(\bk)H(\bk)=H(-\bk)I(\bk)\\
\cline{1-5}\cline{7-7}
{\rm CII} & n=5 
&T^2=-1&TC_4(\bk)=-C_4(-\bk)T&TI(\bk)=I(-\bk)T&&C_4(\bk)H(\bk)=H(\bk)C_4(\bk)\\
&&C^2=-1&CC_4(\bk)=C_4(-\bk)C&CI(\bk)=I(-\bk)C&&I(\bk)H(\bk)=H(-\bk)I(\bk)\\
\cline{1-5}\cline{7-7}
{\rm C} & n=6
&C^2=-1&CC_4(\bk)=C_4(-\bk)C&CI(\bk)=I(-\bk)C&&C_4(\bk)H(\bk)=-H(\bk)C_4(\bk)\\
&&&&&&I(\bk)H(\bk)=H(-\bk)I(\bk)\\
\cline{1-5}\cline{7-7}
{\rm CI} & n=7 
&T^2=1&TC_4(\bk)=C_4(-\bk)T&TI(\bk)=I(-\bk)T&&C_4(\bk)H(\bk)=H(\bk)C_4(\bk)\\
&&C^2=-1&CC_4(\bk)=-C_4(-\bk)C&CI(\bk)=I(-\bk)C&&I(\bk)H(\bk)=H(-\bk)I(\bk)\\
\hline
\end{array}
\renewcommand{\arraystretch}{1}
}
$$
\end{center}
\end{table*}

In this section, we illustrate how the AHSS describes superconducting nodal structures. 
Consider $2d$ superconductors with TRS, inversion and 4-fold rotation symmetries. 
We assume that the gap function obeys the $B_g$ representation under the point group $4/m$. 
Namely, 
$I(\bk) \Delta(\bk) I(-\bk)^T= \Delta(-\bk)$, and 
$C_4(\bk) \Delta(k_x, k_y) C_4(-\bk)^{T} = - \Delta(-k_y, k_x)$.
In the Nambu space, the inversion and 4-fold rotation operators are given as 
\begin{align}
&\tilde I(\bk)
=
\left(
\begin{array}{cc}
I(\bk) & 0 \\
 0& I(-\bk)^*
\end{array}
\right), \\
&\tilde{C}_4(\bk)=\left(
\begin{array}{cc}
C_4(\bk) & 0 \\
 0& -C_4(-\bk)^*
\end{array}
\right),
\end{align}
by which we can determine the factor systems for class DIII ($n=3$).
According to Table~\ref{tab:shift_sym} in Appendix~\ref{sec:factor}, the  factor systems for other gradings $n$ are summarized as in Table~\ref{tab:fac_exam4}.

Consider the cell decomposition in Fig.~\ref{fig:c4}. 
Then the $E_1$-page is found to be 
\begin{align}
\begin{array}{cc|cccc}
{\rm AI} & n=0 & 0+\Z^2+0&\Z+\Z&\Z\\
{\rm BDI} & n=1 & \Z^2+0+\Z^2&0&0\\
{\rm D} & n=2 & 0+\Z^2+0&\Z+\Z&\Z\\
{\rm DIII} & n=3 & \Z^2+0+\Z^2&0&0\\
{\rm AII} & n=4 & 0+\Z^2+0&\Z+\Z&\Z\\
{\rm CII} & n=5 & \Z^2+0+\Z^2&0&0\\
{\rm C} & n=6 &0+\Z^2+0&\Z+\Z&\Z\\
{\rm CI} & n=7 & \Z^2+0+\Z^2&0&0\\
\hline 
& E_1^{p,-n} & \{\Gamma,X,M\} & \{a,b\} & \{\alpha\} \\
&& p=0 & p=1 & p=2 \\
\end{array}.
\end{align}
As we described in Sec.~\ref{sec:ahss_real}, the $E_1$-page $E_1^{p,-n}$ can be regarded  as the space of topological gapless points inside $p$-cells with the symmetry class $(n+1-p)$. 
For example, $E_1^{2,-4}=\Z$ means that the 2-cell $\alpha$ can host stable $\Z$ gapless points belonging to class DIII (i.e., gapless Majorana cones). 

The compatibility relation (incorporating PHS) gives the first differentials.
The non-trivial ones are listed below.
\begin{align}
&d_1^{0,-n}=
\begin{array}{|cc|c}
X & &  \\ 
I=1 & I=-1 & \\ \hline
-1 & -1  & a \\
1 & 1 & b \\ \hline
\end{array}, \\
&d_1^{1,-n}=
\begin{array}{|cc|c}
a & b & \\ \hline
2 & 2 & \alpha \\ \hline 
\end{array},
\end{align}
with even $n$.
It should be noted that $d_1^{1,-n}$ for even $n$ is non-trivial, even though the 1-cells $a$ ($b$) and $a'$ ($b'$) in the same $C_4$ orbit have opposite directions with respect to the 2-cell $\alpha$.
This is because $C_4$ for even $n$ acts as antisymmetry, $C_4 (\bk)H(\bk)=-H(-k_y,k_x)C_4(\bk)$, as shown in  Table~\ref{tab:fac_exam4}.
Occupied bands and empty ones are interchanged after the $C_4$-rotation, so the mismatch of the orientation between $a'$ ($b'$) and $\alpha$ is compensated in the compatibility relation.
As a result, $a$ ($b$) and $a'$ ($b'$) equally contribute to $\alpha$ in $d_1^{1,0}$ as $1+1=2$.
We get the $E_2$-page
\begin{align}
\begin{array}{cc|cccc}
{\rm AI} & n=0 & \Z&0&\Z_2\\
{\rm BDI} & n=1 & \Z^4&0&0\\
{\rm D} & n=2 & \Z&0&\Z_2\\
{\rm DIII} & n=3 & \Z^4&0&0\\
{\rm AII} & n=4 & \Z&0&\Z_2\\
{\rm CII} & n=5 & \Z^4&0&0\\
{\rm C} & n=6 & \Z&0&\Z_2\\
{\rm CI} & n=7 & \Z^4&0&0\\
\hline 
& E_2^{p,-n} & p=0 & p=1 & p=2 \\
\end{array}.
\label{eq:E2example4}
\end{align}

\begin{figure}[!]
	\begin{center}
	\includegraphics[width=\linewidth, trim=0cm 0cm 0cm 0cm]{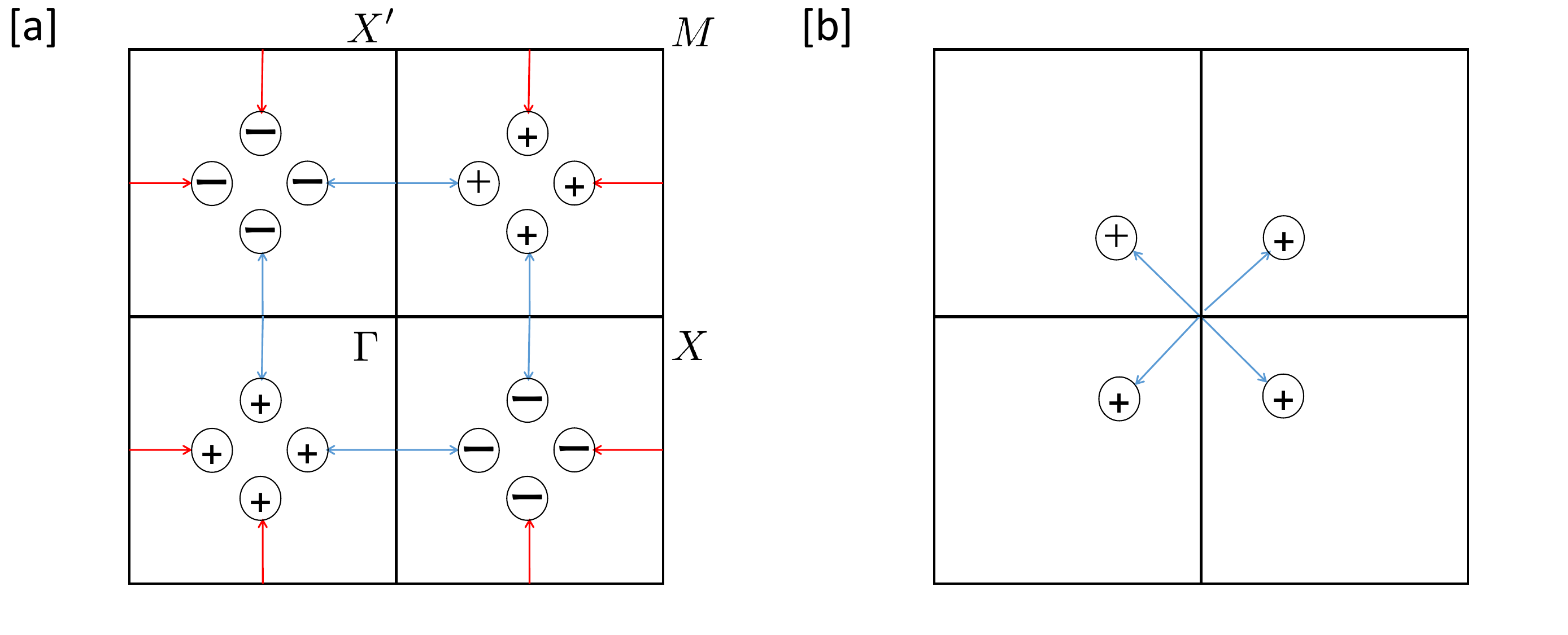}
	\end{center}
	\caption{[a]
	The first differential $d_1^{1,{\rm even}}$. 
	[b]
	The second differential $d_2^{2,{\rm odd}}$.}
	\label{fig:2d_exam4}
\end{figure}

For example, $d_1^{1,-4} = (2,2)$ represents a pair creation of Dirac points from 1-cells to the 2-cell $\alpha$ in class DIII as shown in Fig.~\ref{fig:2d_exam4} [a]. 
Here, two Dirac points related by $C_4$ rotation have opposite charges since
$C_4$ changes the chirality $\Gamma = iTC$. 
The above process changes the total charges in the 2-cell $\alpha$ by an even number,  so only the parity of the charges is topologically stable. 
In other words, we have $E_2^{2,-4}=\Z_2$.

Next, we ask if $E_2^{2,-n}$ with even $n$ is trivialized by the creation of Dirac points from 0-cells. 
The corresponding second differential is $d_2^{0,-(n-1)}$, which maps $ E_2^{0,-(n-1)}$ to $E_2^{2,-n}$.
For even $n$, $E_2^{0,-(n-1)}$ is nothing but $E_1^{0,-(n-1)}=\Z^4$,  which counts (occupied) states with a definite eigenvalue of $C_4$, say $C_4=e^{i\pi/4}$, in the $I=\pm 1$ sectors at $\Gamma$ and $M$, respectively.
Since we have two independent eigensectors of $I$ and two independent 0-cells $\Gamma$ and $M$, 
we need four independent integers $(n^{I=1}_{\Gamma}, n^{I=-1}_{\Gamma}, n^{I=1}_{M},n^{I=-1}_{M})\in \Z^4$.
The second differential arising from the $I=\pm 1$ sector at $\Gamma$ (or $M$) is modeled by the Hamiltonian with the $d_{x^2-y^2}$-wave type gap functions: 
\begin{align}
&{\rm BDI}: 
\left\{
\begin{array}{l}
T=\tau_x K, \quad C=\sigma_x K, \\
H(\bk)=(\bk^2-\mu) \sigma_z + \Delta(k_x^2-k_y^2) \tau_y \sigma_x, \\
\end{array}\right., \\
&{\rm DIII}: 
\left\{
\begin{array}{l}
T=\sigma_y K, \quad 
C=\tau_x K, \\
H(\bk)=(\bk^2-\mu) \tau_z + \Delta(k_x^2-k_y^2) \sigma_y \tau_y, \\
\end{array}\right., \\
&{\rm CII}: \left\{
\begin{array}{l}
T=\tau_y K, \quad
C=\sigma_y K, \\
H(\bk)=(\bk^2-\mu) \sigma_z + \Delta(k_x^2-k_y^2) \tau_x \sigma_y, \\
\end{array}\right., \\
&{\rm CI}: \left\{
\begin{array}{l}
T=\sigma_x K, \quad
C=\tau_y K, \\
H(\bk)=(\bk^2-\mu) \tau_z + \Delta(k_x^2-k_y^2) \sigma_x \tau_x, 
\\
\end{array}\right., 
\end{align}
with 
\begin{align}
C_4=
\begin{pmatrix}
e^{\frac{\pi i}{4} \sigma_z} & 0 \\
0 & -(e^{\frac{\pi i}{4} \sigma_z})^*\\
\end{pmatrix}_{\tau}, \qquad 
I=\pm {\bf 1}_{4\times 4}.
\end{align}
These Hamiltonians describe the creation of Dirac points: when $\mu$ changes the sign from negative to positive, a band inversion occurs at $\Gamma$ (or $M$) and four equivalent Dirac points are pumped from $\Gamma$ (or $M$) to the adjacent $C_4$-equivalent 2-cells (see Fig.~\ref{fig:2d_exam4} [b]).
This process gives  the following  nontrivial $d_2^{0,-(n-1)}$ for even $n$,  
\begin{align}
d_2^{0,-(n-1)}=
\begin{array}{|cc|cc|c}
\Gamma&&M&\\
I=1&I=-1&
I=1&I=-1& 
\\
\hline
1&1&1&1&\alpha\\
\hline
\end{array}.
\end{align}
From this,
we arrive at the limiting page $E_{\infty}=E_3$ as
\begin{align}
\begin{array}{cc|cccc}
{\rm AI} & n=0& \Z&0&0\\
{\rm BDI} & n=1 & \Z^3+2\Z&0&0\\
{\rm D} & n=2 & \Z&0&0\\
{\rm DIII} & n=3 & \Z^3+2\Z&0&0\\
{\rm AII} & n=4 & \Z&0&0\\
{\rm CII} & n=5 & \Z^3+2\Z&0&0\\
{\rm C} & n=6 & \Z&0&0\\
{\rm CI} & n=7 & \Z^3+2\Z&0&0\\
\hline 
& E_3^{p,-n} & p=0 & p=1 & p=2 \\
\end{array}.
\end{align}

As discussed in Sec.~\ref{sec:ind}, a nontrivial element of $E_2^{0,-n}/E_3^{0,-n}$ is the indicator of Dirac points in the 2-cell from topological invariants at $0$-cells. 
For example, $E_2^{0,-3}/E_3^{0,-3} = \Z_2$ for class DIII relates the number of Dirac points $\nu_{\rm ind}$ in the 2-cell to the numbers of occupied states at 0-cells as 
\begin{align}
(-1)^{\nu_{\rm ind}} = (-1)^{n^{I=1}_{\Gamma}+n^{I=-1}_{\Gamma}+n^{I=1}_{M}+n^{I=-1}_{M}},
\end{align}
where $n^{I=\pm 1}_{\bk}$  is the number of occupied states at $\bk=\Gamma, M$ with $C_4=e^{i\pi/4}$ and  $I= \pm 1$.
In general, the high symmetry points $\Gamma$ and $M$ are not located on the Fermi surface,  
so we can neglect the gap function in the caluculation of the latter numbers under the weak pairing assumption.
As a result,  the above equation is recast into
\begin{align}
(-1)^{\nu_{\rm ind}}
= (-1)^{n_{0,\Gamma}^{I=1}+n_{0,\Gamma}^{I=-1}+n_{0,M}^{I=1}+n_{0,M}^{I=-1}},
\end{align}
where $n_{0,\bk}^{I=\pm 1}$ is  the number of bands at $\bk$ in the normal state that are below the Fermi level and with the eigen values $C_4=e^{i\pi/4}$ and $I=\pm 1$. 
For a single band superconductor
\begin{align}
&H(\bk)
= \varepsilon(\bk) \tau_z + \Delta(\bk)\sigma_y \tau_y, \nonumber\\
&\varepsilon(-k_y, k_x)=\varepsilon(k_x, k_y),\nonumber\\
&
\Delta(-k_y, k_x)=-\Delta(k_x, k_y) 
\end{align}
with the weak pairing assumption  $|\Delta(\bk)|\ll |\varepsilon(\bk)|$ at $\bk=\Gamma, M$, the indicator is evaluated as
\begin{align}
(-1)^{\nu_{{\rm ind}}} = {\rm sgn}[\varepsilon(\Gamma)\varepsilon(M)]. 
\end{align}
If the normal state satisfies $\varepsilon(\Gamma)\varepsilon(M)<0$, there must be  point nodes (=Dirac points), irrespective of details of $\varepsilon(\bk)$ and $\psi(\bk)$ if the gap function is in $B_g$ representation.

\section{$E_3$ pages for 230 space groups for class A and AIII}
\label{sec:e2}
In Tables~\ref{tab:230}, \ref{tab:e2_2}, \ref{tab:e2_3}, \ref{tab:e2_4}, \ref{tab:e2_5}, \ref{tab:e2_6}, and \ref{tab:e2_7}, we present the $E_{\infty} = E_4$ pages for all the 230 space groups in the AZ classes A and AIII. 
The $E_3$-pages are the same as
the $E_2$-pages since $d_1^{p,-n}=0$ for any $p$ and $n$ in the abosence
of anti-unitary symmetry. 
We have refered the database~\cite{Bilbao_I} to identify the nonprimitive
lattice translation of each group element.
In Tables \ref{tab:230} to \ref{tab:e2_7}, we use the discrete torsion phase $\epsilon^{\bk=\Gamma}(g,h)$ defined by
\begin{align}
&U_g(\bk=\Gamma) U_h(\bk=\Gamma) \nonumber\\
&= \epsilon^{\bk=\Gamma}(g,h) U_h(\bk=\Gamma) U_g(\bk=\Gamma),
\label{eq:dis_pg}
\end{align}
for $g,h \in G$ with $gh=hg$ to specify the equivalence class of the group cocycle for projective representations for each space group with point group $G$.
(In Table~\ref{tab:pointgroup}, we summarize the group cohomology $H^2(G,U(1))$ for the point groups.)

\begin{table*}
\begin{center}
\caption{
The list of crystallographic point groups and their group cohomologies.
$D_n$ is the dihedral groups, $A_4$ is the alternating group on four letters, and $S_4$ is the symmetric group on four letters. 
Here, $|G|$ is the order of group, $H^2(G, U(1))$ is the group cohomology of $G$, and Sch\"{o}n and Intl are shortened forms of Sch\"{o}flies notation and international notation, respectively. 
}
\label{tab:pointgroup}
$$
\begin{array}{cccll}
{\rm Group} & |G| & H^2(G,U(1)) & \mbox{Sch\"{o}n.} & {\rm Intl}  \\
\hline
\Z_1 & 1 & 0 & C_1 & 1 \\
\Z_2 & 2 & 0 & C_i,C_2,C_S & \bar 1,2, m \\
\Z_3 & 3 & 0 & C_3 & 3 \\
\Z_4 & 4 & 0 & C_4, S_4 & 4, \bar 4 \\
\Z_2^2 & 4 & \Z_2 & C_{2h}, D_2,C_{2v}  & 2/m,222,mm2 \\
\Z_6 & 6 & 0 & C_{3i},C_6, C_{3h} & \bar 3,6, \bar 6 \\
D_3 & 6 & 0 & D_3,C_{3v}  & 32,3m \\
D_4 & 8 & \Z_2 & D_4, C_{4v}, D_{2d} & 422, 4mm, \bar 42m\\
\Z_2^3 & 8 & \Z_2^3 & D_{2h} & mmm\\
\Z_4\times \Z_2 & 8 & \Z_2 & C_{4h}&4/m\\
A_4&12&\Z_2&T&23\\
D_6&12&\Z_2&D_{3d},D_6,C_{6v},D_{3h}&\bar 3m,622,6mm,\bar62m\\
\Z_6\times \Z_2&12&\Z_2&C_{6h}&6/m\\
D_4\times \Z_2&16&\Z_2^3&D_{4h}&4/mmm\\
S_4&24&\Z_2&O,T_d&432,\bar43m\\
A_4\times \Z_2&24&\Z_2&T_h&m\bar3\\
D_6\times \Z_2&24&\Z_2^3&D_{6h}&6/mmm\\
S_4\times \Z_2&48&\Z_2^2&O_h&m\bar3m\\
\end{array}
$$
\end{center}
\end{table*}

It is found that only the space group Nos. 2, 81, 82, 147, and 148 have a torsion group $\Z_2$ in $E_3^{3,0}$.
Except for $E_3^{0,0}$ and $E_3^{3,0}$ in these space groups, the $E_3$-pages in Tables~\ref{tab:230} to \ref{tab:e2_7} coincides with the $E_\infty$ ones:
For space groups in classes A and AIII, only $E^{0,0}_3$ and $E_3^{3,0}$
can be different 
from those in the $E_\infty$ page, since $d_3^{0,0}:E_3^{0,0}\rightarrow
E_3^{3,-2}=E_3^{0,0}$ can be nontrivial. 
However, if $E_3^{0,0}=\Z$, one can argue that $d_3^{0,0}=0$ as follows.
Because no space group symmetry is left as a little group in 3-cells,  
$E_3^{3,0}=\Z$ corresponds to the trivial representation of space group.
Therefore, $d_3^{0,0}$ relates a band inversion between
non-trivial irreps at 0-cells to the trivial representation in 3-cells,
but there is no such a homomorphism.
As a result, the $E_3$-page reduces to the $E_4$-page, which is the
$E_\infty$-page in three dimensions.
  
For the space group Nos. 2, 81, 82, 147, and 148, on the other hand, we have checked that $d_3^{0,0}\neq 0$ and the torsion $\Z_2$ in $E_3^{3,0}$ is trivialized by $d_3^{0,0}$.
This result implies representation enforced Weyl semimetals in these space groups. 
We have also computed the $K$-group $K^{\tau-1}_G(T^3)$ directly via the Mayer-Vietoris sequence.

In the rest of this section, we present some technical details to compute the $E_2$-page.

For a given Bravais lattice $BL (\cong \Z^3)$ and a point group $G$, inequivalent classes of nonprimitive lattice translations $\bm{a}_g$ associated with the point group action $g \in G$ are classified by the group cohomology $H^2(G,BL)$, where $G$ acts on the $BL$ as a usual point group action on the real space. 
A representative $\bm{\nu} \in Z^2(G,BL)$ is given as 
\begin{align}
&\{g|\bm{a}_g\} \{h|\bm{a}_h\}
= \{e|\bm{\nu}_{g,h}\}\{gh|\bm{a}_{gh}\}, \\
&\bm{\nu}_{g,h} = p_g \bm{a}_h + \bm{a}_g - \bm{a}_{gh}, \qquad (g,h \in G).
\end{align}
Here, one nonprimitive lattice translation $\bm{a}_g$ is fixed for each group element of $G$.
In the $\bk$-space, the point group $G$ acts on the Bloch state projectively due to the two origins: (i) space group is nonsymmorphic; (ii) the fundamental degrees of freedom can obey a nontrivial projective representation of the point group $G$.
Let $\{ \ket{\bk,i} \}_i$ be a basis of Bloch states at $\bk \in BZ$. 
The point group acts on $\ket{\bk,i}$ as 
\begin{align}
&\hat g \ket{\bk,i} = \ket{g \bk,j} [U_g(\bk)]_{ji}, \\
&U_g(h\bk)U_h(\bk) = z_{g,h} e^{-i g h \bk \cdot \bm{\nu}_{g,h}} U_{gh}(\bk), \label{eq:app1}
\end{align}
where $(z_{g,h}) \in Z^2(G,U(1))$ is a factor system for a projective representation of the fundamental degrees of freedom. 
At a fixed $\bk \in BZ$, the equivalence class $[(z_{g,h} e^{-i \bk \cdot \bm{\nu}_{g,h}})]$ of the factor system belongs to the group cohomology $H^2(G_{\bk},U(1))$, where $G_{\bk} = \{g \in G | g \bk = \bk\}$ is the little group.
The equivalence class is immediately determined by the discrete torsion phase 
\begin{align}
\epsilon^{\bk}_{g,h} := \frac{z_{g,h} e^{-i \bk \cdot \bm{\nu}_{g,h}}}{z_{h,g} e^{-i \bk \cdot \bm{\nu}_{h,g}}}, \nonumber\\
(g,h \in G_{\bk}, gh=hg).
\end{align}
There is a one-to-one correspondence between an element of group cohomology $H(G_{\bk},U(1))$ and a discrete torsion phase $\epsilon^{\bk}$.~\footnote{
Let $G$ be a finite group and $H^2(G,U(1))$ the second group cohomology. 
It holds that $H^2(G,U(1)) \cong {\rm Hom}(M(G),U(1))$, where $M(G)$ is the Schur multiplier.~\cite{Miller}
}
Once we get a representation at $\bk \in BZ$, by using (\ref{eq:app1}), the representation at other points $g \bk \ (g \notin G_{\bk})$ connected by the point group is given by 
\begin{align}
U_{h \in G_{g\bk}}(g\bk) = \frac{z_{h,g} e^{-i g \bk \cdot \bm{\nu}_{h,g}}}{z_{g,g^{-1} h g} e^{-i g \bk \cdot \bm{\nu}_{g,g^{-1} h g}}} U_{g^{-1}h g}(\bk).
\label{eq:induced_rep}
\end{align}
Especially, the character $\chi^{g \bk}(h) = \tr U_h(g \bk)$ compatible with the factor system at $\bk$ is given in this way.
Let $\chi^{\bk}_{\alpha}(g) (g \in G_{\bk}, \alpha \in {\rm irreps})$ be the irreducible characters with the factor system $z_{g,h} e^{-i \bk \cdot \bm{\nu}_{g,h}}$. 
For a given representation $\rho$ with the same factor system, the irreducible decomposition is given as $\rho = \bigoplus_{i \in {\rm irreps}} n_{\alpha} \rho_{\alpha}$ with the nonnegative integer 
\begin{align}
n_{\alpha} = \frac{1}{|G_{\bk}|} \sum_{g \in G_{\bk}} \chi^{\bk}_{\alpha}(g)^* \chi^{\bk}_{\rho}(g).
\label{eq:irrep_decom}
\end{align}
Combining Eqs.\ (\ref{eq:induced_rep}) and (\ref{eq:irrep_decom}), we can determine the first differentials $d_1^{p,0}$.

\begin{table*}
\begin{center}
\caption{
$E_{\infty}$ pages for 230 space groups. The discrete torsion phase $\epsilon(g,h) \in \{+,-\}$ specifies the algebra among elements in the point group (see eq.(\ref{eq:dis_pg})). 
The discrete torsion phases for spinless and spinful electrons are indicated with the subscripts ``{0}'' and ``{1/2}'', respectively.  
The torsion groups represented by $``(\Z_2)"$ in the columns of $E_{\infty}^{3,0}$ are the $E_3^{3,0}$ groups that disappear after taking the cohomology of the differential $d_3^{0,0}:E_3^{0,0} \to E_3^{3,-2}$. 
I.e.,\ $``(\Z_2)"$ group means the existence of the symmetry-based indicator~\cite{Haruki230} for Weyl semimetals. 
}
\label{tab:230}
\begin{align*}
\begin{array}{llllllllll}
 \text{SG} & \text{Intl} & & E_{\infty}^{0,0} & E_{\infty}^{1,0} & E_{\infty}^{2,0} & E_{\infty}^{3,0} \\
 \hline
 1 & P1 & & \Z & \Z^3 & \Z^3 & \Z \\
 2 & P\bar 1 & & \Z^9 & 0 & \Z^3 & 0(\Z_2) \\
 3 &  P2 & & \Z^5 & \Z^5 & \Z & \Z \\
 4 & P2_1 & & \Z & \Z+\Z_2^3 & \Z & \Z \\
 5 &  C2 & & \Z^3 & \Z^3 & \Z & \Z \\
 6 &  Pm & & \Z^3 & \Z^6 & \Z^3 & 0 \\
 7 & Pc  & & \Z & \Z^2+\Z_2 & \Z+\Z_2 & 0 \\
 8 & Cm  & & \Z^2 & \Z^4 & \Z^2 & 0 \\
 9 & Cc  & & \Z & \Z^2 & \Z+\Z_2 & 0 \\
 \hline
 \\
 \text{SG} & \text{Intl} & \epsilon(2_{001},m_{001}) & E_{\infty}^{0,0} & E_{\infty}^{1,0} & E_{\infty}^{2,0} & E_{\infty}^{3,0} \\
 \hline
 10 &  P2/m & +_{0,1/2} & \Z^{15} & 0 & \Z^3 & 0 \\
  &   & - & \Z & \Z^8 & \Z & 0 \\
 11 & P2_1/m  & +_{0,1/2},- & \Z^6 & \Z^2 & \Z^2 & 0 \\
 12 &  C2/m & +_{0,1/2} & \Z^{10} & 0 & \Z^2 & 0 \\
  &   & - & \Z^3 & \Z^4 & \Z & 0 \\
 13 &  P2/c & +_{0,1/2},- & \Z^7 & \Z^2 & \Z & 0 \\
 14 &  P2_1/c & +_{0,1/2},- & \Z^5 & \Z_2 & \Z & 0 \\
 15 &  C2/c & +_{0,1/2},- & \Z^6 & \Z & \Z & 0 \\
 \hline 
 \\
 \text{SG} & \text{Intl} & \epsilon(2_{100},2_{010}) & E_{\infty}^{0,0} & E_{\infty}^{1,0} & E_{\infty}^{2,0} & E_{\infty}^{3,0} \\
 \hline
 16 &  P222 & +_{0} & \Z^{13} & \Z_2 & 0 & \Z \\
  &   & -_{1/2} & \Z & \Z^{12} & 0 & \Z \\
 17 &  P222_1 & +_{0},-_{1/2} & \Z^5 & \Z^4+\Z_2 & 0 & \Z \\
 18 & P2_12_12  & +_{0},-_{1/2} & \Z^3 & \Z^2+\Z_2^3 & 0 & \Z \\
 19 &  P2_12_12_1 & +_{0},-_{1/2} & \Z & \Z_4^3 & 0 & \Z \\
 20 & C222_1  & +_{0},-_{1/2} & \Z^3 & \Z^2+\Z_2^2 & 0 & \Z \\
 21 &  C222 & +_{0} & \Z^8 & \Z+\Z_2 & 0 & \Z \\
  &   & -_{1/2} & \Z^2 & \Z^7 & 0 & \Z \\
 22 &  F222 & +_{0} & \Z^7 & \Z_2 & \Z_2 & \Z \\
  &   & -_{1/2} & \Z & \Z^6 & \Z_2 & \Z \\
 23 &  I222 & +_{0} & \Z^7 & \Z_2^2 & 0 & \Z \\
  &   & -_{1/2} & \Z & \Z^6+\Z_2 & 0 & \Z \\
 24 &  I2_12_12_1 & +_{0},-_{1/2} & \Z^4 & \Z^3+\Z_2 & 0 & \Z \\ 
 \hline
\end{array}
&& && 
\begin{array}{llllllll}
 \text{SG} & \text{Intl} & \epsilon(m_{100},m_{010}) & E_{\infty}^{0,0} & E_{\infty}^{1,0} & E_{\infty}^{2,0} & E_{\infty}^{3,0} \\
 \hline
 25 &  Pmm2 & +_{0} & \Z^9 & \Z^9 & 0 & 0 \\
  &   & -_{1/2} & \Z & \Z^5 & \Z^4 & 0 \\
 26 & Pmc2_1  & +_{0} & \Z^3 & \Z^3+\Z_2^3 & 0 & 0 \\
  &   & -_{1/2} & \Z & \Z^3 & \Z^2+\Z_2 & 0 \\
 27 & Pcc2  & +_{0} & \Z^5 & \Z^5 & 0 & 0 \\
  &   & -_{1/2} & \Z & \Z & \Z_2^4 & 0 \\
 28 & Pma2  & +_{0},-_{1/2} & \Z^4 & \Z^5 & \Z & 0 \\
 29 & Pca2_1  & +_{0},-_{1/2} & \Z & \Z+\Z_2^2 & \Z_2 & 0 \\
 30 &  Pna2_1 & +_{0},-_{1/2} & \Z^3 & \Z^3 & \Z_2 & 0 \\
 31 &  Pmn2_1 & +_{0},-_{1/2} & \Z^2 & \Z^3+\Z_2 & \Z & 0 \\
 32 & Pba2  & +_{0},-_{1/2} & \Z^3 & \Z^3+\Z_2 & \Z_2 & 0 \\
 33 &  Pna2_1 & +_{0},-_{1/2} & \Z & \Z+\Z_4 & \Z_2 & 0 \\
  34 & Pnn2  & +_{0},-_{1/2} & \Z^3 & \Z^3 & \Z_2 & 0 \\
 35 &  Cmm2 & +_{0} & \Z^6 & \Z^6 & 0 & 0 \\
  &   & -_{1/2} & \Z^2 & \Z^4 & \Z^2 & 0 \\
 36 &  Cmc2_1 & +_{0} & \Z^2 & \Z^2+\Z_2^2 & 0 & 0 \\
  &   & -_{1/2} & \Z & \Z^2+\Z_2 & \Z & 0 \\
 37 & Ccc2  & +_{0} & \Z^4 & \Z^4 & 0 & 0 \\
  &   & -_{1/2} & \Z^2 & \Z^2 & \Z_2^2 & 0 \\
 38 & Amm2  & +_{0} & \Z^6 & \Z^6 & 0 & 0 \\
  &   & -_{1/2} & \Z & \Z^4 & \Z^3 & 0 \\
 39 &  Abm2 & +_{0} & \Z^4 & \Z^4+\Z_2 & 0 & 0 \\
  &   & -_{1/2} & \Z & \Z^2 & \Z+\Z_2^2 & 0 \\
 40 &  Ama2 & +_{0},-_{1/2} & \Z^3 & \Z^4 & \Z & 0 \\
 41 &  Aba2 & +_{0},-_{1/2} & \Z^2 & \Z^2+\Z_2 & \Z_2 & 0 \\
 42 &  Fmm2 & +_{0} & \Z^5 & \Z^5 & 0 & 0 \\
  &   & -_{1/2} & \Z & \Z^3 & \Z^2+\Z_2 & 0 \\
 43 &  Fdd2 & +_{0},-_{1/2} & \Z^2 & \Z^2 & \Z_2 & 0 \\
 44 & Imm2  & +_{0} & \Z^5 & \Z^5 & 0 & 0 \\
  &   & -_{1/2} & \Z & \Z^3+\Z_2 & \Z^2 & 0 \\
 45 &  Iba2 & +_{0} & \Z^3 & \Z^3+\Z_2 & 0 & 0 \\
  &   & -_{1/2} & \Z & \Z+\Z_2 & \Z_2^2 & 0 \\
 46 & Ima2  & +_{0} & \Z^3 & \Z^3+\Z_2 & 0 & 0 \\
  &   & -_{1/2} & \Z^2 & \Z^3 & \Z & 0 \\
\hline
\end{array}
\end{align*}
\begin{align*}
\begin{array}{llllllll}
 \text{SG} & \text{Intl} & \left(\epsilon(m_{100},m_{010}),\epsilon(m_{100},m_{001}),\epsilon(m_{010},m_{001})\right) & E_{\infty}^{0,0} & E_{\infty}^{1,0} & E_{\infty}^{2,0} & E_{\infty}^{3,0} \\
 \hline
 47 &  Pmmm & (+,+,+)_{0} & \Z^{27} & 0 & 0 & 0 \\
  &   & (-,-,-)_{1/2} & \Z^9 & 0 & \Z^6 & 0 \\
  &   & (-,+,+),(+,-,+),(+,+,-) & \Z^3 & \Z^{12} & 0 & 0 \\
  &   & (+,-,-),(-,+,-),(-,-,+) & \Z^5 & \Z^4 & \Z^2 & 0 \\
 48 & Pnnn  & (+,+,+)_{0},(+,-,-),(-,+,-),(-,-,+) & \Z^9 & 0 & \Z_2 & 0 \\
  &   & (-,-,-)_{1/2},(-,+,+),(+,-,+),(+,+,-) & \Z^3 & \Z^6 & 0 & 0 \\
 49 & Pccm  & (+,+,+)_{0},(+,-,-) & \Z^{14} & 0 & \Z & 0 \\
  &   & (-,-,-)_{1/2},(-,+,+) & \Z^6 & \Z^4 & \Z & 0 \\
  &   & (+,-,+),(+,+,-) & \Z & \Z^{10} & 0 & 0 \\
  &   & (-,+,-),(-,-,+) & \Z^5 & \Z^2 & \Z_2 & 0 \\
 50 &  Pban & (+,+,+)_{0},(+,-,-),(-,+,-),(-,-,+) & \Z^9 & 0 & \Z_2 & 0 \\
  &   & (-,-,-)_{1/2},(-,+,+),(+,-,+),(+,+,-) & \Z^3 & \Z^6 & 0 & 0 \\
 51 & Pmma  & (+,+,+)_{0},(+,-,+) & \Z^{12} & \Z^3 & 0 & 0 \\
  &   & (-,-,-)_{1/2},(-,+,-) & \Z^7 & \Z & \Z^3 & 0 \\
  &   & (+,+,-),(+,-,-) & \Z^4 & \Z^7 & 0 & 0 \\
  &   & (-,+,+),(-,-,+) & \Z & \Z^5+\Z_2 & \Z & 0 \\
 52 &  Pnna & {\rm all} & \Z^5 & \Z^2 & 0 & 0 \\
 53 &  Pmna & (+,+,+)_{0},(-,-,-)_{1/2},(+,+,-),(-,-,+) & \Z^9 & \Z & \Z & 0 \\
  &   & (+,-,+),(-,+,+),(+,-,-),(-,+,-) & \Z^2 & \Z^5 & 0 & 0 \\
  \hline
 \end{array}
\end{align*}
\end{center}
\end{table*}

\begin{table*}
\begin{center}
\caption{(Continued from the previous page)}
\label{tab:e2_2}
\begin{align*}
\begin{array}{llllllll}
 \text{SG} & \text{Intl} & \left(\epsilon(m_{100},m_{010}),\epsilon(m_{100},m_{001}),\epsilon(m_{010},m_{001})\right) & E_{\infty}^{0,0} & E_{\infty}^{1,0} & E_{\infty}^{2,0} & E_{\infty}^{3,0} \\
 \hline
 54 &  Pcca & (+,+,+)_{0},(+,+,-),(+,-,+),(+,-,-) & \Z^6 & \Z^3 & 0 & 0 \\
  &   & (-,-,-)_{1/2},(-,+,+),(-,+,-),(-,-,+) & \Z^4 & \Z & \Z_2 & 0 \\
 55 &  Pbam & (+,+,+)_{0},(-,+,+) & \Z^9 & \Z_2^3 & 0 & 0 \\
  &   & (-,-,-)_{1/2},(+,-,-) & \Z^7 & 0 & \Z^2+\Z_2 & 0 \\
  &   & (+,-,+),(+,+,-),(-,+,-),(-,-,+) & \Z & \Z^4+\Z_2 & 0 & 0 \\
 56 &  Pccn & (+,+,+)_{0},(+,+,-),(+,-,+),(+,-,-) & \Z^5 & \Z^2+\Z_2 & 0 & 0 \\
  &   & (-,-,-)_{1/2},(-,+,+),(-,+,-),(-,-,+) & \Z^3 & \Z_2 & \Z_2 & 0 \\
 57 &  Pbcm & (+,+,+)_{0},(+,+,-),(-,+,+),(-,+,-) & \Z^5 & \Z^2+\Z_2 & 0 & 0 \\
  &   & (-,-,-)_{1/2},(+,-,+),(+,-,-),(-,-,+) & \Z^4 & \Z^2 & \Z & 0 \\
 58 & Pnnm  & (+,+,+)_{0},(-,-,-)_{1/2},(-,+,+),(+,-,-) & \Z^8 & \Z_2 & \Z & 0 \\
  &   & (+,+,-),(+,-,+),(-,+,-),(-,-,+) & \Z & \Z^4+\Z_2 & 0 & 0 \\
 59 &  Pmmn & (+,+,+)_{0},(+,+,-),(+,-,+),(+,-,-) & \Z^7 & \Z^4 & 0 & 0 \\
  &   & (-,-,-)_{1/2},(-,+,+),(-,+,-),(-,-,+) & \Z^3 & \Z^2+\Z_2 & \Z^2 & 0 \\
 60 &  Pbcn & {\rm all} & \Z^4 & \Z+\Z_2 & 0 & 0 \\
 61 & Pbca  & {\rm all} & \Z^3 & \Z_2^2 & 0 & 0 \\
 62 &  Pnma & (+,+,+)_{0},(+,-,+),(-,+,+),(-,-,+) & \Z^4 & \Z+\Z_2^2 & 0 & 0 \\
  &   & (-,-,-)_{1/2},(+,-,-),(-,+,-),(+,+,-) & \Z^3 & \Z+\Z_2 & \Z & 0 \\
 63 &  Cmcm & (+,+,+)_{0},(+,+,-) & \Z^8 & \Z^2 & 0 & 0 \\
  &   & (-,-,-)_{1/2},(-,-,+) & \Z^5 & \Z & \Z^2 & 0 \\
  &   & (+,-,+),(+,-,-) & \Z^2 & \Z^3 & \Z & 0 \\
  &   & (-,+,+),(-,+,-) & \Z^4 & \Z^4 & 0 & 0 \\
 64 & Cmca  & (+,+,+)_{0},(+,+,-) & \Z^7 & \Z+\Z_2 & 0 & 0 \\
  &   & (-,-,-)_{1/2},(-,-,+) & \Z^5 & 0 & \Z+\Z_2 & 0 \\
  &   & (+,-,+),(+,-,-) & \Z^2 & \Z^2 & \Z_2 & 0 \\
  &   & (-,+,+),(-,+,-) & \Z^3 & \Z^3 & 0 & 0 \\
 65 & Cmmm  & (+,+,+)_{0} & \Z^{18} & 0 & 0 & 0 \\
  &   & (-,-,-)_{1/2} & \Z^8 & 0 & \Z^4 & 0 \\
  &   & (+,+,-),(+,-,+) & \Z^2 & \Z^8 & 0 & 0 \\
  &   & (-,+,+) & \Z^6 & \Z^6 & 0 & 0 \\
  &   & (+,-,-) & \Z^6 & \Z^2 & \Z^2 & 0 \\
  &   & (-,+,-),(-,-,+) & \Z^3 & \Z^4 & \Z & 0 \\
 66 &  Cccm & (+,+,+)_{0},(+,-,-) & \Z^{11} & 0 & \Z & 0 \\
  &   & (-,-,-)_{1/2},(-,+,+) & \Z^7 & \Z^2 & \Z & 0 \\
  &   & (+,+,-),(+,-,+) & \Z & \Z^7 & 0 & 0 \\
  &   & (-,+,-),(-,-,+) & \Z^3 & \Z^3 & \Z_2 & 0 \\
   67 & Cmma  & (+,+,+)_{0} & \Z^{13} & \Z & 0 & 0 \\
  &   & (-,-,-)_{1/2} & \Z^5 & \Z & \Z^2+\Z_2 & 0 \\
  &   & (+,+,-),(+,-,+) & \Z^5 & \Z^5 & 0 & 0 \\
  &   & (-,+,+) & \Z & \Z^7+\Z_2 & 0 & 0 \\
  &   & (+,-,-) & \Z^3 & \Z^3 & \Z_2^2 & 0 \\
  &   & (-,+,-),(-,-,+) & \Z^6 & \Z & \Z & 0 \\
 68 &  Ccca & (+,+,+)_{0},(+,-,-) & \Z^7 & \Z & \Z_2 & 0 \\
  &   & (-,-,-)_{1/2},(-,+,+) & \Z^3 & \Z^3 & \Z_2 & 0 \\
  &   & (+,+,-),(+,-,+) & \Z^4 & \Z^4 & 0 & 0 \\
  &   & (-,+,-),(-,-,+) & \Z^6 & 0 & \Z_2 & 0 \\
 69 &  Fmmm & (+,+,+)_{0} & \Z^{15} & 0 & 0 & 0 \\
  &   & (-,-,-)_{1/2} & \Z^6 & 0 & \Z^3+\Z_2 & 0 \\
  &   & (+,+,-),(+,-,+),(-,+,+) & \Z^3 & \Z^6 & 0 & 0 \\
  &   & (+,-,-),(-,+,-),(-,-,+) & \Z^4 & \Z^2 & \Z & 0 \\
 70 &  Fddd & (+,+,+)_{0},(+,-,-),(-,+,-),(-,-,+) & \Z^6 & 0 & \Z_2 & 0 \\
  &   & (-,-,-)_{1/2},(+,+,-),(+,-,+),(-,+,+) & \Z^3 & \Z^3 & 0 & 0 \\
 71 & Immm  & (+,+,+)_{0} & \Z^{15} & 0 & 0 & 0 \\
  &   & (-,-,-)_{1/2} & \Z^6 & \Z_2 & \Z^3 & 0 \\
  &   & (+,+,-),(+,-,+),(-,+,+) & \Z^3 & \Z^6 & 0 & 0 \\
  &   & (+,-,-),(-,+,-),(-,-,+) & \Z^4 & \Z^2 & \Z & 0 \\ 
  \hline
 \end{array}
\end{align*}
\end{center}
\end{table*}

\begin{table*}
\begin{center}
\caption{(Continued from the previous page)}
\label{tab:e2_3}
\begin{align*}
\begin{array}{llllllll}
 \text{SG} & \text{Intl} & \left(\epsilon(m_{100},m_{010}),\epsilon(m_{100},m_{001}),\epsilon(m_{010},m_{001})\right) & E_{\infty}^{0,0} & E_{\infty}^{1,0} & E_{\infty}^{2,0} & E_{\infty}^{3,0} \\
 \hline
  72 &  Ibam & (+,+,+)_{0} & \Z^9 & \Z_2 & 0 & 0 \\
  &   & (-,-,-)_{1/2} & \Z^4 & \Z^2 & \Z+\Z_2 & 0 \\
  &   & (+,+,-),(+,-,+) & \Z^2 & \Z^5 & 0 & 0 \\
  &   & (-,+,+) & \Z^5 & \Z^2+\Z_2 & 0 & 0 \\
  &   & (+,-,-) & \Z^8 & 0 & \Z & 0 \\
  &   & (-,+,-),(-,-,+) & \Z^4 & \Z & \Z_2 & 0 \\
 73 &  Ibca & (+,+,+)_{0} & \Z^6 & \Z^3 & 0 & 0 \\
  &   & (-,-,-)_{1/2} & \Z^3 & 0 & \Z_2^2 & 0 \\
  &   & (+,+,-),(+,-,+),(-,+,+) & \Z^5 & \Z^2 & 0 & 0 \\
  &   & (+,-,-),(-,+,-),(-,-,+) & \Z^4 & \Z & \Z_2 & 0 \\
 74 &  Imma & (+,+,+)_{0} & \Z^{10} & \Z & 0 & 0 \\
  &   & (-,-,-)_{1/2} & \Z^7 & 0 & \Z^2 & 0 \\
  &   & (+,+,-),(+,-,+) & \Z^6 & \Z^3 & 0 & 0 \\
  &   & (-,+,+) & \Z & \Z^4+\Z_2 & 0 & 0 \\
  &   & (+,-,-) & \Z^2 & \Z^5 & \Z_2 & 0 \\
  &   & (-,+,-),(-,-,+) & \Z^4 & \Z^2 & \Z & 0 \\
  \hline
\end{array}
\end{align*}
\begin{align*}
\begin{array}{lllllllll}
 \text{SG} & \text{Intl} & & E_{\infty}^{0,0} & E_{\infty}^{1,0} & E_{\infty}^{2,0} & E_{\infty}^{3,0} \\
 \hline
 75 &  P4 & & \Z^8 & \Z^8 & \Z & \Z \\
 76 & P4_1 & & \Z & \Z+\Z_2+\Z_4 & \Z & \Z \\
 77 &  P4_2 & & \Z^4 & \Z^4+\Z_2 & \Z & \Z \\
 78 & P4_3 & & \Z & \Z+\Z_2+\Z_4 & \Z & \Z \\
 79 & I4 & & \Z^5 & \Z^5 & \Z & \Z \\
 80 &  I4_1 & & \Z^2 & \Z^2+\Z_2 & \Z & \Z \\
 81 &  P\bar 4 & & \Z^{12} & \Z & \Z & 0(\Z_2) \\
 82 & I\bar 4 & & \Z^{11} & 0 & \Z & 0(\Z_2) \\
\hline
\\
 \text{SG} & \text{Intl} & \epsilon(2_{001},m_{001}) & E_{\infty}^{0,0} & E_{\infty}^{1,0} & E_{\infty}^{2,0} & E_{\infty}^{3,0} \\
 \hline
 83 &  P4/m & +_{0,1/2} & \Z^{24} & 0 & \Z^3 & 0 \\
  &   & - & \Z^6 & \Z^4 & \Z & 0 \\
 84 & P4_2/m  & +_{0,1/2},- & \Z^{13} & 0 & \Z^2 & 0 \\
 85 & P4/n  & +_{0,1/2},- & \Z^{11} & \Z^3 & \Z & 0 \\
 86 &  P4_2/n & +_{0,1/2},- & \Z^9 & \Z & \Z & 0 \\
 87 & I4/m  & +_{0,1/2} & \Z^{16} & 0 & \Z^2 & 0 \\
  &   & - & \Z^7 & \Z^2 & \Z & 0 \\
 88 &  I4_2/a & +_{0,1/2},- & \Z^8 & 0 & \Z & 0 \\
 \hline
 \\
  \text{SG} & \text{Intl} & \epsilon(2_{001},2_{100}) & E_{\infty}^{0,0} & E_{\infty}^{1,0} & E_{\infty}^{2,0} & E_{\infty}^{3,0} \\
 \hline
 89 & P422  & +_{0} & \Z^{12} & \Z^2+\Z_2 & 0 & \Z \\
  &   & -_{1/2} & \Z^3 & \Z^{11} & 0 & \Z \\
 90 & P42_12  & +_{0} & \Z^7 & \Z^3+\Z_2 & 0 & \Z \\
  &   & -_{1/2} & \Z^4 & \Z^6+\Z_2 & 0 & \Z \\
 91 &  P4_122 & +_{0},-_{1/2} & \Z^4 & \Z^3+\Z_4 & 0 & \Z \\
 92 &  P4_12_12 & +_{0},-_{1/2} & \Z^2 & \Z+\Z_4^2 & 0 & \Z \\
 93 &  P4_222 & +_{0} & \Z^{10} & \Z_4 & 0 & \Z \\
  &   & -_{1/2} & \Z & \Z^9+\Z_2 & 0 & \Z \\
 94 & P4_22_12  & +_{0} & \Z^5 & \Z+\Z_2+\Z_4 & 0 & \Z \\
  &   & -_{1/2} & \Z^2 & \Z^4+\Z_2^2 & 0 & \Z \\
 95 & P4_322  & +_{0},-_{1/2} & \Z^4 & \Z^3+\Z_4 & 0 & \Z \\
 96 &  P4_32_12 & +_{0},-_{1/2} & \Z^2 & \Z+\Z_4^2 & 0 & \Z \\
 97 &  I422 & +_{0} & \Z^8 & \Z+\Z_2 & 0 & \Z \\
  &   & -_{1/2} & \Z^2 & \Z^7 & 0 & \Z \\
 98 &  I4_122 & +_{0} & \Z^5 & \Z+\Z_4 & 0 & \Z \\
  &   & -_{1/2} & \Z^2 & \Z^4+\Z_2 & 0 & \Z \\
\hline \\
\end{array}
&& &&
\begin{array}{lllllllll}
 \text{SG} & \text{Intl} & \epsilon(2_{001},m_{010}) & E_{\infty}^{0,0} & E_{\infty}^{1,0} & E_{\infty}^{2,0} & E_{\infty}^{3,0} \\
 \hline
 99 &  P4mm & +_{0} & \Z^9 & \Z^9 & 0 & 0 \\
  &   & -_{1/2} & \Z^3 & \Z^6 & \Z^3 & 0 \\
 100 &  P4bm & +_{0} & \Z^6 & \Z^6 & 0 & 0 \\
  &   & -_{1/2} & \Z^4 & \Z^5 & \Z & 0 \\
 101 &  P4_2cm & +_{0} & \Z^5 & \Z^5 & 0 & 0 \\
  &   & -_{1/2} & \Z & \Z^2+\Z_2 & \Z+\Z_2 & 0 \\
 102 &  P4_2nm & +_{0} & \Z^4 & \Z^4 & 0 & 0 \\
  &   & -_{1/2} & \Z^2 & \Z^3+\Z_2 & \Z & 0 \\
 103 &  P4cc & +_{0} & \Z^6 & \Z^6 & 0 & 0 \\
  &   & -_{1/2} & \Z^3 & \Z^3 & \Z_2^3 & 0 \\
 104 &  P4nc & +_{0} & \Z^5 & \Z^5 & 0 & 0 \\
  &   & -_{1/2} & \Z^4 & \Z^4 & \Z_2 & 0 \\
 105 &  P4_2mc & +_{0} & \Z^6 & \Z^6 & 0 & 0 \\
  &   & -_{1/2} & \Z & \Z^3+\Z_2 & \Z^2 & 0 \\
 106 &  P4_2bc & +_{0} & \Z^3 & \Z^3+\Z_2 & 0 & 0 \\
  &   & -_{1/2} & \Z^2 & \Z^2+\Z_2 & \Z_2 & 0 \\
 107 &  I4mm & +_{0} & \Z^6 & \Z^6 & 0 & 0 \\
  &   & -_{1/2} & \Z^2 & \Z^4 & \Z^2 & 0 \\
 108 &  I4cm & +_{0} & \Z^5 & \Z^5 & 0 & 0 \\
  &   & -_{1/2} & \Z^2 & \Z^3 & \Z+\Z_2 & 0 \\
 109 &  I4_1md & +_{0} & \Z^3 & \Z^3 & 0 & 0 \\
  &   & -_{1/2} & \Z & \Z^2+\Z_2 & \Z & 0 \\
 110 &  I4_1cd & +_{0} & \Z^2 & \Z^2+\Z_2 & 0 & 0 \\
  &   & -_{1/2} & \Z & \Z+\Z_2 & \Z_2 & 0 \\
\hline
\\
 \text{SG} & \text{Intl} & \epsilon(2_{001},2_{010}) & E_{\infty}^{0,0} & E_{\infty}^{1,0} & E_{\infty}^{2,0} & E_{\infty}^{3,0} \\
 \hline
 111 &  P\bar 42m & +_{0} & \Z^{13} & \Z & 0 & 0 \\
  &   & -_{1/2} & \Z^5 & \Z^6 & \Z & 0 \\
 112 &  P\bar 42c & +_{0} & \Z^{12} & 0 & 0 & 0 \\
  &   & -_{1/2} & \Z^5 & \Z^5 & 0 & 0 \\
 113 &  P\bar 42_1m & +_{0} & \Z^8 & \Z^2 & 0 & 0 \\
  &   & -_{1/2} & \Z^6 & \Z+\Z_2 & \Z & 0 \\
 114 &  P\bar 42_1c & +_{0} & \Z^7 & \Z+\Z_2 & 0 & 0 \\
  &   & -_{1/2} & \Z^6 & \Z_2 & 0 & 0 \\
 121 &  I\bar 42m & +_{0} & \Z^{10} & \Z & 0 & 0 \\
  &   & -_{1/2} & \Z^5 & \Z^3 & \Z & 0 \\
 122 & I\bar 42d  & +_{0},-_{1/2} & \Z^7 & \Z & 0 & 0 \\
\hline
\end{array}
\end{align*}
\end{center}
\end{table*}

\begin{table*}
\begin{center}
\caption{(Continued from the previous page)}
\label{tab:e2_4}
\begin{align*}
\begin{array}{llllllll}
 \text{SG} & \text{Intl} & \epsilon(2_{001},m_{010}) & E_{\infty}^{0,0} & E_{\infty}^{1,0} & E_{\infty}^{2,0} & E_{\infty}^{3,0} \\
 \hline
  115 &  P\bar 4m2 & +_{0} & \Z^{12} & \Z^3 & 0 & 0 \\
  &   & -_{1/2} & \Z^5 & \Z^4 & \Z^2 & 0 \\
 116 &  P\bar 4c2 & +_{0} & \Z^{10} & \Z & 0 & 0 \\
  &   & -_{1/2} & \Z^5 & \Z^2 & \Z_2 & 0 \\
 117 & P\bar 4b2  & +_{0} & \Z^9 & 0 & \Z_2 & 0 \\
  &   & -_{1/2} & \Z^6 & \Z^3 & 0 & 0 \\
 118 &  P\bar 4n2 & +_{0} & \Z^9 & 0 & \Z_2 & 0 \\
  &   & -_{1/2} & \Z^6 & \Z^3 & 0 & 0 \\
 119 &  I\bar 4m2 & +_{0} & \Z^{10} & \Z & 0 & 0 \\
  &   & -_{1/2} & \Z^5 & \Z^3 & \Z & 0 \\
 120 &  I\bar 4c2 & +_{0} & \Z^9 & 0 & 0 & 0 \\
  &   & -_{1/2} & \Z^5 & \Z^2 & \Z_2 & 0 \\
\hline
\end{array}
\end{align*}
\begin{align*}
\begin{array}{llllllll}
 \text{SG} & \text{Intl} & \left(\epsilon(m_{100},m_{010}),\epsilon(m_{100},m_{001}),\epsilon(4^+_{001},m_{001})\right) & E_{\infty}^{0,0} & E_{\infty}^{1,0} & E_{\infty}^{2,0} & E_{\infty}^{3,0} \\
 \hline
 123 & P4/mmm  & (+,+,+)_{0} & \Z^{27} & 0 & 0 & 0 \\
  &   & (-,-,+)_{1/2} & \Z^{13} & 0 & \Z^5 & 0 \\
  &   & (+,-,+) & \Z^{10} & \Z^3 & \Z^2 & 0 \\
  &   & (+,-,-) & \Z^7 & \Z^4 & 0 & 0 \\
  &   & (+,+,-) & \Z^{12} & \Z^3 & 0 & 0 \\
  &   & (-,+,+) & \Z^9 & \Z^9 & 0 & 0 \\
  &   & (-,-,-) & \Z^3 & \Z^5 & \Z^2 & 0 \\
  &   & (-,+,-) & \Z & \Z^8 & \Z & 0 \\
 124 & P4/mcc  & (+,+,+)_{0},(+,-,+) & \Z^{17} & 0 & \Z & 0 \\
  &   & (-,-,+)_{1/2},(-,+,+) & \Z^{11} & \Z^3 & \Z & 0 \\
  &   & (+,-,-),(+,+,-) & \Z^8 & \Z^2 & 0 & 0 \\
  &   & (-,-,-),(-,+,-) & \Z^2 & \Z^5 & \Z_2 & 0 \\
 125 &  P4/nbm & (+,+,+)_{0},(+,-,-) & \Z^{13} & \Z & 0 & 0 \\
  &   & (-,-,+)_{1/2},(-,+,-) & \Z^6 & \Z^4 & \Z & 0 \\
  &   & (+,-,+),(+,+,-) & \Z^8 & \Z^2 & 0 & 0 \\
  &   & (-,+,+),(-,-,-) & \Z^4 & \Z^7 & 0 & 0 \\
 126 &  P4/nnc & (+,+,+)_{0},(+,-,+),(+,-,-),(+,+,-) & \Z^{10} & \Z & 0 & 0 \\
  &   & (-,-,+)_{1/2},(-,+,+),(-,-,-),(-,+,-) & \Z^5 & \Z^5 & 0 & 0 \\
 127 &  P4/mbm & (+,+,+)_{0} & \Z^{18} & 0 & 0 & 0 \\
  &   & (-,-,+)_{1/2} & \Z^{12} & 0 & \Z^3 & 0 \\
  &   & (+,-,+) & \Z^{11} & \Z & \Z^2 & 0 \\
  &   & (+,-,-) & \Z^8 & \Z^2 & 0 & 0 \\
  &   & (+,+,-) & \Z^3 & \Z^3+\Z_2^2 & 0 & 0 \\
  &   & (-,+,+) & \Z^{12} & \Z^3 & 0 & 0 \\
  &   & (-,-,-) & \Z^2 & \Z^5 & \Z_2 & 0 \\
  &   & (-,+,-) & \Z^4 & \Z^2+\Z_2 & \Z & 0 \\
 128 &  P4/mnc & (+,+,+)_{0},(+,-,+) & \Z^{14} & 0 & \Z & 0 \\
  &   & (-,-,+)_{1/2},(-,+,+) & \Z^{12} & \Z & \Z & 0 \\
  &   & (+,-,-),(+,+,-) & \Z^5 & \Z^2 & 0 & 0 \\
  &   & (-,-,-),(-,+,-) & \Z^3 & \Z^3 & \Z_2 & 0 \\
 129 &  P4/nmm & (+,+,+)_{0},(+,-,-) & \Z^{12} & \Z^3 & 0 & 0 \\
  &   & (-,-,+)_{1/2},(-,+,-) & \Z^6 & \Z^2 & \Z^2 & 0 \\
  &   & (+,-,+),(+,+,-) & \Z^7 & \Z^4 & 0 & 0 \\
  &   & (-,+,+),(-,-,-) & \Z^4 & \Z^5 & \Z & 0 \\
 130 &  P4/ncc & (+,+,+)_{0},(+,-,+),(+,-,-),(+,+,-) & \Z^8 & \Z^2 & 0 & 0 \\
  &   & (-,-,+)_{1/2},(-,+,+),(-,-,-),(-,+,-) & \Z^5 & \Z^2 & \Z_2 & 0 \\
 131 & P4_2/mmc  & (+,+,+)_{0},(+,+,-) & \Z^{18} & 0 & 0 & 0 \\
  &   & (-,-,+)_{1/2},(-,-,-) & \Z^7 & \Z & \Z^3 & 0 \\
  &   & (+,-,+),(+,-,-) & \Z^7 & \Z^2 & \Z & 0 \\
  &   & (-,+,+),(-,+,-) & \Z^4 & \Z^7 & 0 & 0 \\
 132 & P4_2/mcm  & (+,+,+)_{0},(+,-,-) & \Z^{15} & 0 & 0 & 0 \\
  &   & (-,-,+)_{1/2},(-,+,-) & \Z^6 & \Z^2 & \Z^2 & 0 \\
  &   & (+,-,+),(+,+,-) & \Z^9 & \Z & \Z & 0 \\
  &   & (-,+,+),(-,-,-) & \Z^5 & \Z^5 & 0 & 0 \\
  \hline
 \end{array}
\end{align*}
\end{center}
\end{table*}

\begin{table*}
\begin{center}
\caption{(Continued from the previous page)}
\label{tab:e2_5}
\begin{align*}
\begin{array}{llllllll}
 \text{SG} & \text{Intl} & \left(\epsilon(m_{100},m_{010}),\epsilon(m_{100},m_{001}),\epsilon(4^+_{001},m_{001})\right) & E_{\infty}^{0,0} & E_{\infty}^{1,0} & E_{\infty}^{2,0} & E_{\infty}^{3,0} \\
 \hline
 133 &  P4_2/nbc & (+,+,+)_{0},(+,-,+),(+,-,-),(+,+,-) & \Z^9 & 0 & 0 & 0 \\
  &   & (-,-,+)_{1/2},(-,+,+),(-,-,-),(-,+,-) & \Z^4 & \Z^4 & 0 & 0 \\
 134 & P4_2/nnm  & (+,+,+)_{0},(+,-,-) & \Z^{12} & 0 & 0 & 0 \\
  &   & (-,-,+)_{1/2},(-,+,-) & \Z^5 & \Z^3 & \Z & 0 \\
  &   & (+,-,+),(+,+,-) & \Z^7 & \Z & 0 & 0 \\
  &   & (-,+,+),(-,-,-) & \Z^3 & \Z^6 & 0 & 0 \\
 135 &  P4_2/mbc & (+,+,+)_{0},(+,+,-) & \Z^9 & \Z_2 & 0 & 0 \\
  &   & (-,-,+)_{1/2},(-,-,-) & \Z^6 & \Z & \Z+\Z_2 & 0 \\
  &   & (+,-,+),(+,-,-) & \Z^8 & 0 & \Z & 0 \\
  &   & (-,+,+),(-,+,-) & \Z^7 & \Z+\Z_2 & 0 & 0 \\
 136 & P4_2/mnm  & (+,+,+)_{0},(+,-,-) & \Z^{12} & 0 & 0 & 0 \\
  &   & (-,-,+)_{1/2},(-,+,-) & \Z^7 & \Z_2 & \Z^2 & 0 \\
  &   & (+,-,+),(+,+,-) & \Z^6 & \Z & \Z & 0 \\
  &   & (-,+,+),(-,-,-) & \Z^6 & \Z^3 & 0 & 0 \\
 137 &  P4_2/nmc & (+,+,+)_{0},(+,-,+),(+,-,-),(+,+,-) & \Z^8 & \Z^2 & 0 & 0 \\
  &   & (-,-,+)_{1/2},(-,+,+),(-,-,-),(-,+,-) & \Z^4 & \Z^2+\Z_2 & \Z & 0 \\
 138 & P4_2/ncm  & (+,+,+)_{0},(+,-,-) & \Z^{10} & \Z & 0 & 0 \\
  &   & (-,-,+)_{1/2},(-,+,-) & \Z^5 & \Z_2 & \Z+\Z_2 & 0 \\
  &   & (+,-,+),(+,+,-) & \Z^5 & \Z^2 & 0 & 0 \\
  &   & (-,+,+),(-,-,-) & \Z^3 & \Z^3+\Z_2 & 0 & 0 \\
 139 &  I4/mmm & (+,+,+)_{0} & \Z^{18} & 0 & 0 & 0 \\
  &   & (-,-,+)_{1/2} & \Z^9 & 0 & \Z^3 & 0 \\
  &   & (+,-,+) & \Z^7 & \Z^2 & \Z & 0 \\
  &   & (+,-,-),(+,+,-) & \Z^8 & \Z^2 & 0 & 0 \\
  &   & (-,+,+) & \Z^6 & \Z^6 & 0 & 0 \\
  &   & (-,-,-),(-,+,-) & \Z^3 & \Z^4 & \Z & 0 \\
 140 & I4/mcm  & (+,+,+)_{0} & \Z^{15} & 0 & 0 & 0 \\
  &   & (-,-,+)_{1/2} & \Z^8 & \Z & \Z^2 & 0 \\
  &   & (+,-,+) & \Z^9 & \Z & \Z & 0 \\
  &   & (+,-,-) & \Z^{10} & \Z & 0 & 0 \\
  &   & (+,+,-) & \Z^5 & \Z^2+\Z_2 & 0 & 0 \\
  &   & (-,+,+) & \Z^7 & \Z^4 & 0 & 0 \\
  &   & (-,-,-) & \Z^2 & \Z^5 & \Z_2 & 0 \\
  &   & (-,+,-) & \Z^4 & \Z^2 & \Z & 0 \\
 141 &  I4_1/amd & (+,+,+)_{0},(+,+,-) & \Z^9 & 0 & 0 & 0 \\
  &   & (-,-,+)_{1/2},(-,-,-) & \Z^6 & \Z & \Z & 0 \\
  &   & (+,-,+),(+,-,-) & \Z^5 & \Z^2 & 0 & 0 \\
  &   & (-,+,+),(-,+,-) & \Z^3 & \Z^3 & 0 & 0 \\
 142 &  I4_1/acd & (+,+,+)_{0},(+,+,-) & \Z^7 & \Z & 0 & 0 \\
  &   & (-,-,+)_{1/2},(-,-,-) & \Z^4 & \Z & \Z_2 & 0 \\
  &   & (+,-,+),(+,-,-) & \Z^6 & 0 & 0 & 0 \\
  &   & (-,+,+),(-,+,-) & \Z^5 & \Z^2 & 0 & 0 \\
  \hline
\end{array}
\end{align*}
\begin{align*}
\begin{array}{llllllll}
 \text{SG} & \text{Intl} & & E_{\infty}^{0,0} & E_{\infty}^{1,0} & E_{\infty}^{2,0} & E_{\infty}^{3,0} \\
 \hline
 143 & P3 & & \Z^7 & \Z^7 & \Z & \Z \\
 144 &  P3_1 & & \Z & \Z+\Z_3^2 & \Z & \Z \\
 145 &  P3_2 & & \Z & \Z+\Z_3^2 & \Z & \Z \\
 146 &  R3 & & \Z^3 & \Z^3 & \Z & \Z \\
 147 &  P\bar 3 & & \Z^{13} & \Z^2 & \Z & 0(\Z_2)\\
 148 & R \bar 3 & & \Z^{11} & 0 & \Z & 0(\Z_2)\\
 149 &  P312 & & \Z^6 & \Z^5 & 0 & \Z \\
 150 &  P321 & & \Z^6 & \Z^5 & 0 & \Z \\
 151 &  P3_112 & & \Z^3 & \Z^2+\Z_3^2 & 0 & \Z \\
 152 &  P3_121 & & \Z^3 & \Z^2+\Z_3 & 0 & \Z \\
\hline
\end{array}
&& &&
\begin{array}{llllllll}
 \text{SG} & \text{Intl} & & E_{\infty}^{0,0} & E_{\infty}^{1,0} & E_{\infty}^{2,0} & E_{\infty}^{3,0} \\
\hline
 153 &  P3_212 & & \Z^3 & \Z^2+\Z_3^2 & 0 & \Z \\
 154 &  P3_221 & & \Z^3 & \Z^2+\Z_3 & 0 & \Z \\
 155 &  R32 & & \Z^4 & \Z^3 & 0 & \Z \\
 156 &  P3m1 & & \Z^5 & \Z^6 & \Z & 0 \\
 157 &  P31m & & \Z^5 & \Z^6 & \Z & 0 \\
 158 &  P3c1 & & \Z^4 & \Z^4 & \Z_2 & 0 \\
 159 &  P31c & & \Z^4 & \Z^4 & \Z_2 & 0 \\
 160 &  R3m & & \Z^3 & \Z^4 & \Z & 0 \\
 161 &  R3c & & \Z^2 & \Z^2 & \Z_2 & 0 \\
 \hline
\end{array}
\end{align*}
\end{center}
\end{table*}

\begin{table*}
\begin{center}
\caption{(Continued from the previous page)}
\label{tab:e2_6}
\begin{align*}
\begin{array}{llllllll}
 \text{SG} & \text{Intl} & \epsilon(-1,m_{1\bar 10}) & E_{\infty}^{0,0} & E_{\infty}^{1,0} & E_{\infty}^{2,0} & E_{\infty}^{3,0} \\
 \hline
 162 &  P\bar 31m & +_{0,1/2} & \Z^{12} & \Z & \Z & 0 \\
  &   & - & \Z^5 & \Z^5 & 0 & 0 \\
 163 &  P\bar 31c & +_{0,1/2},- & \Z^8 & \Z^2 & 0 & 0 \\
\hline \\ 
 \text{SG} & \text{Intl} & \epsilon(-1,m_{010}) & E_{\infty}^{0,0} & E_{\infty}^{1,0} & E_{\infty}^{2,0} & E_{\infty}^{3,0} \\
 \hline
 164 &  P\bar 3m1 & +_{0,1/2} & \Z^{12} & \Z & \Z & 0 \\
  &   & - & \Z^5 & \Z^5 & 0 & 0 \\
 165 &  P\bar 3c1 & +_{0,1/2},- & \Z^8 & \Z^2 & 0 & 0 \\
 166 &  R\bar 3m & +_{0,1/2} & \Z^{11} & 0 & \Z & 0 \\
  &   & - & \Z^4 & \Z^4 & 0 & 0 \\
 167 &  R\bar 3c & +_{0,1/2},- & \Z^7 & \Z & 0 & 0 \\
\hline \\
 \text{SG} & \text{Intl} & & E_{\infty}^{0,0} & E_{\infty}^{1,0} & E_{\infty}^{2,0} & E_{\infty}^{3,0} \\
 \hline
 168 &  P6 & & \Z^9 & \Z^9 & \Z & \Z \\
 169 &  P6_1 & & \Z & \Z+\Z_6 & \Z & \Z \\
 170 &  P6_5 & & \Z & \Z+\Z_6 & \Z & \Z \\
 171 &  P6_2 & & \Z^3 & \Z^3+\Z_3 & \Z & \Z \\
 172 &  P6_4 & & \Z^3 & \Z^3+\Z_3 & \Z & \Z \\
 173 &  P6_3 & & \Z^5 & \Z^5+\Z_2 & \Z & \Z \\
 174 &  P\bar 6 & & \Z^{21} & 0 & \Z^3 & 0 \\
 \hline \\
 \text{SG} & \text{Intl} & \epsilon(2_{001},m_{001}) & E_{\infty}^{0,0} & E_{\infty}^{1,0} & E_{\infty}^{2,0} & E_{\infty}^{3,0} \\
 \hline
 175 &  P6/m & +_{0,1/2} & \Z^{27} & 0 & \Z^3 & 0 \\
  &   & - & \Z^9 & \Z^4 & \Z & 0 \\
 176 &  P6_3/m & +_{0,1/2},- & \Z^{16} & 0 & \Z^2 & 0 \\
\hline \\
\\
\\
\\
\\
\end{array}
&& &&
\begin{array}{llllllll}
 \text{SG} & \text{Intl} & \epsilon(2_{001},2_{110}) & E_{\infty}^{0,0} & E_{\infty}^{1,0} & E_{\infty}^{2,0} & E_{\infty}^{3,0} \\
 \hline
 177 &  P622 & +_{0} & \Z^{10} & \Z^3+\Z_2 & 0 & \Z \\
  &  & -_{1/2} & \Z^4 & \Z^9 & 0 & \Z \\
 178 &  P6_122 & +_{0},-_{1/2} & \Z^3 & \Z^2+\Z_6 & 0 & \Z \\
 179 &  P6_522 & +_{0},-_{1/2} & \Z^3 & \Z^2+\Z_6 & 0 & \Z \\
 180 &  P6_222 & +_{0} & \Z^7 & \Z_6 & 0 & \Z \\
  &   & -_{1/2} & \Z & \Z^6+\Z_3 & 0 & \Z \\
 181 &  P6_422 & +_{0} & \Z^7 & \Z_6 & 0 & \Z \\
  &   & -_{1/2} & \Z & \Z^6+\Z_3 & 0 & \Z \\
 182 &  P6_322 & +_{0},-_{1/2} & \Z^5 & \Z^4+\Z_2 & 0 & \Z \\
\hline \\
 \text{SG} & \text{Intl} & \epsilon(2_{001},m_{1\bar 10}) & E_{\infty}^{0,0} & E_{\infty}^{1,0} & E_{\infty}^{2,0} & E_{\infty}^{3,0} \\
 \hline
 183 &  P6mm & +_{0} & \Z^8 & \Z^8 & 0 & 0 \\
  &   & -_{1/2} & \Z^4 & \Z^6 & \Z^2 & 0 \\
 184 &  P6cc & +_{0} & \Z^6 & \Z^6 & 0 & 0 \\
  &   & -_{1/2} & \Z^4 & \Z^4 & \Z_2^2 & 0 \\
 185 &  P6_3cm & +_{0} & \Z^4 & \Z^4+\Z_2 & 0 & 0 \\
  &   & -_{1/2} & \Z^3 & \Z^4 & \Z & 0 \\
 186 &  P6_3mc & +_{0} & \Z^4 & \Z^4+\Z_2 & 0 & 0 \\
  &   & -_{1/2} & \Z^3 & \Z^4 & \Z & 0 \\
\hline \\
 \text{SG} & \text{Intl} & \epsilon(m_{010},m_{001}) & E_{\infty}^{0,0} & E_{\infty}^{1,0} & E_{\infty}^{2,0} & E_{\infty}^{3,0} \\
 \hline
 187 &  P\bar 6m2 & +_{0} & \Z^{15} & \Z^3 & 0 & 0 \\
  &   & -_{1/2} & \Z^{10} & \Z & \Z^3 & 0 \\
 188 &  P\bar 6c2 & +_{0},-_{1/2} & \Z^{12} & \Z & \Z & 0 \\
\hline \\
 \text{SG} & \text{Intl} & \epsilon(m_{1\bar 10},m_{001}) & E_{\infty}^{0,0} & E_{\infty}^{1,0} & E_{\infty}^{2,0} & E_{\infty}^{3,0} \\
 \hline
 189 &  P\bar 62m & +_{0} & \Z^{15} & \Z^3 & 0 & 0 \\
  &   & -_{1/2} & \Z^{10} & \Z & \Z^3 & 0 \\
 190 &  P\bar 62c & +_{0},-_{1/2} & \Z^{12} & \Z & \Z & 0 \\
\hline \\
\end{array}
\end{align*}
\begin{align*}
\begin{array}{llllllll}
 \text{SG} & \text{Intl} & \left(\epsilon(6^+_{001},m_{001}),\epsilon(m_{1\bar 10},m_{001}),\epsilon(m_{1\bar 10},2_{001})\right) & E_{\infty}^{0,0} & E_{\infty}^{1,0} & E_{\infty}^{2,0} & E_{\infty}^{3,0} \\
 \hline
 191 &  P6/mmm & (+,+,+)_{0} & \Z^{24} & 0 & 0 & 0 \\
  &   & (+,-,-)_{1/2} & \Z^{14} & 0 & \Z^4 & 0 \\
  &   & (+,+,-) & \Z^{12} & \Z^6 & 0 & 0 \\
  &   & (+,-,+) & \Z^{12} & \Z^2 & \Z^2 & 0 \\
  &   & (-,+,+),(-,-,+) & \Z^6 & \Z^6 & 0 & 0 \\
  &   & (-,-,-),(-,+,-) & \Z^7 & \Z^2 & \Z & 0 \\
 192 &  P6/mcc & (+,+,+)_{0},(+,-,+) & \Z^{17} & 0 & \Z & 0 \\
  &   & (+,-,-)_{1/2},(+,+,-) & \Z^{13} & \Z^2 & \Z & 0 \\
  &   & (-,+,+),(-,-,+) & \Z^5 & \Z^5 & 0 & 0 \\
  &   & (-,-,-),(-,+,-) & \Z^7 & \Z & \Z_2 & 0 \\
 193 &  P6_3/mcm & (+,+,+)_{0},(-,+,+) & \Z^{13} & \Z & 0 & 0 \\
  &   & (+,-,-)_{1/2},(-,-,-) & \Z^{10} & 0 & \Z^2 & 0 \\
  &   & (+,+,-),(-,+,-) & \Z^9 & \Z^3 & 0 & 0 \\
  &   & (+,-,+),(-,-,+) & \Z^7 & \Z^2 & \Z & 0 \\
 194 &  P6_3/mmc & (+,+,+)_{0},(-,-,+) & \Z^{13} & \Z & 0 & 0 \\
  &   & (+,-,-)_{1/2},(-,+,-) & \Z^{10} & 0 & \Z^2 & 0 \\
  &   & (+,+,-),(-,-,-) & \Z^9 & \Z^3 & 0 & 0 \\
  &   & (+,-,+),(-,+,+) & \Z^7 & \Z^2 & \Z & 0 \\
  \hline
\end{array}
\end{align*}
\end{center}
\end{table*}

\begin{table*}
\begin{center}
\caption{(Continued from the previous page)}
\label{tab:e2_7}
\begin{align*}
\begin{array}{llllllll}
 \text{SG} & \text{Intl} & \epsilon(2_{001},2_{010}) & E_{\infty}^{0,0} & E_{\infty}^{1,0} & E_{\infty}^{2,0} & E_{\infty}^{3,0} \\
 \hline
 195 &  P23 & +_{0} & \Z^7 & \Z^2+\Z_2 & 0 & \Z \\
  &   & -_{1/2} & \Z^3 & \Z^6 & 0 & \Z \\
 196 &  F23 & +_{0} & \Z^5 & \Z^2+\Z_2 & \Z_2 & \Z \\
  &   & -_{1/2} & \Z^3 & \Z^4 & \Z_2 & \Z \\
 197 &  I23 & +_{0} & \Z^5 & \Z^2+\Z_2^2 & 0 & \Z \\
  &   & -_{1/2} & \Z^3 & \Z^4+\Z_2 & 0 & \Z \\
 198 &  P2_13 & +_{0},-_{1/2} & \Z^3 & \Z^2+\Z_4 & 0 & \Z \\
 199 & I2_13  & +_{0},-_{1/2} & \Z^4 & \Z^3+\Z_2 & 0 & \Z \\
  200 & Pm\bar 3  & +_{0} & \Z^{17} & 0 & 0 & 0 \\
  &   & -_{1/2} & \Z^{11} & 0 & \Z^2 & 0 \\
 201 &  Pn\bar 3 & +_{0} & \Z^{11} & 0 & \Z_2 & 0 \\
  &   & -_{1/2} & \Z^9 & \Z^2 & 0 & 0 \\
 202 &  Fm\bar 3 & +_{0} & \Z^{13} & 0 & 0 & 0 \\
  &   & -_{1/2} & \Z^{10} & 0 & \Z+\Z_2 & 0 \\
 203 &  Fd\bar 3 & +_{0} & \Z^{10} & 0 & \Z_2 & 0 \\
  &   & -_{1/2} & \Z^9 & \Z & 0 & 0 \\
 204 &  Im\bar 3 & +_{0} & \Z^{13} & 0 & 0 & 0 \\
  &   & -_{1/2} & \Z^{10} & \Z_2 & \Z & 0 \\
 205 & Pa\bar 3  & +_{0},-_{1/2} & \Z^9 & 0 & 0 & 0 \\
 206 &  Ia\bar 3 & +_{0} & \Z^{10} & \Z & 0 & 0 \\
  &   & -_{1/2} & \Z^9 & 0 & 0 & 0 \\
 207 &  P432 & +_{0} & \Z^9 & \Z^3+\Z_2 & 0 & \Z \\
  &   & -_{1/2} & \Z^4 & \Z^8 & 0 & \Z \\
 208 &  P4_232 & +_{0} & \Z^7 & \Z+\Z_4 & 0 & \Z \\
  &   & -_{1/2} & \Z^2 & \Z^6+\Z_2 & 0 & \Z \\
 209 &  F432 & +_{0} & \Z^7 & \Z^2+\Z_2 & 0 & \Z \\
  &   & -_{1/2} & \Z^3 & \Z^6 & 0 & \Z \\
 210 &  F4_132 & +_{0} & \Z^4 & \Z^2+\Z_4 & 0 & \Z \\
  &   & -_{1/2} & \Z^3 & \Z^3+\Z_2 & 0 & \Z \\
 211 &  I432 & +_{0} & \Z^7 & \Z^2+\Z_2 & 0 & \Z \\
  &   & -_{1/2} & \Z^3 & \Z^6 & 0 & \Z \\
 212 &  P4_332 & +_{0},-_{1/2} & \Z^3 & \Z^2+\Z_4 & 0 & \Z \\
 213 &  P4_132 & +_{0},-_{1/2} & \Z^3 & \Z^2+\Z_4 & 0 & \Z \\
 214 &  I4_132 & +_{0} & \Z^5 & \Z+\Z_4 & 0 & \Z \\
  &   & -_{1/2} & \Z^2 & \Z^4+\Z_2 & 0 & \Z \\
 215 &  P\bar 43m & +_{0} & \Z^{10} & \Z^2 & 0 & 0 \\
  &   & -_{1/2} & \Z^6 & \Z^3 & \Z & 0 \\
 216 &  F\bar 43m & +_{0} & \Z^9 & \Z^2 & 0 & 0 \\
  &   & -_{1/2} & \Z^6 & \Z^2 & \Z & 0 \\
 217 &  I\bar 43m & +_{0} & \Z^9 & \Z^2 & 0 & 0 \\
  &   & -_{1/2} & \Z^6 & \Z^2 & \Z & 0 \\
 218 &  P\bar 43n & +_{0} & \Z^9 & \Z & 0 & 0 \\
  &   & -_{1/2} & \Z^6 & \Z^2 & 0 & 0 \\
 219 &  F\bar 43c & +_{0} & \Z^8 & \Z & 0 & 0 \\
  &   & -_{1/2} & \Z^6 & \Z & \Z_2 & 0 \\
 220 &  I\bar 43d & +_{0},-_{1/2} & \Z^7 & \Z & 0 & 0 \\
\hline
\end{array}
&& &&
\begin{array}{llllllll}
 \text{SG} & \text{Intl} & \left( \epsilon(2_{001},2_{010}),\epsilon(3^+_{111},-1) \right) & E_{\infty}^{0,0} & E_{\infty}^{1,0} & E_{\infty}^{2,0} & E_{\infty}^{3,0} \\
 \hline
 221 &  Pm\bar 3m & (+,+)_{0} & \Z^{22} & 0 & 0 & 0 \\
  &   & (-,+)_{1/2} & \Z^{14} & 0 & \Z^3 & 0 \\
  &   & (+,-) & \Z^7 & \Z^3 & 0 & 0 \\
  &   & (-,-) & \Z^4 & \Z^5 & 0 & 0 \\
 222 & Pn\bar 3n  & (+,+)_{0},(+,-) & \Z^{11} & \Z & 0 & 0 \\
  &   & (-,+)_{1/2},(-,-) & \Z^8 & \Z^3 & 0 & 0 \\
 223 & Pm\bar 3n  & (+,+)_{0},(+,-) & \Z^{13} & 0 & 0 & 0 \\
  &   & (-,+)_{1/2},(-,-) & \Z^8 & \Z & \Z & 0 \\
 224 &  Pn\bar 3m & (+,+)_{0} & \Z^{13} & 0 & 0 & 0 \\
  &   & (-,+)_{1/2} & \Z^8 & \Z & \Z & 0 \\
  &   & (+,-) & \Z^8 & \Z & 0 & 0 \\
  &   & (-,-) & \Z^6 & \Z^4 & 0 & 0 \\
 225 &  Fm\bar 3m & (+,+)_{0} & \Z^{17} & 0 & 0 & 0 \\
  &   & (-,+)_{1/2} & \Z^{11} & 0 & \Z^2 & 0 \\
  &   & (+,-) & \Z^7 & \Z^2 & 0 & 0 \\
  &   & (-,-) & \Z^5 & \Z^4 & 0 & 0 \\
 226 &  Fm\bar 3c & (+,+)_{0} & \Z^{14} & 0 & 0 & 0 \\
  &   & (-,+)_{1/2} & \Z^{10} & \Z & \Z & 0 \\
  &   & (+,-) & \Z^9 & \Z & 0 & 0 \\
  &   & (-,-) & \Z^6 & \Z^2 & 0 & 0 \\
 227 &  Fd\bar 3m & (+,+)_{0} & \Z^{11} & 0 & 0 & 0 \\
  &   & (-,+)_{1/2} & \Z^9 & 0 & \Z & 0 \\
  &   & (+,-) & \Z^7 & \Z^2 & 0 & 0 \\
  &   & (-,-) & \Z^6 & \Z^2 & 0 & 0 \\
 228 &  Fd\bar 3c & (+,+)_{0} & \Z^9 & \Z & 0 & 0 \\
  &   & (-,+)_{1/2} & \Z^7 & 0 & \Z_2 & 0 \\
  &   & (+,-) & \Z^8 & 0 & 0 & 0 \\
  &   & (-,-) & \Z^8 & \Z & 0 & 0 \\
 229 &  Im\bar 3m & (+,+)_{0} & \Z^{17} & 0 & 0 & 0 \\
  &   & (-,+)_{1/2} & \Z^{11} & 0 & \Z^2 & 0 \\
  &   & (+,-) & \Z^7 & \Z^2 & 0 & 0 \\
  &   & (-,-) & \Z^5 & \Z^4 & 0 & 0 \\
 230 &  Ia\bar 3d & (+,+)_{0},(+,-) & \Z^9 & 0 & 0 & 0 \\
  &   & (-,+)_{1/2},(-,-) & \Z^7 & \Z & 0 & 0 \\
\hline 
\medskip \\
\medskip \\
\medskip \\
\medskip \\
\medskip \\
\medskip \\
\medskip \\
\medskip \\
\end{array}
\end{align*}
\end{center}
\end{table*}

\section{Conclusion and outlook}
\label{sec:conc}
In this paper, we have studied the AHSS for twisted equivariant $K$-theory in the view of band theory. 
As an application, we present the complete classification of topological invariants in A and AIII AZ symmetry classes for 230 space groups, summarized in Tables~\ref{tab:230} to \ref{tab:e2_7}. 
We found that various torsion topological invariants appear even for symmorphic space groups. 

As we have shown in Secs.~\ref{sec:AHSS} and \ref{The AHSS with antiunitary symmetry}, all the ingredients in the AHSS suitably fit into band theory. 
The $E_1$-page has the data of irreps at $p$-cells, i.e., \ high-symmetry points, lines, planes, and volumes. 
At the same time, the $E_1$-page can be thought of as the space of (i) topological insulators on $p$-spheres, each of which is defined by identifying the boundary of a $p$-cell to a point, (ii) topological gapless states in $p$-cells, and (iii) topological singular points in $p$-cells. 
The differentials $d_r (r \geq 1)$ in the AHSS represent the creation of topological gapless states (topological singular points) from $p$-cells to adjacent $(p+r)$-cells.
Especially, the first differential $d_1^{0,-n}: E_1^{0,-n} \to E_1^{1,-n}$ is identified with the compatibility relation in the literature of band theory. 
The $E_{r+1} $-page $(r \geq 1)$ in the AHSS is defined as the cohomology of the $r$th differential $d_r$ as $E_{r+1}^{p,-n} = \ker (d_r^{p,-n}) / \im (d_r^{p-r,-(n-r+1)})$, where $\im (d_r^{p-r,-(n-r+1)})$ is understood as the trivialization of topological gapless states in the $E_r$-page by $(p-r)$-cells, and $\ker (d_r^{p,-n})$ means a generalized compatibility relation for that the topological gapless states in $p$-cells can extend to adjacent $(p+r)$-cells without a singularity. 
Here, a nontrivial $r$th differential $d_r^{0,-n}: E_r^{0,-n} \to E_r^{r,-(n+r-1)}$ serves as the indicator of bulk gapless phases characterized by the high-symmetry points.  
Iterating the cohomology of $d_r$ yields the limiting page $E_{\infty}$. 
Since topological gapless states represented by the $E_{\infty}$-page can not be trivialized by low-dimensional cells, an element of $E_{\infty}$-page is considered an anomalous gapless phase in the sense that it can not be realized as a stand-alone lattice system. 
Moreover, the compatibility with higher-dimensional cells for topological gapless states of $E_{\infty}$ implies that there must be a representative anomalous gapless phase in the whole BZ torus, which leads to the exact sequences (\ref{eq:e_inf}) for the $K$-group. 
In this sense, the $E_{\infty}$-page approximates the $K$-group. 
From the bulk-boundary correspondence, the $E_{\infty}$-page approximates the classification of bulk gapped phases as well as anomalous gapless phases.

We close the paper by mentioning some future directions. 

---
Although we showed the complete list of topological invariants for class A and AIII AZ classes for 230 space groups in Tables~\ref{tab:230} to \ref{tab:e2_7}, the explicit formulas of topological invariants remain undetermined for many space groups, which we left as the future work. 

---
The quick construction of the higher-differentials $d_r (r \geq 2)$ is not known yet. 
Once an efficient method to derive the higher differentials is given, one can finish the computation of the $E_{\infty}$-page for all the magnetic space groups. 
This may be an important step in completely classifying topological crystalline insulators and superconductors.  

---
The $E_{\infty}$-page gives an approximation of the $K$-group in the form of exact sequences (\ref{eq:e_inf}). 
In some cases, the $K$-group as an Abelian group is not settled only by the $E_{\infty}$-page. 
A complementary method is required to compute the exact sequences' extension problem (\ref{eq:e_inf}). 

---
The relationship between bulk topological invariants and physical observables should be clarified. 
Applying the ``gauging crystalline symmetry'' argued in Ref.~\cite{ThorngrenElse} to the torsion topological invariants listed in Tables~\ref{tab:230} to \ref{tab:e2_7} is interesting. 

---
Tables~\ref{tab:230} to \ref{tab:e2_7} indicates various torsion topological invariants in the column of $E_2^{1,0}$. 
In addition to the meaning of the one-dimensional bulk class AIII invariant, $E_2^{1,0}$ is interpreted as the spectral flow index for class A anomalous gapless spectra. 
It is interesting to see the implication of torsion topological invariants in view of the interplay of the chiral anomaly and space group symmetry in the many-body Hilbert space.

\medskip
{\it Acknowledgment---}
K.S. thanks 
Takuya Nomoto, 
Akira Furusaki, 
Takahiro Morimoto, 
Ryo Takahashi and 
Youichi Yanase 
for helpful discussions. 
K.S. is supported by RIKEN Special Postdoctoral Researcher Program. 
M.S. is supported by the JSPS KAKENHI (Nos.JP15H05855, JP15K13498, JP17H02922). 
K.G. is supported by JSPS KAKENHI (JP15K04871). 
K.S. and M.S. thank the Yukawa Institute for Theoretical Physics at Kyoto University, where part of this work was done during the workshop “Novel Quantum States in Condensed Matter
2017” (NQS2017, YITP-T-17-01).

\appendix

\section{Wigner criteria}
\label{app:wigner}
In this Appendix, we explain the Wigner criterion and its generalization in the presence of PHS.

Let $G$ be a finite group and $\phi,c : G \to \Z_2$ the indicators for
unitary/antiunitary and symmetry/antisymmetry, respectively:
\begin{align}
&\phi(g)=\left\{
\begin{array}{cl}
1 & (\mbox{unitary $g\in G$})\\
-1 & (\mbox{antiuniary $g\in G$})
\end{array}
\right.,
\\
&c(g)=\left\{
\begin{array}{cl}
1 & (\mbox{symmetry $g\in G$})\\
-1 & (\mbox{antisymmetry $g\in G$})
\end{array}
\right..
\end{align}
Let $z = (z_{g,h}) \in Z^2(G,U(1)_{\phi})$ be the factor system associated with a projective representation
\begin{align}
z_{g,h} U_{gh} = 
U_g U_h,\quad g,h \in G.
\label{eq:app_factor_system}
\end{align}
The factor system $z_{g,h}$ satisfies the 2-cocycle condition 
\begin{align}
z_{g,hk} z_{h,k}^{\phi(g)} =z_{g,h} z_{gh,k} 
 \qquad (g,h,k \in G). 
\label{eq:app_2cocyclecond}
\end{align}
Consider $G_0= \{g \in G | \phi(g)=c(g)=1\}$, the subgroup consisting of elements of unitary and symmetry, and let $\{\alpha,\beta,\dots\}$ be irreps of $G_0$ with the factor system
$z|_{G_0}$. 
The group $G$ is the disjoint union of cosets $G = G_0 \sqcup a G_0 \sqcup b
G_0 \sqcup ab G_0$, where $a,b,ab$ are elements with $\phi(a)=-c(a)=-1, \phi(b)=c(b)=-1$, and $\phi(ab)=-c(ab)=1$, respectively. 

Let $\ket{i}, i=1,\dots,\dim \alpha$, be a basis of the irrep $\alpha$, i.e., 
\begin{align}
&\hat g \ket{i} = \ket{j} [D_g^{\alpha}]_{ji}, \nonumber\\ 
&D_g^\alpha D_h^\alpha=z_{g,h}D^\alpha_{gh}, \qquad g,h \in G_0.
\end{align}
Then, we formally introduce a conjugate representation $\hat a \ket{i}$ as 
\begin{align}
\hat g (\hat a \ket{i}) 
= (\hat a \ket{j}) \frac{z_{g,a}}{z_{a,a^{-1}ga}} [D^\alpha_{a^{-1}ga}]^*_{ji} 
\end{align}
for $g \in G_0$. 
Here we have used the relations
\begin{align}
&\widehat{a^{-1}ga}|i\rangle=|j\rangle [D^\alpha_{a^{-1}g a}]_{ji},\nonumber\\
&\hat{a} \widehat{a^{-1}ga}=z_{a,a^{-1}ga}\widehat{ga}=\frac{z_{a, a^{-1}ga}}{z_{g,a}}
\hat{g}\hat{a}, 
\end{align}
where the coefficient in the latter equation is determined by
(\ref{eq:app_factor_system}). 
Now we ask if $\hat a \ket{i}$ is unitary equivalent to $\ket{i}$, and
if so, $\hat{a}|i\rangle$ and $|i\rangle$ are a Kramers pair or not.
First, let us assume that $\hat a \ket{i}$ is unitary equivalent to $\ket{i}$, i.e. there exists a unitary matrix $V$ such that 
\begin{align}
\frac{z_{g,a}}{z_{a,a^{-1}ga}} D_{a^{-1}ga}^{\alpha *} = V^{\dag} D^{\alpha}_g V, \qquad (g \in G_0). 
\label{eq:app_unitary_equ_D}
\end{align}
This relation leads to
\begin{align}
&D^\alpha_g (VV^*)^{-1}D^{\alpha}_{a^2}\nonumber\\
&=D^\alpha_g V^T V^{\dagger}D^{\alpha}_{a^2} \nonumber \\
&=\frac{z_{aga^{-1},a}}{z_{a,g}}
V^TD^{\alpha *}_{aga^{-1}}V^{\dagger}D^{\alpha}_{a^2}\nonumber\\
&=\frac{z_{aga^{-1},a}z_{a,aga^{-1}}}{z_{a,g}z_{a^2ga^{-2},a}}V^TV^{\dagger}
D^{\alpha}_{a^2ga^{-2}}D^{\alpha}_{a^2},
\end{align}
which is recast into
\begin{align}
&D^{\alpha}_g (VV^*)^{-1} D^{\alpha}_{a^2}\nonumber\\
&= 
\frac{z_{aga^{-1},a} z_{a,aga^{-1}} z_{a^2 g a^{-2},a^2}}{z_{a,g} z_{a^2 g a^{-2},a} z_{a^2,g}} (VV^*)^{-1} D^{\alpha}_{a^2} D^{\alpha}_g.
\label{eq:app_a10}
\end{align} 
From the 2-cocycle condition (\ref{eq:app_2cocyclecond}), 
the prefactor in the r.h.s.\ of (\ref{eq:app_a10}) is unity, 
and thus we have
$D^\alpha_g (VV^*)^{-1} D^\alpha_{a^2} = (VV^*)^{-1} D^\alpha_{a^2} D^\alpha_g$.
Since the irrep $\alpha$ is irreducible, the Schur's lemma yields that $(VV^*)^{-1}D_{a^2}^\alpha = \xi$ with $\xi$ a constant. 
Substituting this into Eq. (\ref{eq:app_unitary_equ_D}) with $g=a^2$, we have 
\begin{align}
\frac{z_{a^2,a}}{z_{a,a^2}}V^* V/\xi=V^{\dagger}(VV^*)\xi V, 
\end{align}
which leads to $z_{a,a}\xi =\pm 1$, using the 2-cocycle condition $z_{a,a}^{-1} z_{a^2,a}^{-1} z_{a,a^2} z_{a,a}^{-1} = 1$. 
Now introduce a new basis $\wt{\ket{i}} = (\hat a \ket{j}) V^{\dag}_{ji}$ obeying the same representation of $\ket{i}$, i.e.\ $\hat g \wt{\ket{i}} = \wt{\ket{j}} [D^\alpha_g]_{ji}$.
Taking the same transformation twice, we have 
\begin{align}
\wt{(\wt{\ket{i}})} &= (\hat{a}\wt{\ket j})V^{\dagger}_{ji}\nonumber\\
&=(\hat{a}\hat{a}\ket{k})V^T_{kj}V^{\dagger}_{ji}\nonumber\\
&=z_{a,a}(\widehat{a^2}\ket{k})V^T_{kj}V^{\dagger}_{ji}\nonumber\\
&=z_{a,a}\ket{l}[D^\alpha_{a^2}]_{lk}V^T_{kj}V^{\dagger}_{ji}\nonumber\\
&=z_{a,a}\xi\ket{i}
\end{align}
Therefore, when $z_{a,a} \xi = -1 (+1)$, $\ket{i}$ and $\hat a\ket{i}$ are (not) a Kramers pair.

The sign $\xi z_{a,a}$ is computed as follows. 
Using the orthogonality condition $\sum_{g \in G_0} [D^{\alpha}_g]^*_{ij} [D^{\beta}_g]_{kl} = \frac{|G_0|}{{\rm dim}(\alpha)} \delta_{ik} \delta_{jl} \delta^{\alpha,\beta}$ between irreps $\alpha$ and $\beta$, 
we find that 
\begin{align}
&\frac{1}{|G_0|} \sum_{g \in G_0} 
\left[ \frac{z_{g,a}}{z_{a,a^{-1}ga}} D^{\alpha *}_{a^{-1}ga}\right]^* D^\alpha_g\nonumber\\
&= 
\left\{\begin{array}{ll}
\frac{\xi}{{\rm dim}(\alpha)} D_{a^2}^{\alpha\dag} & (\ket{i} \ {\rm and\ } \hat a \ket{i} \ {\rm are\ equivalent}), \\
0 & (\ket{i} \ {\rm and\ } \hat a \ket{i} \ {\rm are\ inequivalent}). 
\end{array}\right.
\end{align}
Moreover, using the 2-cocycle condition (\ref{eq:app_2cocyclecond}), we have 
\begin{align}
W^T_{\alpha}
&:=\frac{1}{|G_0|} \sum_{g \in G_0} z_{ag,ag} \chi_{\alpha}((ag)^2) \nonumber\\
&= 
\left\{\begin{array}{ll}
z_{a,a}\xi =\pm 1 & (\ket{i} \ {\rm and\ } \hat a \ket{i} \ {\rm are\ equivalent}), \\
0 & (\ket{i} \ {\rm and\ } \hat a \ket{i} \ {\rm are\ inequivalent}), 
\end{array}\right.
\end{align}
where $\chi_{\alpha}(g \in G_0) = \tr D^{\alpha}_g$ is the character of the irrep $\alpha$. 

In the same way, for the group element $b \in G$ (a representative element of PHS), we define
\begin{align}
W^C_{\alpha}
&:=\frac{1}{|G_0|} \sum_{g \in G_0} z_{bg,bg} \chi_{\alpha}((bg)^2) \nonumber\\
&= 
\left\{\begin{array}{ll}
\pm 1 & (\ket{i} \ {\rm and\ } \hat b \ket{i} \ {\rm are\ equivalent}), \\
0 & (\ket{i} \ {\rm and\ } \hat b \ket{i} \ {\rm are\ inequivalent}). 
\end{array}\right.
\end{align}
We also introduce for $ab \in G$ (a representative element of unitary PHS) the bit-valued quantity 
\begin{align}
&W^{\Gamma}_{\alpha}\nonumber\\
&:=\frac{1}{|G_0|} \sum_{g \in G_0} \left[\frac{z_{g,ab}}{z_{ab,(ab)^{-1}gab}} \chi_{\alpha}((ab)^{-1}gab)\right]^* \chi_{\alpha}(g) \nonumber\\
&= 
\left\{\begin{array}{ll}
1 & (\ket{i} \ {\rm and\ } \hat ab \ket{i} \ {\rm are\ equivalent}), \\
0 & (\ket{i} \ {\rm and\ } \hat ab \ket{i} \ {\rm are\ inequivalent}). 
\end{array}\right.
\end{align}
Using the datum $(W^T_{\alpha}, W^C_{\alpha},W^{\Gamma}_{\alpha})$, one can determine the emergent AZ symmetry class of the irrep $\alpha$ as in Table.~\ref{tab:wigner_az}.

\section{On the classification of singular points}
\label{app:On the classification of singular points}
This Appendix presents the classification of topologically stable singular points strictly inside a $p$-cell ($p\geq 1$).
Here, a singular point means a point in the $k$-space where the Hamiltonian is not single-valued. 
Let us focus on a $p$-cell $D^p$, where a little group $G_{D^p}$ and a factor system on $D^p$ are given. 
According to Table~\ref{tab:wigner_az}, the Wigner indices (\ref{eq:wt_trs}-\ref{eq:wt_cs}) determines the emergent AZ class realized in the $p$-cell $D^p$. 
Let $s$ be the integer indicating the emergent AZ class as shown in Table~\ref{tab:singular}. 
We observe that a singular point inside the $p$-cell $D^p$ should be the end point of a gapless Dirac line described by the Hamiltonian 
\begin{align}
H_{\rm gapless}=\sum_{\mu=1}^{p-1} k_{\mu} \gamma_{\mu}. 
\label{app:massless_dirac}
\end{align}
The gapless points of this Hamiltonian (\ref{app:massless_dirac}) form a straight line along the $k_p$-axis. 
To have the topological gapless state (\ref{app:massless_dirac}), there should exist the topological invariant on the $(p-2)$-dimensional sphere surrounding the Dirac line of (\ref{app:massless_dirac}), which is classified by the homotopy group $\pi_{p-2}(R_{s}) = \pi_0(R_{s+p-2})$ ($\pi_{p-2}(C_{s}) = \pi_0(C_{s+p-2})$) for emergent real (complex) AZ classes. 
Here, $R_s$ ($C_s$) is the classifying space of the real (complex) AZ class $s$.~\cite{Kitaev}
The homotopy groups for $1$-, $2$- and $3$-cells is summarized in Table~\ref{tab:singular}. 
(The same formula holds true for 1-cells.) 
Since the singular point is the end point of the massless Dirac line, the classification of stable singular points is the same as that for stable massless Dirac lines, which implies that Table~\ref{tab:singular} also gives the classification of the singular points. 
In fact, using the massless Dirac line (\ref{app:massless_dirac}), we have an explicit model for the $p$-dimensional Hamiltonian describing the singular point 
\begin{align}
H_{\rm singular} 
= 
\im \ln \left[ k_p + i \sum_{\mu=1}^{p-1} k_{\mu} \gamma_{\mu} \right]. 
\label{app:singular_dirac}
\end{align}
We see that the Hamiltonian (\ref{app:singular_dirac}) is recast as the Dirac line $H_{\rm singular} \sim \sum_{\mu=1}^{p-1} k_{\mu} \gamma_{\mu}$ on the $k_p$-axis with $k_p>0$, whereas the Hamiltonian (\ref{app:singular_dirac}) has a finite energy gap as $H_{\rm singular} \sim \pm \pi$ on the $k_p$-axis with $k_p<0$.

\begin{table}
	\caption{The classification of singular points inside $p$-cells. 
	The emergent AZ class $s$ is obtained by the Wigner test (\ref{eq:wt_trs}-\ref{eq:wt_cs}). }
	\label{tab:singular}
$$
\begin{array}{ccccc}
\mbox{Emergent AZ class} & s & p=1 & p=2 & p=3 \\
\hline 
\mbox{A} & 0 & 0 & \Z & 0 \\
\mbox{AIII} & 1 & \Z & 0 & \Z \\
\hline 
\mbox{AI} & 0 & 0 & \Z & \Z_2 \\
\mbox{BDI} & 1 & \Z & \Z_2 & \Z_2 \\
\mbox{D} & 2 & \Z_2 & \Z_2 & 0 \\
\mbox{DIII} & 3 & \Z_2 & 0 & \Z \\
\mbox{AII} & 4 & 0 & \Z & 0 \\
\mbox{CII} & 5 & \Z & 0 & 0 \\
\mbox{C} & 6 & 0 & 0 & 0 \\
\mbox{CI} & 7 & 0 & 0 & \Z \\
\hline 
\end{array}
$$
\end{table}

\begin{widetext}
\section{Factor systems}
\label{sec:factor}
In this Appendix, we tabulate factor systems appearing in electronic systems.

\subsection{Time-reversal symmetric spinless/spinful systems}
A peculiarity in electron systems is that the factor system (at the $\Gamma$ point) is determined by a spin representation of continuum rotation group $O(3)$ and TRS. 
Let $G \times \Z_2^T$ be the symmetry group composed by the point group $G$ and the TRS $\Z_2^T$. 
For spin integer systems, the factor system is trivial $z_{g,h} \equiv 1$ for $g,h \in G \times \Z_2^T$. 
For spin half integer systems, the factor system obeys (i) $T^2=-1$, (ii) $T U_g = U_g T (g \in G)$, and the factor system $z_{g,h}$ in $U_gU_h=z_{g,h}U_{gh} (g,h \in G)$ follows the $Pin_-(3)$ group (spin $1/2$ representation of the $O(3)$ group), which is known as the double group in the literature.~\cite{InuiGroup} 
It should be noted that the inversion $I$ always commutes with other symmetry group operators, since the inversion does not affect the internal degrees of freedom of spin.~\footnote{Mathematically, the inversion operator can anticommutes with other symmetry operators.}
The factor system in the whole BZ is summarized as 
\begin{align}
{\rm AI/AII}: \qquad 
\left\{\begin{array}{ll}
T^2=\pm 1, \\
U_g(h\bk) U_h(\bk) = z_{g,h} e^{-i g h \bk \cdot \bm{\nu}_{g,h}} U_{gh}(\bk), \\
TU_g(\bk)= U_g(-\bk)T, 
\end{array}\right.
\label{eq:factor_system_trs}
\end{align}
where $T=U_T K$ with $K$ the complex conjugation, $\pm 1$ is for spinless/spinful, and $z_{g,h}$ is the factor system of the point group. 
We have specified the data $\phi,c$ of elements by symbols $T$ and $U_g$, i.e.\ $\phi(T)=-c(T)=-1, \phi(g)=c(g)=1 (g \in G)$. 
By adding the chiral symmetries with the rule (\ref{eq:shift_n}), the factor systems for other AZ classes are given, which are summarized in Table (\ref{tab:shift_sym}) (with setting $e^{i \theta_g}=1$ and $z_{g,h}=1$ for AI and $z_{g,h}$ being the spin half integer representation for AII). 
%Table~\ref{tab:shift_sym} is the same one as the table II in Ref.~\cite{SSG17}. 
For a derivation of Table (\ref{tab:shift_sym}), see the next section. 

\begin{table}
	\begin{center}
	\caption{The factor systems for time-reversal symmetric systems. 
	 In the table, $T$ and $C$ are TRS and PHS, respectively, $z_{g,h}$ represents the factor system of the point group, and $e^{i \theta_g}$ takes values in $\{\pm 1\}$. }
\label{tab:shift_sym}
{\small
\renewcommand{\arraystretch}{1.5}
$$
\begin{array}{l|l|l|l|l|l}
{\rm AZ} & n & z_{T,T}, z_{C,C} & z_{T,g}/z_{g,T}, z_{C,g}/z_{g,C} & z_{g,h} & c(g) \\
\hline 
{\rm AI} & n=0 & T^2=1 & TU_g(\bk) = U_g(-\bk)T & U_g(h\bk) U_h(\bk) & U_g(\bk)H(\bk)=e^{i\theta_g} H(g\bk)U_g(\bk)\\
&&&&= z_{g,h} e^{-i g h \bk \cdot \bm{\nu}_{g,h}} U_{gh}(\bk)\\
\hline
{\rm BDI} & n=1 & T^2=1 & TU_g(\bk) = e^{i \theta_g} U_g(-\bk)T & U_g(h\bk) U_h(\bk) & U_g(\bk)H(\bk)=H(g\bk)U_g(\bk)\\
&&C^2=1&CU_g(\bk) = U_g(-\bk)C&= z_{g,h} e^{-i g h \bk \cdot \bm{\nu}_{g,h}} U_{gh}(\bk)\\
\hline
{\rm D} & n=2 & C^2=1& CU_g(\bk) = U_g(-\bk)C & U_g(h\bk) U_h(\bk) & U_g(\bk)H(\bk)=e^{i \theta_g}H(g\bk)U_g(\bk)\\
&&&&= z_{g,h} e^{-i g h \bk \cdot \bm{\nu}_{g,h}} U_{gh}(\bk)\\
\hline
{\rm DIII} & n=3 & T^2=-1 & TU_g(\bk) = U_g(-\bk)T & U_g(h\bk) U_h(\bk) & U_g(\bk)H(\bk)=H(g\bk)U_g(\bk)\\
&&C^2=1& CU_g(\bk) = e^{i \theta_g} U_g(-\bk)C&= z_{g,h} e^{-i g h \bk \cdot \bm{\nu}_{g,h}} U_{gh}(\bk)\\
\hline
{\rm AII} & n=4 & T^2=-1 & TU_g(\bk) = U_g(-\bk)T & U_g(h\bk) U_h(\bk) & U_g(\bk)H(\bk)=e^{i\theta_g} H(g\bk)U_g(\bk)\\
&&&&= z_{g,h} e^{-i g h \bk \cdot \bm{\nu}_{g,h}} U_{gh}(\bk)\\
\hline
{\rm CII} & n=5 & T^2=-1 & TU_g(\bk) = e^{i \theta_g} U_g(-\bk)T & U_g(h\bk) U_h(\bk) & U_g(\bk)H(\bk)=H(g\bk)U_g(\bk)\\
&&C^2=-1&CU_g(\bk) = U_g(-\bk)C&= z_{g,h} e^{-i g h \bk \cdot \bm{\nu}_{g,h}} U_{gh}(\bk)\\
\hline
{\rm C} & n=6 &C^2=-1&CU_g(\bk) = U_g(-\bk)C & U_g(h\bk) U_h(\bk) & U_g(\bk)H(\bk)=e^{i \theta_g}H(g\bk)U_g(\bk)\\
&&&&= z_{g,h} e^{-i g h \bk \cdot \bm{\nu}_{g,h}} U_{gh}(\bk)& \\
\hline
{\rm CI} & n=7 & T^2=1& TU_g(\bk) = U_g(-\bk)T & U_g(h\bk) U_h(\bk) & U_g(\bk)H(\bk)=H(g\bk)U_g(\bk)\\
&& C^2=-1&CU_g(\bk) = e^{i \theta_g} U_g(-\bk)C&= z_{g,h} e^{-i g h \bk \cdot \bm{\nu}_{g,h}} U_{gh}(\bk)\\
\hline
\end{array}
$$
\renewcommand{\arraystretch}{1}
}
	\end{center}
\end{table}

\subsection{Time-reversal invariant superconductors}
\label{sec:fs_trs_sc}
In this section we formulate the factor system for spinful time-reversal invariant superconductors (i.e.\ class DIII) with space group symmetry. 
The factor systems for spinless time-reversal invariant superconductors (class BDI) and spinful time-reversal invariant superconductors with $SU(2)$ spin rotation symmetry (class CI) are constructed similarly. 
A peculiarity of superconducting systems is that the factor system
depends on representations of the gap function under the point group. 

Consider the BdG Hamiltonian 
\begin{align}
H(\bk)
= \begin{pmatrix}
{\cal E}(\bk) & \Delta(\bk) \\
\Delta(\bk)^{\dag} & -{\cal E}^T(-\bk)
\end{pmatrix}, 
\end{align}
where ${\cal E}(\bk)$ is the normal Hamiltonian and $\Delta(\bk)$ is the
gap function.
We assume that the normal part is invariant under space group symmetry
$G$ and TRS,
\begin{align}
U_g(\bk){\cal E}(\bk)U^{\dagger}_g(\bk)={\cal E}(g\bk),
\quad
T{\cal E}(\bk)T^{-1}=H(-\bk), 
\end{align}
with the factor system
\begin{align}
U_g(h\bk)U_h(\bk)=z_{g,h}e^{-igh\bk\cdot{\bm \nu}_{g,h}}U_{gh}(\bk), 
\quad
T^2=-1, 
\quad
TU_g(\bk)=U_g(-\bk)T.
\label{eq:factorDIIIspace}
\end{align}
We also assume that the gap function $\Delta(\bk)$ does not break TRS,
so by taking a proper phase of $T$, we have
\begin{align}
T\Delta(\bk)T^{T}=\Delta(-\bk),
\end{align}
without changing the factor system Eq. (\ref{eq:factorDIIIspace}). 
In order for the gap function $\Delta(\bk)$ to keep the space group
symmetry $G$, it should be a 1-dimensional representation of $G$, 
\begin{align}
U_g(\bk) \Delta(\bk) U_g(-\bk)^T = e^{i\theta_g} \Delta(g \bk), \qquad e^{i(\theta_g + \theta_h)} = e^{i \theta_{gh}}, \qquad g,h \in G.
\label{eq:1dimrep_sc}
\end{align}
While the original space group symmetry is
broken when $e^{i \theta_g} \neq 1$,
we can restore it by combining $U(1)$ gauge symmetry. 
However, not all 1$d$ represenations are compatible with TRS:
Applying 
$
T^{-1}U_g^{-1}(-\bk)TU_g(\bk) 
$
to the gap function, we have 
\begin{align}
&[T^{-1}U_g^{-1}(-\bk)TU_g(\bk)]
\Delta(\bk) 
[T^{-1}U_g^{-1}(\bk)TU_g(-\bk)]^T
\nonumber\\
&=T^{-1}U_g^{-1}(-\bk)TU_g(\bk)
\Delta(\bk) 
U_g^T(-\bk)T^T U_g^{*}(\bk)[T^{-1}]^T
\nonumber\\
&=e^{i\theta_g}T^{-1}U_g^{-1}(-\bk)T
\Delta(g\bk)
T^T U_g^{*}(\bk)[T^{-1}]^T
\nonumber\\
&=e^{i\theta_g}T^{-1}U_g^{-1}(-\bk)
\Delta(-g\bk)
U_g^{*}(\bk)[T^{-1}]^T
\nonumber\\
&=e^{2i\theta_g}T^{-1}
\Delta(-\bk)[T^{-1}]^T
\nonumber\\
&=e^{2i\theta_g}\Delta(\bk).
\end{align}
On the other hand, the factor system in Eq.(\ref{eq:factorDIIIspace})
implies $T^{-1}U_g^{-1}(-\bk)TU_g(\bk)=1$, leading to $e^{2i\theta_g}=1$.
Therefore, only when $e^{i\theta_g}=\pm 1$, TRS and the space group $G$ can coexist.

For the BdG Hamiltonian $H(\bk)$, the space group symmetry $G$ and TRS is given as 
\begin{align}
\wt U_g(\bk) H(\bk)\wt U_g(\bk)=H(g\bk),
\quad
\wt TH(\bk){\wt T}^{-1}=H(-\bk),
\end{align}
with
\begin{align}
\wt U_g(\bk):= \begin{pmatrix}
U_g(\bk) & 0 \\
0 & e^{-i \theta_g} U_g(-\bk)^* \\
\end{pmatrix},
\quad
\wt T:=\begin{pmatrix}
T&0\\	
0& (T^{-1})^T
       \end{pmatrix}.
\end{align}
Since $T=U_TK$ with a unitary matrix $U_T$ and
the complex conjugation operator $K$, $\wt T$ is also written as
\begin{align}
\wt T=
\begin{pmatrix}
U_T & 0\\
0& U_T^*      
\end{pmatrix}K. 
\end{align} 
The BdG Hamiltonian also has the inherent PHS 
\begin{align}
CH(\bk)C^{-1}=-H(-\bk),
\quad
C = \begin{pmatrix}
0 & 1 \\
1 & 0 \\
\end{pmatrix} K.
\end{align}
From the above relations, we can identify the factor system among TRS,
PHS, and point group symmetry: 
\begin{align}
{\rm DIII} :\qquad 
\left\{\begin{array}{ll}
T^2=-1, \qquad C^2=1, \qquad  U_g(h\bk)  U_h(\bk)
= z_{g,h} e^{-i g h \bk \cdot \bm{\nu}_{g,h}}  U_{gh}(\bk), \\
T  U_g(\bk) =  U_g(-\bk) T, \qquad 
C  U_g(\bk) = e^{i \theta_g}  U_g(-\bk) C, \qquad e^{i \theta_g} \in \{\pm 1\}, \\
TH(\bk)=H(-\bk)T, \qquad CH(\bk)=-H(-\bk)C, \qquad
 U_g(\bk)H(\bk)=H(-\bk)U_g(\bk).
\end{array}\right. ,
\end{align}
where we have omitted the tilde in $T$ and $U_g$ for the simplicity of
notation. 
This gives the factor system for $n=3$ in Table~\ref{tab:shift_sym}.

By imposing an additional chiral symmetry $\Gamma_1$ on $H(\bk)$,
we can obtain the factor system 
for $n=4$ (class AII) in Table~\ref{tab:shift_sym}. 
For this purpose, we use the chiral operator $\Gamma=iTC$ in the factor
system of class DIII, instead of
PHS: Assuming that $\Gamma_1$ satisfies Eq.(\ref{eq:shift_n}), we have
\begin{align}
{\rm DIII} \to {\rm AII}:\qquad 
\left\{\begin{array}{ll}
\Gamma^2=1, \qquad T^2=-1, \qquad T\Gamma=-\Gamma T, \\
 U_g(h\bk)  U_h(\bk)
= z_{g,h} e^{-i g h \bk \cdot \bm{\nu}_{g,h}} U_{gh}(\bk), \\
T  U_g(\bk) =  U_g(-\bk) T, \qquad 
\Gamma  U_g(\bk) = e^{i \theta_g}  U_g(-\bk) \Gamma, \qquad e^{i \theta_g} \in \{\pm 1\}, \\
\Gamma_1^2=1, \qquad \Gamma \Gamma_1=-\Gamma_1 \Gamma, \qquad  T\Gamma_1=\Gamma_1 T, \qquad \Gamma_1  U_g(\bk) =  U_g(-\bk) \Gamma_1. 
\end{array}\right.
\end{align}
This algebra is indeed that in class AII as follows. 
Without loss of generality, the symmetry operators and the Hamiltonian $H(\bk)$ can be written as 
\begin{align}
\left\{\begin{array}{ll}
\Gamma = \sigma_x \otimes {\bf 1}, \qquad 
\Gamma_1=\sigma_z \otimes {\bf 1}, \qquad 
T=\sigma_z \otimes T', \\
U_g(\bk)=
\left\{
\begin{array}{ll}
\sigma_0 \otimes  U'_g(\bk)& (e^{i \theta_g}=1) \\
\sigma_z \otimes  U'_g(\bk)& (e^{i \theta_g}=-1)
\end{array}
\right.
, \\
H(\bk)=\sigma_y \otimes H'(\bk), 
\end{array}\right. 
\end{align}
with $\sigma_{\mu} (\mu=x,y,z)$ the Pauli matrices.
Then, we have the algebra in class AII: 
\begin{align}
{\rm AII}: \qquad 
\left\{\begin{array}{ll}
T'^2=-1, \qquad T'  U_g'(\bk) =  U_g'(-\bk) T', \\
\ U_g'(h\bk)  U_h'(\bk)= z_{g,h} e^{-i g h \bk \cdot \bm{\nu}_{g,h}}  U_{gh}'(\bk), \\
T'H(\bk)=H(-\bk)T', \qquad U_g'(\bk)H(\bk)=e^{i \theta_g}H(g\bk)\wt U_g'(\bk).
\end{array}\right.
\end{align}
An important point is that the obtained symmetry $U_g'(\bk)$ behaves as
a PHS if $e^{i \theta_g}=-1$.  
In a similar way, the factor systems for other AZ classes are
constructed by imposing additional chiral symmetries.

\subsection{Superconductors with broken TRS}
In a way similar to the previous section, we formulate the factor system for superconductors with broken TRS. 
The only difference from the previous section is the absence of TRS. 
There are no constraint relations among the TRS and space group symmetry, which means any 1-dimensional representations of the superconducting gap function in (\ref{eq:1dimrep_sc}) are allowed. 
We have the factor system for $n=2$ (class D)
\begin{align}
{\rm D}: \qquad 
\left\{\begin{array}{ll}
C^2=1, \qquad  U_g(h\bk)  U_h(\bk)
= z_{g,h} e^{-i g h \bk \cdot \bm{\nu}_{g,h}}  U_{gh}(\bk), 
\qquad C U_g(\bk) = e^{i \theta_g} U_g(-\bk) C, \\
CH(\bk)=-H(-\bk)C, \qquad  U_g(\bk)H(\bk)=H(-\bk) U_g(\bk).
\end{array}\right.
\end{align}
Adding chiral symmetries with Eq. (\ref{eq:shift_n}), we get the
factor systems for other AZ classes as summarized in
Table~\ref{tab:shift_sym_sc}. 
 
\begin{table}
	\begin{center}
	\caption{The factor systems for superconductors with broken TRS.
	 In the table, $T$ and $C$ are TRS and PHS, respectively, $z_{g,h}$ represents the factor system for the point group. 
	 $e^{i \theta_g}: G \to U(1)$ is a 1-dimensional irrep.}
\label{tab:shift_sym_sc}
{\small
\renewcommand{\arraystretch}{1.5}
$$
\begin{array}{l|l|l|l|l|l}
{\rm AZ} & n & z_{T,T}, z_{C,C} & z_{T,g}/z_{g,T}, z_{C,g}/z_{g,C} & z_{g,h} & c(g) \\
\hline 
{\rm AI} & n=0 & T^2=1 & TU_g(\bk) = e^{i \theta_g} U_g(-\bk)T & & \\
\cline{1-4}
{\rm BDI} & n=1 & T^2=1 & TU_g(\bk) = e^{i \theta_g} U_g(-\bk)T & \\
&&C^2=1&CU_g(\bk) = e^{i \theta_g}U_g(-\bk)C&\\
\cline{1-4}
{\rm D} & n=2 & C^2=1& CU_g(\bk) = e^{i \theta_g}U_g(-\bk)C & \\
\cline{1-4}
{\rm DIII} & n=3 & T^2=-1 & TU_g(\bk) = e^{i \theta_g}U_g(-\bk)T &  & \\
&&C^2=1& CU_g(\bk) = e^{i \theta_g} U_g(-\bk)C&U_g(h\bk) U_h(\bk)&U_g(\bk)H(\bk)=H(g\bk)U_g(\bk) \\
\cline{1-4}
{\rm AII} & n=4 & T^2=-1 & TU_g(\bk) =e^{i \theta_g} U_g(-\bk)T & = z_{g,h} e^{-i g h \bk \cdot \bm{\nu}_{g,h}} U_{gh}(\bk) & \\
\cline{1-4}
{\rm CII} & n=5 & T^2=-1 & TU_g(\bk) = e^{i \theta_g} U_g(-\bk)T & &\\
&&C^2=-1&CU_g(\bk) = e^{i \theta_g}U_g(-\bk)C&\\
\cline{1-4}
{\rm C} & n=6 &C^2=-1&CU_g(\bk) =e^{i \theta_g} U_g(-\bk)C & & \\
\cline{1-4}
{\rm CI} & n=7 & T^2=1& TU_g(\bk) = e^{i \theta_g}U_g(-\bk)T & & \\
&& C^2=-1&CU_g(\bk) = e^{i \theta_g} U_g(-\bk)C&\\
\hline
\end{array}
$$
\renewcommand{\arraystretch}{1}
}
	\end{center}
\end{table}

\subsection{Type III and IV magnetic space group symmetry}
Let us consider the factor system for magnetic space (Shubnikov) group symmetry of type III and IV. 
In such groups, there is no TRS itself. 
An antiunitary symmetry appears as a combined symmetry with the TRS and a space group element. 
Let us denote unitary and antiunitary symmetries by $U_g(\bk)$ and $A_g(\bk)$, respectively, where $A_g(\bk)$ includes the complex conjugation. 
Adding chiral symmetries, we have the factor systems for $n>0$ as shown in Table~\ref{tab:shift_sym_mag}.

\begin{table}
	\begin{center}
	\caption{The factor systems for insulators with type III or IV magnetic space group symmetry. 
	$\Gamma$ is refereed as the chiral symmetry.
	We denote the unitary (antiunitary) symmetry by $U_g(\bk)$ ($A_g(\bk)$).
	}
\label{tab:shift_sym_mag}
{\small
\renewcommand{\arraystretch}{1.5}
$$
\begin{array}{l|l|l|l|l}
n & {\rm chiral\ sym} & z_{g,\Gamma}/z_{\Gamma,g} & z_{g,h} & c(g) \\
\hline 
n=0 
& & & &  \\
& & & & U_g(\bk)H(\bk)=H(g\bk)U_g(\bk) \\
\cline{1-3} 
n=1
& \Gamma^2=1 & U_g(\bk)\Gamma=\Gamma U_g(\bk) & U_g(h\bk)U_h(\bk)=z_{g,h}e^{-i gh\bk\cdot \bm{\nu}_{g,h}}U_{gh}(\bk) & A_g(\bk)H(\bk)=H(g\bk)A_g(\bk) \\
& & A_g(\bk)\Gamma=\Gamma A_g(\bk)& U_g(h\bk)A_h(\bk)=z_{g,h}e^{-i gh\bk\cdot \bm{\nu}_{g,h}}A_{gh}(\bk) & \\
\cline{1-3} \cline{5-5}
n=2
& & & A_g(h\bk)U_h(\bk)=z_{g,h}e^{-i gh\bk\cdot \bm{\nu}_{g,h}}A_{gh}(\bk) &  \\
& & & A_g(h\bk)A_h(\bk)=z_{g,h}e^{-i gh\bk\cdot \bm{\nu}_{g,h}}U_{gh}(\bk) & U_g(\bk)H(\bk)=H(g\bk)U_g(\bk) \\
\cline{1-3} 
n=3
& \Gamma^2=1 & U_g(\bk)\Gamma=\Gamma U_g(\bk) & & A_g(\bk)H(\bk)=-H(g\bk)A_g(\bk) \\
& & A_g(\bk)\Gamma=-\Gamma A_g(\bk) & & \\
\hline
n=4
& & & &  \\
& & & & U_g(\bk)H(\bk)=H(g\bk)U_g(\bk) \\
\cline{1-3} 
n=5
& \Gamma^2=1 & U_g(\bk)\Gamma=\Gamma U_g(\bk) & U_g(h\bk)U_h(\bk)=z_{g,h}e^{-i gh\bk\cdot \bm{\nu}_{g,h}}U_{gh}(\bk) & A_g(\bk)H(\bk)=H(g\bk)A_g(\bk) \\
& & A_g(\bk)\Gamma=\Gamma A_g(\bk)& U_g(h\bk)A_h(\bk)=z_{g,h}e^{-i gh\bk\cdot \bm{\nu}_{g,h}}A_{gh}(\bk) & \\
\cline{1-3} \cline{5-5}
n=6
& & & A_g(h\bk)U_h(\bk)=z_{g,h}e^{-i gh\bk\cdot \bm{\nu}_{g,h}}A_{gh}(\bk) &  \\
& & & A_g(h\bk)A_h(\bk)=-z_{g,h}e^{-i gh\bk\cdot \bm{\nu}_{g,h}}U_{gh}(\bk) & U_g(\bk)H(\bk)=H(g\bk)U_g(\bk) \\
\cline{1-3} 
n=7
& \Gamma^2=1 & U_g(\bk)\Gamma=\Gamma U_g(\bk) & & A_g(\bk)H(\bk)=-H(g\bk)A_g(\bk) \\
& & A_g(\bk)\Gamma=-\Gamma A_g(\bk) & & \\
\hline
\end{array}
$$
\renewcommand{\arraystretch}{1}
}
	\end{center}
\end{table}

\subsection{Superconductors with magnetic space group symmetry of type III and IV}
Let us consider superconductors without TRS but preserving a combined symmetry between TRS and a space group symmetry, i.e.\ magnetic space group symmetry of type III and IV. 
The derivation of the factor system is parallel to Sec.~\ref{sec:fs_trs_sc}. 
Let $G = G_0+Tg G_0$ with $Tg$ an antiunitary symmetry. 
Using a $U(1)$ phase rotation, one can assume the gap function is invariant under $Tg$, $U_{Tg}(\bk) \Delta(\bk)^* U_{Tg}(-\bk)^T=\Delta(-g\bk)$. 
For unitary subgroup $G_0$, the gap function obeys a 1-dimensional irrep of $G_0$, $e^{i \theta_h} U_h \Delta(\bk) U_h(-\bk)^T = \Delta(h\bk)$ for $h \in G_0$. 
We introduce $\hat{\tilde h} = \hat h e^{-\theta_h \hat N/2}$ to restore the $G_0$ symmetry. 
The compatibility between $Tg$ and $h \in G_0$ leads to the condition $e^{i \theta_h} \in \{\pm 1\}$. 
Then, the 1-dimensional irrep $e^{i \theta_h} (h \in G_0)$ can be extended to the irrep for $G$ by putting $e^{i \theta_{Tg}}=1$.
With this 1-dimensional irrep for $G$, we introduce the symmetry operators acting on the BdG Hamiltonian as 
\begin{align}
&\wt U_g(\bk)= \begin{pmatrix}
U_g(\bk) & 0 \\
0 & e^{i\theta_g} U_g(-\bk)^* \\
\end{pmatrix} \qquad g \in G_0, \\
&\wt A_{g}(\bk)
= 
\begin{pmatrix}
U_g(\bk) & 0 \\
0 & e^{i \theta_g} U_g(-\bk)^* \\
\end{pmatrix}K, \qquad g \notin G_0.
\end{align}
We find that $CU_g(\bk)=e^{i\theta_g}U_g(-\bk)C$ and $CA_g(\bk)=e^{i\theta_g}A_g(-\bk)C$. 
The factor system for class D as well as other AZ classes is summarized in Table.~\ref{tab:shift_sym_mag_sc}. 

\begin{table}
	\begin{center}
	\caption{The factor systems for superconductors with type III or IV magnetic space group symmetry.  
	We denote the unitary (antiunitary) symmetry by $U_g(\bk)$ ($A_g(\bk)$).
	$e^{i \theta_g} \in \{\pm 1\}$ is a 1-dimensional irrep of $G$. 
	}
\label{tab:shift_sym_mag_sc}
{\scriptsize
\renewcommand{\arraystretch}{1.5}
$$
\begin{array}{l|l|l|l|l|l|l}
{\rm AZ} & n & T,C &&& z_{g,h} & c(g) \\
\hline 
{\rm D}&n=2
&C^2=1 &CU_g(\bk)=e^{i\theta_g}U_g(-\bk)C&CA_g(\bk)=e^{i\theta_g}A_g(-\bk)C&&U_g(\bk)H(\bk)=H(g\bk)U_g(\bk)\\
\cline{1-5}
{\rm DIII}&n=3
&T^2=-1&TU_g(\bk)=e^{i\theta_g}U_g(-\bk)T&TA_g(\bk)=e^{i\theta_g}A_g(-\bk)T&U_g(h\bk)U_h(\bk)=z_{g,h}e^{-i gh\bk\cdot \bm{\nu}_{g,h}}U_{gh}(\bk)&A_g(\bk)H(\bk)=H(g\bk)A_g(\bk)\\
&&C^2=1&CU_g(\bk)=e^{i\theta_g}U_g(-\bk)C&CA_g(\bk)=e^{i\theta_g}A_g(-\bk)C&U_g(h\bk)A_h(\bk)=z_{g,h}e^{-i gh\bk\cdot \bm{\nu}_{g,h}}A_{gh}(\bk)&\\
\cline{1-5}\cline{7-7}
{\rm AII}&n=4
&T^2=-1&TU_g(\bk)=e^{i\theta_g}U_g(-\bk)T&TA_g(\bk)=e^{i\theta_g}A_g(-\bk)T&U_g(h\bk)A_h(\bk)=z_{g,h}e^{-i gh\bk\cdot \bm{\nu}_{g,h}}A_{gh}(\bk)&U_g(\bk)H(\bk)=H(g\bk)U_g(\bk)\\
\cline{1-5}
{\rm CII}&n=5
&T^2=-1 &TU_g(\bk)=e^{i\theta_g}U_g(-\bk)T&TA_g(\bk)=e^{i\theta_g}A_g(-\bk)T&A_g(h\bk)A_h(\bk)=z_{g,h}e^{-i gh\bk\cdot \bm{\nu}_{g,h}}U_{gh}(\bk)&A_g(\bk)H(\bk)=-H(g\bk)A_g(\bk)\\
&&C^2=-1&CU_g(\bk)=e^{i\theta_g}U_g(-\bk)C&CA_g(\bk)=-e^{i\theta_g}A_g(-\bk)C&&\\
\hline
{\rm C}&n=6
&C^2=-1 &CU_g(\bk)=e^{i\theta_g}U_g(-\bk)C&CA_g(\bk)=-e^{i\theta_g}A_g(-\bk)C&&U_g(\bk)H(\bk)=H(g\bk)U_g(\bk)\\
\cline{1-5}
{\rm CI}&n=7
&T^2=1&TU_g(\bk)=e^{i\theta_g}U_g(-\bk)T&TA_g(\bk)=-e^{i\theta_g}A_g(-\bk)T&U_g(h\bk)U_h(\bk)=z_{g,h}e^{-i gh\bk\cdot \bm{\nu}_{g,h}}U_{gh}(\bk)&A_g(\bk)H(\bk)=H(g\bk)A_g(\bk)\\
&&C^2=-1&CU_g(\bk)=e^{i\theta_g}U_g(-\bk)C&CA_g(\bk)=-e^{i\theta_g}A_g(-\bk)C&U_g(h\bk)A_h(\bk)=z_{g,h}e^{-i gh\bk\cdot \bm{\nu}_{g,h}}A_{gh}(\bk)&\\
\cline{1-5}\cline{7-7}
{\rm AI}&n=0
&T^2=1&TU_g(\bk)=e^{i\theta_g}U_g(-\bk)T&TA_g(\bk)=-e^{i\theta_g}A_g(-\bk)T&U_g(h\bk)A_h(\bk)=z_{g,h}e^{-i gh\bk\cdot \bm{\nu}_{g,h}}A_{gh}(\bk)&U_g(\bk)H(\bk)=H(g\bk)U_g(\bk)\\
\cline{1-5}
{\rm BDI}&n=1
&T^2=1 &TU_g(\bk)=e^{i\theta_g}U_g(-\bk)T&TA_g(\bk)=-e^{i\theta_g}A_g(-\bk)T&A_g(h\bk)A_h(\bk)=-z_{g,h}e^{-i gh\bk\cdot \bm{\nu}_{g,h}}U_{gh}(\bk)&A_g(\bk)H(\bk)=-H(g\bk)A_g(\bk)\\
&&C^2=1&CU_g(\bk)=e^{i\theta_g}U_g(-\bk)C&CA_g(\bk)=e^{i\theta_g}A_g(-\bk)C&&\\
\hline
\end{array}
$$
\renewcommand{\arraystretch}{1}
}
	\end{center}
\end{table}

\end{widetext}

\bibliography{ref}

\end{document}